\documentclass[letter,12pt,oneside,spanish,english]{book}
\pdfoutput=1

\usepackage[round]{natbib}
\usepackage[latin1]{inputenc}
\usepackage[english,spanish]{babel}
\usepackage[pdftex]{graphicx}
\usepackage{color}
\usepackage{array}
\usepackage{colortbl}
\usepackage{longtable}
\usepackage{enumerate}
\usepackage{amsmath}
\usepackage{amssymb}
\usepackage{wasysym}
\usepackage{pifont}
\usepackage[pdftex, bookmarks, colorlinks=true, linkcolor=blue, plainpages = false, citecolor = blue, urlcolor = blue, filecolor = blue]{hyperref}
\usepackage[small]{caption}
\usepackage{pdfpages}
\usepackage{fancybox}
\usepackage{amssymb}

\newcommand{\citeopcit}[1]{%
	\citeauthor{#1}\defcitealias{#1}{op. cit.}\citepalias{#1}%
}

\newcommand{\opcitp}[2][]{\defcitealias{#2}{op. cit.#1}\citepalias{#2} }
\reversemarginpar
\newcommand{\revolucion}[1]{$\textsc{s}_{#1}$}
\newcommand{\sementera}{$\pitchfork$}

\newcommand{\siglozocam}[1]{$\textsc{zc}_{#1}$}
\newcommand{\sigloacrot}[1]{$\textsc{ac}_{#1}$}
\newcommand{\bxogonoa}[1]{$\textsc{b}_{#1}$}
\newcommand{\zocam}[1]{$z_{#1}$}
\newcommand{\acrot}[1]{$a_{#1}$}
\newcommand{\grados}{${}^\circ$}
\begin{document}
\frontmatter
%%%%%%%%%%%%%%%%%%%%%%%%%%%%%%%%%%%%%%
%% See reference sheet of plainnat in:
%% http://merkel.zoneo.net/Latex/natbib.php
\bibliographystyle{chicago}
%%%%%%%%%%%%%%%%%%%%%%%%%%%%%%%%%%%%%%
%% COVER

\title{The Muisca Calendar: An approximation to the timekeeping system of the ancient native people of the northeastern Andes of Colombia\footnote{Dissertation presented to the Départament d'Anthropologie, Faculté des études supérieures, Universit\'e de Montr\'eal, as prequisite to obtain the diploma of \textit{Maître \`es Sciences en Anthropologie.} v.3. \copyright{} Manuel Arturo Izquierdo Pe\~na, 2008} \footnote{Note: This version contains corrections and additions to the text presented to the Universit\'e de Montr\'eal. For the original version (v.2) see: \url{http://arxiv.org/pdf/0812.0574v2}.}}
\author{Manuel Arturo Izquierdo Pe\~na\footnote{ma.izquierdo@umontreal.ca, aizquier@gmail.com}}

\date{November 2008}

\thispagestyle{empty}

\maketitle

\newpage
\pagestyle{myheadings}

%%%%%%%%%%%%%%%%%%%%%%%%%%%%%%%%%%%%%%%%%%%%%%%%%%%%%%%%%%%%%%%%%%%%%%%%%%%%
\newpage\null\vspace{1in}
\begin{huge}
\noindent\textbf{Résumé}
\end{huge}
\vspace{1cm}

Le but de ce travail est d'examiner et de compléter le modèle proposé
par le prêtre José Domingo Duquesne de Madrid (1745-1821), concernant
le calendrier de l'ancienne culture Muisca, située en Colombie
centrale. Ce modèle, écarté par les chercheurs à la fin du 19ème
siècle, a longtemps été considéré comme une belle invention.
Cependant, une analyse détaillée de son travail montre que
l'interprétation du système de gestion du temps lui fût rapporté par
les autochtones. Ceci peut-être démontré par le biais de données
ethnohistoriques et archéologiques confirmant que Duquesne n'est pas à
l'origine d'un tel système.
Nous examinerons cette hypothèse à travers, notamment, des cérémonies
Muisca relatées par des chroniqueurs espagnols du 15ème siècles. Ces
cérémonies correspondent à un cycle astronomique de conjonctions des
planètes Jupiter et Saturne, ce même cycle pouvant se superposer au
cycle de 60 ans décrit par Duquesne à propos du siècle acrotome Muisca.
Il en est de même pour les outils de calculs des prêtres Muisca dont
parle Duquesne. Une pierre sculptée, découverte dans le village de
Choachi (Cundinamarca) témoigne de tels éléments numériques.
\vspace{0.5cm}

\noindent\textbf{Mots clés:} 
\begin{small}
Archeoastronomie, Muiscas, Colombie, Calendrier
\end{small}
%%%%%%%%%%%%%%%%%%%%%%%%%%%%%%%%%%%%%%%%%%%%%%%%%%%%%%%%%%%%%%%%%%%%%%%%%%%
% abstract
\newpage
\null\vspace{1in}
\begin{huge}
\noindent\textbf{Abstract}
\end{huge}
\vspace{1cm}

The aim of this project is to review and expand upon the model proposed by Father José Domingo Duquesne de la Madrid (1745-1821) regarding the calendar of the ancient Muisca culture of the central Colombia. This model was dismissed by scholars in the late 19th century, calling it just a simple invention of a clergyman; however, a detailed analysis of Duquesne's work shows that his  interpretation of the timekeeping system was based on information given to him by indigenous informers. Based on his work, we can be derive somewhat indirectly, some aspects of the calendar that apparently were not understood by the priest. This confirms that such a system was not his own invention. Ethnohistorical and archaeological evidence provide support for Duquesne's calendar model. Massive Muisca ceremonies described by 15th century Spanish chroniclers, is examined, and the occurrence of such ceremonies seem to match the astronomical cycle of conjunctions of the planets Jupiter and Saturn, wich also agrees  with the 60-year span described by Duquesne as the Muisca Acrotom Century. Archaeological artifacts, such as a carved stone found in the village of Choachi (Cundinamarca) that shows numerical elements supports Duquesne's model that suggests this stone was a  calendar calculation tool for  Muisca priests.
\vspace{0.5cm}

\noindent\textbf{Keywords:} 
\begin{small}
Archaeoastronomy, Muiscas, Colombia, Calendar
\end{small}

%%%%%%%%%%%%%%%%%%%%%%%%%%%%%%%%%%%%%%%%%%%%%%%%%%%%%%%%%%%%%%%%%%%%%%%%%%%
\newpage\null\vspace{1in}
\begin{huge}
\noindent\textbf{Resumen}
\end{huge}
\vspace{1cm}

El objetivo del presente trabajo es reexaminar y complementar el modelo propuesto por el padre José Domingo Duquesne de la Madrid (1745-1821) acerca del calendario de la antígua cultura Muisca del centro de Colombia. Tal modelo había sido descartado por investigadores de finales del siglo \textsc{xix}, consideránosele simplemente como una ingeniosa invención del sacerdote. Un análisis detallado de su trabajo muestra, sin embargo, que Duquesne estaba únicamente interpretando un sistema de manejo del tiempo que le fué comunicado por sus informantes indígenas. Con base en su trabajo puede derivarse de manera indirecta varios aspectos del calendario que aparentemente no fueron notados por el religioso, siendo esto una confirmación que tal sistema no era de su invención. Se presenta evidencia etnohistórica y arqueológica que soporta dicho modelo del calendario. Se estudia el caso de ceremonias masivas muiscas descritas por cronistas españoles del siglo \textsc{xvi} cuya ocurrencia parece concordar con un ciclo astronómico de conjunciones de los planetas Júpiter y Saturno, el cual esta también de acuerdo con un lapso de sesenta años descrito por Duquesne para el Siglo Acrótomo Muisca. Artefactos arqueológicos como una piedra labrada encontrada en el municipio de Choachí (Cundinamarca) muestra elementos numéricos que concuerdan con el modelo de Duquesne, sugiriendo que se trataba de una herramienta de cálculo para el sacerdote Muisca.
\vspace{0.5cm}

\noindent\textbf{Palabras claves:} 
\begin{small}
Arqueoastronomía, Muiscas, Colombia, Calendario
\end{small}
%%%%%%%%%%%%%%%%%%%%%%%%%%%%%%%%%%%%%%%%%%%%%%%%%%%%%%%%%%%%%%%%%%%%%%%%%%%

\tableofcontents
\listoftables
\listoffigures

\newpage
\null\vspace{2in}
\begin{center}
For\\
my father, Gustavo A. Izquierdo\\
and\\
my mother, Celina Peña
\end{center}

%%%%%%%%%%%%%%%%%%%%%%%%%%%%%%%%%%%%%%%%%%%%%%%%%%%%%%%%%%%%%%%%%%%%%%%%%%%
\newpage\null\vspace{1.8in}
\begin{huge}
\noindent\textbf{Acknowledgments}
\end{huge}
\vspace{1cm}

I wish to thank  professors Louise Paradis and Ariane Burke from the Université de Montréal, for their direction and support along my research these years in Canada. In Colombia, I have to thank to Margarita Reyes, from the Museo Nacional de Colombia and Margarita Silva from the Museo Arqueológico de Sogamoso, who allowed me to examine the archaeological collections of their museums. A special mention goes to my colleagues of the Observatorio Astronómico Nacional, of the Universidad Nacional de Colombia, with whom I initiated my professional career. I also thank  the kind commentaries of    Breen Murray,  from the Universidad de Monterrey, Mexico.

I thank the Ministère de l'éducation de Québec and the FICSUM, for their fellowships that facilitated my studies. To the internet sites www.assembla.com for providing me a free SVN hosting for the files of my project, the www.wikipedia.org project, who always provided me information at hand, the free software projects  Stellarium\footnote{http://www.stellarium.org/} and  the Swiss Ephemeris\footnote{http://www.astro.com/swisseph/}, invaluable tools for the archaeoastronomer, and to the overall project GNU Linux, who provided me with a free and virus clean computing environment to perform my work.

A big acknowledgment to the support and commentaries from my dear friends, Diana Carvajal,  Juan David Morales, Luis Francisco López, Federico \textit{el primo} Ariza,   Luz Marina Bedoya,  Juan Carlos Martínez, Fanny Guyon, and  Basile Sohet. To Benjamin Patenaude and Caroline Chochol to their help correcting my english in the manuscript. To all my family, specially my son Gabriel Alejandro, and  the love of my life, my wife Patricia whose love and  support has been for me a lighthouse all these months. 

\newpage

%%%%%%%%%%%%%%%%%%%%%%%%%%%%%%%%%%%%%%
%%                        C  H  A  P  T  E  R  S
%%%%%%%%%%%%%%%%%%%%%%%%%%%%%%%%%%%%%%
%% INTRO

\chapter{Introduction}\label{intro}\hyphenation{Muis-ca}
I first became interested in the Muisca Calendar in the mid 1990s, when I was a bachelor's student in Anthropology at the Universidad Nacional de Colombia. However, it was not until  2004, when I taught archaeoastronomy at the Astronomical Observatory at the same university, that I began to seriously research this topic. As result of several discussions with my colleagues Francisco López and Juan David Morales about the Muisca calendar and astronomy, I took on the arduous task to carefully read and analyze the texts of José Domingo Duquesne de la Madrid, a late 18th century (1795) priest who was seemingly in close contact with the descendants of the Muisca Indians. From these texts, I obtained a plethora of information about the Muisca timekeeping system. Despite the fact that Duquesne's work was, however, dismissed by later generations of scholars, who considered it a product of the priest's imagination, I started to inquire whether such opinion was fully justified. After four years of work analyzing his writings, checking for clues in the ethnohistorical, archaeological, and astronomical sources,  my current opinion is that Duquesne's work has to be reconsidered as a an important source of information about this topic, and that the dismissal of his work seriously slowed our understanding of the calendar and astronomy of the Muiscas. This model cannot be considered, however, as a definitive description of the calendar of the precolumbian Muisca, but as a legitimate piece of the whole jigsaw, result of the Duquensne's effort to interpret and to describe an ethnographically-obtained timekeeping system.  The main goal of this project is to introduce, analyse and complement the Duquesne's model of the Muisca Calendar, providing preliminarly information supporting the model, from ethnohistorical and archaeological sources.

In Chapter \ref{ch1}, I introduce the theory of archaeoastronomy, and especially the framework that will allow me to frame the hypothesis that will be described throughout the course of this work. In chapter \ref{muisca-chapter}, I provide a succinct introduction to the Muisca culture, helping the reader who is unfamiliar with such a culture.

Due to the fact that Chapter \ref{apuntes-duquesne} is the product of a careful analysis of Duquesne's description of the Muisca Calendar, it is the  densest part of this dissertation.  Numerous explanations of the texts are given in the chapter in order to give the reader a better understanding of the priest's work, which is, in many parts, difficult to interpret. New calendar elements, not explicitly described  by Duquesne can be deduced from his work are shown, for example the \textit{Cycle of Ata} and the \textit{Zocam Extended Centuries}. Furthermore, the analysis links the described calendar with the descriptions of the earlier chronicler Fray Pedro Simón (1625),  and aims to construct a general model of the calendar, that  comprises mythical time-spans up to a duration of almost six thousand years.

In Chapter \ref{ubaque-chapter}, ethnohistorical sources are explained, in particular those that relate episodes associated to the Muis\-ca religion provide clues supporting the analysis started in Chapter \ref{apuntes-duquesne}. The study of a massive ceremony held by the Muiscas in 1563 in a village known as Ubaque, led to the discovery that this  ceremony occurred during an astronomical conjunction of the planets Jupiter and Saturn,  phenomenon that significantly fits with the Acrotom Century described by Duquesne. Another massive ceremony described by Juan Rodríguez Freyle (1636) is also studied, leading to the possibility that it was performed an Acrotom Century before the Ubaque ceremony.

Chapter \ref{stones-chapter} addresses the existence of some archaeological artifacts that suggests the possibility of the presence of a basic system of arithmetic notation. The Choachí Stone in particular stands out; it is a small stone slab that shows an arrangement of figures that represent values that can be associated to the calendar. It is suggested that such a stone could be considered as a `pocket device' to perform calendar computations. Examples of other archaeological artifacts are also discussed, as in the case of a conch trumpet and spindle whorls showing, in their decoration, the graphical theme of a bird-head, already present in the Choachi Stone, which were possibly  used as  containers of numerical information. Future research may wish to analyze this topic further.

Appendix \ref{duquesne-work} includes a complete transcription of Duquesne's writings, in their original Spanish version. Appendix \ref{muisca-centuries-table} presents several tables showing the \textit{Muisca Centuries} (\textit{Acrotom} and \textit{Zocam}) and the \textit{Cycle of Ata}. Tables for the higher mythical time spans are also presented. These tables were generated by a computer program developed conforming to Duquesne's descriptions, which I wrote in C language for the purpose of this research.

\mainmatter
%% Part I

\part{Theoretical Background}

\chapter{The archaeoastronomy}\label{ch1}
\section{Definition}
Archaeoastronomy is an auxiliary discipline of  Archaeology, and can be defined as the ``study of the practice of astronomy in ancient cultures using both written and unwritten records'' \citep{Aveni2003}, with special emphasis on it being a ``study of the sky-watching practices made with a cultural purpose''\citep{Belmonte2006} in order to understand the way ancient people ``understood the phenomena in the sky, how they used phenomena in the sky and what role the sky played in their cultures'' \citep{Sinclair2006}. This approach has lead to the proposition of alternative names for the discipline, for example Astronomy in the Culture or Cultural Astronomy \citep{Ruggles1993}. 

Although closely related,  Archaeoastronomy is not the study of the ancient astronomy, since \textit{astronomy} ``is a culturally specific concept and ancient peoples may have related to the sky in a different way'' \citep{Ruggles2005}. Hence, the study of ancient astronomy and the historical process that lead to the formation of modern astronomy is part of the History of Science, by means of its methods and goals, and is formally a sub-discipline of  History.

\section{History and research trends}

The first studies in archaeoastronomy can be dated back to the 19th century, Norman Lockyer's classic book \textit{Dawn of Astronomy} \citeyearpar{Lockyer1893}, discusses astronomical elements associated with the orientation of Egyptian monuments. However, it is not until the 1960's  that the works of Alexander
Thom \citeyearpar{Thom1974, Thom1975} and Gerald Hawkins \citeyearpar{Hawkins1965, Hawkins1974} on the astronomical orientation of the megalithic monuments of  the British Isles attracted the interest of both archaeologists and astronom\-ers to inquire whether astronomical phenomena played a role in ancient cultures.  

The first meeting to unite many researchers in archaeoastronomy was held in 1981 in Oxford, England. This meeting brought to light the staggering differences in methodology and research objectives between the scholars of  Europe and America, leading to the publication (in two volumes) of the meeting's proceedings which  defined the  two current main trends of research in this field: The \textit{Brown Archaeoastronomy} and the \textit{Green Archaeoastronomy}. The former was named for the color of the paper cover of the volume describing the work of the American researchers, who concentrated
on  the study of the Amerindian societies,  and included works that dealt with archaeological, ethnographic and ethnohistorical sources in order to inquire about the ancient astronomical knowledge of the studied cultures,  (as in the case of the  Maya and their calendar, which  deserve special attention \citep{Aveni1980} ). From this publication, it is evident that, the study of  timekeeping systems had been an active topic of research in this branch of the discipline.  The latter publication, the \textit{Green Archaeoastronomy}, corresponds to the European research, and it focuses on the astronomical alignments of monuments, specially the megalithic archaeological sites of Europe. Its methodology relies on rigorous statistical analysis  to determine if the studied alignments are a product of chance, or whether they are  intentional phenomena. 

Both sides criticized each other. the criticism of the Brown Archaeoastronomy lies in its apparent disregard for statistical rigor where astronomical orientations in ancient monuments  is concerned. The criticism of Green Archaeoastronomy lies in in the fact that it does not provide a comprehensive enough interpretation of components of ancient societies  \citep{Kintigh1992}.

 In general,  Green Archaeoastronomy is best at answering the question: \textit{there  was really an astronomic intention in the construction of a given site?}, whereas the Brown Archaeoastronomy does better with the question: \textit{how it was the astronomic intention of a given past culture?}. Most of the work done in archaeoastronomy today could be classified under one of these approaches, although currently many new works  integrate both approaches.

\section{The issues of a maturing discipline}
 Unfortunately,  archaeoastronomy has been viewed with skepticism by the archaeological community,  due in part to the many works that have been conduced by researchers from non humanistic studies, especially astronomy.  Although they are very rigorous in the astronomical part of their research, they lose  credibility when they generate wrong interpretations from an archaeological point of view. Similarly, researchers without an astronomical formation have proposed archaeological theories based on erroneous astronomical postulates, (see for example \citealt{Milla1983}). Juan Antonio Belmonte addresses a key element of this problem, when he argues that  archaeoastronomy is defined as a `no-man's land', where both astronomers do not recognize it as their own, nor do anthropologists fully understand its usefulness. Any individual without  academic formation  could attribute himself the title of archaeoastronomer in order to posit wild theories that are absolutely unfounded, horrifying the social scholars who  will as a consequence consider  archaeoastronomy as a `lunatic fringe' expression \citep{Belmonte2006}. 

Despite this problem, archaeoastronomy, in the same way that disciplines as Archaeology, Anthropology and Astronomy  arose after a period of immaturity,  is living such a process, working to get rid of these `fringe aspects' and thus become a well-established academic discipline.   Methodological foundations as the interdisciplinary effort of anthropologists and astronomers, have turned into the definition of a new kind of professional designation, as \citeopcit{Belmonte2006} indicates: ``After more than one decade of field work, my current opinion, and the one of many of my colleagues is:  what is really needed is an almost complete recycling of the current astronomer or anthropologist, who must become  into a truly  archaeoastronomer, forgetting many of the epistemological references learnt during his long years of study and learning new ones that are completely unknown to him''. Despite these  efforts to establish a  general theory on archaeoastronomy, the creation of such a theory is still a work-in-progress\citep{Iwaniszewski2001,Iwaniszewki2003,Belmonte2006}.

Belmonte  provides theoretical elements to help the discipline avoid the ``fringe'' trap, and thus gain credibility in the academic community. I consider these elements very important, because I  intend to frame the ideas  I discuss in this dissertation into this theoretical  `safety zone'.

\section{The Belmonte's classification of archaeoastronomical research }
\label{belmote-classification}

\citeopcit{Belmonte2006} proposes a concise framework to classify the current work in archaeoastronomy, which  divides the research into five categories of credibility (from the more to the less credible) and are defined as: Formal work, Serious speculation, Endearing speculation, Wild speculation, and \textit{Making money}\footnote{Term in english in the original text in spanish.}. 

As a scholar coming from the exact sciences (Astrophysics), Belmonte emphasizes the use of the principle named in latin  \textit{Testis unus, testis nullus}, which states that ``an unique case of something... is not enough evidence to elaborate a hypothesis, and much less, a theory since it cannot be falsifiable'' \opcitp{Belmonte2006}. He also  proposes the use, as far as possible, of the rule of  \textit{Occam's razor} and the \textit{Principle of Economy} to distinguish the boundary between what can be considered science and not. Therefore, in his opinion, such a boundary ``could be located in some point between the Endearing speculation and the Wild speculation'' \opcitp{Belmonte2006}. Succinctly, these categories can be described as follows:

\begin{description}
\item  [Formal work] When a research produces a theory that is supported by a rigorous methodology and solid evidence coming from astronomical, archaeological, ethnohistorical and/or ethnographic sources.
\item  [Serious speculation] When a work, based on a  rigorous methodology, proposes a very reasonable hypothesis, but lacks of complete supporting evidence from the mentioned sources.
\item  [Endearing speculation] When a work only can propose a very reasonable hypothesis, lacking of any supporting evidence. 
\item  [Wild speculation] When the researcher, faced to the uncertainties, fill the holes of evidence with his own speculations  and falls in the temptation to convince himself of the truth of such hypothesis. Thus, in an act of enthusiasm (and pareidolia), his proposed model turns into a `truth'.
\item  [\textit{Making money}] When the archaeoastronomer turns to the dark side. The researcher discovers that such fabricated truths sell very well in the market of the `fringe literature', and his work becomes a best seller aside akin titles of UFOlogy, Astrology,  New Age. This is quite good for his budget, but certainly \textit{not} for  Archaeoastronomy, as an academic discipline.
\end{description}

Formally, an hypothesis is a conjeture with supporting data, and a conjecture is simply an a priori explanation. Belmonte coloquially describes these ranges of `speculations' to  address these differences. Although the freedom to speculate (to conjeture, to hypothesize) is useful in Science to get new ideas around a problem, these have to be treated only as working tools, not truths. The validation or invalidation of  several hypothesis around a problem will give to the researcher  a  map indicating what could be the solution to a problem, and of equal importance,  what is \textbf{not} the solution. Furthermore,  Belmonte's framework is valuable because it provides a basis to classify a given explanation in such a  range, in order to bring the researcher the capability to know the scope of such idea, allowing him to use it correctly as a methodological tool, and not letting him  fall in the dangerous fringe trap. 

\chapter{The Muisca Culture}\label{muisca-chapter}

\section{Origin and chronology}

\begin{figure}[h]
 \centering
\includegraphics[width=6.5cm]{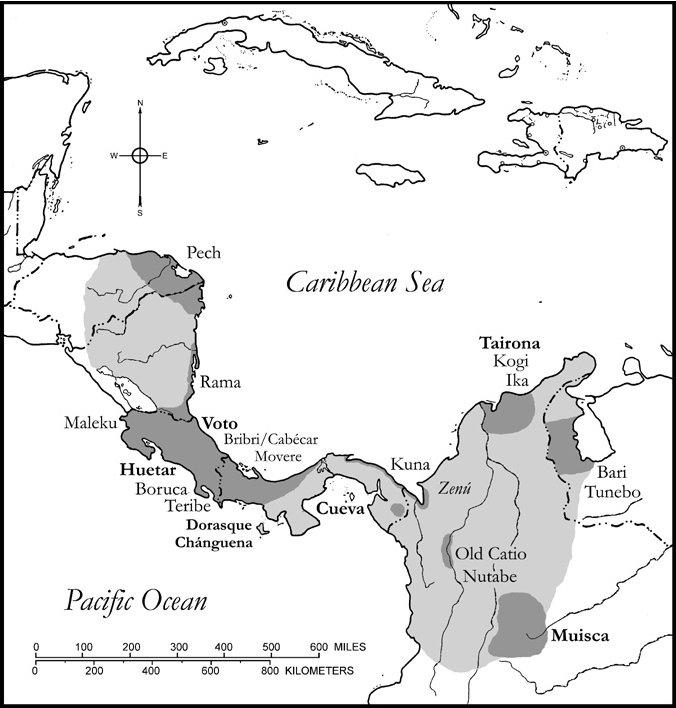} (a)
\includegraphics[width=6.5cm]{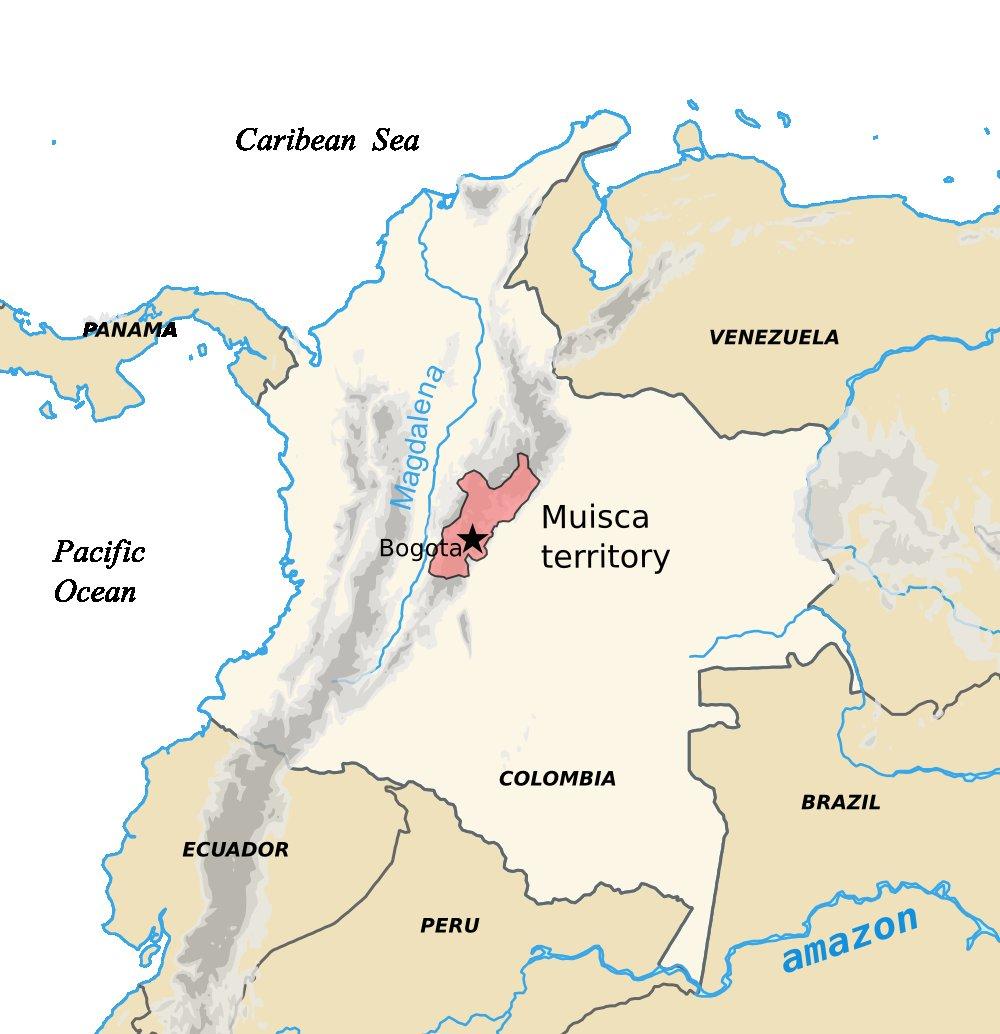} (b)
 % muisca-territory.png: 1000x1034 pixel, 179dpi, 14.20x14.69 cm, bb=0 0 403 416 
 \caption{(a) Map of the Chibcha world. Figure from \cite{Hoopes2005}. (b) Location of the ancient Muisca territory on the current Colombia.}
 \label{muisca-territory}
\end{figure}

 \textit{Muisca}  culture developed in the \textit{altiplano}  located in the Northeastern Colombian Andes (figure \ref{muisca-territory}b). The origins of Muisca culture are attributed to migrations of Chibcha-speaking groups coming from the caribean coast during the 8th century AD, who either replaced or fused (which is an on-going topic of dicussion \citep{Rodriguez1999}) with  former populations of the \textit{altiplano} known as the \textit{Herrera} Culture. Nevertheless, since this century, several distinct changes in the archaeological record have been observed suggesting a tendency  towards increasing social complexity in the region, which contrasts sharply with the simple material culture of the \textit{Herrera} \citep{Lleras1987,Lleras1995}. The appareance of the Muisca culture seems to belong to a series of social phenomena leading to the social complexity of cultures along the Intermediate Area since as early  as the 4th century AD \citep{Hoopes2005}. As most of these social changes involved groups belonging to the Chibcha linguistic family and  some researchers \citep{Hoopes2005,Warwick1997} have proposed the regional concept of a `Chibcha World' (figure \ref{muisca-territory}a), in order to approach such groups and their associated cultural development from a wide perspective.

A succint chronology for the \textit{altiplano} can be therefore proposed as follows \citep{Lleras1995,Rodriguez1999}:

\begin{itemize}
\item Preceramic: $\pm$20000 BC --- 1500 BC	
\item Herrera: 1500 BC --- 800 BC
\item Muisca: 800 AD --- 1536 AD
\end{itemize}

It is widely accepted that Preceramic populations arrived in this region from the Magdalena River Valley via previous ancient migrations attributed to the pursuit of megafauna found in the \textit{altiplano}. The \textit{Herrera} culture can be asociated with a formative stage, however, very little is known about this period. Following the precolumbian Muisca period, the 16th century marked the arrival of Europeans, this beggining a colonial period that spanned from the 16th to 18th centuries, in it, assimilated a lot of cultural elements from new power, but also achieved to maintain most of its native cultural traits.

\section{Political organization}
\begin{figure}[t]
 \centering
 \includegraphics[width=10cm]{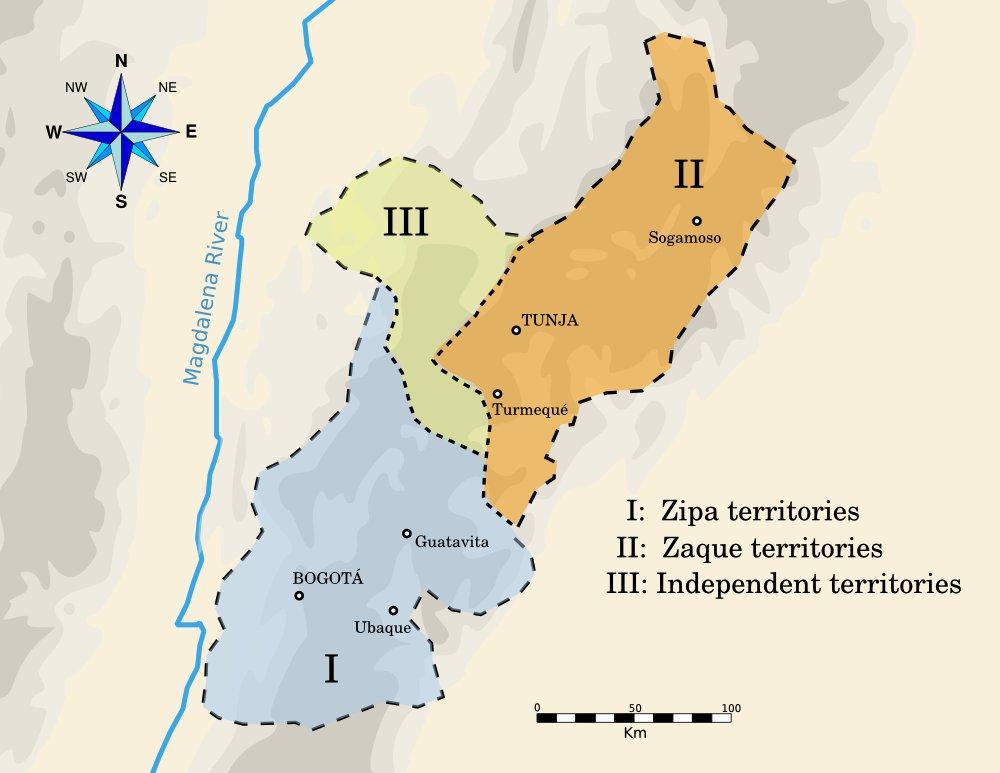}
 % muisca-territory.png: 1000x1034 pixel, 179dpi, 14.20x14.69 cm, bb=0 0 403 416 
 \caption{Territories of the main Muisca chiefdoms in the times of the European conquest. Based on the map published by \cite{Falchetti1973}.}
 \label{muisca-territory-zoom}
\end{figure}

The Muisca were  characterized politically by a complex system of chiefdoms, which during the first contact with the Europeans formed into two main divisions occupying the northern and southern parts of the Muisca territory, composed by chiefdoms and subject to  two Rulers: the \textit{Zaque} in the north and the \textit{Zipa} in the south. The \textit{Zaque} controled the chiefdoms of \textit{Sogamoso},  \textit{Turmeque} and \textit{Tunja}, which was his capital. The \textit{Zipa} ruled over the chiefdoms of \textit{Guatavita}, Ubaque and Bogotá, with the current day city of Funza as his capital (see figure \ref{muisca-territory-zoom}). At the same time, each one of these subjected chiefdoms was composed of other smaller chiefdoms, which were also subdivided  until a mininum political unit of the Muisca society, known as the \textit{uta}.   According the chroniclers, these two confederations of chiefdoms were in conflict, due to the expansionist tendencies of both rulers. However, some  chiefdoms to the northeast of the territory, maintained an independient existence, sometimes intervening, depending the circumstances, in favor of the Zaque or the Zipa.  In the eye the medieval-european Spaniards, they viewed  in the highlands  an ashaming very organized set of `kingdoms', in contrast to the `savage hordes' of the low lands; however, these kingdoms were  ruled by cruel and despotic kings wishing to control and subjugate those territories. The Europeans were so impressed with the new discovered society, that associated it with a burgeaning empire. In fact,  declaring heroic victories over vast, despotic and cruel empires was a very attractive idea to the Spanish conquerors, since they needed to gain prestige when justifing  their conquests  before the Spanish crown  in order to legitimize newly acquired property and the  profits derived from it. Consequenlty, this notion of the `Muisca empire', as depicted by the european chroniclers since the 16th century, must to be considered in light of their agenda, and  has been contested by current scholars \citep{Correa2004}, who have established that the Muisca never reached  state organization, rather it was organized as a complex system of subjected chiefdoms governed by series of matrilineal descended lineages of chiefs named  \textit{cacique}  by the spaniards and \textit{sijipcua} in the Muisca language \citep{Villamarin1999}. The realm of each \textit{cacique} took on the name of its ruler who was considered  divine by his people. Consequently, the \textit{cacique}'s power trascended the political domain of Muisca society into the religious sphere, consequently, this created the effect of being legitimately linked with the sacred aspects of the Muisca world in the eyes of his people. The duties of the \textit{cacique} were aimed to warrant, from both in the secular and sacred domains, the reproduction and the health of his society \citep{Correa2004}. In order to achieve this end, the \textit{caciques} followed a rigorous training regimen since early childhood, aimed at the proper fulfillment of their duties, whose performance  extended after death, becoming their mummified bodies as part of the collective memory of the ancestors. These were kept in such sacred places as temples or caves, under specialized care of  priests. 

\section{Economy}

\begin{figure}[t]
  \begin{center}
    \includegraphics[height=6cm]{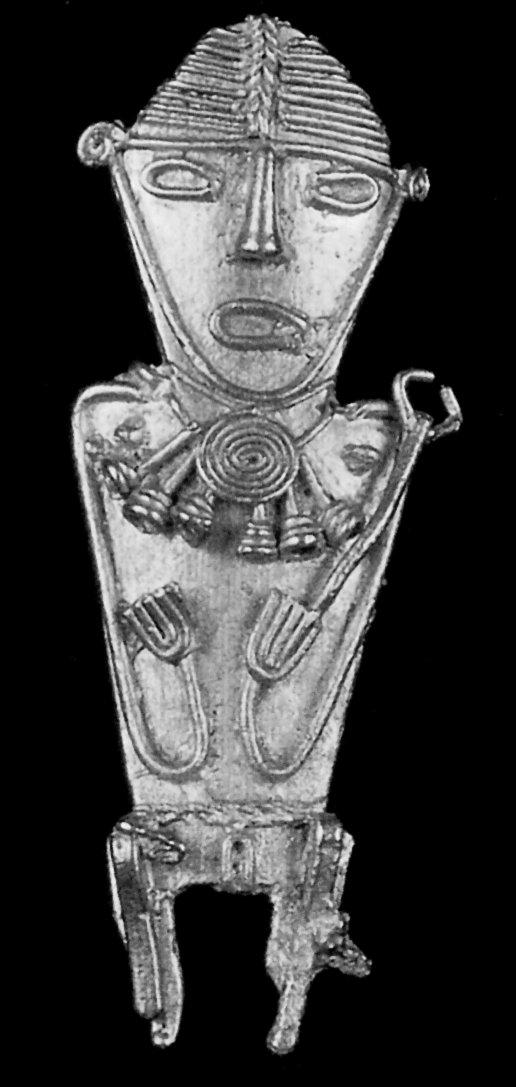} (a)
    \includegraphics[height=6cm]{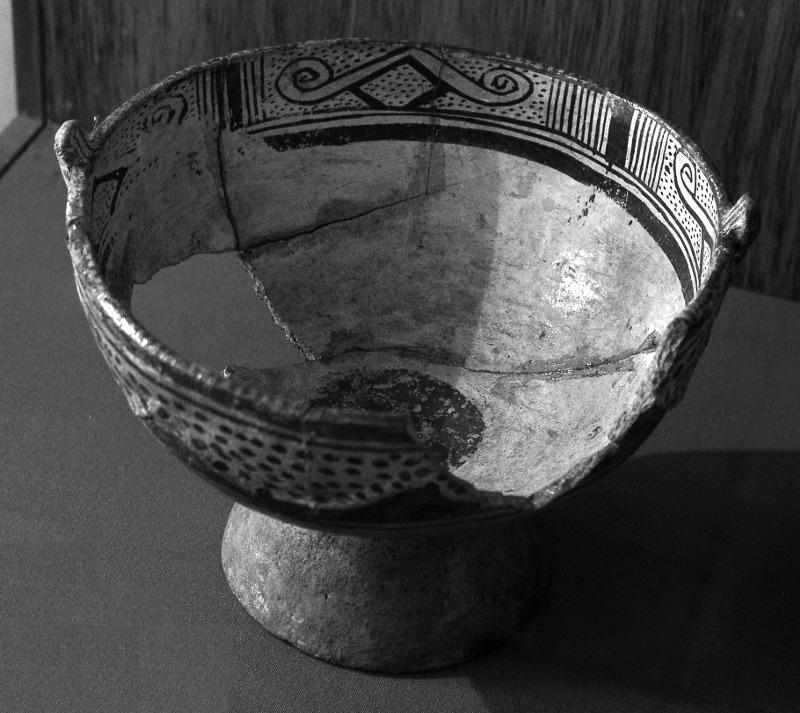} (b)
  \end{center}
  \caption{(a) Gold \textit{tunjo} \textit{(Museo del Oro)}. (b) Muisca ceremonial ceramics \textit{(Museo Arqueológico de Sogamoso)}.}
  \label{tunjo}
\end{figure}

The economy of the Muisca was based principally on the agriculture.  Their crops consisted mainly of high-land plants as corn, potatoes and quinua; however, some Muisca chiefdoms had territories the low lands profiting of its natural resources. In an effort to control the lacustrine terrains peculiar to the \textit{altiplano} the Muisca constructed channels which allowed them to not only irrigate their fields, but to fish, which contributed to their diet. An alternative source of animal protein was the \textit{curi}\footnote{Also known as \textit{cuy}, \textit{Fuquy} in Muisca language.} (Guinea Pig), which was domesticated. Deer was also hunted, however its consumption was restricted to the religious practices of the elite.

Mining and mineral extraction was practiced by the Muisca, especially the explotaion of emeralds and salt. These resources were traded with neighbouring groups, for such available commodities as cotton and gold, highly appreciated by the Muiscas. Gold was needed to supply the artisan goldsmiths (\textit{plateros}) who covered the demand for votive gold figurines (\textit{tunjos}) used to perform religious offerings (figure \ref{tunjo}a).

\begin{figure}[t]
 \centering
 \includegraphics[height=7cm]{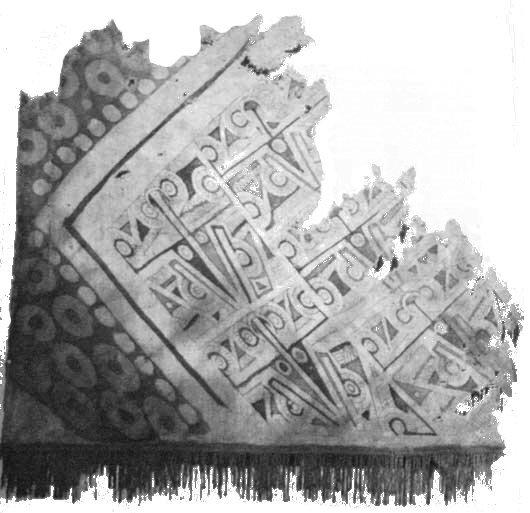}\hspace{0.5cm} (a)
 \includegraphics[height=7cm]{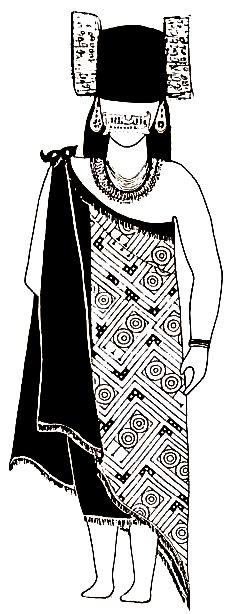} (b)
 % blancket.jpg: 530x513 pixel, 72dpi, 18.70x18.10 cm, bb=0 0 530 513
 \caption{(a) Muisca painted blancket fragment \textit{(Museo del Oro)}. (b) Wearing of blanckets.  From \cite{Cortez1990}.}
 \label{blanckets}
\end{figure}

Imported cotton was used to manufacture blankets (which were also exported),  which was a highly specialized industry in Muisca society (see figure \ref{blanckets}). These blankets were commonly used as clothes, and, special painted blankets were considered  highly valued goods. Unfortunately, a few scraps have been conserved in museum collections. However, numerous decorated stone spindle whorls made in stone  (see chapter~\ref{stones-chapter}), are better represented, thus attesting to the large scale of this industry. 

Another important industry was  pottery, which was manufactured at on a very large scale and mostly produced utilitarian and domestic vessels, more specifically the manufacture of special big pots used in the extraction of salt, which were used to evaporate salty water until obtaining a block of salt which was then extracted after cracking the ceramics.  However, special ceremonial vessels were produced, which differred from the utilitarian ones in terms of their quality and decoration the elaborated geometrical designs in its decoration (see figure \ref{tunjo}b) .
 
Other imported goods were  marine strombus conchs, used for ceremonial ocasions; in addition to another goods associated with the religion, as hayo (coca), yopo, tobacco and exotic feathers.

\section{Religion}

Religion was managed by specialists known as the \textit{Xeques}, \textit{Mohanes} or \textit{Ojques}, who alongside the caciques and enjoyed a similar level of prestige and power, including the right to intervene in political affairs (see chapter \ref{ubaque-chapter}).  The historical accounts refer to them as specific type of priests, who underwent a special training from an early age in places named as the \textit{Cucas} \citep{CorreaAguirre2001}. This training was very similar to that undertaken by cacique initiates, and consisted of long periods of fasting, locked up inside a ceremonial house. The iniciates were forbidden to leave the house or have any contact with  sunlight during their seclusion, for that reason, the ceremonial houses were built with teh intention of having a permanent semidark interior. The initiates were instructed by older Xeques during the nights, and taught about the  skills required by priesthood.  Such fasting lasted for about four to seven years, and included a restricted lean diet, and the restriction of the sexual habits of the young iniciates, which were only allowed once the iniciate passed a series of tests completing the fast. In an similar fashion to the the caciques, the priesthood was transmited through matrilineal lineages, and normally each \textit{cacique} had several \textit{Xeques} at his service.

The details of the religious rites performed by the Muisca priests are not well known, since the \textit{Xeques} were strongly persecuted and killed during the Spanish Conquest and the early colonial period. The historical archives provide descriptions of ceremonial activities known as \textit{borracheras} (see chapter \ref{ubaque-chapter}), which involved the use of psychotropic plants, as used in most of the Native American religions. Chroniclers also emphasize the existence of special children used in sacrifice rituals, brought from  foreign lands in the eastern plateau of Colombia at very young age and were raised by the priests following a similar training as the \textit{Cacique}s and \textit{Xeque}s. These children were sacrificed in special ceremonies at the set of their adolescence. 

As I will dicuss in chapter \ref{ubaque-chapter}, historical records provide data referring to possible astronomical elements conforming to the Muisca religion, suggesting that the astronomical and timekeeping knowledge was a mandatory skill of the \textit{Xeques}. Their systematic extermination by the Spaniards only served to form a large gap in our understanding of the these aspects of  Muisca culture.

The main religious center of the Muisca culture was located in the present-day city of Sogamoso\footnote{Whose name is a variation of the Muisca word \textit{Suamox} that means `House of the Sun'.}, where the Spaniards were impressed by the sizeable  ceremonial structure found there, described by the chroniclers as the `Temple of the Sun', later destroyed during the conquest \citep{Simon1625}. Although described as a temple by the Europeans, it was not a public place, rather it was accesible only to the priesthood and served as a place to iter and care for the mummies of distinguished ancestors as well as the storage of valuable objects that served as religious paraphernalia.

The Muisca worshiped figurines made of gold, copper or wood and wrapped in cotton cloths, named by the spaniards as \textit{tunjos}, \textit{santillos}, or \textit{ídolos}. These were possesed by both the common people and elite and were usually kept on homes, fields of crops, caves and ceremonial houses. These figurines  were the preferred plunder of the spaniard's idolatry eradication offensives, since they were made of gold \citep{Correa2004}.

\begin{figure}[t]
 \centering
 \includegraphics[height=7.5cm]{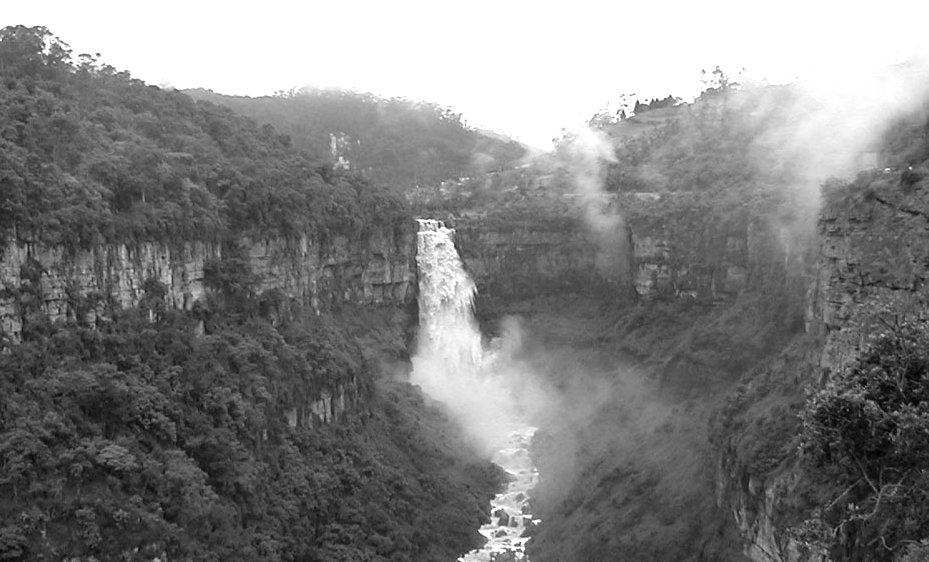}
 % tequendama.jpg: 628x930 pixel, 72dpi, 22.15x32.81 cm, bb=
 \caption{\textit{Salto de Tequendama}. Photograph by Francisco A. Zea.}
 \label{tequendama}
\end{figure}

\begin{figure}
 \centering
 \includegraphics[width=13cm]{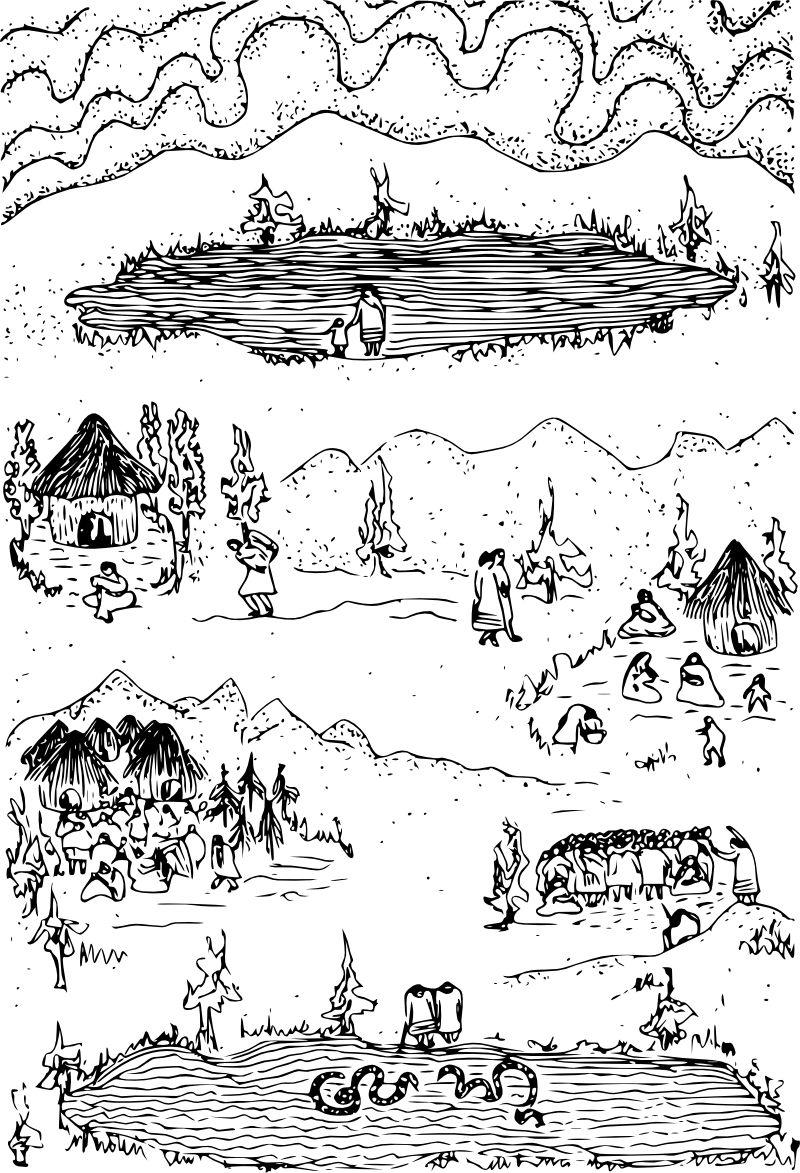}
 % bachue.png: 800x1173 pixel, 131dpi, 15.53x22.77 cm, bb=0 0 440 646
 \caption[Artistic representation of the Bachue's myth]{Artistic representation of the Bachue's myth.  From \cite{Pena1972}.}
 \label{bachue}
\end{figure}

\begin{figure}
 \centering
 \includegraphics[width=13cm]{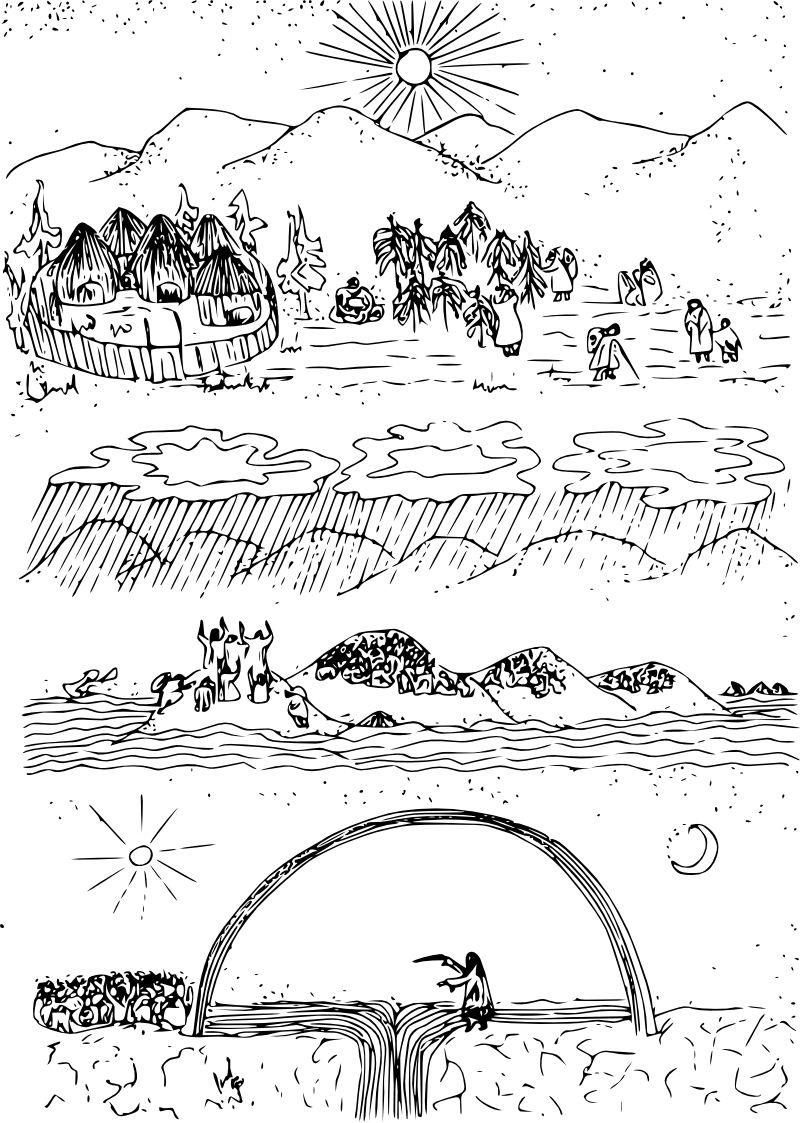}
 % bachue.png: 800x1173 pixel, 131dpi, 15.53x22.77 cm, bb=0 0 440 646
 \caption[Artistic representation of the Bochica's myth]{Artistic representation of the Bochica's myth.  \cite{Pena1972}.}
 \label{bochica}
\end{figure}

Among the deities that made up the Muisca religion described by the chroniclers, two stand out \textit{Bachue} and \textit{Bochica}. \textit{Bachue} was said to be the mother of mankind. She emerged from a sacred lake named \textit{Iguaque}, carrying a very young boy who, at the set of adulthood married her and thus they spawned all people into the world. When they reached old age they transformed into snakes and returned to the sacred lake from which they had emerged (see figure \ref{bachue}). Bochica is their civilizor hero, it is said that long ago he arrived into the Muisca region  and  traveled across the territory teaching the people such skills as agriculture,  weaving, and timekeeping. After this, his image became associated with the sun. According to tyhe myths, he performed fantastic feats, such as  creating the \textit{Salto de Tequendama}, a high water fall currently located in the southern end of the \textit{Sabana de Bogotá} (see figures \ref{tequendama} and \ref{bochica}) in order to save people from a massive flooding that covered part of the Muisca World. It is said that Bochica disapeared in the town of Iza, in the northeastern part of the territory. This personage will be important in our analysis of the Muisca calendar (see chapter \ref{apuntes-duquesne}). 

\section{The Muisca after the colony period.}

Although the Muiscas survived along the colony period (1537-1810), during the processes of Colombian independence in the 19th century they lost their rights to  communal lands \textit{(resguardos)} granted to them by the former Spanish administration. Under the new establishment, they were progressively forced to sell their lands to landowners. Thus they culturally  vanished in the 19th century as a recognizable ethnic group, mixing with Creoles and becoming the majority of agricultural laborers in today's provinces of Boyacá and Cundinamarca. Most of the original  Muisca cultural features are now dispersed and mixed with European traditions in the current population. In 1991, with the proclaiming of a new Political Constitution, the indigenous groups acquired more political rights, so, there exist currently some efforts of small populations claiming  direct descendant of the Muiscas to rescue and keep the ancient traditions, seeking the reconstruction of a lost identity, after more than 470 years of heavy assimilation \citep{Duran2005}.

%%% Notes
%% ASCII ENCODING: ISO8859-1

\hyphenation{Len-gua-za-que Du-ques-ne Mo-nu-men-tos}

\part{The Calendar}\label{part-notes}
\section*{Introduction}
This part \ref{part-notes} is the product of my analysis of the description of the Muisca Calendar written by the priest José Domingo Duquesne in the end of 18th century. As Duquesne's work is not easy to read, my research was centered to understand his work, in order to evaluate its validity as a ethnohistorical source. This analysis process not only allowed me to verify the reliability of his work, but, with the aid of alternate information coming from other chroniclers, to deduce additional inherent features  that were not originally described by the priest  . 

Chapter \ref{apuntes-duquesne} must to be seen as a \textit{reconstruction of the Duquesne's model}, but \textbf{not} as a definitive model of the Muisca Calendar. As  the product of a reflection of a colonial clergyman about the memories of the late 18th century Muisca indians, we cannot concede it all the truthfulness due to our ignorance of all the circumstances about the obtaining of his data, and how the indian's timekeeping vision could have been affected after three centuries of contact with Europeans. However, as it will be shown here, Duquesne's work is a entirely valid starting point in the process of understanding the timekeeping customs of this culture. A lot of archaeoastronomical, ethnohistorical, archaeological, and perhaps ethnographical work is yet needed in order to arrive, consequently, to a final conclusion about this issue. Part \ref{part-connections} will show a preliminary effort towards this goal.

\chapter{Considerations regarding the analysis of José Domingo Duquesne's Interpretation of the  Muisca Calendar}%
\label{apuntes-duquesne}

\section{The work of José Domingo Duquesne}
José Domingo Duquesne has obtained a place in Colombian history for being one of the first researchers to pursue the study of the Muisca culture. Born in Santafé de Bogotá in 1745, he excelled in his studies in the Colegio Mayor y Seminario de San Bartolomé, where he earned his doctorate in Theology and Canonic Law \citep{Vargas1991}. Since 1778, Duquesne served for fifteen years as the parish priest of the village of Lenguazaque. After this period,  he was transferred to another village named Gachancipá, where he remained until the year 1800 when he received an appointment at the cathedral of Santa Fé de Bogotá as a overseer of the Catholic Church's affairs in the city in the early years of the Colombian Republic, office he held until the year of his death, 1821. 

\begin{figure}[h]
 \centering
 \includegraphics[width=5cm]{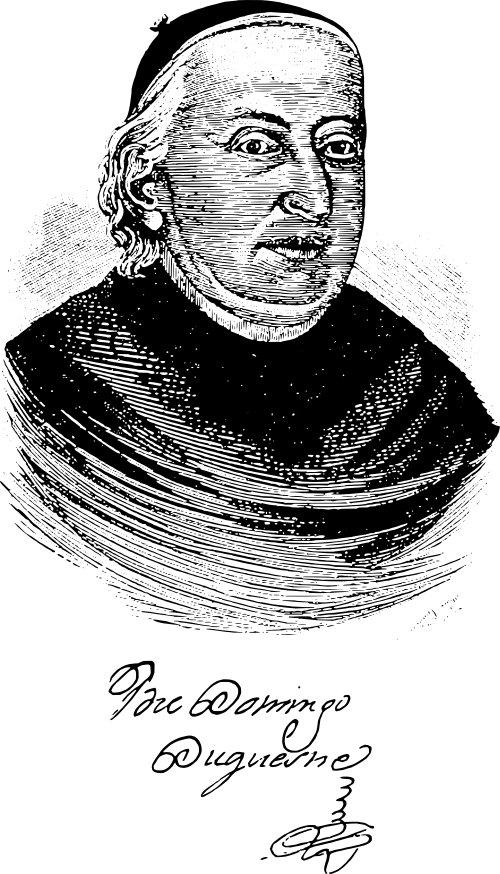}
 \caption{Jos\'e Domingo Duquesne. Engraving of Antonio Rodriguez \citeyearpar{Zerda1882}.}
 \label{duq-photo}
\end{figure}

As a product of his years of interaction with the indian communities of Lenguazaque and Gachancipá, descendants of the ancient Muisca, Duquesne was profoundly interested in this culture, its history, language and especially, its calendar, which details he knew presumedly from indigenous sources. 
According to a version accounted by some relatives of Duquesne to the author Liborio Zerda, his curiosity about the native antiquities, added to the kindness of priest with the native community, gained the sympathy of the indians, who frequently brought him archaeological objects, which he studied with a lot of interest.  Duquesne was then offered to be taken in secret to a hidden cave containing ancient objects,  kept under the indians' care since old times. Duquesne was also demanded  by an indian elder  to keep the secret of the existence of that place. Once there, the indians explained Duquesne  the meaning of some objects and probably, the calendar \citep[p.243]{Zerda1882}\footnote{It is unknown so far in what of the two villages served by Duquesne could happened this episode.}.

In 1795 Duquesne addressed (and dedicated) to José Celestino Mutis, general chief of the \textit{Real Expedición Botánica}, a paper which title translates into: \textit{Dissertation on the calendar of the Muyscas, natural Indians of this New Kingdom of Granada, dedicated to the Doctor José Celestino Mutis, chief of  the Botanical Expedition. By Doctor José Domingo Duquesne de la Madrid, priest of the church of Gachancipá of the same Indians, year of 1795.}\footnote{\textit{``Disertaci\'{o}n sobre el calendario de los muyscas, indios naturales de este nuevo Reino de Granada, dedicada al Se\~{n}or Doctor Jos\'{e} Celestino Mutis, director general
de la Expedici\'{o}n Bot\'{a}nica. Por el Doctor Jos\'{e} Domingo Duquesne de la Madrid, cura de la iglesia de Gachancip\'{a} de los mismos 
indios, a\~{n}o de 1795''}} This text was compiled by Colonel Joaquín Acosta as an appendix of his \textit{``Compendio Histórico del descubrimiento y colonización de la Nueva Granada''}, published in \citeyear{Acosta1848}. Afterwards, towards the end of the nineteen century, Liborio Zerda \citeyearpar{Zerda1882}, published a series of articles in the \textit{Papel Periódico Ilustrado} magazine, titled \textit{``El Dorado''}\footnote{Which will be published posteriorly as a book with the same name.}, and showcases  another  Duquesne's texts on the same subject, titled  \textit{``Dissertation on the origin from the calendar and hieroglyphs of the Moscas'', ``The Astronomical Ring of the Moscas'', ``Explanation of the Symbols of the Century or Calendar of the Muiscas'', ``Table of the Muisca's years''.}\footnote{\textit{``Disertaci\'{o}n sobre el origen del calendario''} y
\textit{``Jerogl\'{\i}ficos de los moscas''}, \textit{``Anillo astron\'{o}mico de los
moscas''}, \textit{``Explicaci\'{o}n de los s\'{\i}mbolos del siglo o calendario de
los moscas''}, \textit{``Tabla de los a\~{n}os muiscas''}} These manuscripts apparently were owned by a certain Alberto Quijano \citep{Zerda1882}; like the previous paper, it is also dated to 1795. As a methodological convention, along this work I will refer to \textit{Dissertation} as the text published by Acosta, and to the \textit{Astronomical Ring} as the set of writings published by Zerda\footnote{In the Appendix \ref{duquesne-work} of the present work is included a transcription of the whole work of Father Duquesne (\textit{Dissertation} and \textit{Astronomical Ring}) in its original version in spanish.}.

At first glance, both the \textit{Dissertation} and the \textit{Astronomical Ring}  seems to be the same text, but a closer look reveals important differences. It is possible that the \textit{Astronomical Ring} was a preliminary draft to the final \textit{Dissertation} sent to Mutis. This fact was noted by Zerda, who argued that the \textit{Dissertation} is just a summary of the results described in detail in the \textit{Astronomical Ring}:

\begin{quote}
...he  limited himself to to present only the summary of his manuscripts to Mr. Mutis, director of the Botanical Expedition of the Viceroyalty; this summary was published by Colonel Acosta... prior to this memoir or dissertation, he wrote other texts that we also have in hand, which were his first writings on this topic and served as base to write the memoir that he addressed to Mr. Mutis, and in which he improved upon the interpretation of the Muisca Calendar.%
\footnote{``... se limit\'{o} unicamente a presentar el resumen de sus 
manuscritos al Se\~{n}or Mutis, director de la Expedici\'{o}n Bot\'{a}nica 
del Virreinato, este resumen fu\'{e} el publicado por el coronel 
Acosta... Antes de esta Memoria o disertaci\'{o}n, escribi\'{o} otras 
que tenemos a la vista, que fueron sus primeros escritos en \'{e}sta 
materia y que le sirvieron de base para confeccionar la Memoria 
que dirigi\'{o} al se\~{n}or Mutis, en la que perfeccion\'{o} la
interpretaci\'{o}n del calendario Muisca...'' \citep{Zerda1882}}.         
\end{quote} 

\noindent Nevertheless, this difference is not too emphasized by 19th and 20th century researchers, as the \textit{Astronomical Ring} is the more frequently cited version of Duquesne's works \citep{Restrepo1892,Rozo1997}.

\section{Criticisms to Duquesne's work}
Although throughout the 19th century  Duquesne's model was widely accepted,  in 1892 this model suffered a setback when the scholar Vicente Restrepo published a very harsh attack against Duquesne's work entitled \textit{Crítica de los trabajos arqueológicos del Dr. José Domingo Duquesne}, which  discredited his thesis, arguing that he \textit{"was swept away by his brilliant imagination"} and concluding that his model was  \textit{"an invention of  honourable fantasy"}\citep{Restrepo1892}. The main arguments proposed by Restrepo against Duquesne basically were: (a) The impossibility to find any surviving aspects of Muisca culture  in the late 18th century indian communities from the villages where Duquesne served as parish priest. (b)  His only valid sources would have been the descriptions and accounts of the Spanish chroniclers, which consequently, would have matched these of Duquesne's, what does not happen. An example of this is his description of the year, which is not found in any other chronicler's work, and differs from the well known description by  Fray Pedro Simón, and (d) if the indians were his actual informants, Father Duquesne should not have had no reason to hide that fact. All of these arguments shaped the view of mainstream academia, and since then it has been the prevailing opinion of the scholars regarding this topic \citep{Lloreda1992}.
  
Nevertheless, in reading Restrepo's analysis, I find it discriminatory, outdated, and hardly acceptable by the standars of modern anthropology. For example, in the first pages of his work, he opens by dismissing the validity of oral tradition: 

\begin{quote}
``The Spanish conquest ended in the complete subjugation of Muisca Nation. No parts of it remained autonomous in order to retain its language, beliefs and traditions. Everybody, either forced or voluntarily adopted the victor's language and religion (...) When Doctor Duquesne served as parish priest in some indian villages, he just found a poor-ignorant people that could not teach him anything about their ancestors' knowledge (...) the generations that followed two and a half centuries had forgotten everything, even their own language (...) What the illiterated  indians from Turmequé, Lenguazaque and Gachancipá could teach him about such convoluted subjects as etymology, astronomy and theogony, which his studies were about?''\footnote{``...La conquista española terminó con el sometimiento completo de la nación chibcha. Ninguna fracción de ésta logró permanecer en un aislamiento tál que le hubiera permitido conservar su idioma, sus creencias y sus tradiciones. Todos, de grado o por fuerza, adoptaron la lengua y la religión del vencedor (...) Cuando el Doctor Duquesne sirvió como Cura de almas en algunas poblaciones de indios, no encontró en ellas sino pobres gentes ignorantes que nada podían enseñarle de los conocimientos de sus antepasados, (...) Las generaciones que se habrían sucedido en el transcurso de dos siglos y medio habían acabado por olvidarlo todo, hasta su propia lengua (...) ¿Qué podían enseñarle los indios iletrados de Turmequé, Lenguazaque y Gacahancipá, sobre las ar\-duas materias que fueron objeto de sus estudios, y que rozan con la etimología, la astronomía y la teogonía?'' \citep[p.4,6]{Restrepo1892}}
\end{quote}

This argument is clearly \textit{a priori} and biased. Although Spanish conquest intervened in the culture of native communities, it was not a process of deletion-and-replacement as Restrepo alludes. It is worth noting that Muisca communities retained most of their cultural identity through out the colonial period, partly due to the right for communal lands \textit{(resguardos)} granted by the colonial administration, and the recognition of the authority of their \textit{caciques} over these territories \citep{Correa2004}, allowed the Muisca to maintain most of their social structures, despite the shock that conquest represented. Archaeological evidence  also suggest for the the retention of native religious rites until the 18th century \citep{Cardenas1990,Cardenas1990a}. In contrast, it seems more plausible that these native communities would have had more to teach to Duquesne than Restrepo imagined.

Restrepo viewed the written record as was the only legitimate foundation for a valid discourse, hence, he argued that Duquesne only had access to \textit{``the printed chronicles available on that time, which were only the  works of Bishop Piedrahita, Father Zamora, and Herrera''}\footnote{``...las crónicas que corrían entonces impresas, que se reducían a las obras del Obispo Piedrahita, del Padre Zamora y de Herrera''\citep[p.5]{Restrepo1892}} and pointed out that the clergyman \textit{``did not have the chance to consult the writings of Jiménez de Quesada, Fray Pedro Simón and other authors, because the library whose manuscripts were kept in, was established in Bogotá years after he did his work''}\footnote{``Ni tuvo (Duquesne) siquiera ocasión de consultar los escritos de Jiménez de Quesada, de Fray Pedro Simón y de otros autores, que se conservaban manuscritos en la Biblioteca, porque no se estableció en Bogotá sino años después de haber dado cima a sus trabajos.''\citep[p.5]{Restrepo1892}}. Consequently, Restrepo systematically rejected all the details given by Duquesne  that were not found in these sources. Restrepo never recognized the opportunity to evaluate the material brought forth by Duquesne as an alternative source of ethnohistorical data, and simply considered dismissed it as incorrect. For example  his criticism of  Alexander von Humboldt's commentary on the ceremonial processions performed by the Muisca during the \textit{Guesa}'s sacrifice \citep{Humboldt1878} , which is directly based on Duquesne's work  (see Appendix \ref{duquesne-work}).  Restrepo argues:

\begin{quote}
...no chronicler says that the Chibcha knew about the gnomonic observations,  represented Bochica with three heads, or confused him with the Sun, who was the Chía's husband, the Moon, or that they weared masks in their processions imitating frogs and customs resembling the monstrous Tomagata (...) This is the way history becomes distorted!\footnote{``...ningún cronista da cuenta de que los Chibchas conocieran las observaciones gnomónicas; ni de que representaran á Bochica con tres cabezas; ni que confundieran á éste con el Sol, que era el esposo de Chía, la Luna; ni de que llevaran en las procesiones caretas imitando ranas y disfraces que recordaran al monstruoso Tomagata (...) ¡Así es como se desfigura la historia!''\citep[p.7-8]{Restrepo1892}}
\end{quote}
\noindent Using similar arguments, Restrepo dismisses the Duquesne's description of the calendar,  since it is not a simple verbatim copy of the   Sim\'on's chronicle.

However, certain details have been confirmed by later ethnohistorical discoveries, as the trial of Cacique Ubaque in 1564 \citep{Casilimas2001,londono2001puu} (see chapter \ref{ubaque-chapter}), and recent studies  confirm the association Bachue-Sun \citep{Correa2004}. Furthermore,  section \ref{simonduquesne} I will show how both Sim\'on and Duquesne's descriptions of the calendar are not in conflict, rather they validate and support each other instead.

Other criticisms to Duquesne's work involves his lack to properly mention his sources. However, the ethnographic origin of his data seems very plausible. As Liborio Zerda indicates: \textit{without the frequent and intimate relationships and communications with the indians, it cannot be understood how Duquesne could interpret the Muisca calendar and collect so much and interesting data.}\footnote{``Sin las relaciones íntimas y frecuentes comunicaciones con los indios, no se puede comprender como Duquesne hubiera podido interpretar el calendario de los muiscas, y recoger tantos y tan interesantes datos.''\citep[p.242]{Zerda1882}}. Restrepo  argued that if Duquesne really obtained  his information from the indians, there would be no reason to keep it quiet. I think that, on the contrary,  he effectively did. Considering the probable circumstances under which Duquesne was introduced to the  traditional knowledge of the Muisca (see previous section), his permissive attitude of tolerating the indian `pagan tradition' would have caused him serious problems at the risk of losing his sacerdotal career. Throughout three centuries of colonianlism, the usual reaction of catholic priests towards slightest evidence of native religious practices was punishment and confiscation of native religious paraphernalia, especially if such paraphernalia was made of gold. Clearly, Duquesne was an unusual  case for that time and consequently, he must have maintained a prudent silence about his relationship with the indians at the risk of his career. This could explain why the clergyman avoided revealing his close relationship with the people he studied, which was only vaguely expressed in this paragraph from the \textit{Astronomical Ring} (the \textit{italics} is mine): 

\begin{quote}
``... I guess that the scholar that would take the work to combine into one idea the remaining historial news of this nation, and \textit{if in any case has treated the indians with some frequency (and not superficially), and has penetrated their mood and their misterious and emphatic character}, will know the  soundness of the foundations over I will establish this interpretation.''\footnote{``Bien que, creo que el lector erudito que se tome el trabajo de combinar bajo una idea las noticias históricas que nos han quedado de esta nación, si por otra parte ha tratado con alguna frecuencia (y no superficialmente) a los indios; si ha penetrado su genio y su carácter misterioso y enfático, conocerá la solidez de los fundamentos sobre que establecemos esta interpretación.''\citep{Zerda1882}}
\end{quote}

The criticisms of Restrepo are, thus,  founded on outdated academic views and biases against the Indians. However some flaws can be  detected in  Duquesne's description, for example his asumption that three solar years exactly equals 37 lunar months (see sections \ref{acrotomyear} and \ref{muisca-centuries}), these are not a solid premise upon which to reject the entirely of the work and declare it the product of a madman, as proposed by the Restrepo's argument.

It seems likely that Duquesne's description is merely an incomplete snapshot, derived from surviving vestiges  of Muisca traditions during the 18th century.  The aim of future research should be to compile a more complete picture of the Muisca Calendar System. The remainder of this chapter will introduce  Duquesne's model of the Muisca Calendar, and will further demonstrate how new structures of this system arise from the deductive analysis of both Duquesne's work and  ethnohistorical sources. 

\section{Duquesne's model}
\subsection{Work's structure}

The fact that the texts of the clergyman are obscure does not diminish their importance.  Duquesne's work deserves a closer look than the one previously accorded him by scholars like Restrepo. When  the \textit{Astronomical Ring} and the \textit{Dissertation} are read simultaneously, one notices that  they provide complementary information, contributing to a greater understanding of their meaning. 
Although both texts share identical paragraphs,  the \textit{Dissertation} tends to show a more compact writing, which could have lead Zerda to consider it just a summary, though he may  not be right at all. Indeed, the \textit{Dissertation} might be considered as a more polished text derived from the \textit{Astronomical Ring}. 

Both the \textit{Astronomical Ring} and the \textit{Dissertation}, show the same thematic structure: 

\begin{enumerate}
\item They begin with a commentary on the manners  the ancient people had in making their calendars according to the Sun and the Moon, and argues that the Muiscas developed similarly. 
\item  Both described the numerical system, giving a etymological interpretation to the names of the numbers. In this description, one finds one of the more convoluted aspects of Duquesne's thesis:  his claim that the Muiscas developed writing and the existence of pictorial signs to represent numbers.
\item  The lunar cycle with the denominations for the lunar phases are described.  Lunar phases are named  using the same terms defined for the numbers. 
\item  Next, they  describe the Acrotom years, defined as a lunisolar system with the intention of fitting both the lunar and the solar movements.
\item  They continue describing the rite of the Guesa's sacrifice, showing certain calendrical cycles associated with ceremonies. 
\item  Then, they provide the description and interpretation of a calendar stone that was given to Duquesne, where supposedly,  the exposed cycles are recorded. 
\item  Finally they describe the \textit{Table of the Muisca years}, a tool apparently designed by the author to find dates in the Muisca chronological system. 
\end{enumerate}

A third  document that alludes to Duquesne's work is the chapter \textit{Monumentos de los indios muiscas}  of the  work \textit{``Sitios de las cordilleras de América''} by the famous naturalist Alexander von Humboldt \citeyearpar{Humboldt1878}, who  met Duquesne during his visit to Santa Fé de Bogot\'a in 1801. In essence, his chapter is a repetition of the \textit{Dissertation}, extended with a personal analysis of Duquesne's model. He also gives some of  Duquesne's opinions on the matter, which are not found in the texts published by Acosta and Zerda.

\subsection{The Muisca numbering system}

This system is fundamental to understand the calendar system proposed by Duquesne. It is based on the use of ten names, or ``labels'', that  represent the values from the number 1 to 10. In order to express the numbers from 11 to 20, these same names are used, but added to them is a \textit{Quihicha} prefix: 

\begin{quote}
 ...they only have names for ten, and once finished, they passed from the hands to the feet, adding to each  word \textit{quihicha}, that means  ``foot'': \textit{quihicha ata}, the one of foot, or eleven, \textit{quihicha bosa}, twelve, etc. (p. 210)
\end{quote} 

Duquesne provides a table with the number names, which, by means of \textit{quihicha} will express the values from 1 to  20: 

\begin{center}
\begin{tabular}{rl||ll}
(1) & Ata & (11) & Quihicha Ata\\
 (2) & Bosa &(12) & Quihicha Bosa \\
 (3) & Mica &(13) & Quihicha Mica\\
 (4) & Muihica &(14) & Quihicha Muihica\\
 (5) & Hisca &(15) & Quihicha Hisca\\
 (6) & Ta &(16) & Quihicha Ta\\
 (7) & Cuhupcua &(17) & Quihicha Cuhupcua\\
 (8) & Suhusa &(18) & Quihicha Suhusa\\
 (9) & Aca &(19) & Quihicha Aca\\
 (10) & Ubchihica &(20) & Quihicha Ubchihica
\end{tabular}
\end{center}
 
\begin{figure}[h]
 \centering
 \includegraphics[width=12cm]{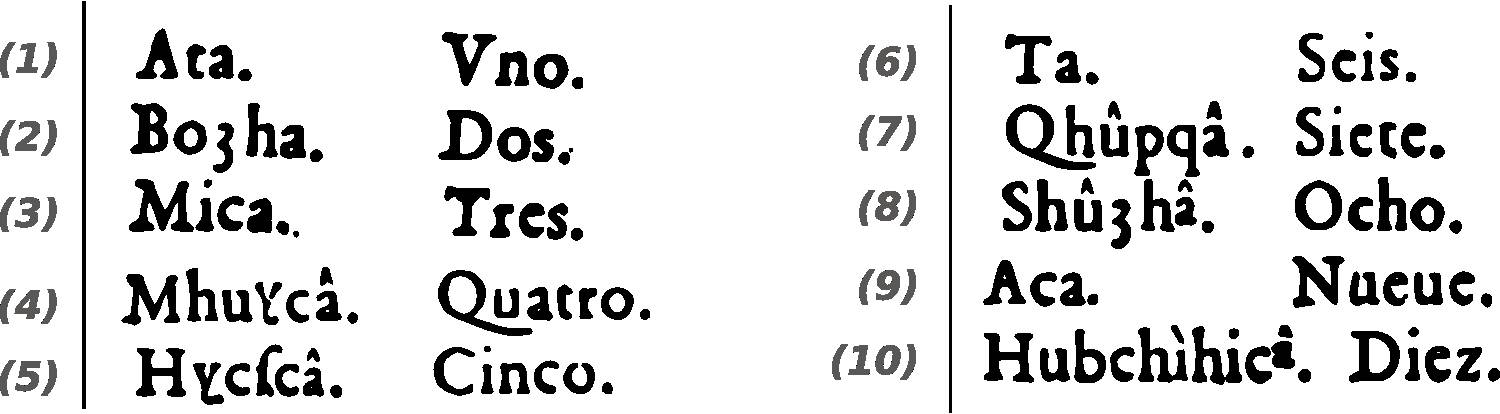}
 % lugo-numbers.png: 800x889 pixel, 80dpi, 25.43x28.26 cm, bb=0 0 721 801
 \caption{Facsimile of Fray Fernando de Lugo's description of the Musica numbers~(\citeyear{Lugo1619}).}
 \label{lugo-numbers}
\end{figure}

Note that the term for twenty, \textit{Quihicha Ubchihica}, has the same meaning as the prefix \textit{Gueta}, which is used to denominate a group of twenty. Added to the previous names, it serves to express any amount. The priest Fray Fernando de Lugo wrote in \citeyear{Lugo1619} a detailed description of the Muisca's system of numeration in his \textit{Gramática en la Lengua General del Nuevo Reyno, llamada Mosca}. Pioneering the concept of the modern phonetic alphabet used by linguists,  Father de Lugo designed some typographical signs in order to match as much as possible the original Muisca phonetics (see figure \ref{lugo-numbers}), which are in essence, the same terms given by Duquesne, two centuries later, in the 18th century.

\newcommand{\iM}{\raisebox{-0.5ex}{\rotatebox[origin=c]{180}{$\lambda$}}}
\newcommand{\zM}{\raisebox{-0.4ex}{3}}

About the use of the term \textit{Quihicha Ubchihica} (20) both authors agree that alternatively the word \textit{gueta} can be used, which apparently was more appropriate to express twenty and other derived amounts. Furthermore, Father Lugo explains in more detail the correct grammatical formation of numerical expressions: 

\begin{quote}
...and to count twenty one, we put between the term gueta and the term ata the particle ``asaq{\iM}'', that means `and more', and for proper pronunciation and sound after the term gueta, we add the letter S... \footnote{ ``...Y para contar veynte y uno, pondremos entre este termino 
gueta, y el termino ata esta part\'{\i}cula asaq{\iM}, que quiere 
dezir y mas, y para la buena pronunciaci\'{o}n y sonido despu\'{e}s 
del termino gueta, a\~{n}adiremos esta letra S''...\citep[p110]{Lugo1619}.}                                                      
\end{quote}

He gives the dictions for the numbers from  21 to 40: 

\begin{center}
\begin{tabular}{rl||ll}
(21)&  Guetas asaq{\iM} ata &(31)&  Guetas asaq{\iM} qhicha ata   \\
(22)&  Guetas asaq{\iM} bo{\zM}ha &(32)&  Guetas asaq{\iM} qhicha bo{\zM}ha  \\
(23)&  Guetas asaq{\iM} mica &(33)&  Guetas asaq{\iM} qhicha mica  \\
(24)&  Guetas asaq{\iM} mhu{\iM}c\^{a} &(34)&  Guetas asaq{\iM} qhicha
mhu{\iM}c\^{a}  \\(25)&  Guetas asaq{\iM} h{\iM}csc\^{a} &(35)&  Guetas
asaq{\iM} qhicha h{\iM}csc\^{a}  \\(26)&  Guetas asaq{\iM} ta &(36)&  Guetas
asaq{\iM} qhicha ta   \\(27)&  Guetas asaq{\iM} qh\^{u}pq\^{a} &(37)&  Guetas
asaq{\iM} qhicha qh\^{u}pq\^{a}   \\(28)&  Guetas asaq{\iM} sh\^{u}{\zM}h\^{a}
&(38)&  Guetas asaq{\iM} qhicha sh\^{u}{\zM}h\^{a}   \\(29)&  Guetas asaq{\iM} aca
&(39)&  Guetas asaq{\iM} qhicha aca   \\(30)&  Guetas asaq{\iM}
hubch\`{\i}hic\^{a} &(40)&  Guetas asaq{\iM} qhicha
hubch\`{\i}hic\^{a}\end{tabular}\end{center}

The value 40 can be replaced by the term \textit{guebo{\zM}ha}, that is, two twenties; therefore, (41) \textit{guebo{\zM}ha asaq{\iM} ata}, (42) \textit{guebo{\zM}ha asaq{\iM} bo{\zM}ha}, and so on. Note that this notation indicates a cognitive base 20 plan of numbering: in order to express any value it is necessary to group it in amounts of twenties. 

In order to understand  Duquesne's writings, it is necesary to take into account  that in situations where he is speaking about amounts higher than ten, he only used the basic name  for the values from 1 to 10, as he was not diligently using the appropiate prefixes, a fact that could cause confusion and consequently, lead to a misunderstanding of his work.

\subsection{Muisca `written ciphers': a hypothetical reinterpretation}\label{muisca-constellations}
\begin{figure}[h]
\begin{center}\renewcommand{\tabcolsep}{0cm}
 \begin{tabular}{m{2cm}cccccccccccc}
 &\rotatebox{90}{\scalebox{0.6}[1]{\textbf{\textsc{ata}}}} &\rotatebox{90}{\scalebox{0.6}[1]{\textbf{\textsc{bosa}}}} &\rotatebox{90}{\scalebox{0.6}[1]{\textbf{\textsc{mica}}}} &\rotatebox{90}{\scalebox{0.6}[1]{\textbf{\textsc{muihica}}}} &\rotatebox{90}{\scalebox{0.6}[1]{\textbf{\textsc{hisca}}}} &\rotatebox{90}{\scalebox{0.6}[1]{\textbf{\textsc{ta}}}} &\rotatebox{90}{\scalebox{0.6}[1]{\textbf{\textsc{cuhupcua}}}} &\rotatebox{90}{\scalebox{0.6}[1]{\textbf{\textsc{suhusa}}}} &\rotatebox{90}{\scalebox{0.6}[1]{\textbf{\textsc{aca}}}} &\rotatebox{90}{\scalebox{0.6}[1]{\textbf{\textsc{ubchihica}}}} &\rotatebox{90}{\scalebox{0.6}[1]{\textbf{\textsc{gueta}}}}\\\hline
Acosta &\includegraphics[width=0.6cm]{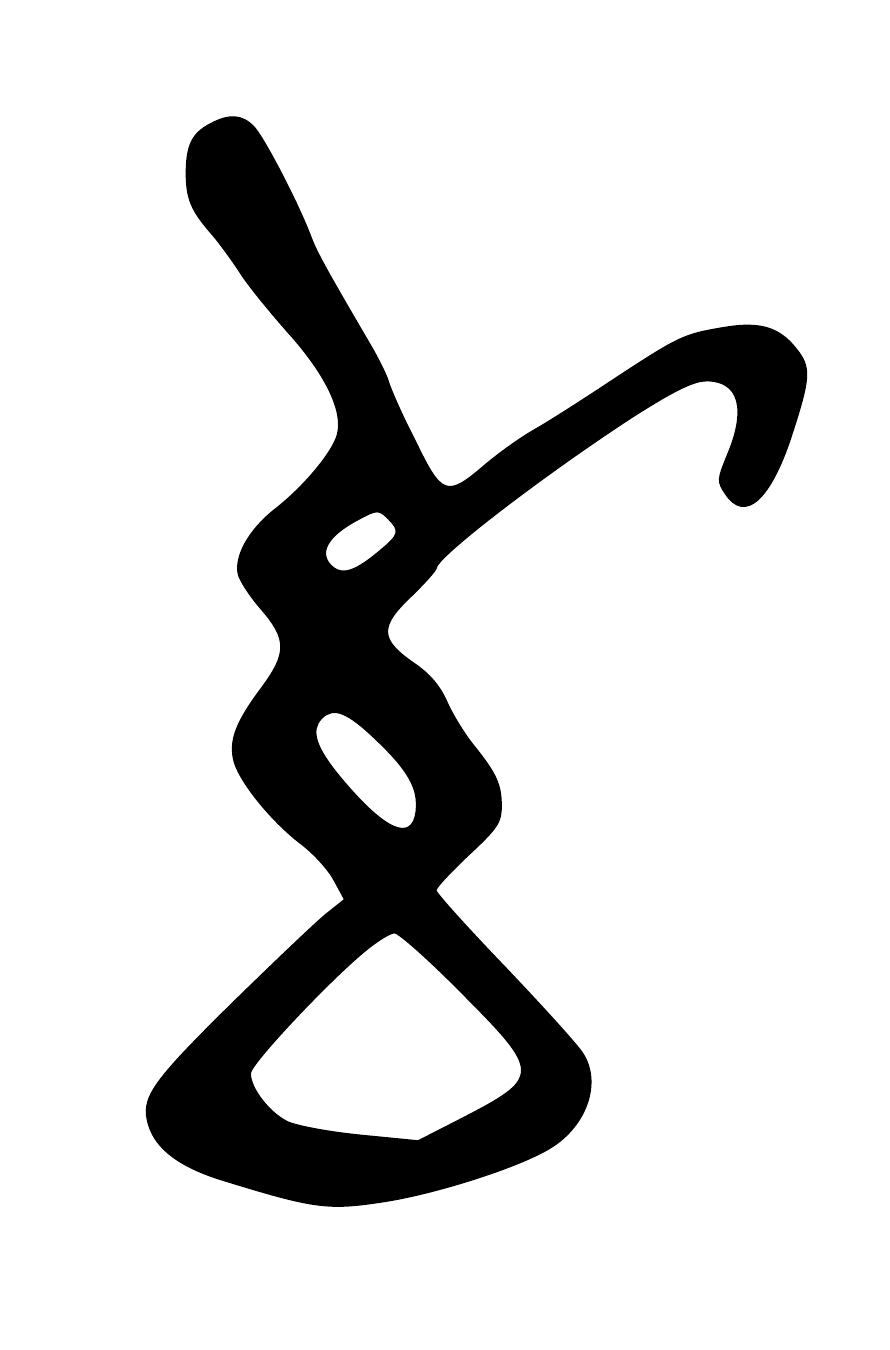} &\includegraphics[width=0.6cm]{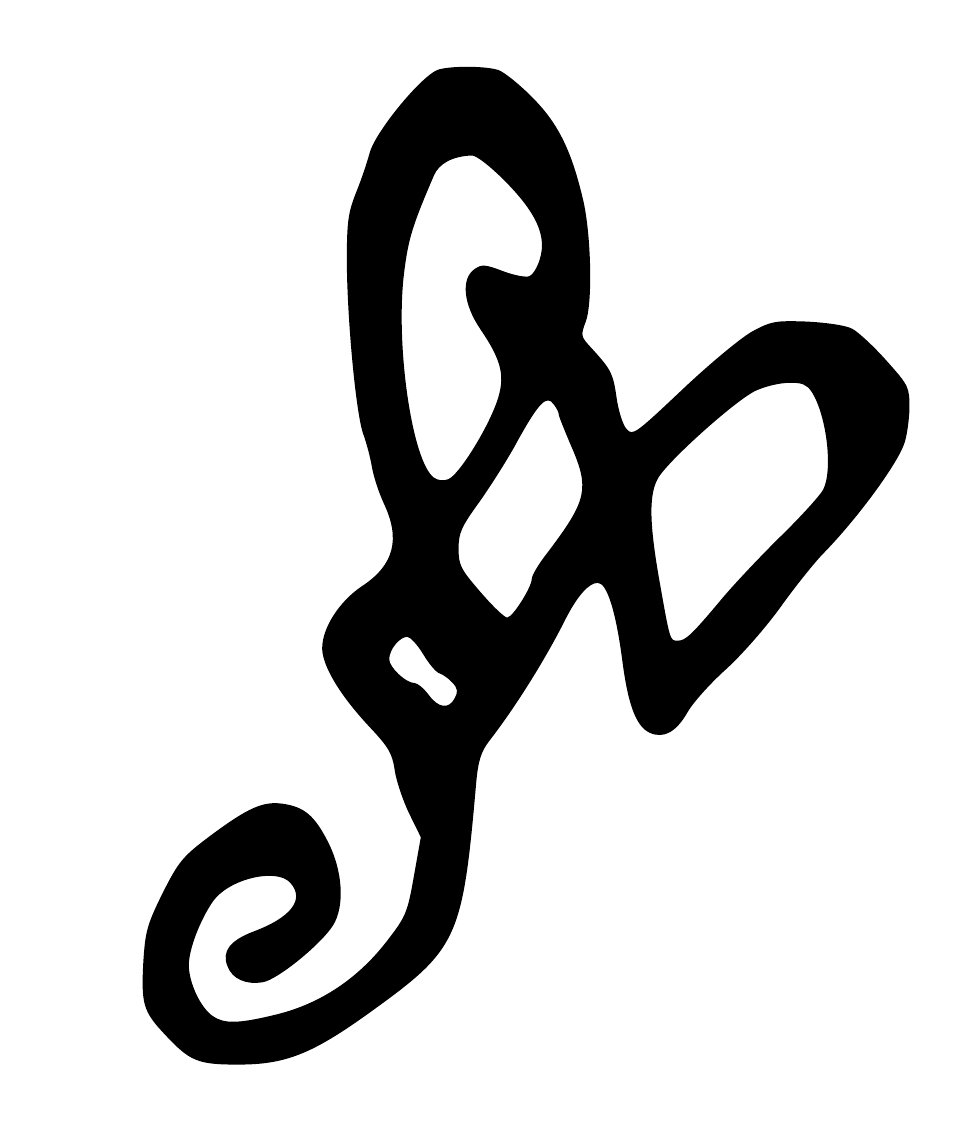} &\includegraphics[width=0.6cm]{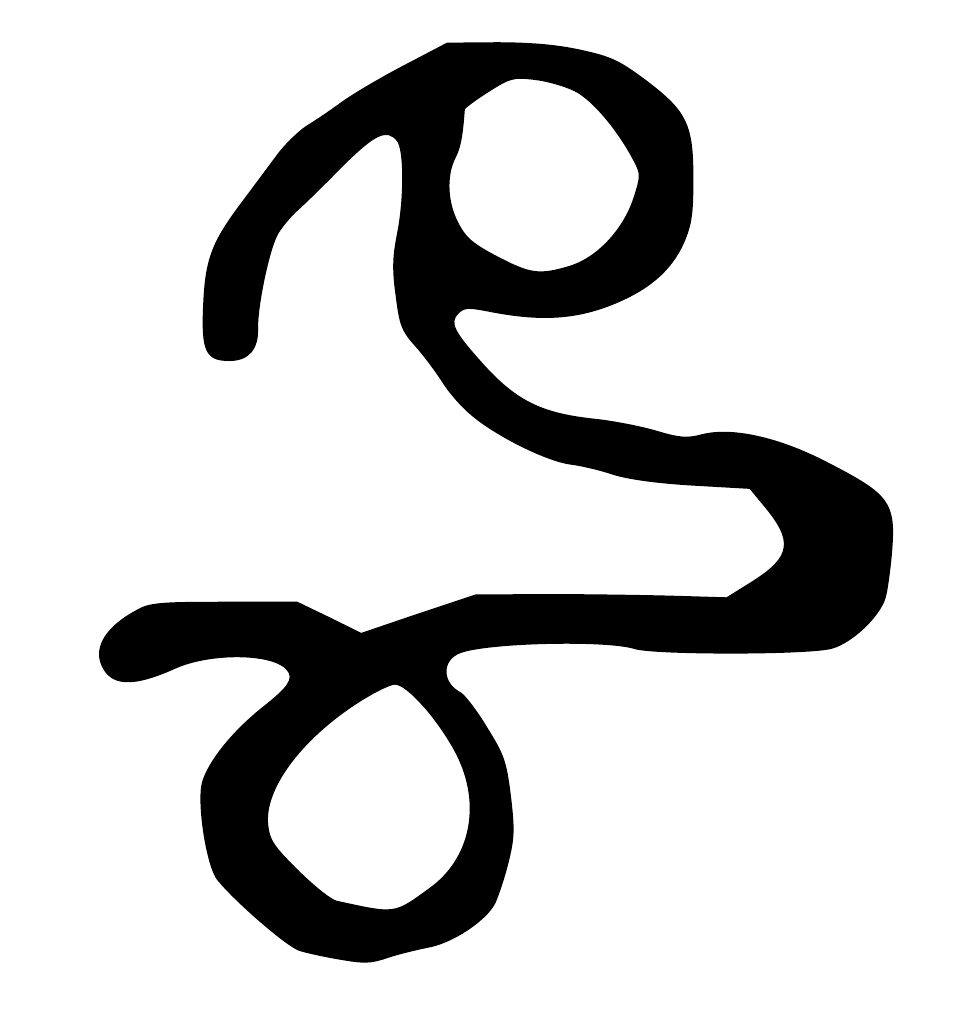} &\includegraphics[width=0.6cm]{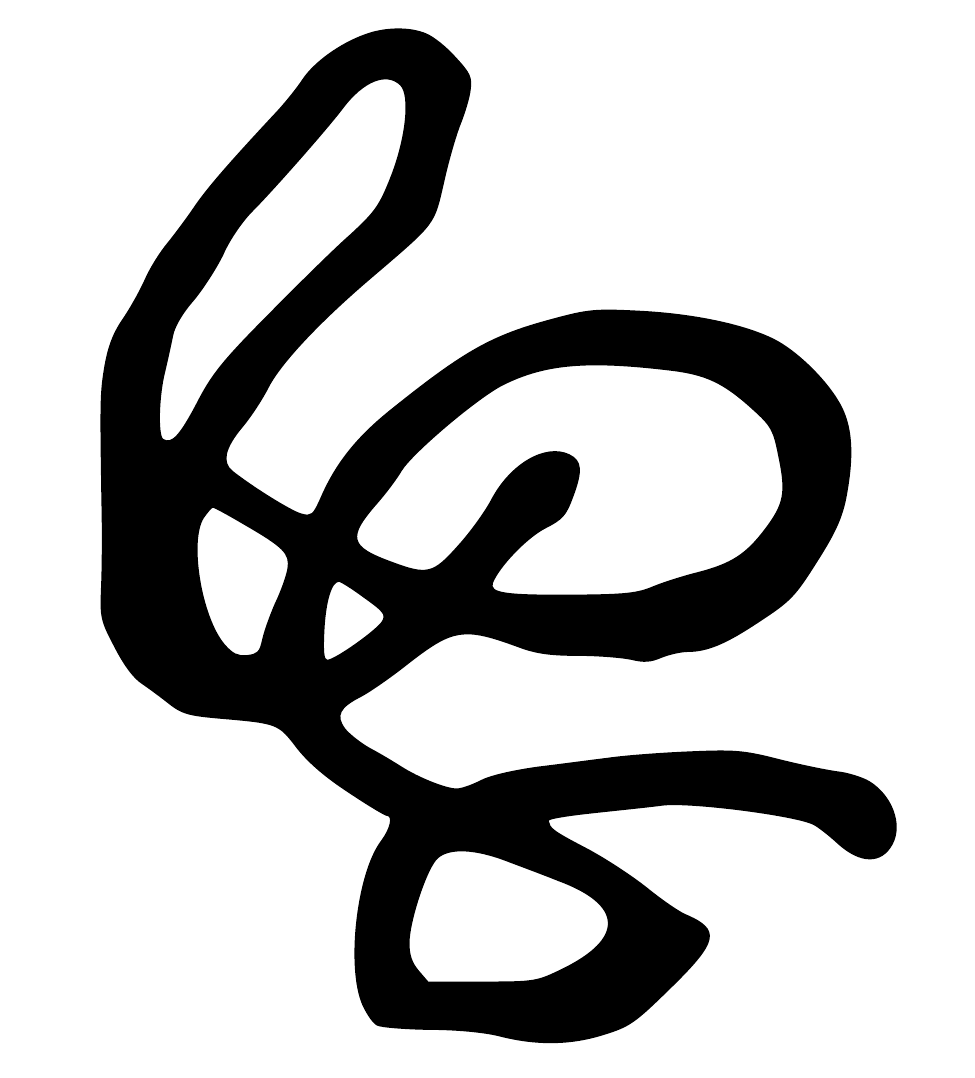} &\includegraphics[width=0.6cm]{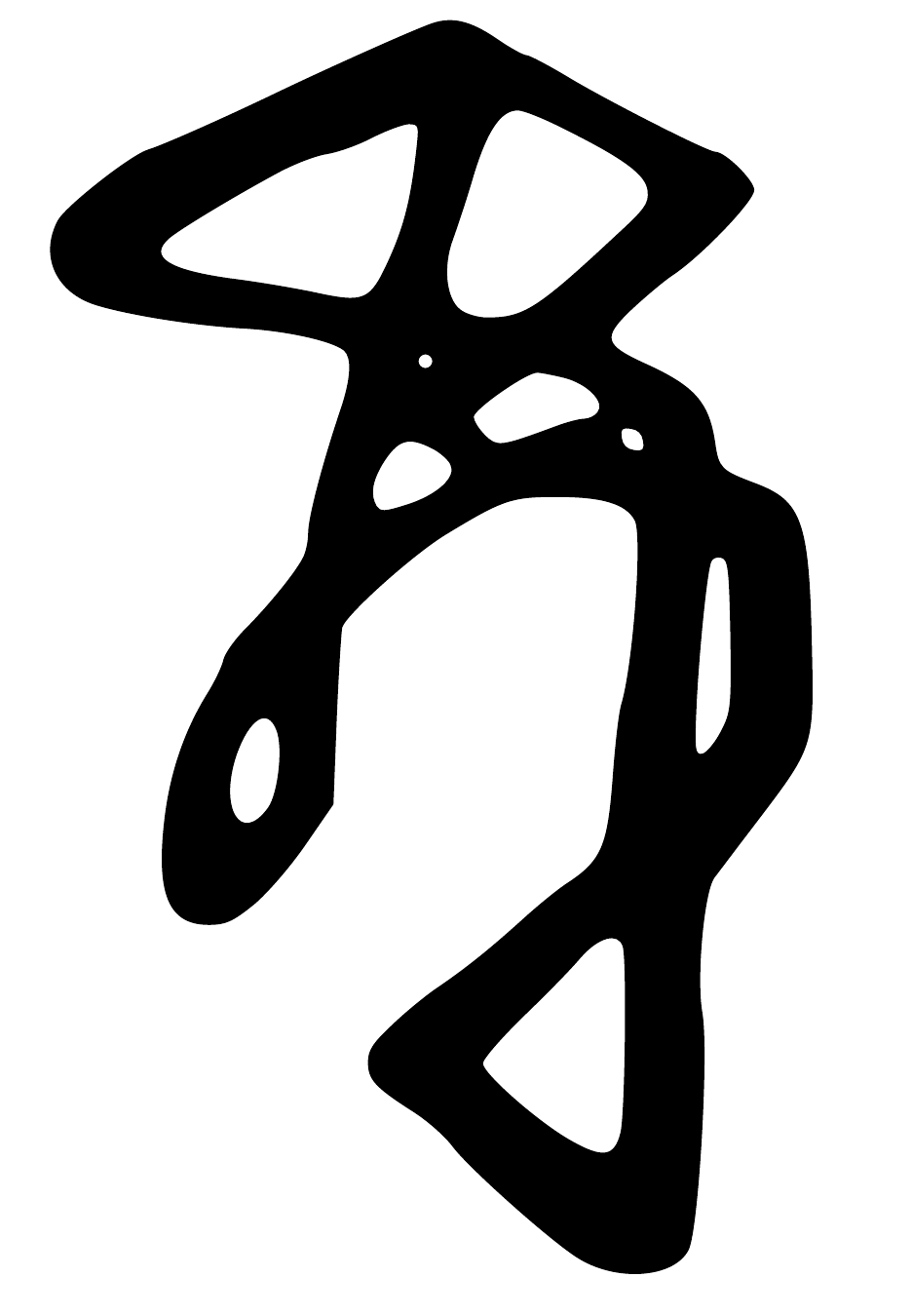} &\includegraphics[width=0.6cm]{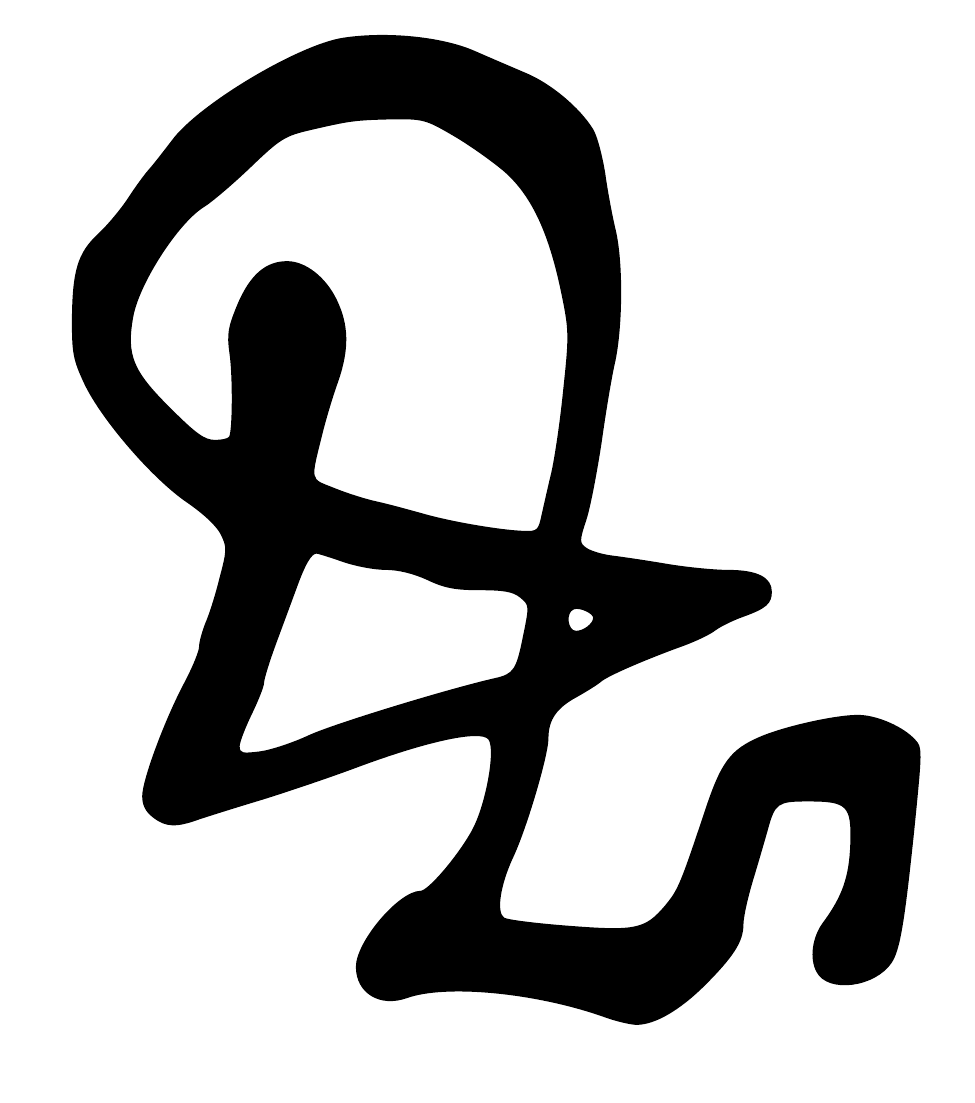} &\includegraphics[width=0.6cm]{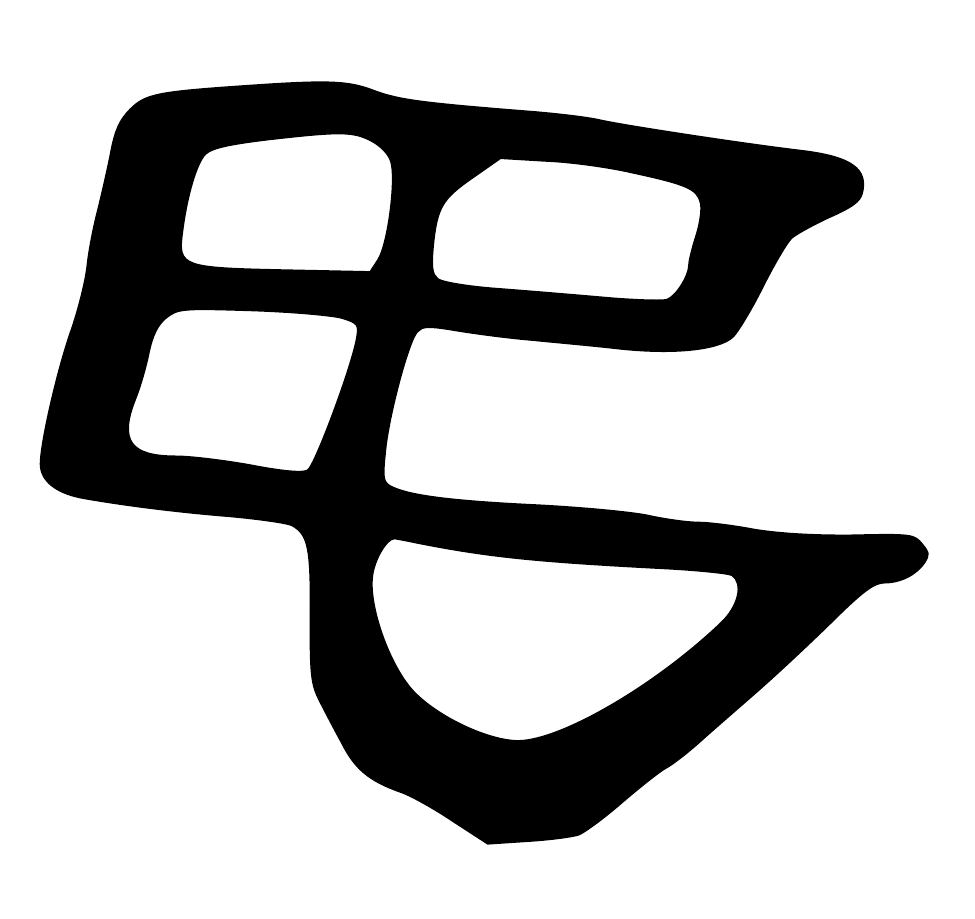} &\includegraphics[width=0.6cm]{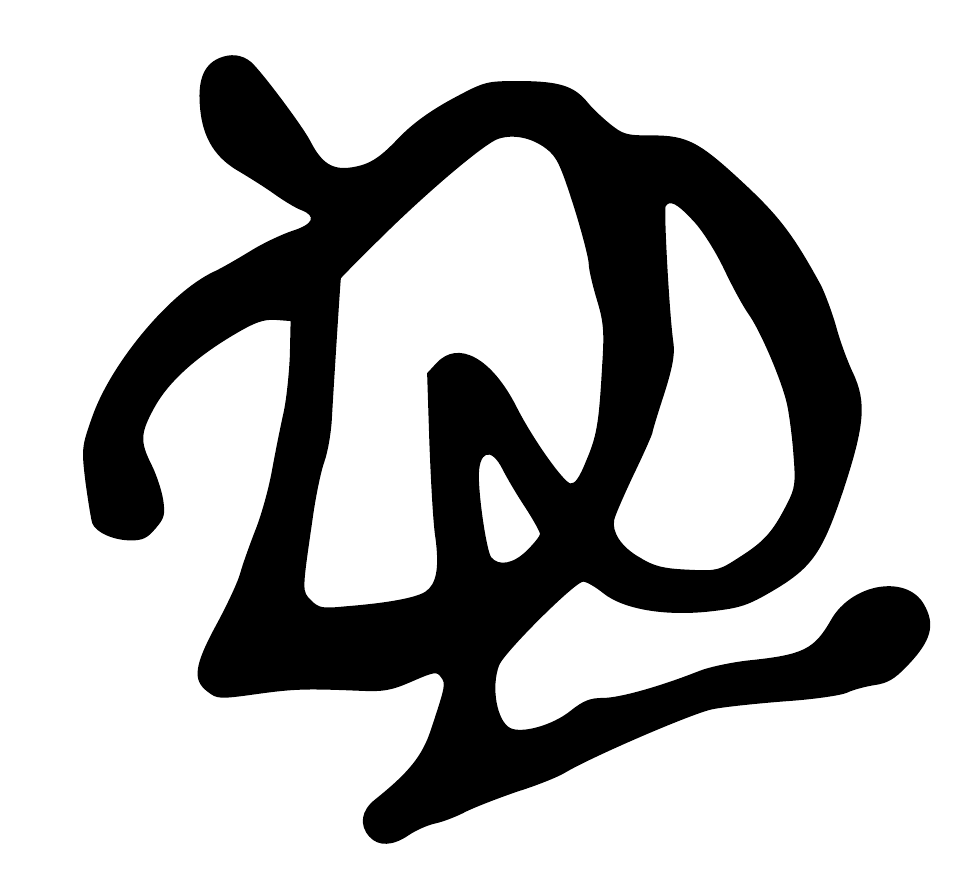} &\includegraphics[width=0.6cm]{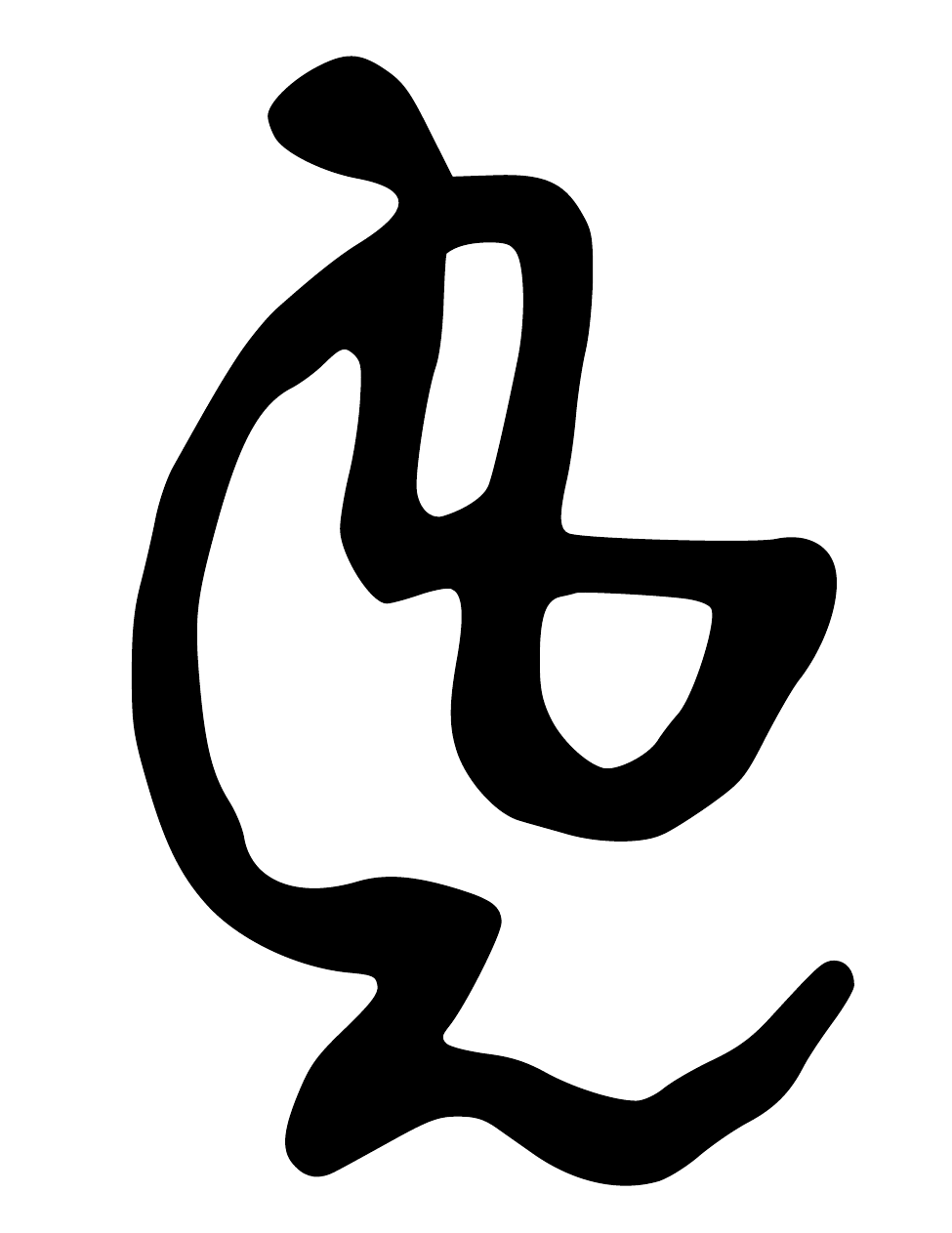} &\includegraphics[width=0.6cm]{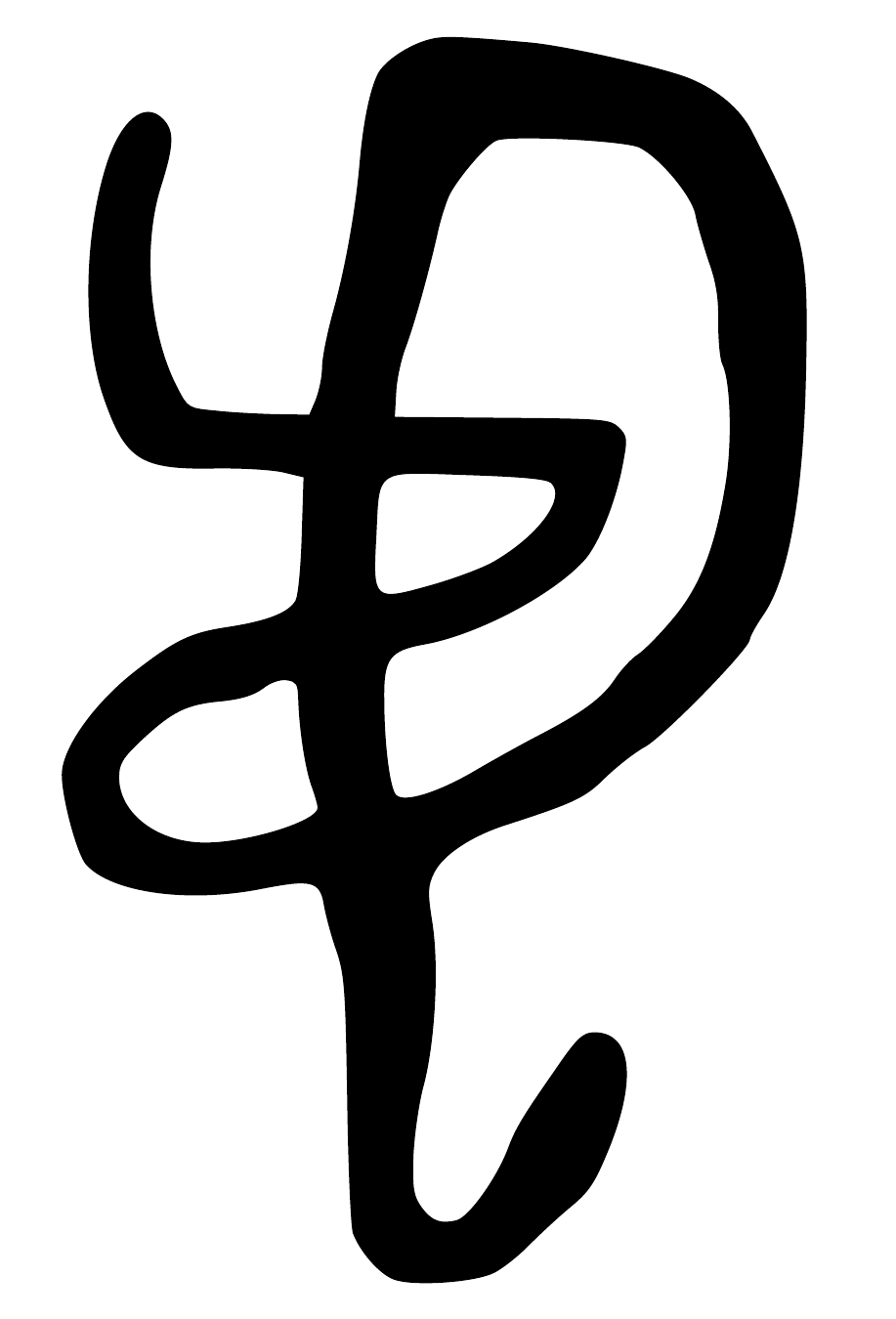} &\includegraphics[width=0.6cm]{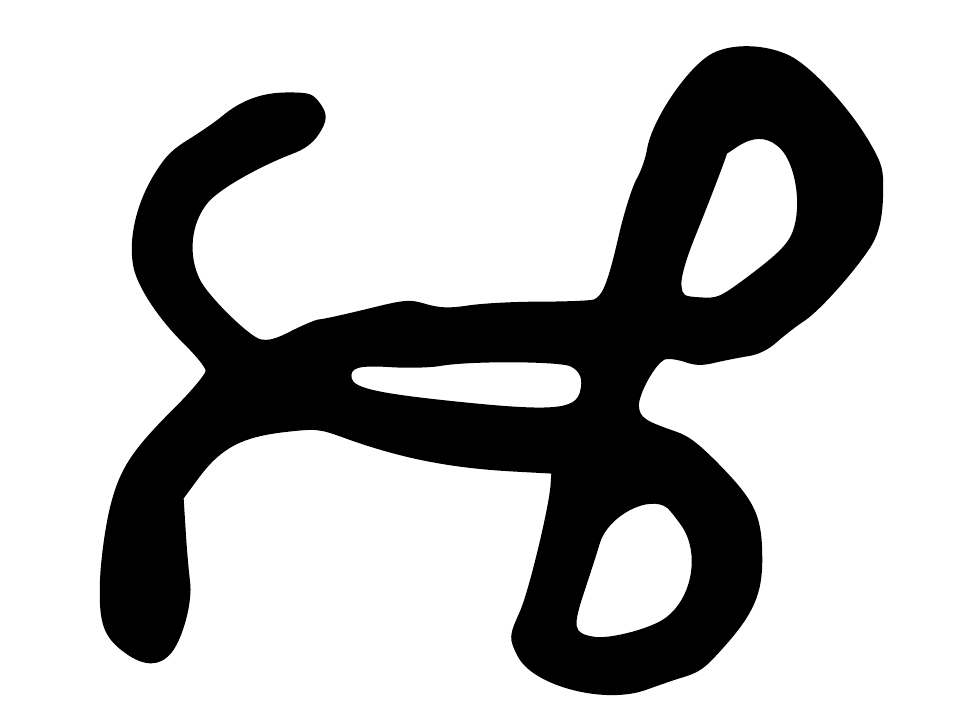} \\
Humboldt& \includegraphics[width=0.6cm]{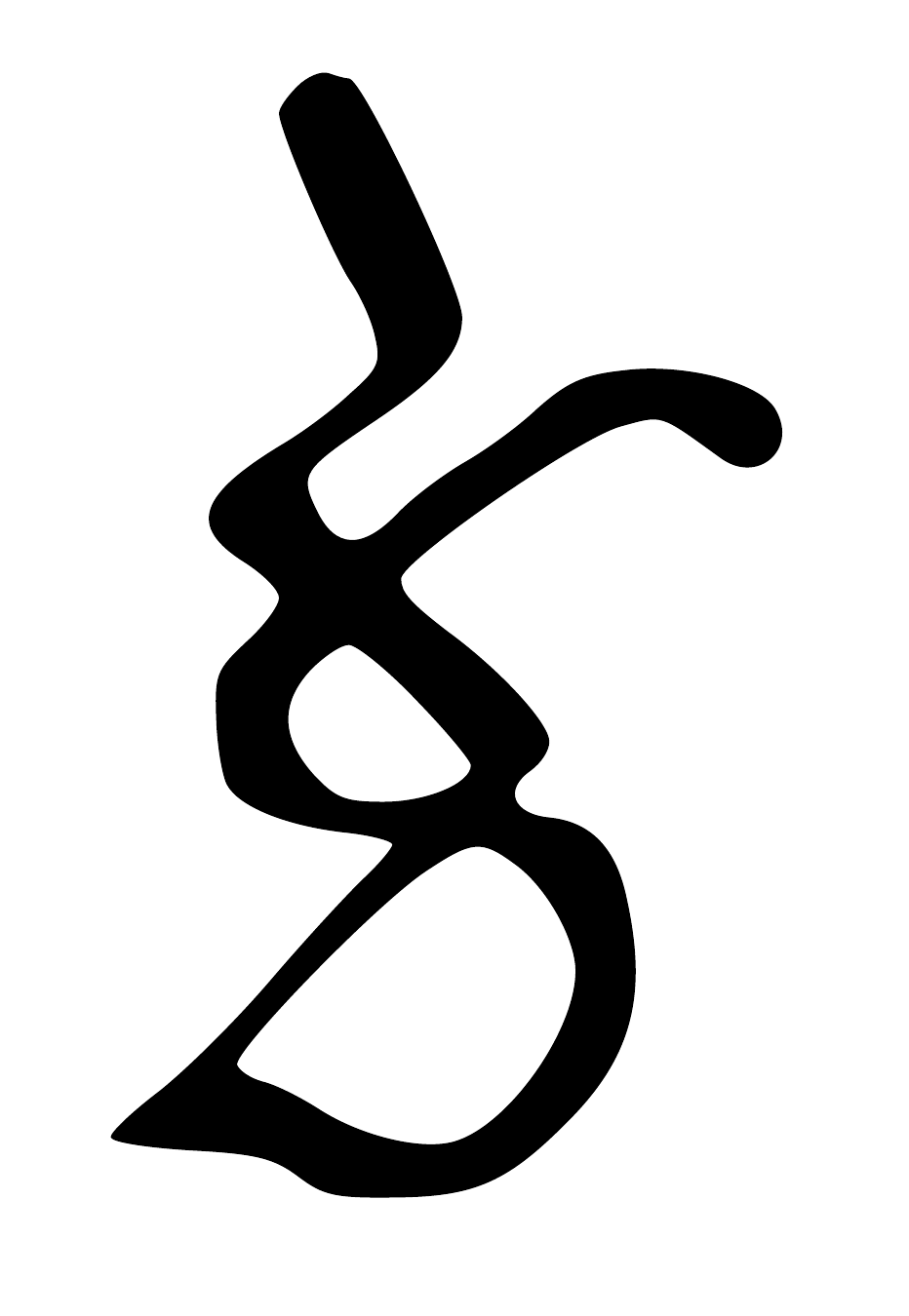} &\includegraphics[width=0.6cm]{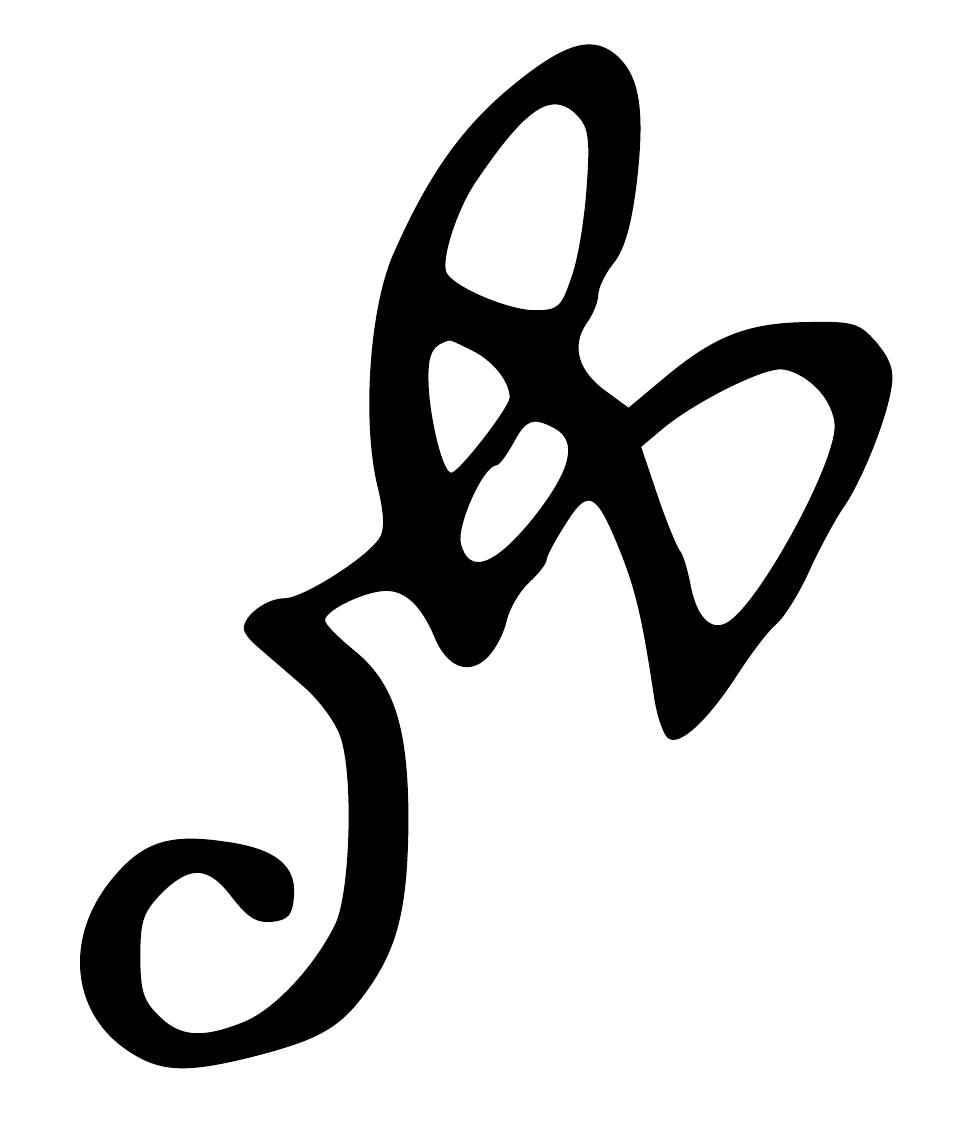} &\includegraphics[width=0.6cm]{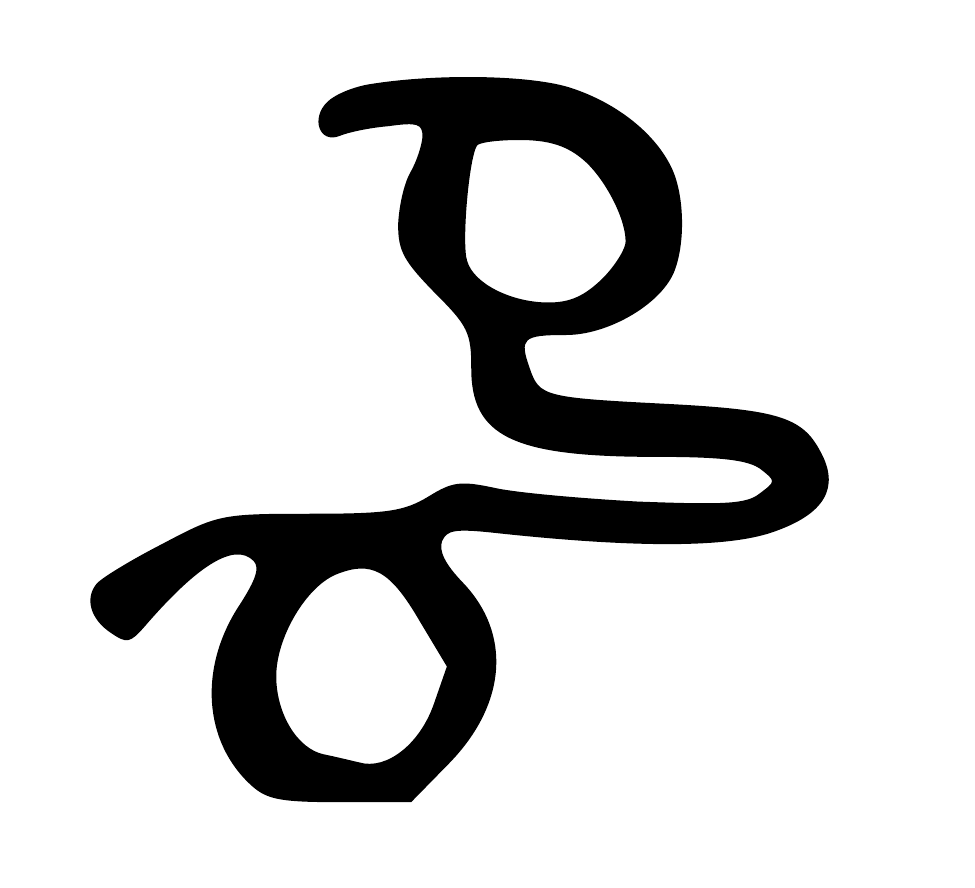} &\includegraphics[width=0.6cm]{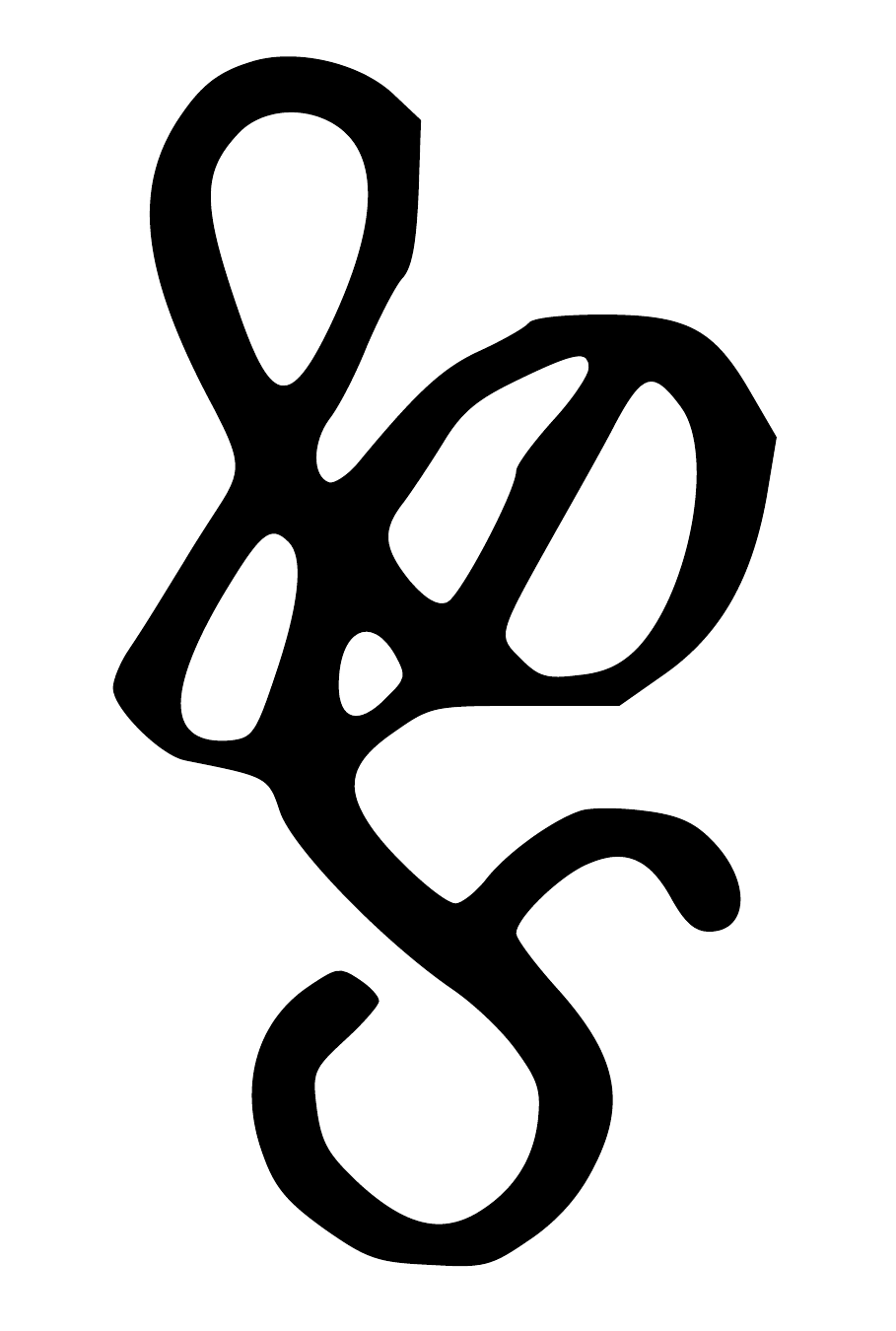} &\includegraphics[width=0.6cm]{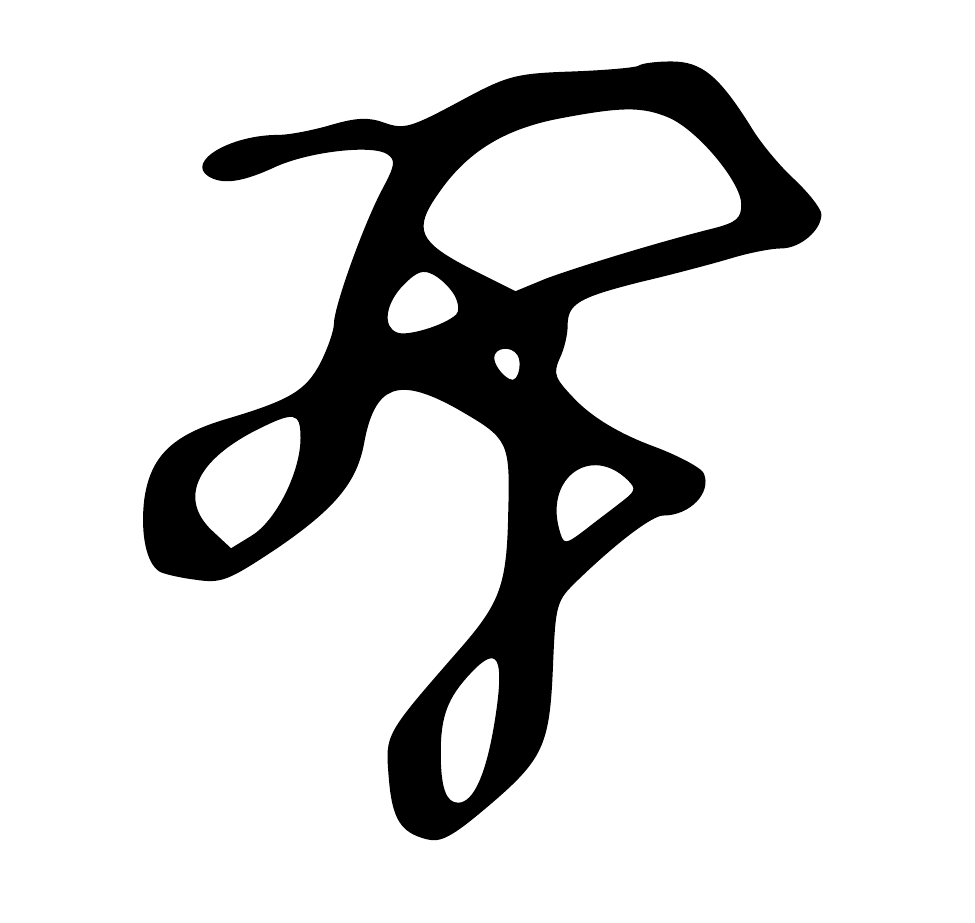} &\includegraphics[width=0.6cm]{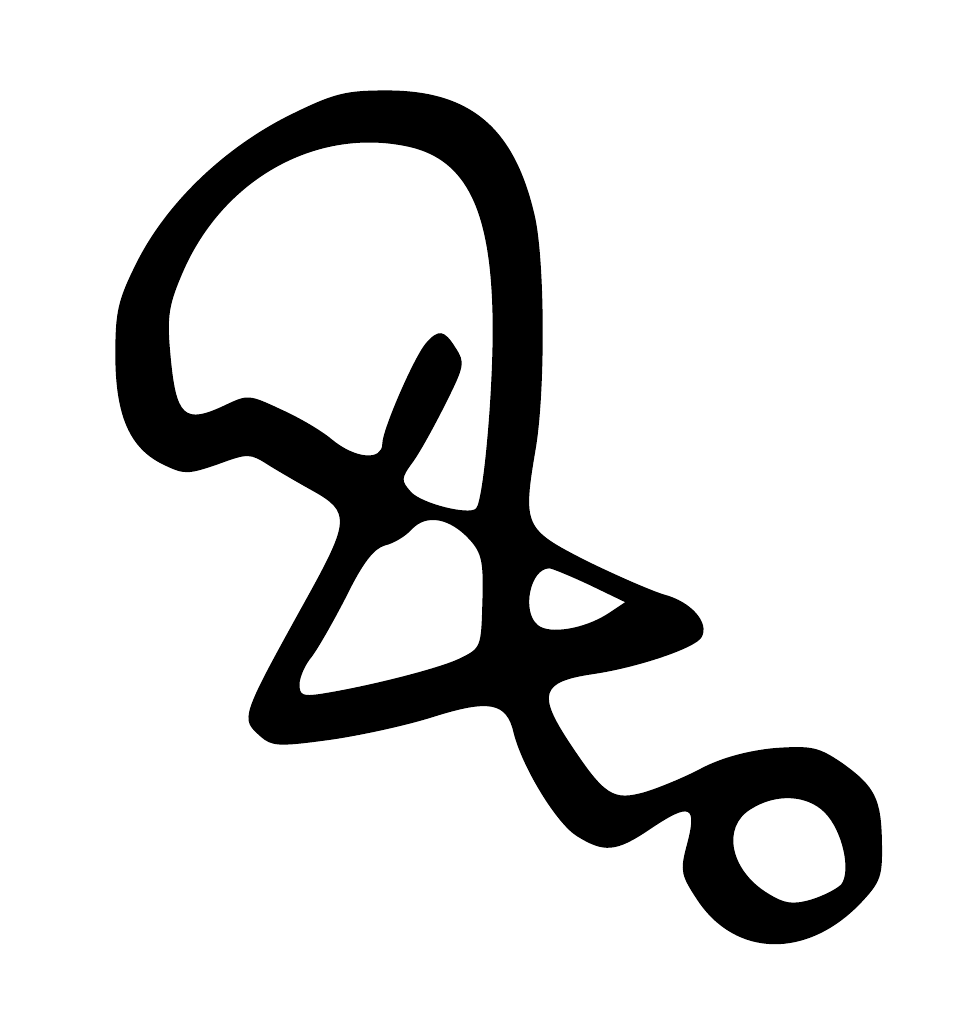} &\includegraphics[width=0.6cm]{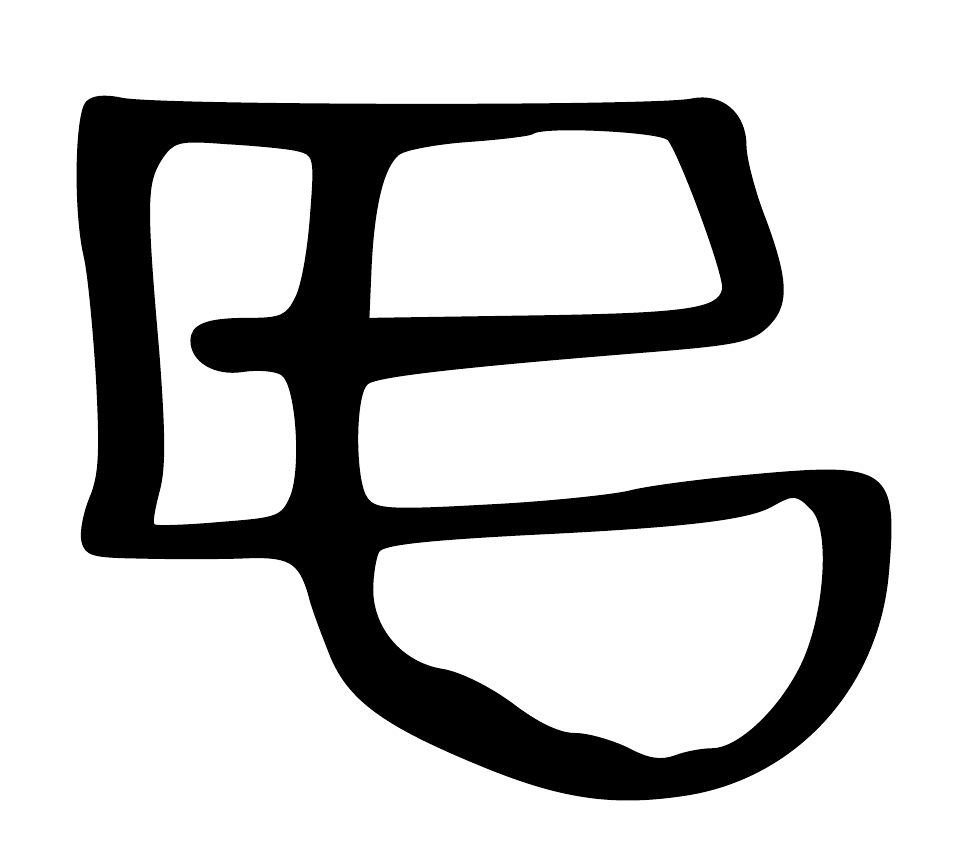} &\includegraphics[width=0.6cm]{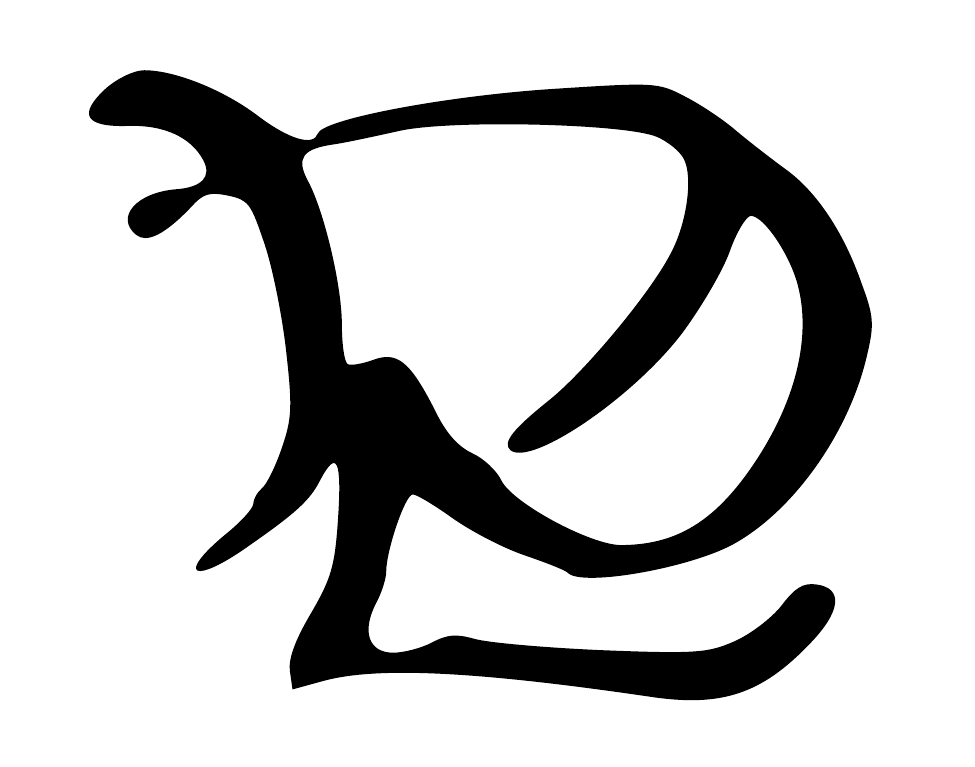} &\includegraphics[width=0.6cm]{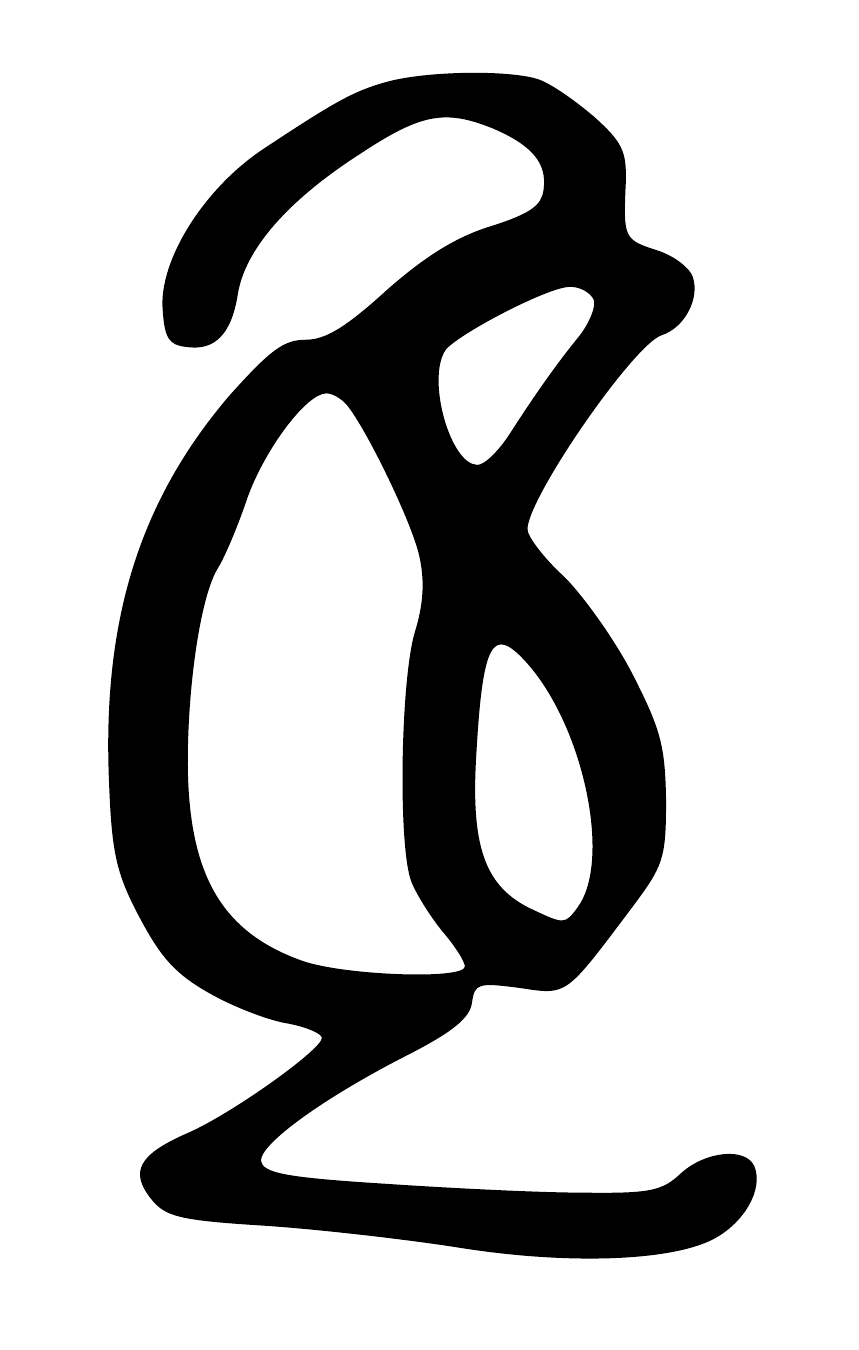} &\includegraphics[width=0.6cm]{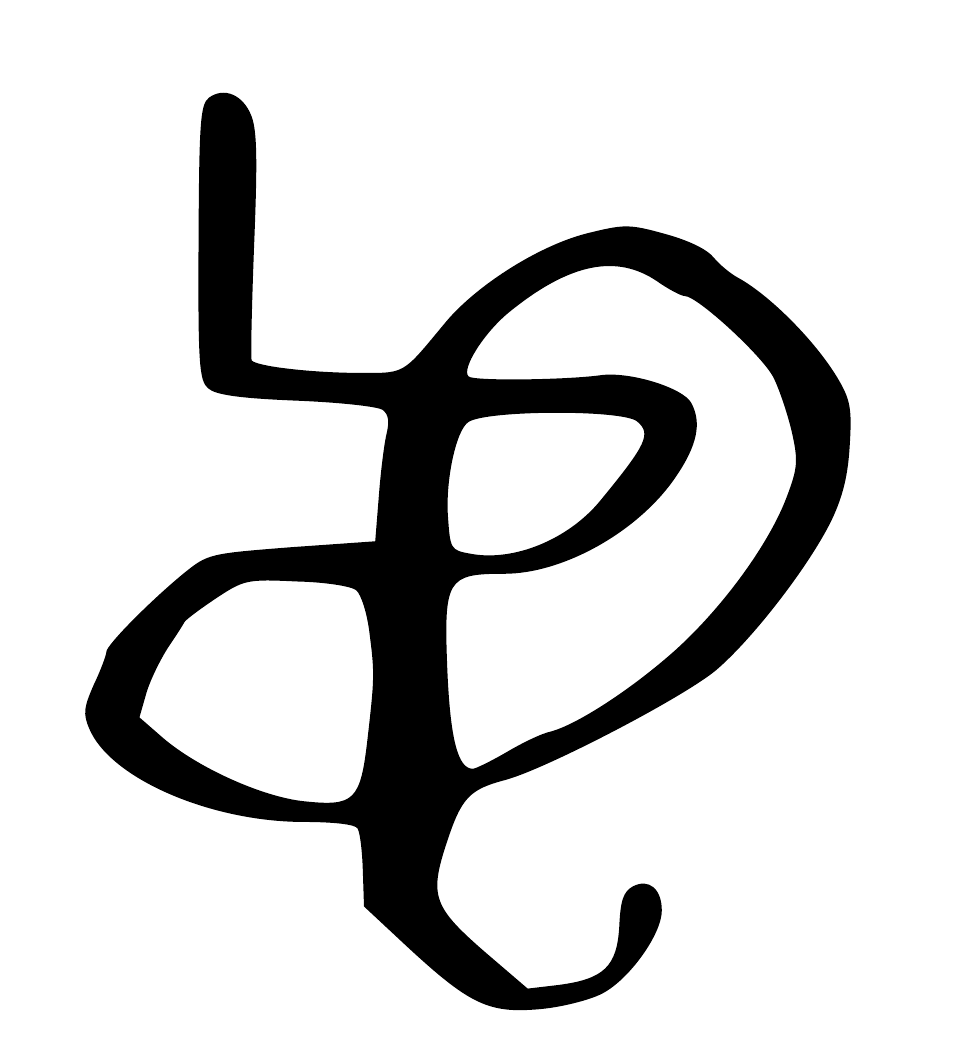} &\includegraphics[width=0.6cm]{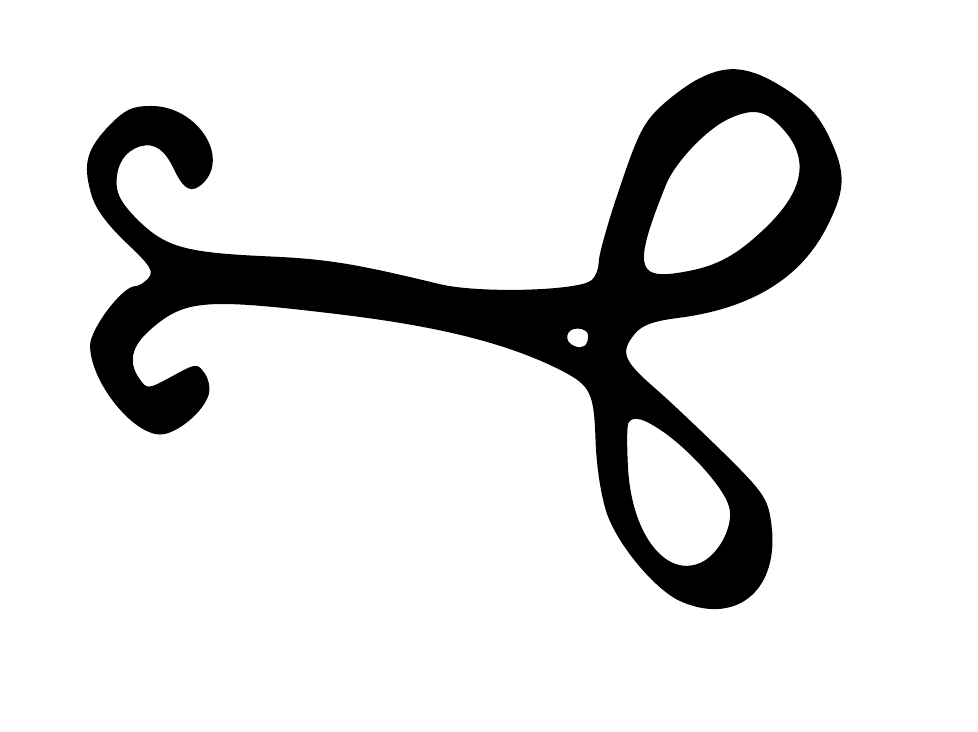} \\
Zerda &\includegraphics[width=1cm]{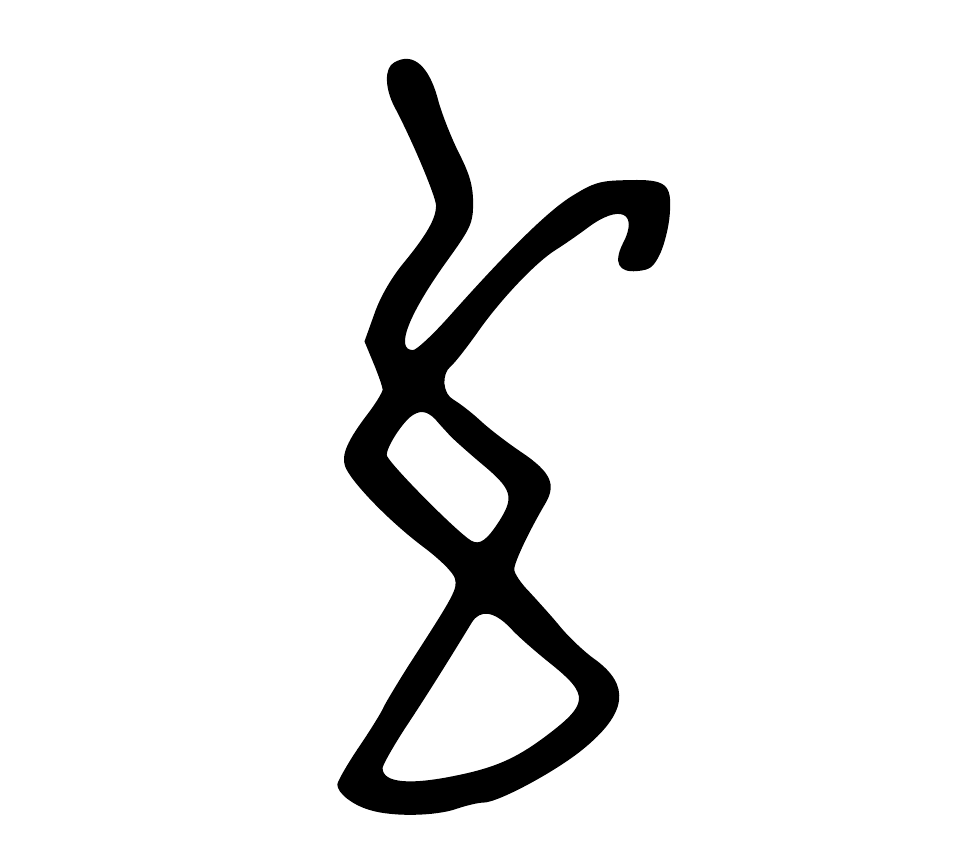} &\includegraphics[width=1cm]{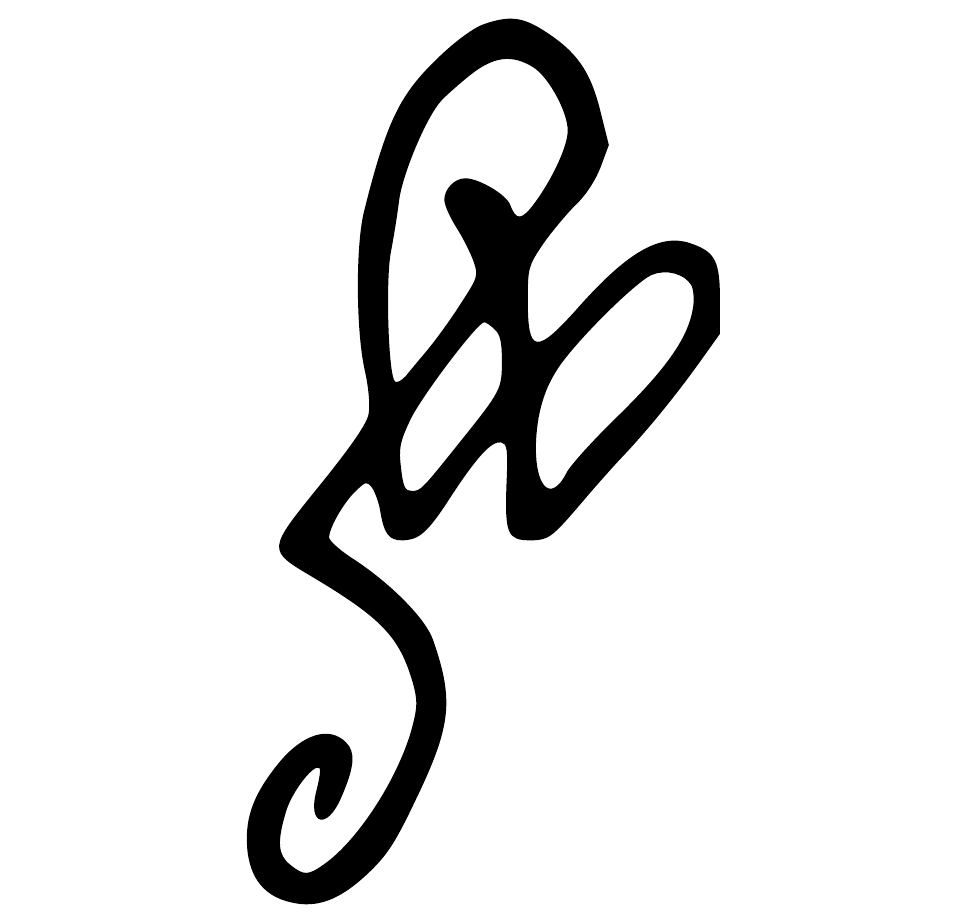} &\includegraphics[width=1cm]{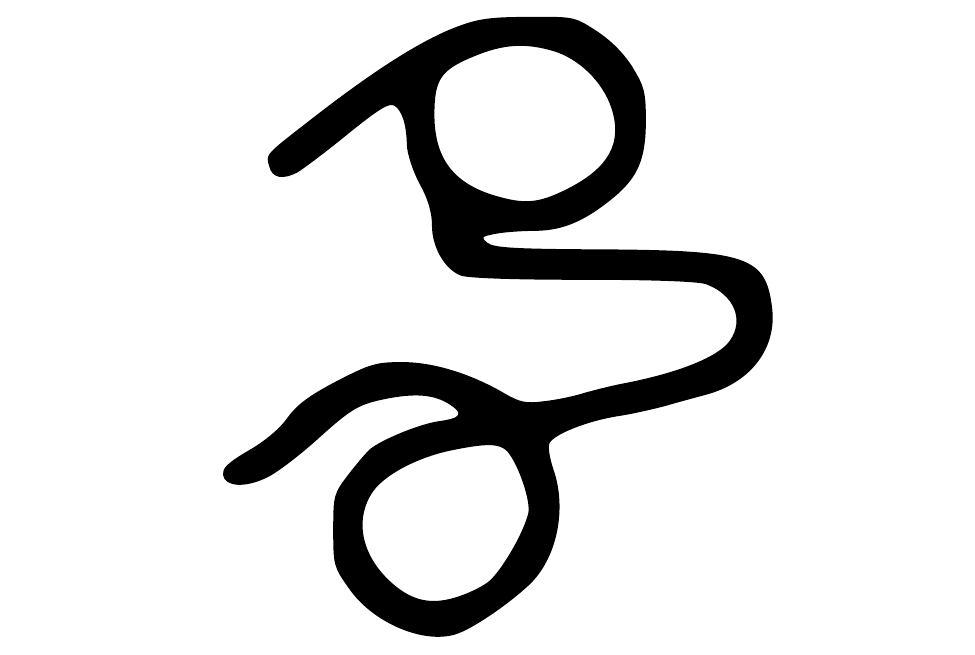} &\includegraphics[width=1cm]{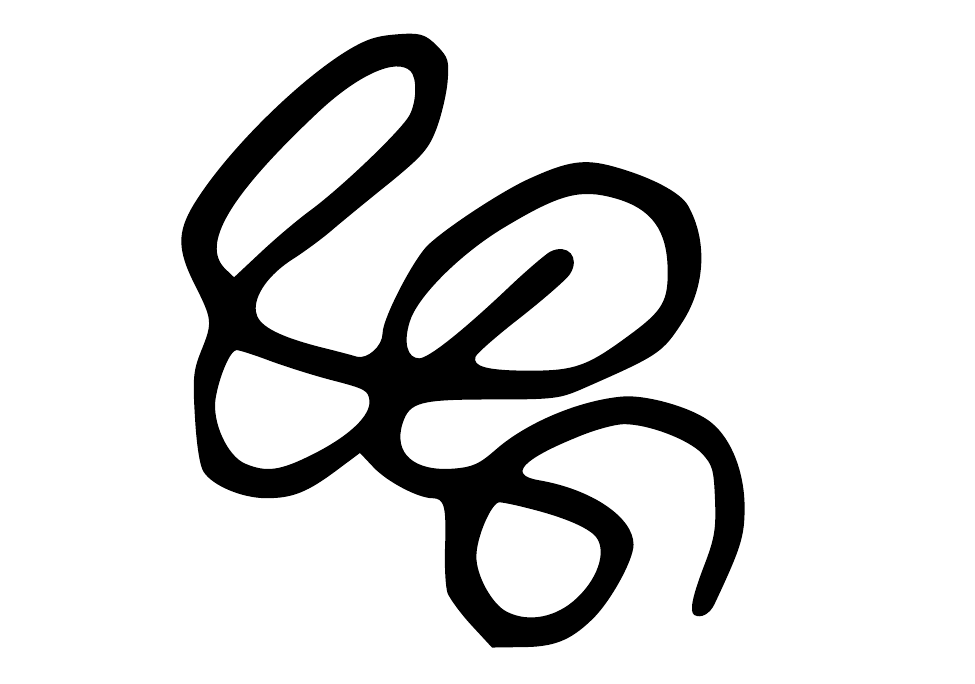} &\includegraphics[width=1cm]{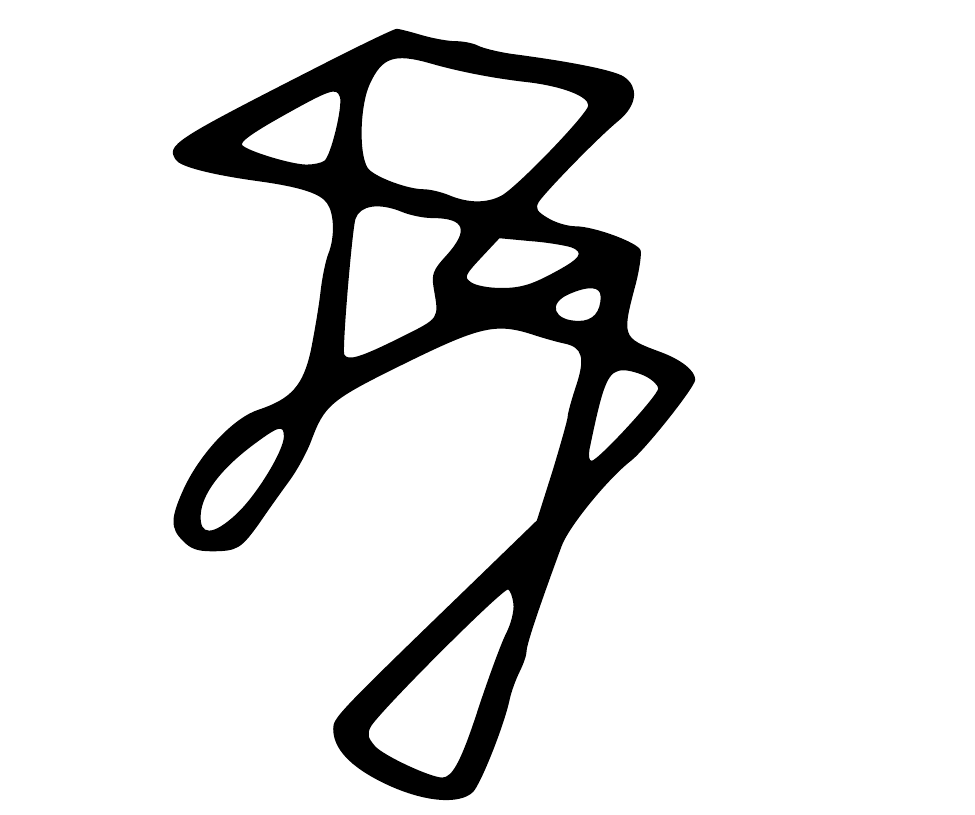} &\includegraphics[width=1cm]{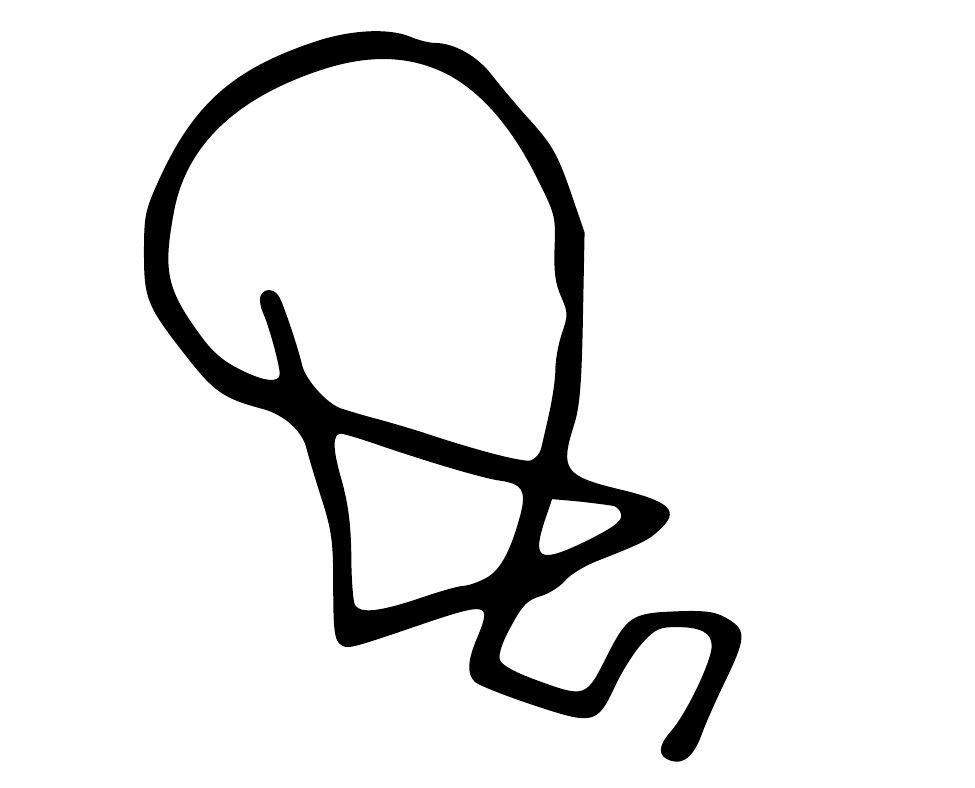} &\includegraphics[width=1cm]{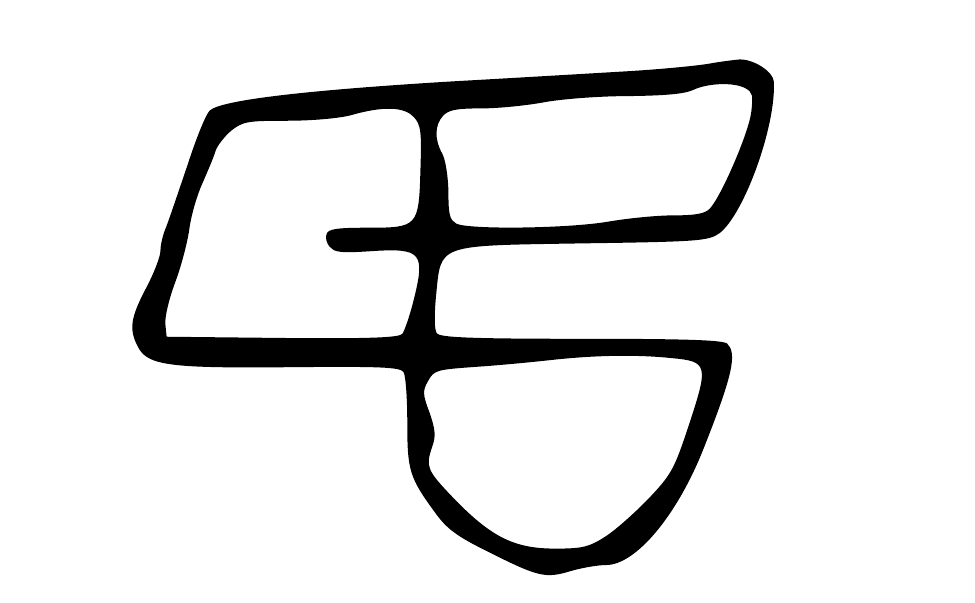} &\includegraphics[width=1cm]{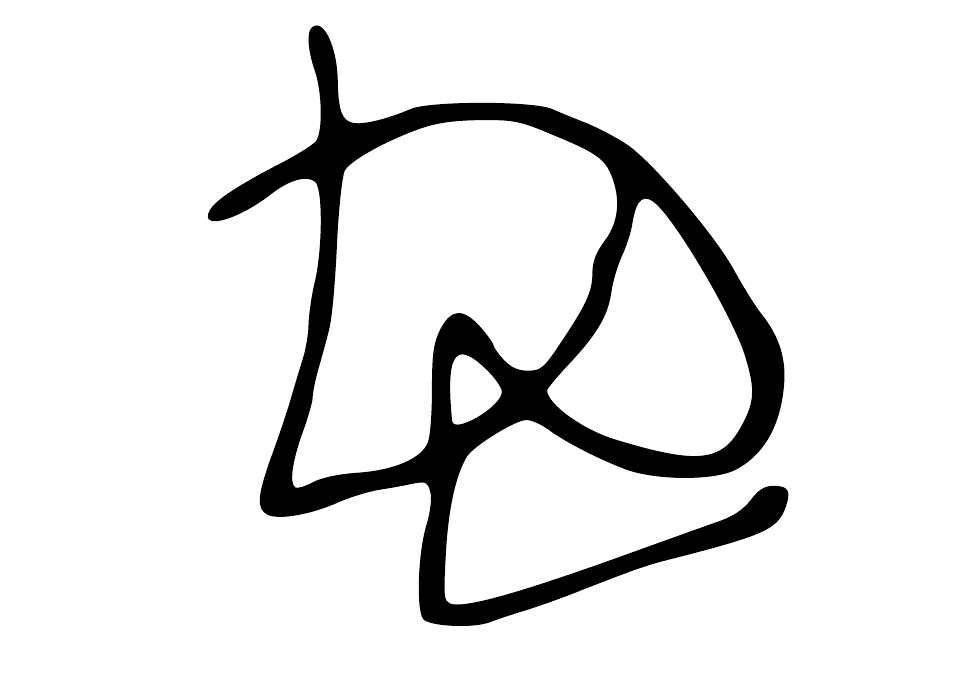} &\includegraphics[width=1cm]{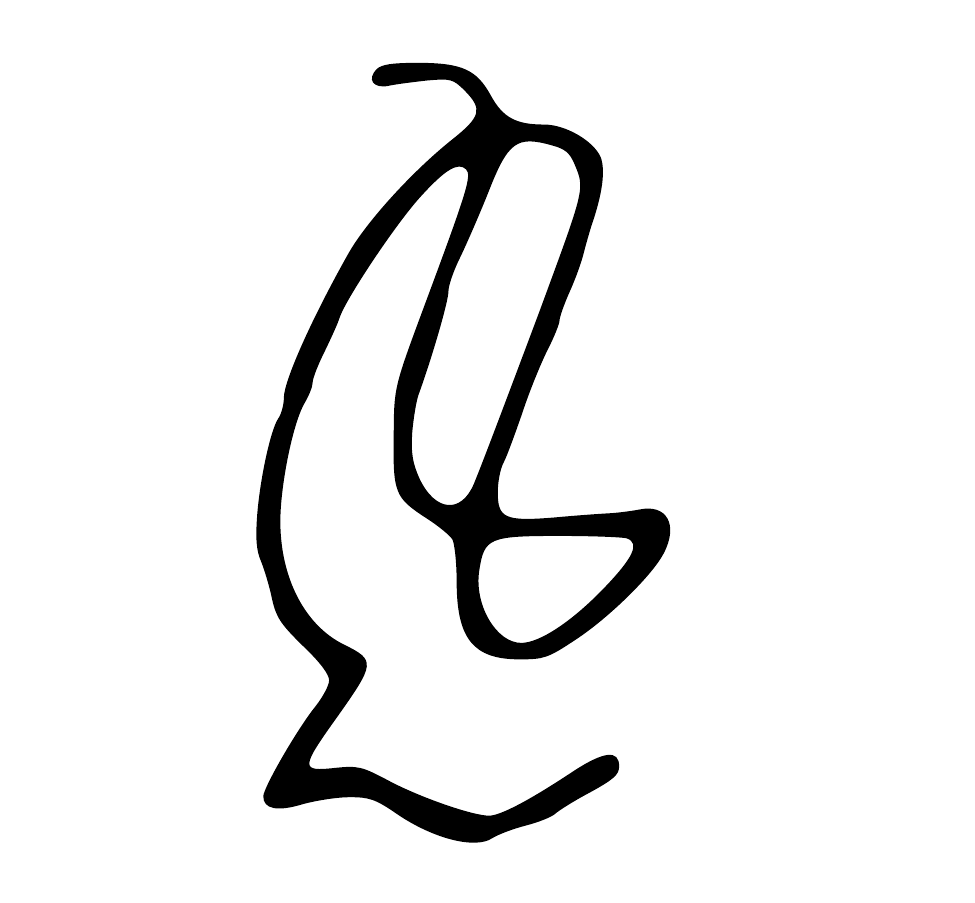} &\includegraphics[width=1cm]{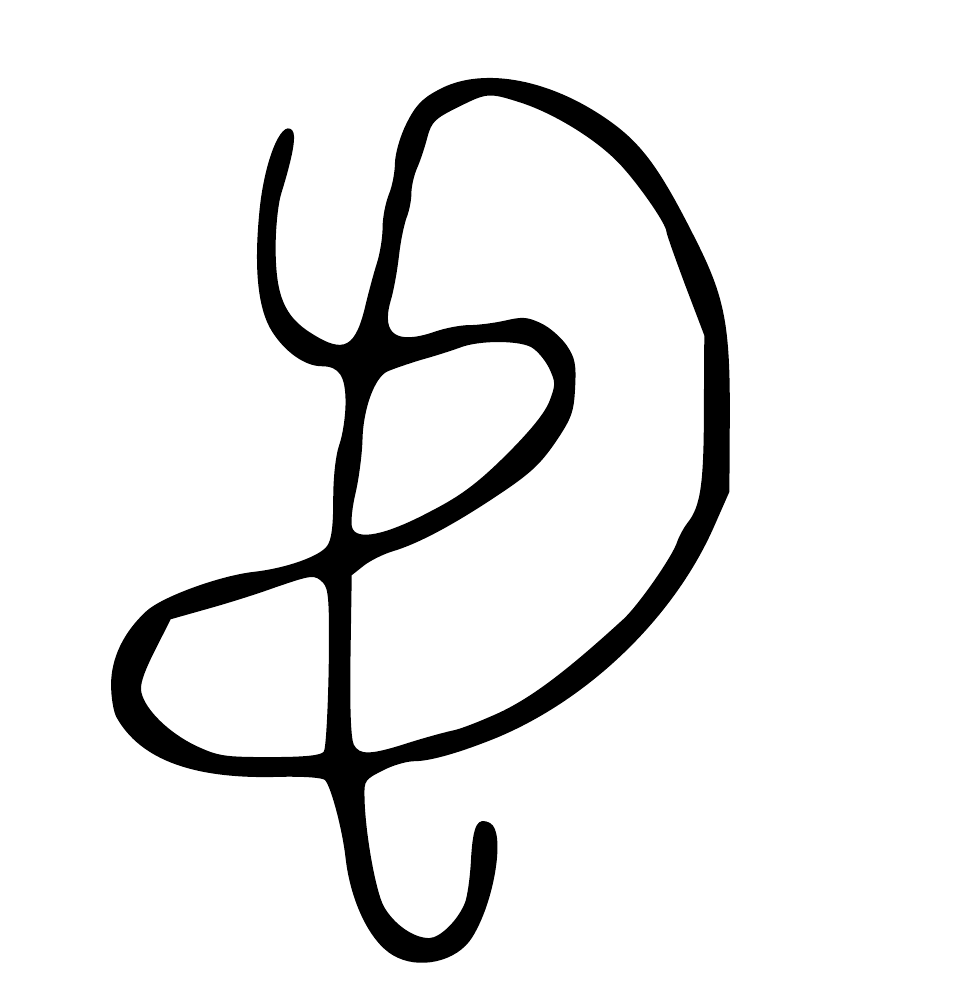} &\includegraphics[width=1cm]{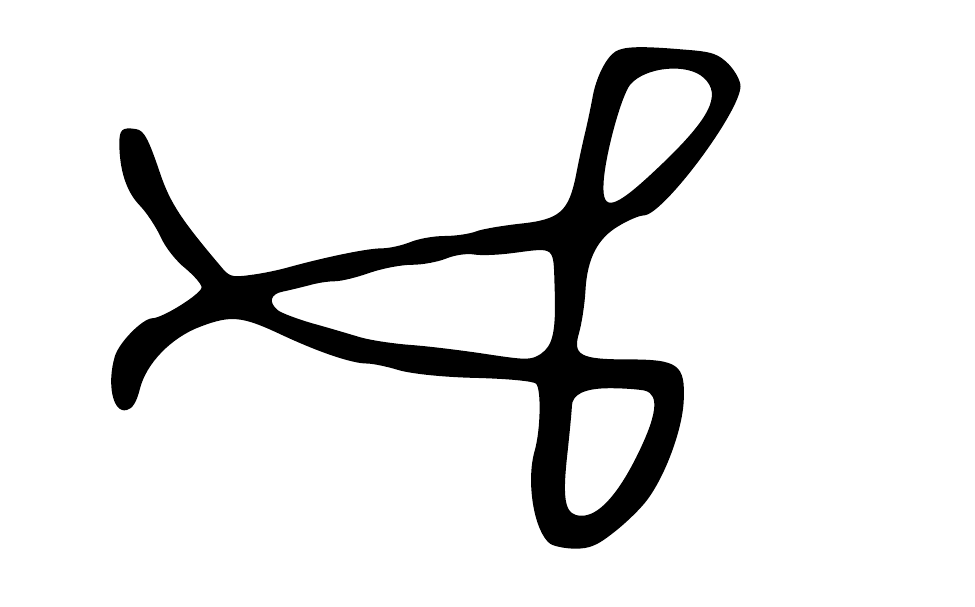}\\\hline
& 1&2&3&4&5&6&7&8&9&10&20\\
 \end{tabular}
\caption[Muisca `written ciphers' according Duquesne.]{Muisca `written ciphers' according Duquesne.  The version of the symbols as published by \cite{Acosta1848}, \cite{Humboldt1878} and \cite{Zerda1882} are shown}
\label{muisca-guarisms}
\end{center}
\end{figure}

In his description of the Muisca numbers, father Duquesne says that they used written signs to denote them, and gives a plate showing the graphic symbols for each name and number. Figure \ref{muisca-guarisms} shows such `ciphers', as they were published by \cite{Acosta1848}, \cite{Zerda1882} and \cite{Humboldt1878}, and the slight differences between them can be explained by the style of the different  artists who prepared the plates for printing. Seeing them together helps us get an idea of the  symbols in the original Duquesne manuscript. 

A problem with these graphic symbols is that they are practically nonexistent in any of the material expressions of this culture, for example  pottery, textile, rock art, gold and stone artifact decoration. Also, several chroniclers emphatically affirm that the Muiscas lacked a writing system\citep{Lugo1619,Simon1625}. These facts have been used as critical arguments against Duquene's credibility  \citep{Restrepo1892}.   

Nevertheless, to state that such `cyphers' are just an invention of the clergyman is just an easy way out, so I consider it plausible that  such symbols were given to the clergyman by his indigenous informants. However, considering that the numerical quantities represented in some some archaeological artifacts of this culture  (i.e. the Choachi stone, see section \ref{choachi-section})
the Muiscas may have opted for a graphically simpler way to represent numbers than the intricate \textit{graffitis} given by Duquesne. How can be explained such symbols? What did the indigenous informant mean when he drew such symbols for the inquiring priest? Perhaps the priest understood as cyphers some graphic symbols that represented a different thing than numbers, but related to them in some way.

\begin{table}[h]
\centering
\begin{tabular}{rlm{10cm}}
\multicolumn{2}{c}{\textbf{Number name}} & \multicolumn{1}{c}{\textbf{Duquesne's description and/or etymologies}}\\\hline
 1 & \textit{Ata} &A toad in the act of jumping, which characterizes the start of the year.\\\hline
 2 & \textit{Bosa} &The nose and the nostrils.\\\hline
3 & \textit{Mica} &Two opened eyes and the nose. To look for, to find, to choose small things.\\\hline
 4 & \textit{Muihica} &Two closed eyes. Home stone, black thing, to grow.\\\hline
 5 & \textit{Hisca} &Union of two figures: fecundity symbol. Green thing. Happiness, to lie one over another, medicine, to enjoy.\\\hline
 6 & \textit{Ta} &The stick and the cord: with it they formed the circle of their houses and their sowing fields. Sowing, harvest.\\\hline
7 & \textit{Cuhupcua} &The two ears covered. Deaf person. The figure of a snail or an ear.\\\hline
 8 & \textit{Suhusa} &The stick and the cord. Tail. Do not pull of another thing. To spread.\\\hline
9 & \textit{Aca} &  A toad from which tail begins to form another toad. The goods.\\\hline
 10 & \textit{Ubchihica} &Ear, it means the moon phases. Shining moon, painted house, to paint.\\\hline
20 & \textit{Gueta} & A lying or stretched toad . House and sowing field, to touch.\\\hline
\end{tabular} 
\caption{Duquesne's description and etymologies of the Muisca numbers.}
\label{number-meanings}
\end{table}

From analyzing Duquesne's interpretation about the meaning of the etymology of the names of number, we came to a possible interpretation. Table \ref{number-meanings} shows a summary of such an analysis. Those definitions were received with distrust by other scholars, who considered them vague and without sense in the context of numbering: \textit{``...but it is not possible to admit that when the ignorant man feels the basic need to count, he would name four to a black thing (muyhica), six to the sowing (ta) and twenty to the home (gueta)''} \citep{Humboldt1878}.  Such apparent lack of sense could be conciliated if those descriptions were not seen as \textit{meanings for numbers}, but as \textit{descriptions of celestial asterisms} instead, which at the same time could have with the same names as the numbers. In this alternative context, it is more convincing to speak about shapes that resembles toads, human face features, and agricultural tools. Note also that  some remarkable `black things' in the sky are  dark zones of the Milky Way, as seen in the southern hemisphere, which  were considered asterisms by Native American cultures of the Andes \citep{Urton1981,Urton1981a}. 

\begin{figure}
 \centering
 \includegraphics[width=9cm,bb=0 0 948 649]{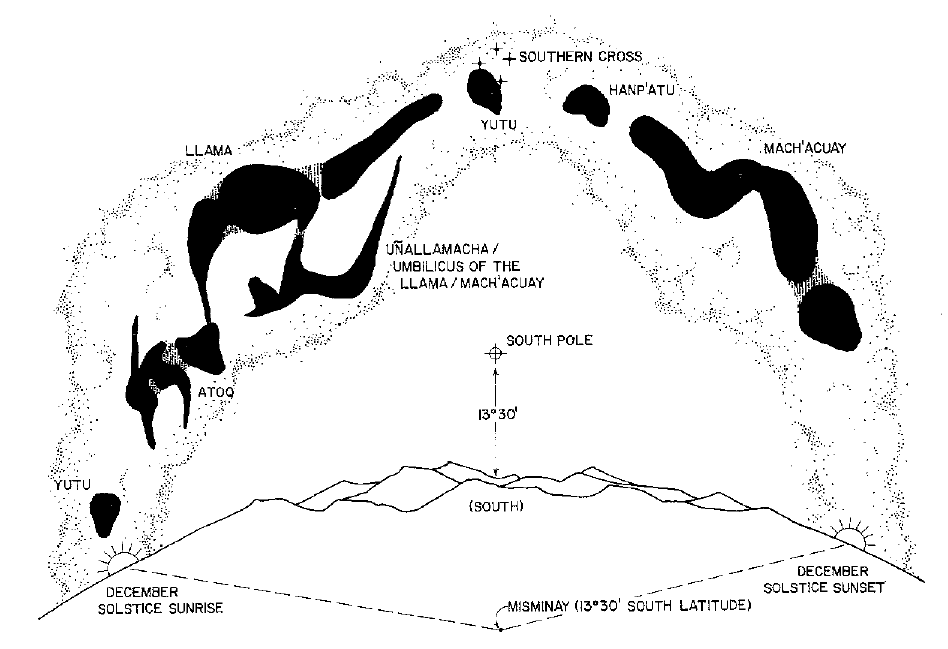}
 % quechua-asterisms.png: 948x649 pixel, 72dpi, 33.44x22.89 cm, bb=0 0 948 649
 \caption{Dark zones of the Milky Way,  seen as constellations by current Andean communities. \cite[p.171]{Urton1981a} }
 \label{black-asterisms}
\end{figure}

Although this interpretation is not directly argued by Duquesne,  his commentary about the possible constellations the Muiscas valued suggests it.  In it he associates the name for one \textit{(Ata)} to the Pisces constellation and the name of five \textit{(Hisca)} to Gemini. Additionally, he provides information about an asterism dedicated to the Cacique Tomagata, a monstrous  mythical personage that  had only an eye, four ears and a long tail, which  Duquesne supposes is the memory of an ancient sighting of a comet: 

\begin{quote}
...his name Tomagata, means fire that burns. [The Muisca] located in the astrological sky this horrifying comet, and I, according to the circumstances of his story, guessed that they will classify him rather as eunuch of the Gleaner Virgin than as Sagittarius' partner\footnote{``...Su nombre Tomagata, significa fuego que hierve. Ellos pasaron al cielo astrológico este espantoso cometa, y yo, según las circunstancias de su historia, creo que le señalarían mas bien por eunuco de la Virgen Espigadora que por compañero de Sagitario...'' (see \textit{Dissertation}, p.\pageref{eunuco-virgen})}.
\end{quote} 

\noindent Which probably relates to the sky area occupied by the western Scorpius constellation.   The association of the name of the numbers 6 \textit{(Ta)} and 8 \textit{(Suhusa)} as a stick and a cord used to trace circles, strongly emphasized by Duquesne as the basis for their `numbers', could be related to the daily observation of the circular translation of the stars around the sky. At the equatorial latitude of central Colombia, the imaginary stick for such movement would correspond to the geographical north and south cardinal points.

Consequently, an hypothetical explanation for these \textit{graffitis} is to consider that instead of being `written' ciphers, they are just stylized drawings of celestial asterisms associated to the number names, as seen in the sky by the informer. This could be considered a \textit{serious speculation},  claiming for further ethnohistorical and ethnographical research in the next future.

\subsection{The Day and the Week}\label{dayweek}

The Muisca considered daylight time and the night as different entities. In his Noticias Historiales, chronicler Fray Pedro Sim\'on \citeyearpar{Simon1625} tell us that \textit{``the days were counted by suns, because it was the cause of them, thus so many suns were so many days; these were divided in three parts, morning, noon, and afternoon''}\footnote{``...los días contaban por soles, viendo que él era la causa de allos, de manera que tantos soles eran tantos días; estos distinguían en sólo tres partes, mañana, medio y tarde...''\citep[p306]{Simon1625}}. According Duquesne, \textit{``...the artificial day was called} \textrm{sua}, \textit{that is, a sun measured from dawn to sunset''}, the daylight span apparently was divided in two parts, one from dawn until noon and another from the noon to sunset. Duquesne gives the terms \textit{suamena} and \textit{suameca} for these spans, which is agreed by Ezequiel Uricoechea's grammar and vocabulary \citep{Uricoechea1871}, however, the 17th century \textit{Diccionario y gramática Chibcha} by anonymous author, indicates that these terms are  synonyms for the `afternoon' span, giving for the `morning' the alternative term $Zacoca$  \citep{Gonzalez1987}. All the sources agree that the word for `night' entity was \textit{Za}, and similarly to the day, it was divided in two: \textit{Zasca} corresponded to the span from sundown to midnight and \textit{Cag\"ui}, from midninght to dawn. \textit{Cag\"ui}, also means `morning star' \cite{Gonzalez1987}, which possibly relates such part of the day with the sighting of bright planets, especially Venus, during the early morning. 

Apparently, the Muisca lacked of a concept of week, at least as we understand it. The descriptions provided by different chorniclers are different, for example in the \textit{Epítome de la conquista del Nuevo Reino de Granada} the Adelantado Gonzalo Jiménez de Quezada is described
\begin{quote}
...They have distributed the times of months and years in a very convenient way. The first ten days of the month, they eat an herb known in the sea coast as hayo, that supports and  purges them their unwellness.  After these days, cleaned of the hayo, they work another ten days in their sows and lands.  And the another remaining ten days of the month, are spent in their homes, talking and resting with their women.\footnote{``...tienen repartidos los tiempos de meses y año, mui al propósito: los diez días primeros del mes, comen una yerba que en la costa de la mar llaman hayo, que los sustenta mucho y les hace purgar sus yndisposiciones. Al cabo de estos días, limpios ya del hayo, tratan otros días en sus labranzas y haciendas, y los otros diez que quedan del mes, los gastan en sus casas, en conversar con sus mujeres y en holgarse con ellas'' \citep{Benavides2001}}
\end{quote}

According this, the 30 day-month would be divided in three `weeks' of ten days each. Sim\'on, however, gave us a different version of the week:

\begin{quote}
They counted the months by moons with its wanings and crescents, dividing each one of these in another two, such that they were four parts of the month or moon, in the same manner we divide it by four weeks.\footnote{``...los meses contaban por lunas con sus menguantes y crecientes, dividiendo cada una de éstas dos en otras dos, con que venían hacer cuatro partes del mes, ó la luna al modo que nosotros lo dividimos por cuatro semanas... \citep[p306]{Simon1625}''}
\end{quote}

The situation is even confuse when in his work, Duquesne affirms that \textit{``the week was of three days, and it is known about its use because each three days they used to make in Turmequ\'e, place belonging to Guatavita, a market.''}\footnote{La semana era de tres días, y se conoce que usaban de ella porque cada tres días hacían en Turmequé, lugar perteneciente al Guatavita, un mercado.}. It can be considered that such different versions are product to the intention to match the european idea of the week to the diverse times associated to rutinary activities of society, which certainly were not regulated by a standarized timespan between day and month. The \textit{Diccionario y gram\'atica Chibcha}, gives us some clues in this direction in its entries for sentences as \textit{Fasinga domingoca}: `the next Sunday' , \textit{Mi\'ercolesca hunga}: `The Wednesday he will come' , \textit{Viernesc inanga}: `I will go this friday', where the Muisca translations for the names of days of the week appeal to the corresponding spanish words (Domingo, Mi\'ercoles, Viernes) integrated into the native grammar, as was usual when the described concept was alien to the Muisca language. As it will be discussed in the next sections, the levels of resolution of the Muisca timekeeping system can be considered as days, months, years and centuries.

\subsection{The Lunar Cycle}\label{lunarcycle}
As  already been said, both the \textit{Astronomical Ring} and the \textit{Dissertation} discuss the arrangement of the lunar cycle as a set of formulas designed to find the time corresponding to the lunar phases. These formulas are based on counts of days done with the fingers. We read in the \textit{Astronomical Ring}: 

\begin{quote}
... The signs that have human shapes represent the different aspects of the moon, that successively vary in each month. 
\textit{Mica} and \textit{Muihica} appear on the syzygies; \textit{Mica}, in the open eyes, the full moon and the opposition; \textit{Muihica} in the closed eyes, the conjunction or new Moon. 
\textit{Bosa} and \textit{Cuhupcua}, the first phases of the Moon, and \textit{Ubchihica} the quadratures... 

...\textit{Cuhupcua} is the neomemia\footnote{The time of the new moon.}...  this is one a common and universally well-known first phase;  as the new moon is brief in some months and in others is more delayed, it was designated as the fourth day so that there were differences, unless we want \textit{Bosa} representing the neonemia of the brief months and \textit{Cuhupcua} the others. 
Counting from \textit{Muihica} they found  the number 7 the first quadrature, in \textit{Ubchihica} after the  neonemia; and counting from \textit{Mica} they found in the number 8  the second one after the opposition, on the same sign...
Finally, \textit{Bosa} gave them the last phase of the moon placed in the third finger before \textit{Muihica}\footnote{``...Los signos que tienen facciones humanas representan los diferentes 
aspectos de la luna, que sucesivamente se var\'{\i}an en cada mes.\\
\textit{Mica} y \textit{Muihica} figuran las dos Zisigias; \textit{Mica}, en 
los ojos abiertos, el plenilunio y la oposici\'{o}n; \textit{Muihica} 
en los ojos cerrados, la conjunci\'{o}n o girante.\\
\textit{Bosa} y \textit{Cuhupcua}, las primeras fases de la luna, y
\textit{Ubchihica} las cuadraturas...\\
(...) \textit{Cuhupcua} es la neonemia... es una primera fase vulgar 
y universalmente conocida; y como el novilunio en unos meses 
es m\'{a}s breve, en otros m\'{a}s tard\'{\i}o, se coloc\'{o} en el d\'{\i}a 
cuarto para que hubiese diferencia, a no ser que queramos que 
\textit{Bosa} represente la neonemia de los unos y \textit{Cuhupcua} la 
de los otros.\\
Contando desde \textit{Muihica} hallaban al n\'{u}mero 7 la primera 
cuadratura, en \textit{Ubchihica} despu\'{e}s de la neonemia ; y contando 
desde \textit{Mica} encontraban al n\'{u}mero 8 la segunda despu\'{e}s 
de la oposici\'{o}n, en el mismo signo...\\
Finalmente \textit{Bosa} les daba la \'{u}ltima fase de la luna colocada 
al tercer dedo antes de \textit{Muihica}''. \citep[p216]{Duquesne1882}}. 
\end{quote}

\begin{figure}
 \centering
 \includegraphics[width=8cm]{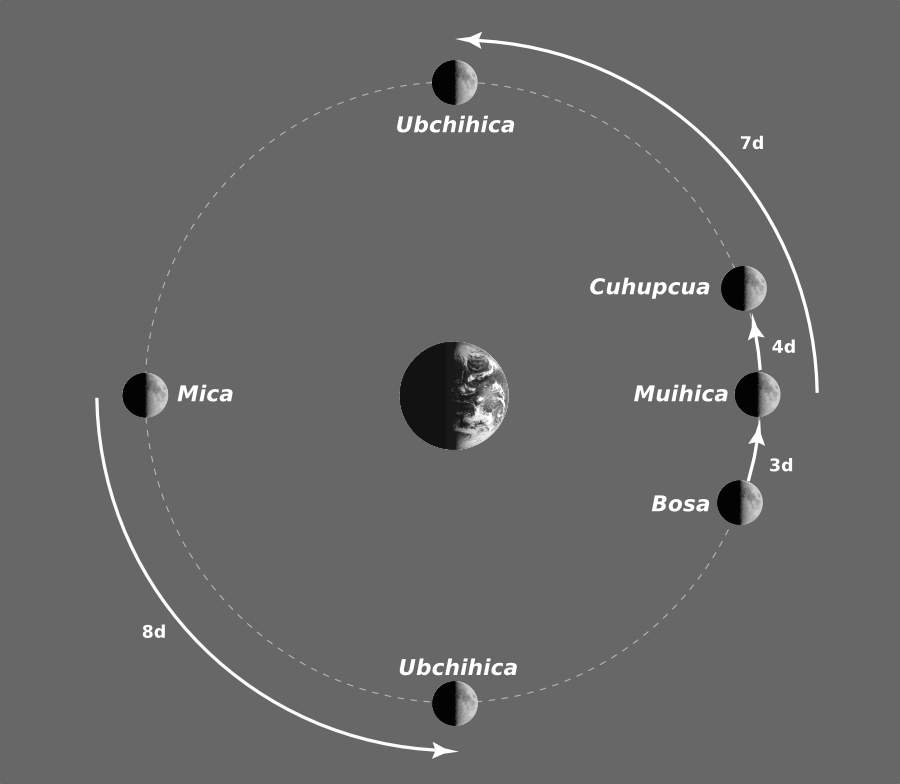}
 % lunas-muiscas-anillo.pdf: 1179666x1179666 pixel, 0dpi, infxinf cm, bb=
 \caption{Lunar phases according the \textit{Astronomical Ring}}
 \label{lunas-anillo}
\end{figure}

The \textit{Astronomical Ring} does not clearly to explain why, in order to determine moments in  time, names of numbers are thus used. An apparent argument should be the etymological relations of the names of numbers and the shape of their lunar phases. Thus, it could be supposed that  the  names are used as descriptive labels for the moments of the lunar cycle. In figure \ref{lunas-anillo}, we can appreciate a scheme of the lunar orbit where these names are located according to the position of the satellite around the Earth. We see that \textit{Ubchihica} corresponds to the quadratures (waxing and waning), \textit{Mica} marks the Full Moon, \textit{Muihica} the New Moon, \textit{Bosa} and \textit{Cuhupcua} the heliacal rising and settings  (neonemies) of our satellite. 

It is however, surprising to find  a radically different description in the \textit{Dissertation}:

\begin{quote}
 ...they commenced to count the month from  the opposition or full-moon in the sign of \textit{Ubchihica}, which signified brilliant moon: counting seven days on the fingers, beginning at \textit{Ata}, which follows \textit{Ubchihica}, they found the quadranture in \textit{Cuhupcua}; counting from this, seven, they found the next immersion of the Moon in \textit{Muyhica}, which meant anything black; and the next day, the conjunction symbolized in \textit{Hisca}, (...), then counting eight days, they found the other quadrature in \textit{Mica}, which meant a changing object, (...), The first aspect of the first phase was in \textit{Cuhupcua}, and as  this symbol fell during the quadrature, they gave it two ears, calling it  deaf for reasons of superstition...\footnote{``...comenzaban \'{a} contar el mes desde la oposici\'{o}n, \'{o} plenilunio 
figurado en \textit{Ubchihica}, que significa luna brillante; contando 
siete dias en los dedos comenzando por \textit{Ata}, que se sigue 
\'{a} \textit{Ubchihica}, hallaban la cuadratura en \textit{Cuhupcua}; contando 
de all\'{\i} siete encontraban la pr\'{o}xima inmersi\'{o}n de la luna 
en \textit{Muyhica}, que significa cosa negra, y al dia siguiente 
la conjunci\'{o}n simbolizada en Hisca, (...), contando despu\'{e}s 
ocho dias hallaban la otra cuadratura en Mica, que significa 
cosa varia,(...) El primer aspecto de la primera faz la se\~{n}alan 
en Cuhupcua, y como en este s\'{\i}mbolo caia la cuadratura le 
daban dos orejas, y le llamaban sordo por otros motivos de superstici\'{o}n''. \citep{Duquesne1848}}
\end{quote} 

\begin{figure}
 \centering
 \includegraphics[width=8cm]{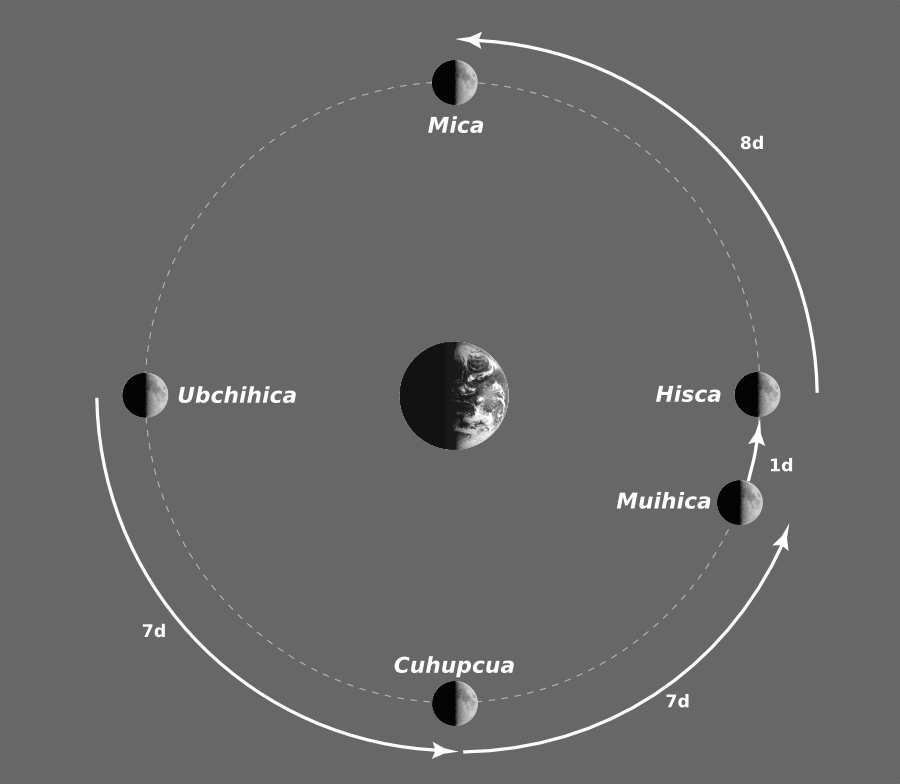}
 % lunas-muiscas-disertacion.pdf: 1179666x1179666 pixel, 0dpi, infxinf cm, bb=
 \caption{Lunar phases according the \textit{Dissertation}}
 \label{lunas-disertacion}
\end{figure}

How come Duquesne contradicts himself as much as he does in his two texts? Can it be attributed to   transcription error by Acosta or Zerda? It does not seem possible; both authors are emphatic in their desire to publish  faithfully  Duquesne's writings, aiming to contribute to future research. Perhaps  the \textit{Astronomical Ring} is imprecise, it being a rough draft. At at first glance it could give this impression. Nevertheless, a detailed analysis reveals that indeed both texts are congruent and complementary to each other. 

Two reasons explain the confusion: first, the assumption made in the \textit{Astronomical Ring} that considers the use of names of numbers as mere labels is erroneous. The \textit{Dissertation} shows that the use of those names  really obeys  a day-to-day rule, when it explains that the lunar month begins ``\textit{...by Ata,} which follows \textit{Ubchihica...}'', that is, a day numbered as 1 (\textit{Ata}), follows after another one numbered as 10 (\textit{Ubchihica}). Consequently, if such a name is really pointing to the ending day of the previous lunar cycle, it would be referring not to the 10, but to the number 30, which is expressed as `\textit{Guetas asaq{\iM}} ubchihica'. The lack of care that Duquesne gives to the rigorous use of prefixes for the amounts greater than ten, results in the second cause of confusion, leading him to put a same number name to different moments of the lunar cycle. The dates 5, 15, and 25 of a certain month, for example, are simply referred to as days whose name is ``Five''. 

Let us check again the texts, step by step, to locate the given symbols in a sequence of thirty days. Let us begin by the \textit{Dissertation}:

\begin{itemize}
\item ``\textit{...they commenced to count the month from the opposition or full-moon in the sign of Ubchihica}''. As mentioned already, it apparently assigns the beginning of the count to the last day of the previous cycle, [Guetas asaq{\iM}] Ubchihica (30). 

\item ``\textit{...counting seven days on the fingers, beginning at Ata, which follows Ubchihica, they found the quadranture in Cuhupcua}'' Here it is clear that it is a wheel of 30 days, where \textit{Ata} (1)  follows \textit{[Gueta asaq{\iM}] Ubchihica} (30). Obviously, \textit{Cuhupcua} is the seventh day of the cycle. However, a peculiar and important detail must be noted: in the case of the \textit{Dissertation}, Duquesne always uses \textit{the next day} to mean the day taken as origin. In this case, the origin is \textit{[Gueta asaq{\iM}] Ubchihica}, but he counts from the next day, Ata. 
\item ``\textit{...counting from this, seven, they found the next immersion of the Moon in \textit{Muyhica}}'' That is, counting from the seventh day \textit{(Cuhupcua)} there would be the heliacal lunar set at the 14 day,  which is \textit{[Quihicha] Muyhica}. 

\item ``\textit{...and the next day, the conjunction symbolized in Hisca}'' that is, the 15 day, \textit{[Quihicha] Hisca}.

\item ``\textit{...then counting eight days, they found the other quadrature in Mica}'' that is, the 23 day, \textit{[Guetas asaq{\iM}] Mica}.

\item ``\textit{...The first aspect of the first phase was in \textit{Cuhupcua}, and as  this symbol fell during the quadrature, they gave it two ears, calling it  deaf for reasons of superstition...}'' This  \textit{Cuhupcua} refers to the 27 day , \textit{[Guetas asaqy] Cuhupcua}. Here the  text is a bit confusing when it mentions a quadrature in a moment corresponding to a full Moon;  Duquesne actually refers to the 7 day, (Cuhupcua), which etymologically matches the date of the full Moon.
\end{itemize}

Let us now examine  the \textit{Astronomical Ring}. It is important to notice that in this case, unlike in the \textit{Dissertation}, Duquesne now counts from the origin number. It is not clear why Duquesne did so in this case.

\begin{itemize}
 \item ``\textit{...\textit{Mica} and \textit{Muihica} appear on the syzygies; \textit{Mica}, in the open eyes, the full moon and the opposition; \textit{Muihica} in the closed eyes, the conjunction or new Moon.}'' That is, the 3rd day, \textit{Mica}, and the 14th day \textit{[Quihicha] Muihica}.

\item ``\textit{...\textit{Bosa} and \textit{Cuhupcua}, the first phases of the Moon (...) \textit{Cuhupcua} is the neonemia (...) is  a common and universally well-known first phase; as the new moon is brief in some months, and in others is more delayed, it was designated as the fourth day so that there were difference, unless we want \textit{Bosa} representing the neonemia of the brief months and \textit{Cuhupcua} the one of the others (...) Finally, \textit{Bosa} gave them the last phase of the Moon placed in the third finger before \textit{Muihica}}'' It refers to the 12th day, \textit{[Quihicha] Bosa}, and the 17th day, \textit{[Quihicha] Cuhupcua}. The day \textit{[Quihicha] Bosa} points to the heliacal lunar setting,  according Duquesne occuring three days before the conjunction, whereas \textit{[Quihicha] Cuhupcua} gives the heliacal lunar rise, four days after the conjunction, which he places on the 14th day, \textit{[Quihicha] Muihica}.

\item ``\textit{...Counting from \textit{Muihica} they found in the number 7 the first quadrature, in \textit{Ubchihica} after the  neonemia}'' It refers to the 14th day, \textit{[Quihicha] Muihica}, the supposed new moon, and from it they counted seven days, arriving to the 20th day, \textit{[Quihicha] Ubchihica}. It is peculiar that Duquesne describes this quadrature as the first one, leading one to think that the month started in the new moon. However, in  the \textit{Dissertation}, the quadratures  are described in an inverse order, and despite of this it is not said which one is the first or the second. In that version, it is implicit that the month began during the full moon.

\item ``\textit{and counting from \textit{Mica} they found in the number 8 the second one in same sign, after the opposition...}'' From the 3th day, \textit{Mica}, they counted eight days in order to arrive to the 10th day, \textit{Ubchihica}, that Du	quesne would designate as ``the second'' conjunction.

\end{itemize}

Let us place in a table the data provided by Duquesne, in order to compare both versions more easily. A table of thirty rows can be used, each one assigned to a day in the sequence from 1 to 30:

For ease of comparison, let us display Duquesne's data in a table. Each of the thirty rows in table \ref{ARvsD} represents one day in a sequence from 1 to 30.

{\footnotesize
\begin{longtable}{|r|l|l|p{1.3in}|p{1.3in}|}
\caption{Comparison between the lunar cycle times according to the \textit{Dissertation} and the \textit{Astronomical Ring}.}\label{ARvsD}\\
\hline
% ROW 1
{\raggedleft \textrm{{ \#Day}}} &
{\raggedright \textrm{{ Prefix}}} &
{\raggedright \textrm{{ Name}}} &
{\centering \textrm{{ Dissertation}}} &
{\centering \textrm{{ Astronomical Ring}}}\\
\hline
\endfirsthead
\hline
{\raggedleft \textrm{{ \#Day}}} &
{\raggedright \textrm{{ Prefix}}} &
{\raggedright \textrm{{ Name}}} &
{\centering \textrm{{ Dissertation}}} &
{\centering \textrm{{ Astronomical Ring}}}\\
\hline
\endhead
% ROW 2
{\raggedleft \textrm{{ 30}}} & 
{\raggedright { Guetas asaq{\iM}}} & 
{\raggedright { Ubchihica}} & 
{\raggedright \textrm{ Opposition:} \textit{``...they commenced to count the month from the opposition or full-moon in the sign of Ubchihica...''}} & {\raggedright }\\
\hline
% ROW 3
{\raggedleft \textrm{{ 1}}} & 
{\raggedright { ---}} & 
{\raggedright { Ata}} & 
{\raggedright \textit{``...beginning at \textit{Ata}, which follows \textit{Ubchihica}...''} } & 
{\raggedright }\\
\hline
% ROW 4
{\raggedleft \textrm{{ 2}}} & 
{\raggedright { ---}} & 
{\raggedright { Bosa}} & 
{\raggedright } & 
{\raggedright }\\
\hline
% ROW 5
{\raggedleft \textrm{{ 3}}} & 
{\raggedright { ---}} & 
{\raggedright { Mica}} & 
{\raggedright } & 
{\raggedright  Opposition: \textit{``...\textit{Mica} and \textit{Muihica} appear on the syzygies; \textit{Mica}, in the open eyes, the full moon and the opposition; \textit{Muihica} in the closed eyes, the conjunction or new Moon....''} }\\
\hline
% ROW 6
{\raggedleft \textrm{{ 4}}} & 
{\raggedright { ---}} & 
{\raggedright { Muihica}} & 
{\raggedright } & 
{\raggedright }\\
\hline
% ROW 7
{\raggedleft \textrm{{ 5}}} & 
{\raggedright { ---}} & 
{\raggedright { Hisca}} & 
{\raggedright } & 
{\raggedright }\\
\hline
% ROW 8
{\raggedleft \textrm{{ 6}}} & 
{\raggedright { ---}} & 
{\raggedright { Ta}} & 
{\raggedright } & 
{\raggedright }\\
\hline
% ROW 9
{\raggedleft \textrm{{ 7}}} & 
{\raggedright { ---}} & 
{\raggedright { Cuhupcua}} & 
{\raggedright  Quadrature: \textit{``...counting seven days on the fingers, beginning at \textit{Ata}, which follows \textit{Ubchihica}, they found the quadranture in \textit{Cuhupcua}...''}} & {\raggedright }\\
\hline
% ROW 10
{\raggedleft \textrm{{ 8}}} & 
{\raggedright { ---}} & 
{\raggedright { Suhusa}} & 
{\raggedright } & 
{\raggedright }\\
\hline
% ROW 11
{\raggedleft \textrm{{ 9}}} & 
{\raggedright { ---}} & 
{\raggedright { Aca}} & 
{\raggedright } & 
{\raggedright }\\
\hline
% ROW 12
{\raggedleft \textrm{{ 10}}} & 
{\raggedright { ---}} & 
{\raggedright { Ubchihica}} & 
{\raggedright } & 
{\raggedright Quadrature: ``\textit{...and counting from \textit{Mica} (3) they found in the number 8 the second one in same sign, after the opposition...}''}\\\hline
% ROW 13
{\raggedleft \textrm{{ 11}}} & 
{\raggedright { Quihicha}} & 
{\raggedright { Ata}} & 
{\raggedright } & 
{\raggedright }\\
\hline
% ROW 14
{\raggedleft \textrm{{ 12}}} & 
{\raggedright { Quihicha}} & 
{\raggedright { Bosa}} & 
{\raggedright } & 
{\raggedright  Heliacal lunar set: \textit{``...representing (Bosa) the neonemia of the brief months (...) Finally, \textit{Bosa} gave them the last phase of the Moon placed in the third finger before \textit{Muihica}...''}}\\
\hline
% ROW 15
{\raggedleft \textrm{{ 13}}} & 
{\raggedright { Quihicha}} & 
{\raggedright { Mica}} & 
{\raggedright } & 
{\raggedright }\\
\hline
% ROW 16
{\raggedleft \textrm{{ 14}}} & 
{\raggedright { Quihicha}} & 
{\raggedright { Muihica}} & 
{\raggedright Heliacal lunar set: \textit{``...counting from this (Cuhupcua [7]), seven, they found the next immersion of the Moon in \textit{Muyhica}...''}} & 
{\raggedright Conjunction: \textit{``...\textit{Mica} and \textit{Muihica} appear on the syzygies; (...) \textit{Muihica} in the closed eyes, the conjunction or new Moon...''}}\\
\hline
% ROW 17
{\raggedleft \textrm{{ 15}}} & 
{\raggedright { Quihicha}} & 
{\raggedright { Hisca}} & 
{\raggedright Conjunction: \textit{``...and the next day, the conjunction symbolized in \textit{Hisca}...''} } & 
{\raggedright }\\
\hline
% ROW 18
{\raggedleft \textrm{{ 16}}} & 
{\raggedright { Quihicha}} & 
{\raggedright { Ta}} & 
{\raggedright } & 
{\raggedright }\\
\hline
% ROW 19
{\raggedleft \textrm{{ 17}}} & 
{\raggedright { Quihicha}} & 
{\raggedright { Cuhupcua}} & 
{\raggedright } & 
{\raggedright Heliacal lunar set: ``...\textit{Cuhupcua is the neonemia} [that is] \textit{the one of the others} [brief months]...''}\\
\hline
% ROW 20
{\raggedleft \textrm{{ 18}}} & 
{\raggedright { Quihicha}} & 
{\raggedright { Suhusa}} & 
{\raggedright } & 
{\raggedright }\\
\hline
% ROW 21
{\raggedleft \textrm{{ 19}}} & 
{\raggedright { Quihicha}} & 
{\raggedright { Aca}} & 
{\raggedright } & 
{\raggedright }\\
\hline
% ROW 22
{\raggedleft \textrm{{ 20}}} & 
{\raggedright { Quihicha}} & 
{\raggedright { Ubchihica}} & 
{\raggedright } & 
{\raggedright Quadrature: \textit{``...Counting from \textit{Muihica} they found in the number 7 the first quadrature, in \textit{Ubchihica} after the  neonemia...''}}\\\hline
% ROW 23
{\raggedleft \textrm{{ 21}}} & 
{\raggedright { Guetas asaq{\iM}}} & 
{\raggedright { Ata}} & 
{\raggedright } & 
{\raggedright }\\
\hline
% ROW 24
{\raggedleft \textrm{{ 22}}} & 
{\raggedright { Guetas asaq{\iM}}} & 
{\raggedright { Bosa}} & 
{\raggedright } & 
{\raggedright }\\
\hline
% ROW 25
{\raggedleft \textrm{{ 23}}} & 
{\raggedright { Guetas asaq{\iM}}} & 
{\raggedright { Mica}} & 
{\raggedright Cuadrature: \textit{``...counting eight days} [from hisca (15)], \textit{they found the other quadrature in \textit{Mica}...''}} &
{\raggedright }\\
\hline
% ROW 26
{\raggedleft \textrm{{ 24}}} & 
{\raggedright { Guetas asaq{\iM}}} & 
{\raggedright { Muihica}} & 
{\raggedright } & 
{\raggedright }\\
\hline
% ROW 27
{\raggedleft \textrm{{ 25}}} & 
{\raggedright { Guetas asaq{\iM}}} & 
{\raggedright { Hisca}} & 
{\raggedright } & 
{\raggedright }\\
\hline
% ROW 28
{\raggedleft \textrm{{ 26}}} & 
{\raggedright { Guetas asaq{\iM}}} & 
{\raggedright { Ta}} & 
{\raggedright } & 
{\raggedright }\\
\hline
% ROW 29
{\raggedleft \textrm{{ 27}}} & 
{\raggedright { Guetas asaq{\iM}}} & 
{\raggedright { Cuhupcua}} & 
{\raggedright Full Moon: \textit{``...The first aspect of the first phase was in \textit{Cuhupcua}...''}} & 
{\raggedright }\\
\hline
% ROW 30
{\raggedleft \textrm{{ 28}}} & 
{\raggedright { Guetas asaq{\iM}}} & 
{\raggedright { Suhusa}} & 
{\raggedright } & 
{\raggedright }\\
\hline
% ROW 31
{\raggedleft \textrm{{ 29}}} & 
{\raggedright { Guetas asaq{\iM}}} & 
{\raggedright { Aca}} & 
{\raggedright } & 
{\raggedright }\\
\hline
\end{longtable}
}

In contrast to the  synodic period of the moon of about 29.5 days, the 30 day account as described by Duquesne suggest that the Muiscas calculated this period using an integer number based arithmetic, as seen in  other precolumbian astronomical systems. Ignoring the differences of time in the moon's translation around the Earth in accordance to the Kepler's second law, we can safely approximate:

$$
 \textrm{Days between each lunar phase} =\frac{29.5}{4} \textrm{days}= 7.375 \textrm{days}
$$

\noindent being 29.5 the timespan of a synodic period of the moon. The times for the waning quarter, new Moon, waxing quarter, and full Moon would be 7.375, 14.75, 22.125 and 29.5 days, respectively (these could be rounded to 7, 15, 22, 30). Thus, the days for each phase, according the two sources can be seen as shown in the table \ref{ARvsD2}.

\begin{table}[h]
\centering
\begin{tabular}{|l|r|c|c|c|}
\hline
% ROW 1
{\raggedright } & 
{\centering {\footnotesize \#Day}} & 
{\centering {\footnotesize \#Day (rounded)}} & 
{\centering {\footnotesize Dissertation}} & 
{\centering {\footnotesize Astr. Ring}}\\
\hline
% ROW 2
{\raggedright {\footnotesize Quadrature}} & 
{\centering {\footnotesize 7.375}} & 
{\centering {\footnotesize 7}} & 
{\centering {\footnotesize 7}} & 
{\centering {\footnotesize 10}}\\
\hline
% ROW 3
{\raggedright {\footnotesize Conjunction}} & 
{\centering {\footnotesize 14.750}} & 
{\centering {\footnotesize 15}} & 
{\centering {\footnotesize 15}} & 
{\centering {\footnotesize 14}}\\
\hline
% ROW 4
{\raggedright {\footnotesize Quadrature}} & 
{\centering {\footnotesize 22.125}} & 
{\centering {\footnotesize 22}} & 
{\centering {\footnotesize 25}} & 
{\centering {\footnotesize 20}}\\
\hline
% ROW 5
{\raggedright {\footnotesize Opposition}} & 
{\centering {\footnotesize 29.500}} & 
{\centering {\footnotesize 30}} & 
{\centering {\footnotesize 30}} & 
{\centering {\footnotesize 3}}\\
\hline
\end{tabular}
\caption{Lunar time errors}\label{ARvsD2}
\end{table}

What is described in both the \textit{Astronomical Ring} as the \textit{Dissertation} does not correspond  exactly to the names for the astronomical days of lunar quarters, varying the Dissertation in the day of the waning quadrature,  and the \textit{Astronomical Ring} vary by as much as three days before and after  astronomical moments, showing an apparent  lack of precision. 

Nevertheless, when regarding the data in the table \ref{ARvsD}, a somewhat interesting trend arises: The cells occupied by citations in the column of the \textit{Dissertation} appear to complement the empty cells of the column of the \textit{Astronomical Ring} and vice versa. Moreover, considering  the information from the two columns as a whole, it becomes evident that each citation is marking the beginning or the end of timespans associated  with each lunar quarter. For example,  according to the \textit{Dissertation}, the waxing quadrature, occurs on the 7th day, whilst in the \textit{Astronomical Ring} it occurs on the 10th day, covering so in the table a lapse of four cells, from  number 7 to  10. From this, the following spans  can be deduced from the texts (see table \ref{lunarlapses}): 

\begin{enumerate}
\item Around the opposition (days 27-3) \textit{(the full Moon span)}.
\item Around the waxing quadrature  (days 7-10) \textit{(the waning Moon span)}.
\item Around the conjunction, including the heliacal sets and rises (days 12-17) \textit{(the new Moon span)}.
\item Around the waning quadrature (days 20-23) \textit{(the waxing Moon span)}.
\end{enumerate}

\begin{longtable}{|rl|c|c|}%{|p{2.3cm}|p{1.5cm}|p{0.5cm}|p{3.2cm}|}
\caption{Lunar lapses as deduced from the \textit{Astronomical Ring} and the \textit{Dissertation} }\label{lunarlapses}\\
\hline
% ROW 1
{\centering {\footnotesize \textbf{Number}}} & 
{\centering {\footnotesize \textbf{name}}} & 
{\centering {\footnotesize \textbf{Day}}} & 
{\centering {\footnotesize \textbf{Time span}}}\\
\hline
% ROW 1
{\centering {\footnotesize }} & 
{\centering {\footnotesize Ata}} & 
{\centering {\footnotesize 1}} & 
{\centering {\footnotesize \textit{the full Moon span}}}\\
\hline
% ROW 2
{\centering {\footnotesize }} & 
{\centering {\footnotesize Bosa}} & 
{\centering {\footnotesize 2}} & 
{\centering {\footnotesize \textit{the full Moon span}}}\\
\hline
% ROW 3
{\centering {\footnotesize }} & 
{\centering {\footnotesize Mica}} & 
{\centering {\footnotesize 3}} & 
{\centering {\footnotesize \textit{the full Moon span}}}\\
\hline
% ROW 4
{\centering {\footnotesize }} & 
{\centering {\footnotesize Muihica}} & 
{\centering {\footnotesize 4}} & 
{\centering {\footnotesize *}}\\
\hline
% ROW 5
{\centering {\footnotesize }} & 
{\centering {\footnotesize Hisca}} & 
{\centering {\footnotesize 5}} & 
{\centering {\footnotesize *}}\\
\hline
% ROW 6
{\centering {\footnotesize }} & 
{\centering {\footnotesize Ta}} & 
{\centering {\footnotesize 6}} & 
{\centering {\footnotesize *}}\\
\hline
% ROW 7
{\centering {\footnotesize }} & 
{\centering {\footnotesize Cuhupcua}} & 
{\centering {\footnotesize 7}} & 
{\centering {\footnotesize \textit{the waning Moon span}}}\\
\hline
% ROW 8
{\centering {\footnotesize }} & 
{\centering {\footnotesize Suhusa}} & 
{\centering {\footnotesize 8}} & 
{\centering {\footnotesize \textit{the waning Moon span}}}\\
\hline
% ROW 9
{\centering {\footnotesize }} & 
{\centering {\footnotesize Aca}} & 
{\centering {\footnotesize 9}} & 
{\centering {\footnotesize \textit{the waning Moon span}}}\\
\hline
% ROW 10
{\centering {\footnotesize }} & 
{\centering {\footnotesize Ubchihica}} & 
{\centering {\footnotesize 10}} & 
{\centering {\footnotesize \textit{the waning Moon span}}}\\
\hline
% ROW 11
{\centering {\footnotesize Quihicha}} & 
{\centering {\footnotesize Ata}} & 
{\centering {\footnotesize 11}} & 
{\centering {\footnotesize *}}\\
\hline
% ROW 12
{\centering {\footnotesize Quihicha}} & 
{\centering {\footnotesize Bosa}} & 
{\centering {\footnotesize 12}} & 
{\centering {\footnotesize \textit{the new Moon span}}}\\
\hline
% ROW 13
{\centering {\footnotesize Quihicha}} & 
{\centering {\footnotesize Mica}} & 
{\centering {\footnotesize 13}} & 
{\centering {\footnotesize \textit{the new Moon span}}}\\
\hline
% ROW 14
{\centering {\footnotesize Quihicha}} & 
{\centering {\footnotesize Muihica}} & 
{\centering {\footnotesize 14}} & 
{\centering {\footnotesize \textit{the new Moon span}}}\\
\hline
% ROW 15
{\centering {\footnotesize Quihicha}} & 
{\centering {\footnotesize Hisca}} & 
{\centering {\footnotesize 15}} & 
{\centering {\footnotesize \textit{the new Moon span}}}\\
\hline
% ROW 16
{\centering {\footnotesize Quihicha}} & 
{\centering {\footnotesize Ta}} & 
{\centering {\footnotesize 16}} & 
{\centering {\footnotesize \textit{the new Moon span}}}\\
\hline
% ROW 17
{\centering {\footnotesize Quihicha}} & 
{\centering {\footnotesize Cuhupcua}} & 
{\centering {\footnotesize 17}} & 
{\centering {\footnotesize \textit{the new Moon span}}}\\
\hline
% ROW 18
{\centering {\footnotesize Quihicha}} & 
{\centering {\footnotesize Suhusa}} & 
{\centering {\footnotesize 18}} & 
{\centering {\footnotesize *}}\\
\hline
% ROW 19
{\centering {\footnotesize Quihicha}} & 
{\centering {\footnotesize Aca}} & 
{\centering {\footnotesize 19}} & 
{\centering {\footnotesize *}}\\
\hline
% ROW 20
{\centering {\footnotesize Quihicha}} & 
{\centering {\footnotesize Ubchihica}} & 
{\centering {\footnotesize 20}} & 
{\centering {\footnotesize \textit{the waxing Moon span}}}\\
\hline
% ROW 21
{\centering {\footnotesize Guetas asaq{\iM}}} & 
{\centering {\footnotesize Ata}} & 
{\centering {\footnotesize 21}} & 
{\centering {\footnotesize \textit{the waxing Moon span}}}\\
\hline
% ROW 22
{\centering {\footnotesize Guetas asaq{\iM}}} & 
{\centering {\footnotesize Bosa}} & 
{\centering {\footnotesize 22}} & 
{\centering {\footnotesize \textit{the waxing Moon span}}}\\
\hline
% ROW 23
{\centering {\footnotesize Guetas asaq{\iM}}} & 
{\centering {\footnotesize Mica}} & 
{\centering {\footnotesize 23}} & 
{\centering {\footnotesize \textit{the waxing Moon span}}}\\
\hline
% ROW 24
{\centering {\footnotesize Guetas asaq{\iM}}} & 
{\centering {\footnotesize Muihica}} & 
{\centering {\footnotesize 24}} & 
{\centering {\footnotesize *}}\\
\hline
% ROW 25
{\centering {\footnotesize Guetas asaq{\iM}}} & 
{\centering {\footnotesize Hisca}} & 
{\centering {\footnotesize 25}} & 
{\centering {\footnotesize *}}\\
\hline
% ROW 26
{\centering {\footnotesize Guetas asaq{\iM}}} & 
{\centering {\footnotesize Ta}} & 
{\centering {\footnotesize 26}} & 
{\centering {\footnotesize *}}\\
\hline
% ROW 27
{\centering {\footnotesize Guetas asaq{\iM}}} & 
{\centering {\footnotesize Cuhupcua}} & 
{\centering {\footnotesize 27}} & 
{\centering {\footnotesize \textit{the full Moon span}}}\\
\hline
% ROW 28
{\centering {\footnotesize Guetas asaq{\iM}}} & 
{\centering {\footnotesize Suhusa}} & 
{\centering {\footnotesize 28}} & 
{\centering {\footnotesize \textit{the full Moon span}}}\\
\hline
% ROW 29
{\centering {\footnotesize Guetas asaq{\iM}}} & 
{\centering {\footnotesize Aca}} & 
{\centering {\footnotesize 29}} & 
{\centering {\footnotesize \textit{the full Moon span}}}\\
\hline
% ROW 30
{\centering {\footnotesize Guetas asaq{\iM}}} & 
{\centering {\footnotesize Ubchihica}} & 
{\centering {\footnotesize 30}} & 
{\centering {\footnotesize \textit{the full Moon span}}}\\
\hline
\end{longtable}

Although these spans seem akin to the `Muisca week' described by Sim\'on \citeyearpar{Simon1625}, it is unlikely the existence of a concept of week among the Muisca, as  discussed in  section \ref{dayweek}. Instead, it could be concluded that both the \textit{Dissertation} and the \textit{Astronomical Ring} simply provide pragmatic recipes to allow an observer to establish timespans around the lunar phases, instead of determining the exact date of each quarter, or weeks,
as the western traditional system employed by Duquesne to understand the Muisca account of the lunar cycle.

\subsubsection{Adjustment of  moon cycle duration}
Given that the synodic lunar period is  about 29.5 (29.530589) days, this would present difficulties to the Muisca, because of their integer-based  30 day month account. It can observed that after two full Moons, a half day difference could accumulate resulting in an error of one day, easily observable. Duquesne did not account for this problem and the eventual adjustment done by the Muiscas. So as a result, we are forced to assume the hypothetical adjustment of one day, leading to the count of thirty days in one month, and twenty nine for the next.

It may be suggested that 17th day of the cycle, \textit{[Quihicha] Cuhupcua}, could be used  to perform the adjustment. Located at the end of the period ``New moon span'',  \textit{Cuhupcua} meant ``deaf person'', which, according to Duquesne, was considered by the Muiscas as propicious to `ignore' some dates in the context of  Muisca years to make similar adjustements possible\footnote{Topic to be covered in the next pages.}. On a monthly level, it would be convenient to include or ignore  a \textit{Cuhupcua} day because it would fall in a period when the Moon was not visible. 

\subsection{The Calendar System}

This is the most complex part of the Duquesne's work, which unfortunately, has not been fully understood. Albeit the \textit{Dissertation} and the \textit{Astronomical Ring} are virtually identical in this part, Duquesne  uses the Muisca number names yet again, this time in the context of accounting lunar months. Similar to the Moon phases, he is not rigurous in the use of the number name prefixes. 

According Father Duquesne, the Muiscas had two types of year, both of which were based upon the lunar cycles. The first one was called the \textit{Zocam}, which corresponded to a timespan of twenty moons. The second one, although unnamed by the Muiscas, had a span of 37 moons, and was therefore was given the name  \textit{Acrotom Year} by Duquesne .

\subsection{The Rural year}\label{ruralyear}

Although Duquesne did not  explicitly mention  this kind of year in his work, it can deduced  that the Muiscas had an account of twelve moons, a timespan that was pragmatically useful in agricultural activities.  In the analysis of the Muisca calendar conducted by alexander von Humboldt
this year is referred to as the ``Rural year'', which is also described by chroniclers as Pedro Sim\'on \citeyearpar{Simon1625}: 

\begin{quote}
...they had also a year of twelve moons or months, that began in January and finished in December, 
as cleverly than us whom use to start it at such a month, unlike the Roman calendar that began in March. So, they began it in January, because, being  a dry season, they started to farm and prepare the land, in order to  cultivate it in the waning moon of March, which is when  the first rains of the first wet season of this land begin. And, from January's moon when they began to sow, until December, when they harvested, there was a span of twelve moons, the so called \textit{Chocan}, which means the same we name year, and in order to refer to the previous years, they said \textit{Chocamana}, and for the current year \textit{Chocamata}...\footnote{``...tenían tambien año de doce meses ó lunas, que comenzaba en
Enero y se acababa en Diciembre, pero por la inteligencia que nosotros tenemos
para comenzarlo en aquel mes, ni como la que tuvieron los romanos de
comenzarlo en el de Marzo; pues sólo lo daban principio desde Enero, porque
desde allí a labrar y disponer la tierra por ser tiempo seco y de verano, para
que ya estuviesen sembradas la menguantes de la luna de Marzo, que es cuando
comienzan las aguas del primer invierno en esta tierra, y como es de la
luna de Enero que comenzaban estas sementeras, hasta la del Diciembre, que
las acababan de coger, hay doce lunas, a este tiempo llamaban con este vocablo
Chocan, que es lo mismo que nosotros llamamos año, y para significar el pasado
decían Chocamana, y al año presente Chocamata'' \cite[p306]{Simon1625}} 
\end{quote} 

The term \textit{Chocan} is found in the known vocabularies of the Muisca language, translated as ``year'', without a further explanation about what kind of year is \citep{Gonzalez1987}. From Simón it can be deduced that such term is associated to a 12 moon year.

\subsection{The \textit{Zocam} year}\label{zocam-year}

It is worth noting that the term \textit{Zocam},  described by  Duquesne, is a variation of the word \textit{Chocan}, however,  the Duquesne's \textit{Zocam} refers to a period of twenty moons, which differs to the description of Simón. As a convenion,  we will use the term \textit{Zocam} along this work to refer to this twenty moon year, opposed to the 12 moon year, which we will prefer to denominate as `Rural' year.

Father Duquesne emphasized that the number twenty, base of their numbering system, was of great importance, in both their religious and civil affairs:

\begin{quote}
...We have said many times that the Moscas considered  the number 20 as sacred. So, they must have designed their year around it, if not, they had  confused all their accounts (...) everything had to be governed by this number: Gueta was the symbol of  happiness, and among this superstitious people, the years that had not been sealed with this character, would have been diminished and laden with misfortune; therefore, among them,  the year of twenty moons was mandatory. (...) When they engaged in warfare, they remained in the battlefield for 20 continuous days, singing and being eager for the prospect of victory. If  they lost, they still remained on the  battlefield for 20 days more, crying and lamenting their misfortune. It is said that Zipa Nemequene and Zaque Michua arraged a solemn truce of 20 moons; and that the mysterious dream of their memorable Bochica\label{bochicasdream} lasted in his fantasy twenty times five twenties of years...\footnote{``\ldots Ya hemos dicho muchas veces que los moscas miraban como sagrado el
número 20. No podían menos que ajustar por él el año, porque de otra suerte se hubieran confundido en
todas sus
cuentas (\ldots) todo se debía gobernar por éste número: \emph{Gueta} era
símbolo de la
felicidad, y entre esta gente supersticiosa hubieran sido menguados e infelices
los años que no se hubiesen
sellado con este carácter; era, pues inexcusable entre ellos el año de veinte
lunas\ldots'' \citep[p217]{Duquesne1882}

``\ldots Cuando denunciaban la guerra asistían por 20 días seguidos en el campo,
cantando y alegrandose
por la esperanza de la victoria, y si perdían la batalla permanecían en el mismo
campo otros 20
días llorando y lamentando su negra y desdichada fortuna. Se dice que el zipa
Nemequene y el zaque Michua
ajustaron una solemne tregua de 20 lunas; y que el misterioso sueño de su
memorable Bochica duró
en su fantasía veinte veces cinco veintes de años \ldots'' \citep[p211]{Duquesne1882}}
\end{quote} 

Duquesne does not give more details about such a 20 day span, but center the description in the years of twenty moons. The priest refers to them as `vulgar years', as opposed to  the \textit{Acrotom Year} (see next section), and presents the term \textit{Zocam} as the Muisca translation for `year'.  Consequently, the exact meaning of \textit{Zocam/Chocam}, is unknown so far. Only as a convention, however, we will use the term \textit{Zocam} to refer the described 20 moon span along this work.

\subsection{The Acrotom year}\label{acrotomyear}
Another cycle of moons, was named \textit{Acrotom} year by Duquesne, and was designed by the Muiscas in order to reconcile the differences between the Rural year (lunar year) and the solar annual cycle (solar year), in terms of moons counted in the \textit{Zocam} year. Twelve lunar months have a duration of 354.37 days, therefore, it is out of phase by approximately 10.8 days given the 365.25 day solar year. Duquesne affirmed that the Muiscas  discovered that after three lunar years, or 36 moons, they could insert or intercalate an  additional month, as defined in their language as ``deaf person'', that  corresponds approximately to the accumulated error,  and compensates for it, after three solar years  the Sun and the Moon became synchronized: 
$$
 	29.530589\textrm{days}\times(36+1) = 1092.631793\textrm{days}\quad\sim\quad 365.25\textrm{days}\times 3 = 1095.75\textrm{days}
$$

Humboldt did not use the term \textit{Acrotom}, instead he preferred to use the term \textit{Sacerdotal} year. He agrees with Duquesne when considering that the calculation of the intercalation was exclusive to the  Muisca religious elite,  the Xeques and Ojques. Apparently, the knowledge of the Acrotom year was of a \textit{esoteric} nature, in contrast with the Rural and Zocam years, which were publicly known,  that is, \textit{exoteric}.

Duquesne described the use of this year as something ``hidden'' in the Zocam account of 20 moons, 

\begin{quote}
... the hidden year of the Moscas will be understood under this assumption, because  20 moons made up a year, in the following one,  the 17th moon arrived, they began to sow, according to the month  from which they had begun, they intercalated it, in that, they let it pass as useless, and they sowed during the following moon that was the 18th...\footnote{ ``...En este supuesto se entenderá el año oculto de los moscas, porque pasadas las 20 lunas de un año, al siguiente, llegando a la 17a en que
les competía sembrar, según el mes por donde habían comenzado, la intercalaban, es decir, la
dejaban pasar por inoficiosa, y sembraban en la siguente que era la
18a\ldots'' \citep[p218]{Duquesne1882}}                            
\end{quote}  

\begin{figure}[t]
 \centering
 \includegraphics[width=10cm]{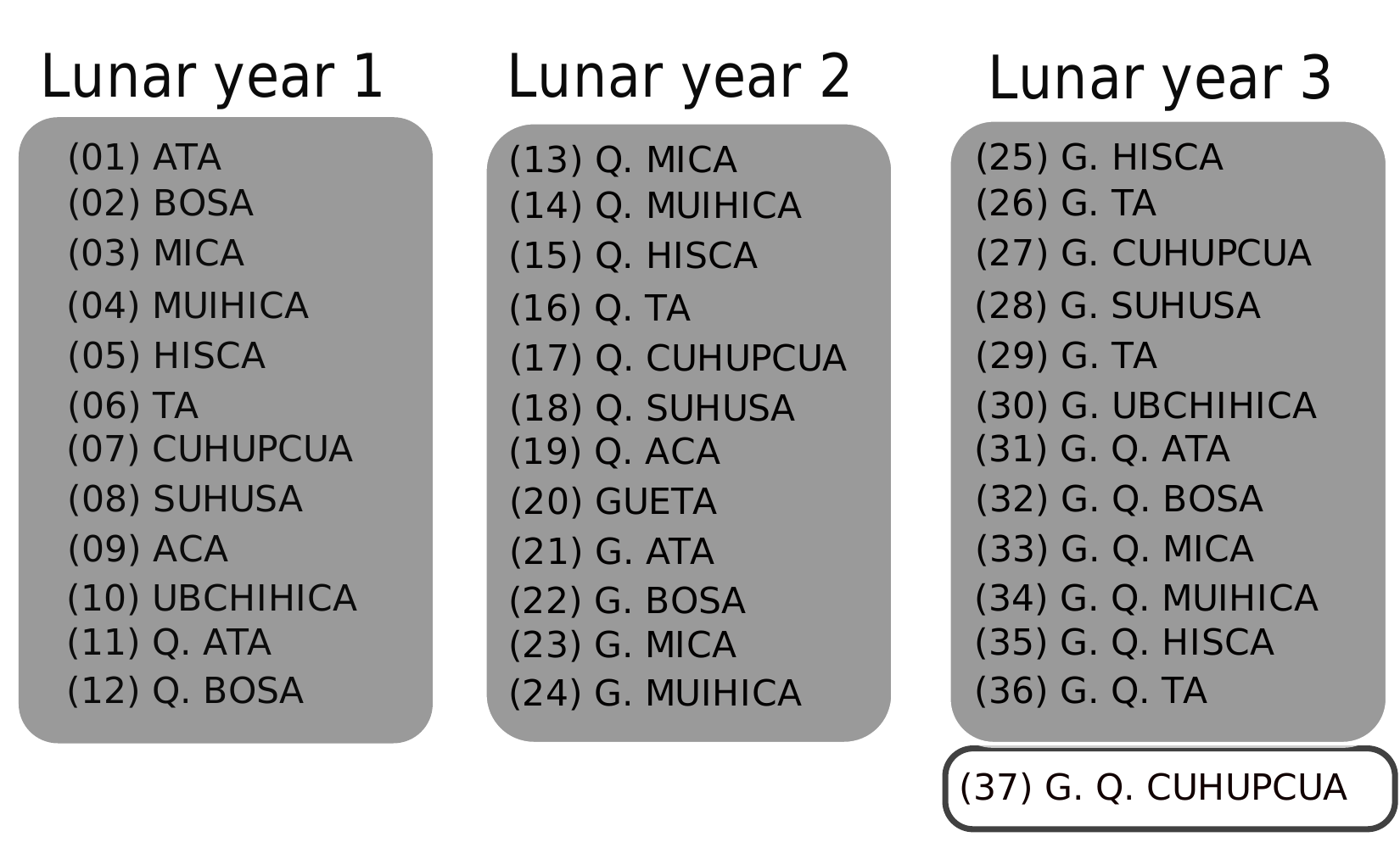}
 % acrotomos.pdf: 576x389 pixel, 72dpi, 20.32x13.72 cm, bb=
 \caption{Acrotom year}
 \label{acrotomos-fig}
\end{figure}

\noindent This gives a series of either $(12+12+12+1)$ or $(20+17)$ months\footnote{The linguistic origin is clear in  $(20+17)$: \emph{Guetas asaq{\iM}} (20) + \emph{quihicha cuhupcua}~(17) $= 37$.}, the result of which is the same: a timespan of 37 moons (see figure \ref{acrotomos-fig}). 

In the \textit{Dissertation}, Duquesne describes a  formula that  was based yet again on counting with one's fingers,  to determine the times to sow according to the Acrotom year sequence:

\begin{quote}
 ...{they made two continuous sowings with a \textit{sign} between them, and the third one with two signs} (...) Let us distribute the Muisca signs among the fingers, and this digital table will give us all the combinations. Let us suppose that \textit{Ata}, which is in the first finger, corresponds to January, and that it is a month suitable to seed. {Runned the fingers, it corresponds to  the second sowing \textit{Mica}, intercepting to Bosa, which is in the middle of \textit{Ata} and \textit{Mica},} in such a way that this sowing is made in the thirteenth moon, with respect to \textit{Ata}. 

{Now running  the fingers from \textit{Mica}, the sow corresponds to \textit{Hisca}, intercepting to \textit{Muyhica}, which is in the middle of \textit{Mica} and \textit{Hisca}. Such that  sowing is made in the thirteenth moon,  with respect to \textit{Mica}.}

{Finally, run the fingers from \textit{Hisca}, and the sowing will be made on \textit{Suhuza}, intercepting two signs: \textit{Ta} and \textit{Cuhupcua}, that are in the middle of \textit{Hisca} and \textit{Suhuza}; this is in the fourteenth moon respect to \textit{Hisca}}...\footnote{``\ldots sembraban dos sementeras seguidas con un \emph{signo} de por medio, y
la tercera con dos (\ldots) Distribuyamos pues los signos muyscos en los dedos, y esta tabla digita
nos dará todas las combinaciones. Supongamos que \emph{Ata}, que está en el primer dedo, corresponde á enero, y que es un mes apto para sembrar. Corridos
los dedos corresponde la segunda sementera en \emph{Mica}, interceptando á \emph{Bosa}, que está en medio de \emph{Ata} y \emph{Mica}. De suerte que esta
sementera se hace en la luna decimatercia, respecto de \emph{Ata}.

Corriendo ahora los dedos desde \emph{Mica}, corresponde la sementera en \emph{Hisca}, interceptando á \emph{Muyhica}, que está en medio de \emph{Mica},
é \emph{Hisca}. De modo que se hace en la luna décima tercia respecto de \emph{Mica}.

Corramos últimamente los dedos desde \emph{Hisca}, y se hará la sementera en \emph{Suhuza}, interceptando dos signos: \emph{Ta} y \emph{Cuhupcua}, que están
en medio de \emph{Hisca} y \emph{Suhuza}; esto es en la luna décima cuarta respecto de \emph{Hisca}'' \citep[p409-410]{Duquesne1848}}
\end{quote}

\begin{figure}[t]
\begin{center}
\centering
\includegraphics[width=15cm]{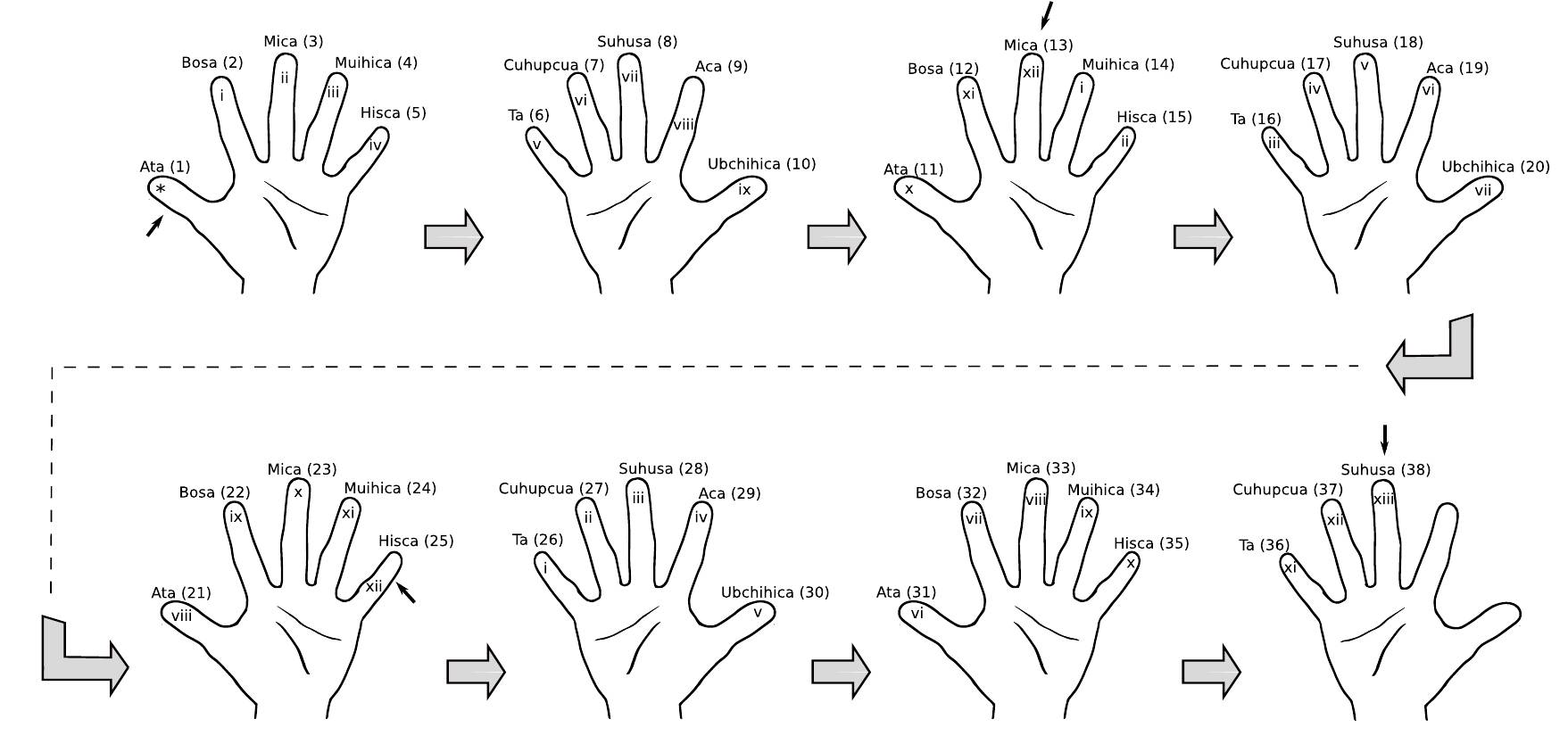}
\caption[Finger accounting to determine the sowings and the intercalary month in the Acrotom year.]{Finger accounting to determine the sowings and the intercalary month in the Acrotom year. The tiny arrows mark the signs where sowing have to be made. Note the last finger, whose number is \textsc{xiii} (13)}\label{manoacrotomos}
\end{center}
\end{figure}

A careful reading reveals that  three intervals of twelve months lunar are being defined, plus an additional month, in order to get thirty seven months. Analyzing now the text in greater depth:

\begin{itemize}
 \item \textit{``...they made two continuous sowings with a \textit{sign} between them, and the third one with two signs''}. Here Duquesne is speaking about the operation by which this calculation was made. The figure \ref{manoacrotomos} shows how the corresponding number names or, ---according to Duquesne--- signs, that can be assigned to the fingers, we could therefore assign the value \textit{Ata} to the left thumb, and \textit{Bosa} to the left index finger, counting the fingers until arriving at the right thumb, which would have the value of \textit{Ubchihica}. The idea is to  count cyclically on the fingers, returning to the  finger marked as \textit{Ata}, whenever the count has arrived at the \textit{Ubchihica} finger. According to this, Duquesne means that the sowing occurs at the signs \textit{Ata, Mica, Hisca,} and \textit{Suhuza}. 

\item \textit{``...Run} [from Ata] \textit{the fingers, it corresponds in \textit{Mica} the second sowing, intercepting to Bosa, that is in the middle of \textit{Ata} and \textit{Mica}''}. First lunar year. Strictly, this \textit{Mica} is \textit{Quihicha Mica}, which is, the thirteenth moon from \textit{Ata}. Again, Duquesne counts from the next sign of the referred one. Reviewing  figure \ref{manoacrotomos}, we see this happens when he counts twelve fingers starting at \textit{Bosa} and not at \textit{Ata} (refer to the Roman numeration). The action of `interception' can be interpreted as  having one or two fingers between the current sign and the previous one. 

\item \textit{``...Now running  the fingers from \textit{Mica}, the sow corresponds to \textit{Hisca}, intercepting to \textit{Muyhica}, which is in the middle of \textit{Mica} and \textit{Hisca}. Such that  sowing is made in the thirteenth moon,  with respect to \textit{Mica}.''}. Second lunar year. This \textit{Hisca} refers to  \textit{Guetas asaq{\iM} hisca}, the 25th moon of the Acrotom year (refer to the arabic numeration in figure \ref{manoacrotomos}). 

\item \textit{``Finally,  run the fingers from \textit{Hisca}, and the sowing will be made on \textit{Suhuza}, intercepting two signs: \textit{Ta} and \textit{Cuhupcua}, that are in the middle of \textit{Hisca} and \textit{Suhuza}; this is in the fourteenth moon respect to \textit{Hisca}''}. Third lunar year and an intercalary month. Strictly, \textit{Suhuza} refers to \textit{Guetas asaq{\iM} quihicha suhusa}, the 38th moon, which is the first moon of the next Acrotom year; effectively, Duquesne describes it as ``true January''. \textit{Cuhupcua}, refers to \textit{Guetas asaq{\iM} quihicha cuhupcua}, the 37th moon, the one to be intercalated, the action entailing the interception of two signs\footnote{Note in the figure \ref{manoacrotomos} that the number \textsc{xiii} corresponds to \emph{Suhusa}, because the count is done from  \emph{Bosa}, not \emph{Ata}.}. Note that the meaning of \textit{Cuhupcua}, is \textit{deaf person}, according to Duquesne. The Muisca linguistic talent arranged to assign a metaphor commanding the user ``to ignore'' or ``not to consider''   such a sign of the intercalary month. 
\end{itemize}

\subsection{The Solar Year}

Although there is no direct reference in the Duquesne's work of a Solar Year, the mere definition of the Acrotom Year implicitly entails that some kind of solar-based account of time was performed by the Muisca.  An important clue is given by Sim\'on when he comments a Muisca myth of creation where, in the beginning, only existed Caciques \textit{Ramiriquí} and \textit{Sogamoso}, who from yellow earth created the men and from a tall-hollow herb the women, afterwards, \textit{Ramiriquí} ascended up to sky becoming  the sun, then \textit{Sogamoso}  followed him converted in the moon. Simón inform us also that this mythical event \textit{``happened in the month of December, so in commemoration of this event, the indians of this province, especially the ones of Sogamoso, used to perform in this month a feast named} \textsl{huan...''}\footnote{``Esto, según se cuenta, sucedió por el mes de Diciembre, y así en recuerdo y memoria de este suceso, hacían los indios de esta provincia, en especial los Sogamosos, en este mes, una fiesta que llamaban huan...''\citep[p312]{Simon1625}}. 

Such ceremony of \textit{Huan}, apparently had a close relationship with Rural Year (see section \ref{ruralyear}), as it can be deduced from the rest of the description:

\begin{quote}
 ...there went out twelve men, all wearing red dresses and garlands with a small bird in the brow. Among these twelve, it was another one dressed in blue, and  together they sang in their tongue about how all of them were mortals and how their bodies will become ash: they said this with so sadly words such  the audience sobbed and cried, so, it was law that in order to comfort them, the Cacique invited and cheer up them with a lot of wine, which made them to abandon the sadness' house and fully entered into the happiness and forgetting of death one.\footnote{``..salían doce, vestidos todos de colorado, con guirnaldas y chasines que cada una de ellas remataba en una cruz, y hacia la frente un pájaro pequeño. En medio de estos doce de librea estaba otro que la tenía azul, y todos éstos juntos cantaban en su lengua cómo todos ellos eran mortales y se habían de convertir los cuerpos en ceniza, sin saber el fin que habían de tener sus almas: decían esto con palabras tan sentidas que hacían mover a lágrimas y llantos los oyentes con la memoria de la muerte, y así era ley que para consolarlos de su aflicción, había de convidar á todos el Cacique y alegrarlos con mucho vino, con que salían de la casa de la tristeza y se entregaban del todo en la de la alegria y olvido de la muerte...''\citep[p313]{Simon1625}}
\end{quote}

\noindent although it is not clear whether this `December' corresponds to a Rural (lunar) Year or a solar one, it have been suggested elsewere \citep{Correa2004}, that the \textit{Huan} was performed during the December's solstice.  Chapter \ref{ubaque-chapter} will show that the same relationship is present in a similar ceremony occurred in 1563 in the village of Ubaque.

It can be concluded so far, that the Muisca were aware of the annual movements of the sun and took care of the solsticial extremes. Probably akin ceremonies were performed during the June solstice and the March and September equinoxes as well, however, we lack of information about them in the ethohistorical sources.

\subsection{The Cycle of \textit{Ata}}\label{cycle-of-ata}
According Father Duquesne,  the Muiscas would count the moons  in the Zocam and  Acrotom accounts, in such a way that a given Moon should have a corresponding number (or sign) referring to the month  in both systems. If both the Zocam and Acrotom years start simultaneously, their numbers will coincide until the 20th moon, such that, the next moon will be the first moon of the next Zocam year and the 21th of the same Acrotom year. Furthermore, the next moon will be the 2nd Zocam month and the 22th Acrotom month, and so on,  both accountings return to the same starting numbers and the cycle is complete. This would lead to span of 740 moons. However,  following the description provided by Duquesne a shorter span of 160 moons appears, instead of 740. The reason for such a difference can be extracted from the same description, and  will be explained.  Such 160 moon timespan will be denominated as the \textit{Cycle of Ata}.  

\begin{figure}[t]
 \centering
 \includegraphics[width=10cm]{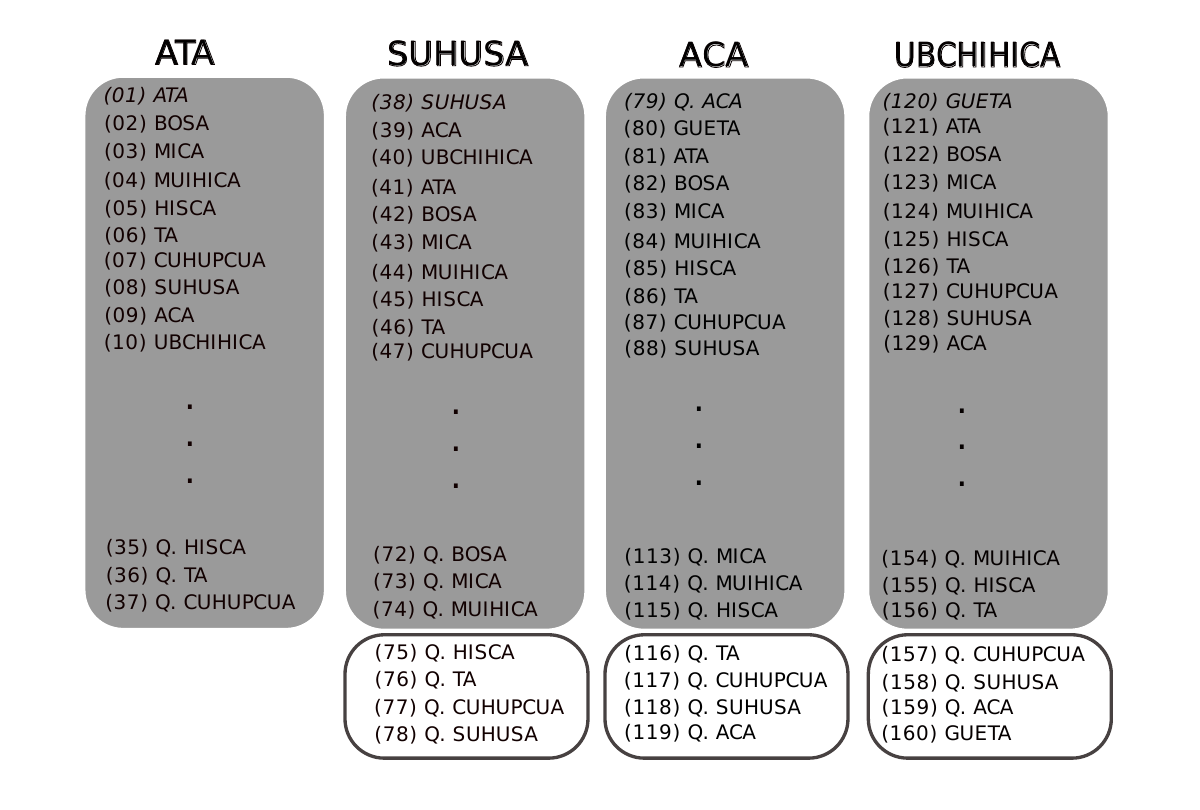}
 \caption{Cycle of \textit{Ata}}\label{atacycle-fig}
\end{figure}

Duquesne describes the Zocam dates that are passing as the Acrotom years goes by, until  they become synchronized again:

\begin{quote}
 ...\textsl{[The Acrotom years method]} would have been confusing, if it had not been established under a fixed procedure with simple and general rules. It had  four signs assigned to achieve such a goal: First \textit{Ata}; then \textit{Suhusa}, in the three last \textsl{[signs]} of the year, which they so-called the tail; \textit{Aca} and \textit{Gueta}, which were those that concluded the twenty moons year and that we therefore can call terminals. %
So, beginning  by \textit{Ata},  corresponded the number 17 of the next \textsl{[Zocam year]} to \textit{Cuhupcua}, who by this intercalation they  called  the `deaf person', and sowing was carried out in the  next one, \textit{Suhusa} 18. %
Spent the year of twenty moons, corresponded the number 17 of the next one to \textit{Muihica}, the `blind person', the sowing and  representation of the first month passed to \textit{Aca}. %
In the biennium of this one corresponded the number 17 to \textit{Hisca}, and \textit{Gueta} entered in the place and   function of \textit{Aca}, %
which 
in its biennium was 17, \textit{Ta}, and the turn came back to the first one, \textit{Ata}.%
\footnote{``...\textsl{[El método de los años acrótomos]} hubiera sido muy expuesto y confuso si no se hubiera
establecido bajo un pie fijo y con reglas fáciles y generales. Tenía pues,
destinados para este objeto, cuatro signos: el primero \emph{Ata}, y los tres
últimos del año \emph{Suhusa}, al cual llamaban por esto, la cola; \emph{Aca} y
\emph{Gueta}, que eran los que fenecían el año de veinte lunas y a los que por
esta razón podemos llamar terminales. Comenzado, pues por \emph{Ata}, tocaba al
número 17 del siguiente a \emph{Cuhupcua}, a quien por esta intercalación
llamaban con toda propiedad el sordo, y se hacía la siembra en el siguiente
\emph{Suhusa} 18. Pasado el año de veinte lunas, al siguiente tocaba el número
17 a \emph{Muihica}, el ciego, y pasaba la siembra y representación del primer
mes a \emph{Aca}. En el bienio de éste correspondía el 17 a \emph{Hisca}, y
entraba \emph{Gueta} en el lugar y en los oficios de \emph{Aca}; en su bienio
era 17, \emph{Ta}, y volvía el turno al primero, \emph{Ata}.'' \citep[p218]{Duquesne1882}}
\end{quote} 

This procedure, may seem confusing at first. However, one must  analyze the text carefully,  using the table of the Appendix \ref{muisca-centuries-table}\footnote{\textbf{Important}: The text from this section (\ref{cycle-of-ata}) to  section (\ref{muisca-centuries}) requires the reader to have in hand the Appendix \ref{muisca-centuries-table} in order to track the described sequences of moons (particularly, their numbers).}. This will help  to understand  the sequence:

\begin{itemize}
 \item \textit{``So, beginning  by \textit{Ata},  corresponded the number 17 of the next \textsl{[Zocam year]} to \textit{Cuhupcua}, who by this intercalation they  called  the `deaf person', and sowing was carried out in the  next one, \textit{Suhusa} 18.''} %
According to this passage the Acrotom year is presented in the form of $20+17$ months. \textit{Cuhupcua} is being related to the 37th moon, which is, \textit{Guetas asaq{\iM} quihicha cuhupcua}. When the sowing is made in \textit{Suhusa} is put forth it refers to \textit{Guetas asaq{\iM} quihicha suhusa}, the 38th  moon. This becomes the first month of the next Acrotom year. Note that this sowing agrees with the description of the figure \ref{manoacrotomos}.

\item \textit{``Spent the year of twenty moons, corresponded the number 17 of the next one to \textit{Muihica}, the `blind person', the sowing and  representation of the first month passed to \textit{Aca}.''} %
\textit{Muihica} represents the 74th moon (\textit{Guemicas asaq{\iM} quihicha}\footnote{From this point I will dispense with giving the whole name of a Muisca number, and I only will use the \textit{Quihicha} prefix, when necessary. The reader is warned about the occurrence of several name repetitions, referring to different Moons. In order to avoid falling in the same confusion of the Duquesne's texts, the respective moon number is indicated between parenthesis.}), and the  end of the second Acrotom year. Using the table in Appendix \ref{muisca-centuries-table}, we see that the 75th moon corresponds to  \textit{Quihicha Hisca}. According to the calendar wheel, should be in the first month of the third Acrotom year. In this point,  something else arises: In the absence of further explanation from Duquesne, he asserts that this first month is \textit{Aca}, and not \textit{Hisca} as could be interpreted. By examining the table, we see the 79th moon, \textit{Quihicha Aca}, which seems to be the one referred to in the text. If this is correct,  a shift of four months, \textit{Quihicha Hisca} (75th), \textit{Quihicha Ta} (76th), \textit{Quihicha Cuhupcua} (77th), \textit{Quihicha Suhusa} (78th) is performed.

\item \textit{``In the biennium of this one} \textsl{[Aca]} \textit{corresponded the number 17 to \textit{Hisca}, and \textit{Gueta} entered in the place and   function of \textit{Aca}''} Effectively,  37 moons after the 79th moon \textit{(Quihicha Aca)}, we arrive at \textit{Quihicha Hisca} (115). Then, another shift of four months occurs: \textit{Quihicha Ta} (116), \textit{Quihicha Cuhupcua} (117), \textit{Quihicha Suhusa} (118) and \textit{Quihicha Aca} (119) are skipped; the 120th moon, \textit{Quihicha Ubchihica} (or \textit{Gueta}), thus becomes the first month of the fourth Acrotom year. 

\item \textit{``...in its biennium} \textsl{[the corresponding to Ubchihica]} \textit{it was 17, \textit{Ta}, and the turn came back to the first one, \textit{Ata}.''} Finally,  another shift of four moons occurs: \textit{Quihicha Cuhupcua} (157), \textit{Quihicha Suhusa} (158), \textit{Quihicha Aca} (159) and \textit{Quihicha Ubchihica} (160). This is done in order to complete a 160 moon timespan, forcing the Zocam and Acrotom  accounts  to synchronize, beginning a new cycle in the 161th month  with the same name \textit{Ata} for  both accounts.
\end{itemize}

This Cycle of Ata is made up of a series consisting of four acrotom years. Each year was headed by the signs  \textit{Ata}, \textit{Suhusa}, \textit{Aca} and \textit{Ubchihica}, but  the end of the last three years it was intercalated by four months for each one, adding a total of 12 moons, or one lunar year (see figure \ref{atacycle-fig}). As a result, this cycle has a span of 160 moons: $$(37\leftmoon\times 4) + (4\leftmoon\times 3) = 160\leftmoon $$ \noindent which is also: $$\textrm{8 Zocam years} = 160\leftmoon$$ \noindent this corresponds also to 12 solar years and 342 days, which is almost thirteen solar years. It is important to note that this intercalation of three series of four moons (12 moons) forces the synchronization of  signs to take place over a 160 moon span; otherwise, such a synchronization would  happen  only after a 740 moon span, which corresponds to the \textit{Acrotom Century}, yet another cycle.   Despite the fact that Duquesne did not mention such intercalated moons directly, we will see that such a play of intercalations is important to manage larger time spans on the order of centuries. Such intercalations are shown in Appendix \ref{muisca-centuries-table},  under the heading \textit{Supplemental Series}.

\subsection{Muisca ``centuries''}\label{muisca-centuries}

According to Duquesne, the Muiscas had two varieties of long term cycles, appart from the \textit{Ata Cycle} and are described as `centuries', based on a series of lunar months as well. 

The first  period consists of twenty by twenty moons $(20\times 20)$ or twenty \textit{Zocam} years, that is four hundred lunar months. It is equal to  11812.236 days, and corresponds to 32 solar years and 124 days\footnote{We consider here a lunar synodic period of 29.530589 days}. The priest also notes a functional analogy between the \textit{Zocam} and the Mexican \textit{Xiuhmolpilli}:

\begin{quote}
...Twenty moons, thus, made one year. Once finished, they counted another twenty, and so on, turning in a continuous circle until reaching a twenty of twenties...\footnote{``\ldots Veinte lunas, pues, hacían el año. Terminadas estas, contaban otras veinte, y así sucesivamente, rodando en un círculo continuo hasta concluír un veinte de veintes\ldots'' \citep[p409]{Duquesne1848}} 

...[this] year of twenty moons is called \textit{Zocam} (...) a very significant term, that expresses the union of  one with the another, this is, from one year of twenty moons spent, to the next twenty-moon year,  following its turn  in a continous circle; an energetic word that corresponds to the \textit{xiuhmolpili} whereupon the Mexicans knew their famous fifty  two years wheel, which means in that language,  fastening of the years.\footnote{
``\ldots [Este] año de veinte lunas llamaban \emph{Zocam} (\ldots) término muy significativo, que expresa la unión del uno con el otro, esto es, del año de veinte lunas que pasó con el de las otras veinte que sigue para seguir su vuelta en un continuado círculo; palabra enérgica que corresponde a la \emph{xiuhmolpili} con que los mexicanos conocían aquella su célebre rueda de cincuenta y dos años, que en aquella lengua quiere decir atadura de los años.'' \citep[p219]{Duquesne1882}}
\end{quote} 

\noindent as a convention,  this kind of century  will be renamed as \textit{Zocam Century}.

The second period consist in a span of twenty by thirty seven moons $(20 \times 37)$ or twenty Acrotom years\footnote{Or 37 Zocam years.}, which corresponds to 740 moons. Duquesne refers to this period as a `Muisca Century':

\begin{quote}
 The Muisca century  consisted of twenty intercalary years of 37 moons. Each one,  corresponds to 60 of our  years, and is composed of four revolutions counted by five in five, each one consists of ten Muisca years, and fifteen of our years, until completing twenty, in which the sign \textit{Ata} takes the turn from where it started. The first revolution was finished in \textit{Hisca}, the second one in \textit{Ubchihica}, the third one  in \textit{Quihicha Hisca} and the fourth one in \textit{Gueta}.\footnote{``El siglo de los Muyscas constaba de veinte años intercalares de 37 lunas cada
uno, que corresponden á 60 años nuestros, y le componían de cuatro revoluciones
contadas de cinco en cinco, cada una de las cuales constaba de diez años
muyscos, y quince nuestros, hasta completar los veinte, en que el signo Ata
vuelve á tomar el turno de donde comenzó la vez primera. La primera revolución
se cerraba en \emph{Hisca}, la segunda en \emph{Ubchihica}, la tercera en
\emph{Quihicha Hisca} y la cuarta en \emph{Gueta}''. \citep[p410]{Duquesne1848}} 
\end{quote}

 \noindent 740 moons correspond to 59.83 solar years, which differs to  60 solar years in about two months, due to the difference of 3.11 days between an Acrotom year (1092.63 days) and three solar years (1095.75~days) (see section \ref{acrotomyear}). However, it is unknown whether the Muiscas were aware of this mismatch and devised a correction.\footnote{A satisfactory correction would be to add a day to each rural year of the Acrotom year plus another day to the intercalary month. In this way, the Acrotom year would have 1096 days, and an Acrotom Century would be 60.0136 solar years long, with a closer difference of about four days.}. Until new evidence comes to light, we have to consider this difference as an inaccuracy of the Muisca system; possibly the Acrotom year was only an approximation to indicate the days that an empirical observation of the sun was mandatory in order to apply a correction, in a similar way that has been suggested for the Inca calendar \citep{Bauer1995}.

Each one of the  four revolutions described covered five Acrotom years, which corresponds to 185 moons:
\begin{center}
\begin{tabular}{rcl} 
$740\leftmoon$ & $=$ & $4 \times (\textrm{5 Acrotom years})$\\
$ 37\leftmoon \times 20 $&$=$& $4 \times \left(5 \times 37\leftmoon \right)$\\
&$=$& $4 \times 185\leftmoon$ 
\end{tabular}
\end{center}
\noindent these are approximately fifteen solar years (14 solar years, 350 days), and correspond to the months 185, 370, 555 and 740; these values match the symbols given by Duquesne. As a convention,  this kind of century  will be renamed as \textit{Acrotom Century}.

At this point, it is interesting to note that the Acrotom Century and its four revolutions do not  account for the Cycle of Ata, which alternatively seems to be associated with the 400 moon period. This period is described by Duquesne as the `Vulgar Century':

\begin{quote}
...after a Muisca vulgar century of  20 moon years plus 17 years, such that a finished century, or astronomical revolution of 20 intercalary years consisted of 37 moons each,  three vulgar years are needed to complete two vulgar centuries. Thus, arriving at this case, they do not account for these three vulgar years that they were not needed for sowing,  religion, or history, and they started a vulgar year in Ata (to which it was arrived the turn), beginning of a new century that completely resembles the first one, which we have described.%
\footnote{
``\ldots despues de un siglo vulgar muysco de años de 20 lunas, y mas 17 años,
de suerte que, terminando el siglo, ó revolución astronómica de 20 años
intercalares de 37 lunas cada uno, les faltan tres años vulgares para completar
dos siglos vulgares. En llegando pues á este caso no hacian mas cuenta de
aquellos tres años vulgares de que no necesitaban para la labranza, ni para la
religión, ni para la historia, y empezaban en Ata (á que habia llegado el turno)
un año vulgar, nuevo principio de un siglo nuevo en todo semejante al primero
que hemos descrito''. \citep[p417]{Duquesne1848}
}
\end{quote} 
\noindent These described `vulgar years' are, in fact, Zocam years. Hence, the `Vulgar Century' refers to the timespan we have renamed as \textit{Zocam Century}. Two Zocam Centuries correspond to 800 moons, while an Acrotom Century is 740 moons. This gives a difference of 60 months, and are the same three Zocam years referred to by Duquesne. It appears that the clergyman was unable to explain how the Muiscas dealt with such a difference, and he assumed that they merely ignored such a timespan and restarted the counting of a new century.  He did not realized, that it was the Cycle of Ata that really fit a 800 moon period, this being the intercalations of a four moon series described as the `Supplemental Series' (see Section \ref{cycle-of-ata} and the Appendix \ref{muisca-centuries-table})
the responsible to accumulate the sixty  moons required to complete the 800 moon span or two Zocam centuries. Therefore, in a similar fashion to  the Acrotom century revolutions, two Zocam centuries can be divided by five consecutive Cycles of Ata:

\begin{center}
\begin{tabular}{rcl} 
$800\leftmoon$ & $=$& $740\leftmoon + 60\leftmoon$\\
$400\leftmoon\times 2$ & $=$ & $5 \times \textrm{Cycles of Ata}$\\
$20\leftmoon\times 20 \times 2 $&$=$& $5\times \left(\left(4 \times \textrm{Acrotom year}) + (\textrm{Lunar year}\right) \right)$  \\
                                &$=$& $5\times \left(\left(4 \times 37\leftmoon) + (3 \times 4\leftmoon\right) \right)$ \\
&$=$&$5 \times 160\leftmoon$
\end{tabular}
\end{center}

\noindent It appears, that multiplying the timespan of the Zocam centuries  by two  gave a  more numerically akin period to an Acrotom century than  400 moon span. Therefore, we could rename this 800 moon span as the  \textit{Extended Zocam Century}. Both the Acrotom century and the Extended Zocam centuries are summarized in the Table \ref{siglomuisca}.
\begin{table}[h]
\begin{center}
% use packages: array
\begin{footnotesize}
\begin{tabular}{|c|c|c|c|c|c|c|}\hline
\textsc{Century type}     & $\leftmoon$  & \textsc{Divisions} & \textsc{name div.} & \textsc{days} & \textsc{years} \\ \hline
Acrotom          & 740         & $4\times 185\leftmoon$     & ``Revolutions'' & 21852.6914 & 59.8306 \\ 
Extended Zocam & 800         & $5\times 160\leftmoon$     & Ata Cycle & 23624.5664 & 64.6819 \\ \hline
\end{tabular}
\caption{Values of the two muisca ``centuries''}
\label{siglomuisca}
\end{footnotesize}
\end{center}
\end{table}

The function of the Cycle of Ata would be to split two zocam Centuries into portions, similar to those of the  Acrotom Century revolutions, giving  an apparent  synchronization  of  month signs of the Zocam and the Acrotom years in a timespan of 160 moons, shorter than 740 moons ---or one Acrotom century--- being the least common multiple of 37 and 20 moons. It is important to note that if in a certain moment both the Acrotom  and  Extended Zocam centuries would begin synchronized, they will return to their common origin after 29600 moons\footnote{Least common multiple of 740 and 800.} (or 2393.2193 solar years). As a consequence,  as that centuries go by, the  number of moon indicating each Acrotom century revolutions, will shift with respect to the \textit{Zocam} centuries, but always conserving their number sign (or name),  \textit{Hisca}, \textit{Ubchihica} and \textit{Gueta}. For example, once  the first Acrotom century is finished, the first revolution of the following Acrotom century, occuring 185 moons later, would finish at the 125th moon of the second Extended Zocam century, or the 925th moon, with respect to the common origin of the Acrotom and Zocam centuries. This differs with the number of moon for the same revolution occuring in the first Extended Zocam,  the moon 185th, being evident such a shift.

\subsection{The \textit{Bxogonoa} and the ``Bochica's dream'' as long term timespans}\label{simonduquesne}

Until this point in the analysis of the Duquesne's calendrical model, the 29600 moon cycle is derived as the mere theoretical result of combining the spans of both the Acrotom and  Extended Zocam Centuries. Duquesne does not mention such a cycle, and apparently does not take its existence into account. However,  Pedro Simón's  historical account provides new information that, when interpreted through Duquesne's calendrical system, gives a clue  that effectively points out that the Muiscas knew of such a cycle, and in addition they apparently had defined some associated long-term periods of time.
In his \textit{Noticias Historiales}, Simón describes the myth of the last coming of the civilizing hero Bochica, in accordance with  the traditions of  Bogotá (south)  and Sogamoso (north). For the Bogotá tradition, he wrote: 

\begin{quote}
...It helps  a very true tradition a lot that everybody has in this reign,  arrived into it twenty ages ago, and each age counts as seventy years, an unknown old man, dressed in wool, with long hair and a beard down to his waist...
\footnote{``...A que ayuda mucho una tradicion certísima que tienen todos los de
este Reino, de haber venido a el, veinte edades, y cuentan con cada edad setenta
años, un hombre no conocido de nadie, ya mayor en años y cargado de lanas, el
cabello y barba larga basta la cintura...''\citep[Not. 4, chap. III, p.284]{Simon1625}}                                                                                \end{quote} 

\noindent and, for Sogamoso:

 \begin{quote}
... [And] four ages ago, so-called as Bxogonoa, a man came with the same shape and dress  we described  in the lands of the Bogota...\footnote{``habra cuatro edades, que las nombran por este vocable Bxogonoa, vino un hombre del mismo talle y vestido que le pintamos tratando de el en estas tierras del Bogota.''\citep[Not. 4, chap. IX, p.314]{Simon1625}} 
\end{quote}

It is clear that both citations refer to two different `ages' but they express the same timespan: the \textit{Coming of Bochica}. \cite{Rozo1997}, estimates these `twenty ages' as  about 1400 years and each \textit{Bxogonoa} as 350 years. However, Rozo considers these spans as unlikely, and concludes that such denominations only indicate the multiple ways to objectivize the time among the northern and southern Muisca traditions. 

The 70 year span described by Simón  for one such `ages'  does not  match any of the periods described by Duquesne. Nevertheless, the context of  Simón's narration shows such ages as century-like periods. Did there exist an alternative period of 70 years, unknown to Duquesne, or did Simon simply give a wrong value?

Considering the astronomical and archaeological factors discussed in chapters \ref{ubaque-chapter} and \ref{stones-chapter}, which support  Duquene's model, and the 20-based arithmetic of the muiscas, it is possible that perhaps Simón, or  a scribe that copied the original manuscript, may have misspelled the original word describing the value of such an age. In Spanish, the number 60 is written \textit{`se\textbf{s}enta'}, whilst 70 is \textit{`se\textbf{t}enta'}, the error of writing an `s' as a `t'  is likely due to the handwritten nature of the transcribed palaeographic document.

Considering  this hypothesis, we can attempt to check whether such `ages' match Duquesne's model. Assuming that misspelled ages correspond to periods of 60 years, it becames quite apparent  that such ages refer to Acrotom Centuries.  We  could therefore establish:

$$\textrm{1 age} = \textrm{(possibly) 60 solar years} \quad\sim\quad \textrm{1 Acrotom Century} = 740\leftmoon$$

\noindent Then, 

\begin{center}
\begin{tabular}{rcl}
 20 ages &$=$& $20 \times 740\leftmoon$ \\
 20 ages  & $=$ & $14800\leftmoon$
\end{tabular}               
\end{center}

\noindent Effectively, the 14800 moon span, is  half  the 29600 moon cycle, and corresponds both to the zocam year number 740, and the acrotom year number 400. Therefore, it marks the 37th Zocam Century  and the 20th Acrotom Century. This match now suggests that, the `transcription error' hypothesis is probably correct.

Continuing in this direction, it can be assumed that the related 20 ages for the northern Muiscas correspond to the 4 \textit{Bxogonoa} of the southern Muiscas. If so, it becomes easy to set up an equation  to determine the duration of a \textit{Bxogonoa}:

\begin{center}
\begin{tabular}{rcl}
 4 \textit{Bxogonoa} &$=$& 20 ages\\
             &$=$& $20 \times 740\leftmoon$ \\
             &$=$& $14800\leftmoon$ \\
 1 \textit{Bxogonoa}  &$=$& $\frac{1}{4}\times 14800\leftmoon$ \\
             &$=$& $3700\leftmoon$ 
\end{tabular}               
\end{center}

\noindent A \textit{Bxogonoa} is composed of 3700 moons, which corresponds to 185 Zocam years (or 100 Acrotom years), which are also equal to 5 Acrotom Centuries.  A \textit{Bxogonoa},  corresponds to a span of 299.1579 solar years. Therefore, the 29600 moon period will correspond to 8 \textit{Bxogonoa}. 

With this point in mind, it is important to turn back to one of Duquesne's citation(previously mentioned on  page \pageref{bochicasdream}), wherein he refers to \textit{``the mysterious dream of their memorable Bochica lasted in his fantasy twenty times five twenties of years''}. How can this timespan be expressed?  If we  again use the Acrotom years we find:

\begin{center}
\begin{tabular}{rcl}
Bochica's dream &$=$& $\left(20\times\left(5\times 20\right)\right) \times 37\leftmoon$\\
                &$=$& $74000\leftmoon$
\end{tabular}               
\end{center}

\noindent this corresponds to 3700 Zocam years or 2000  Acrotom years, which is equal to 185 Zocam centuries and 100 Acrotom centuries, thus giving a time span of 5978.3691 solar years. Alternatively, if we try to express this 74000 moon span in terms of \textit{Bxogonoa}s, we obtain a pleasing result:

$$
	\frac{74000\leftmoon}{3700\leftmoon} = 20\leftmoon
$$

\noindent \textit{Bochica's dream} appears to be the next multiple $(\times 20)$ level of the \textit{Bxogonoa}. 

 It is very interesting to find that the \textit{Diccionario y gramática Chibcha} report four definitions for the term ``formerly'' \textit{(antiguamente)}:

\begin{quote}
\begin{description}
\item \textsc{Formerly}, \textit{Sasia}; it means a long time ago.
\item \textsc{Formerly}, speaking of much more time. \textit{Fanzaquia}.
\item \textsc{Formerly}, yet speaking of more time than the past. \textit{Sasbequia.}
\item \textsc{Formerly}, so that, in the beginning of the world.  \textit{Zaitania;} so, \textit{zaitania} means a very long time ago, the most possible oldness.  \textit{Unquynxie} [or] \textit{unquyquienxie} means, ab initio seculi, in the beginning of everything.
\item \textsc{Antiquity}, the things that were in the beginning of the world. \textit{Zaita} [or] \textit{zaita caguequa}. the indians say: \textit{Dios zaitaz abquy},  God created the antiquity.\footnote{Antiguamente. \textit{Sasia}; çignifica tiempo algo largo.
Antiguamente, ablando de mucho más tiempo. \textit{Fanzaquia}.
Antiguamente, aún ablando de más tiempo qu[e] el pasado. \textit{Sasbequia}.
Antiguamente, esto es, al prinçipio del mundo. \textit{Zaitania}; de suerte que \textit{zaitania} quiere deçir, antequísimamente, todo lo que puede ser. \textit{Unquynxie} [o] \textit{unquyquienxie} quiere deçir, ab initio seculi, ante todas cosas.
Antigüedad, las cosas que ubo al prinçipio del mundo. Zaita [o] zaita caguequa. Dios zaitaz abquy, diçen los yndios, Dios crió lo antiguo.\citep[p.189]{Gonzalez1987}}
\end{description}
\end{quote}

\noindent which support the idea that the Muisca had the concept of long duration spans hierarchically embedded. It could therefore be suggested that the `dates' of both the coming of Bochica and his `misterious dream' are not historical dates, rather they are mythical time spans profoundly  bound to the arithmetical structure of Muisca timekeeping system.

The transcription error hypothesis allows us to understand the information given by Simón in relation Duquesne's model, solving an apparent mismatch between the two sources. This argument was used by scholars to dismiss Duquesne's model \citep{Restrepo1892}.

\section{A general organization of the system}
After having presented  Duquene's model and all the derived time spans, one could have a confused picture of these time spans, as shown in the order of moons as follows:

\begin{center}
\begin{tabular}{lrr}\hline
\multicolumn{1}{c}{Time span}  & \multicolumn{1}{c}{Moons} & \multicolumn{1}{c}{Referred by}\\\hline
Rural Year & 12 & Simón, Humboldt\\
Zocam Year & 20 & Duquesne\\
Acrotom Year &37& Duquesne\\
Ata Cycle &160& Deduced from Duquesne\\
Astr. Revolution &185& Duquesne\\
Zocam Century &400& Duquesne \\
Acrotom Century &740& Duquesne \\
Ext. Zocam Century &800& Deduced from Duquesne\\
Bxogonoa & 3700 & Simón \\
Bochica's coming & 14800& Simón \\
29600 moons cycle &29600& Calculated by Izquierdo\\
Bochica's Dream &74000& Duquesne\\\hline
\end{tabular}
\end{center}

How are they organized? Upon a closer examination it can be stated that the \textit{Acrotom year, Acrotom century, Bxogonoas, Bochica's coming,} and \textit{Bochica's dream} belongs to a sequence $u$ that can be described mathematically as: 
\[
	u_{n,k} = n\textbf{az}^k
\]
\noindent where $\textbf{a}=37$, $\textbf{z}=20$, $n$ and $k$ are natural numbers, whereby $n > 0$. Therefore, such  time spans appear to correspond to the obtained values  for  $n = 1$  and $n = 5$, and $k = 0$  to $k = 2$:

\newcommand{\flechita}{\reflectbox{\ding{219}}}

\begin{center}
{\footnotesize
\begin{tabular}{c|cl|cl|cl}\hline
$n$ & $u_{n,0}$ & & $u_{n,1}$& & $u_{n,2}$ \\\hline
\textbf{1}   & \textbf{37} & \flechita Acr. Year & \textbf{740} & \flechita Acr. Century & \textbf{14800} &\flechita Bochica's Coming \\
2   & 74 &                  &1480 &                     & 29600 &\flechita 29600 moon cycle\\
3   & 111 &                 &2220 &                     & 44400 &\\
4   & 148 &                 &2960 &                     & 59200 &\\
\textbf{5}   & \textbf{185} & \flechita Astr. Revolution  &\textbf{3700} & \flechita Bxogonoa  & \textbf{74000} &\flechita Bochica's Dream\\\hline 
\end{tabular} 
}
\end{center}

According to this, these time spans could be organized as  value couples, wherein the second value of a couple is 5 times the value of the first. Furthermore, these couples are organized  into three levels or orders, based on the value of $k$:

\[
	\begin{array}{p{4.6cm}cr}
	 \textrm{Acrotom year} & = & 37 \times 20^0 \\
	 \textrm{Astr. Revolution} & = & 5\times37 \times 20^0 	
	\end{array}
	\Biggl\} \textrm{1st order}
\]
\[
	\begin{array}{p{4.6cm}cr}
	 \textrm{Acrotom Century} & = & 37 \times 20^1 \\
	 \textrm{Bxogonoa} & = & 5\times37 \times 20^1 	
	\end{array}
	\Biggl\} \textrm{2nd order}
\]
\[
	\begin{array}{p{4.6cm}cr}
	 \textrm{Bochica's coming} & = & 37 \times 20^2 \\
	 \textrm{Bochica's Dream} & = & 5\times37 \times 20^2 	
	\end{array}
	\Biggl\} \textrm{3rd order}
\]

Note that the Ata Cycle and the Zocam Centuries are independent of this sequence. Alternatively, they belong to another sequence $v$ as described by:

\[
	v_{n,k} = \textbf{a}^k \left(4\textbf{a}n + 12 n\right) 
\]

\noindent where $\textbf{a}=37$, wherein the Ata cycle will correspond to $v_{1,0} = 160$. In the Extended Zocam century  $v_{5,0}=800$, and in the 29600 moon cycle  $v_{5,1}=29600$. The occurrence of the 29600 cycle in both sequences  $u_{5,2} = v_{5,1}$ is explained by the fact that such a number is the least common multiple of the two sequences.

\section{The  `Table of the Muisca years'}

\begin{figure}[t]
 \centering
 \includegraphics[width=8cm]{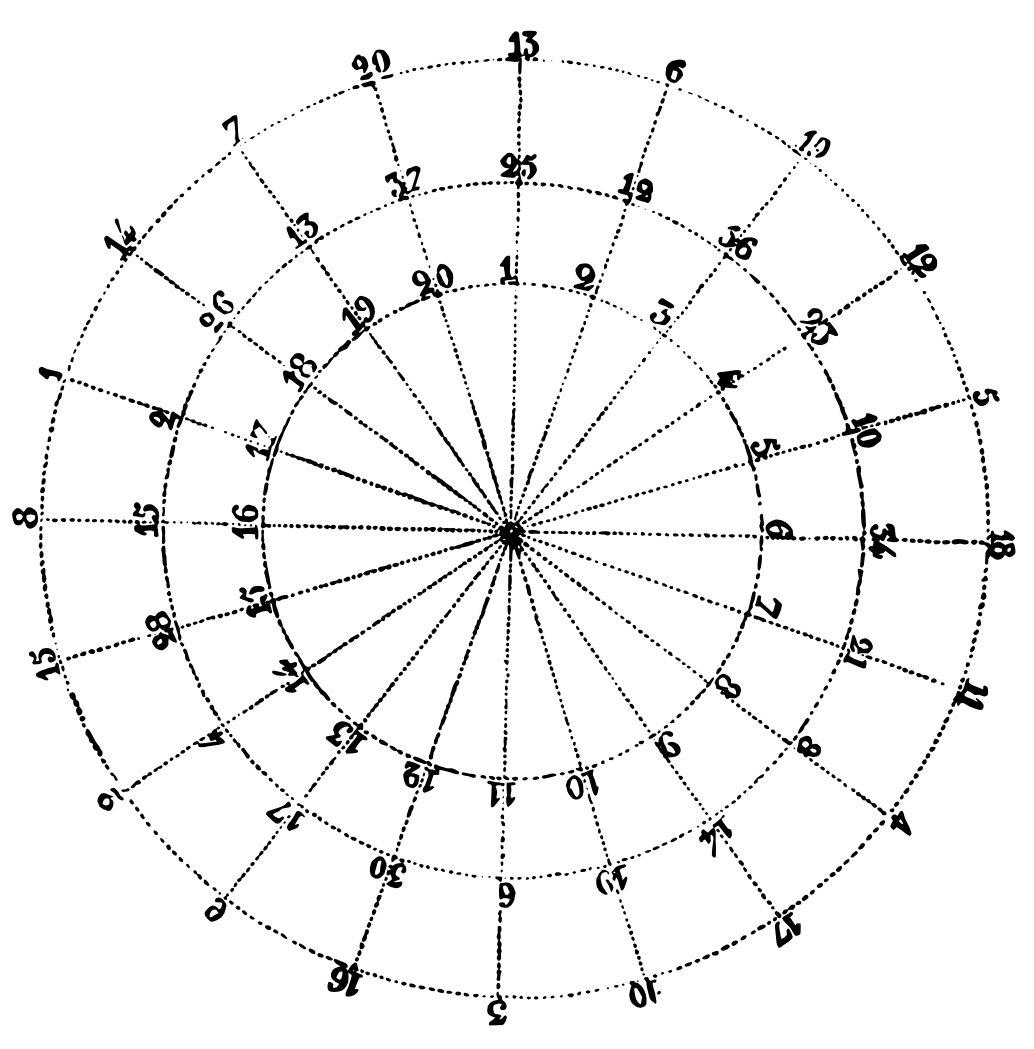}
 % calendar-table-duquesne.png: 1034x1049 pixel, 44dpi, 59.26x60.11 cm, bb=0 0 1680 1704
\vspace{0.5cm}

\resizebox{0.7\width}{\height}{
\begin{tabular}{|l|r|r|r|r|r|r|r|r|r|r|r|r|r|r|r|r|r|r|r|r|}\hline
Outer circle:&13&6&19&12&5&18&11&4&17&10&3&16&9&2&15&8&1&14&7&20\\\hline
Middle circle:&25&12&36&23&10&34&21&8&14&19&6&30&17&4&28&15&2&26&13&37\\\hline
Inner circle:&1&2&3&4&5&6&7&8&9&10&11&12&13&14&15&16&17&18&19&20\\\hline
\end{tabular} 
} 
 \caption{Duquesne's Calendar Table}
 \label{calendar-table}
\end{figure}

In the last part of his work, father Duquesne shows a table that was apparently designed by him, and compiles the number of intercalations along the Acrotom Century. Such a table is composed of three concentric circles, each one having twenty values (see figure \ref{calendar-table}) which are explained by Duquesne as being:

\begin{quote}
...The inner circle represents the 20 moons of the vulgar Muisca year\footnote{a.k.a. the Zocam year}, of which signs are  intercalated throughout in the span of the century.

The second circle represents the Muisca years  corresponding to the intercalation of each sign.

The third circle expresses the order of this intercalation.\footnote{``...El círculo interior representa las veinte lunas del año muysco vulgar, cuyos signos todos se intercalan en el espacio del siglo... El círculo segundo expresa los años muyscas a que corresponde la intercalación de cada signo... El círculo tercero expresa el orden de esta intercalación'' (see \textit{Disertation}; \textit{Astronomical Ring})}
\end{quote}

This table can be checked against the Century table provided in  Appendix \ref{muisca-centuries-table}. For example, when reading  the corresponding values of the first intercalation as shown in Duquesne's table, the following are retrieved when moving from outer to inner circles: 1, 2, 17. The first intercalation, corresponds to the number 1, which will happen during the 37th moon of the Acrotom Century. This is labeled in Appendix \ref{muisca-centuries-table} as \textit{Quihicha Cuhupcua}, thus matching  with the number 17 of  Duquesne's table. Such a moon is also the 17th moon of the \textit{second} Zocam year, which whose start point is labeled by the symbol $z_1$ in the 20th moon . Note that the index of such a symbol shows that the number of Zocam years just finished. Therefore the number of the current Zocam year can be expressed as $z_{n+1}$, which corresponds to the number 2 of Duquene's table. Consequently, it describes that the first intercalation occurs \textit{during} the second Zocam year of the Acrotom Century.

%% Part II

\part{Ethnohistorical and Archaeological connections}\label{part-connections}
\section*{Introduction}

Part \ref{part-connections} of this study seeks to examine and establish potential connections between the ethnohistorical (chapter \ref{ubaque-chapter}) and archaeological (chapter \ref{stones-chapter}) data, which support the model of the Muisca Calendar, previously discussed in chapter \ref{apuntes-duquesne}. The intention here is not to put forth  a complete validation of such a model, rather these chapters are aimed at compiling an initial set of relevant information, which may provide elements to support the future research on this subject.

Due to the hypothetical nature of many  elements covered in  these chapters, a strong emphasis will be placed on the framework proposed by Belmonte (see section \ref{belmote-classification}) to properly frame the proposed suggestions  in a responsable and academic manner.

\chapter{Ethnohistorical clues as to the understanding of the Muisca Century}\label{ubaque-chapter}

%% hypenation of the muisca/spanish word troublesome to LaTeX
\hyphenation{Muis-ca Gua-ta-vi-ta Ca-ci-caz-go U-ba-que}

\section{The Ubaque's ceremony of 1563}

In December of 1563, the colonial authorities of Santa fé de Bogotá were alerted to the organization a massive ceremony in a valley called Ubaque, located  east of the capital of the New Granada, which was named after the ruler \textit{(Cacique)} of thas land.  It was clear that this ceremony appeared to involve the perfomance of  the forbidden rituals of the native religion, which under the european view had many expresions of idolatry,  practice condemned by the Spanish authorities.

Immediately, the local authorities dispatched an inquiry on the spot, in order investigate the idolatrous nature of this ritual, and if that would be the case, to react accordingly. The commission arrived on Ubaque, and observed  part of the ceremony, and was effectively condemned as idolatry. The authorities put on abrupt stop to the ceremony, and proceeded to arrest the main persons involved in such ritual. These included the Cacique Ubaque, and other \textit{caciques} coming from other regions of the ancient muisca territory. A trial, soon followed, especially against the Cacique Ubaque.

The documentation that comes from this  trial is one of the earliest and most detailed historical source of data about the religious activities of the Muiscas. Currently, most of the documentation is conserved in the Archivo General de Indias (Spain).  A detailed transcript of each archived item, accompanied with a detailed ethnohistorical analysis of the trial, can be found in \citet{Casilimas2001,londono2001puu}

The document brings a detailed description of the  performance  of Muisca rituals, or at least of the public activities that the Spaniards witnessed. These activities included massive procession of people organized in groups or delegations by the region from which they came (to assist in this ceremony), the playing of sad and melancholic music, the performing of \textit{borracheras} (drunkenness), a term used by the Spaniards to describe the drinking of \textit{chicha} (corn beer), but actually, such an activity may be associated  with the ingestion of hallucinogen substances also, such as Yopo. Also included the ritual was the use of paraphernalia, for example masks, nets, metal jewerly. The Spaniards were especially concerned whether human sacrifices were done in such ceremonies, because as it was known since before the arrival of the Europeans that such ceremonies sometimes included the sacrifice of boys.

\subsection{Significance of this event}

 This ceremony certainly  had a huge importance for the Muiscas at that moment. Basically, it was in direct defiance of the Christian establishment, done in an era when both the Spanish and the Muisca powers were involved in  a process where one was expanding whilst the other was diminishing, or at least, fading into a subversive existence. It is important to note that the Cacique Ubaque did not intend to hold a secret reunion; on the contrary, he called several caciques to attend the ceremony openly. With the aid of his \textit{encomendero}, Juan de Cespedes, who signed letters to authorize Ubaque's messengers to go across the territory, Ubaque summoned the caciques of several towns in the region to come to his town to perform the ceremony. He sent his  messengers at the beginning of December in order to give  the guests time to prepare for the ceremony. In spite of the clear illegal aspect of the gathering, several caciques answered the summons, sending a delegation of about ten to twenty persons as representatives of their towns. The caciques themselves went with the delegations, or in the cases of such  caciques who feared repercussions, they sent their \textit{capitanes}:

\begin{quotation}
... the caciques and capitanes that came are, Bogotá, Suba and Tuna, Hontibón, Boza, Ciénega, Chía, Tibacuy, Pasca, Sichaque, Queca, Une, Pasusaga, Cáqueza, Susa, Tuche, /(1417r.) Teusacá, Fitatá, Cota, Cajicá, Sopó, and tmore more [caciques] came than the ones that had [already] spoken. [The cacique Ubaque] was asked why the caciques that he had invited did not come and  sent instead their capitanes, and he said it was because they feared that the \textit{Oidor} or another \textit{Zipa} would come to seize them...\footnote{``...son los caciques y capitanes que han venido Bogotá, Suba y Tuna, Hontibón, Boza, Ciénega, Chía, Tibacuy, Pasca, Sichaque, Queca, Une, Pasusaga, Cáqueza, Susa, Tuche, /(1417r.) Teusacá, Fitatá, Cota, Cajicá, Sopó, e que no han venido más de los que han hablado. Preguntado que cómo no vinieron los caciques que ha declarado y enviaron sus capitanes. Dijo que porque se temían que había de venir el dicho señor oidor o otro Çipa a prenderlos.'' \citep[f1417r]{Casilimas2001} }                                                                                        
\end{quotation}

The number of chiefs who attended the ceremony were certainly more than the ones Ubaque confessed, it may amount to thirty names of caciques and capitanes in the documentation of the trial. Essentially, such a ceremony was performed by people coming from the majority of the towns of the southern  part of the Muisca territory, corresponding to the ancient lands of the Zipa Bacata. However, the document mentions that the summon was extended far into the northern lands of Tunja and Sogamoso:

\begin{quotation}
...[They] have brought news of how in the town and land of the Cacique Ubaque a big meeting of Indians is to be held, summoned by the said Cacique and others, who come from the city of Tunja and other places, in order to make parties and races in order to celebrate idolatries and worship of the devil...\footnote{``... se les ha dado noticia cómo en el pueblo y repartimiento del cacique de Ubaque se ordena grande junta de indios convocados por el dicho cacique y por otros, a donde vienen desde la ciudad de Tunja y otras partes para hacer convites y carreras en las cuales celebran idolatrías en culto y veneración del demonio...'' \opcitp[ , f1396]{Casilimas2001} }
\end{quotation} 

\begin{quotation}
... [so, he] went where this witness was, and was told how the Cacique Ubaque summoned him and who also summoned the caciques of all the lands, Sogamoso, and Guatavita, and all the people from the land of the city of Santafé, in order for everyone to meet in the town of Ubaque and make \textit{borrachera}, and that Cacique Fusagasugá asked the witness for his permission  to go to the gathering of  Cacique Ubaque and celebrate \textit{borrachera}...\footnote{``...venido donde este testigo estaba le dijo cómo el dicho cacique de Ubaque lo había enviado a llamar y que también el dicho cacique de Ubaque había hecho llamar a los caciques de toda la tierra y Sogamoso y Guatavita y todos los de la tierra desta ciudad de Santafé y que todos se juntasen en el dicho pueblo de Ubaque a hacer borrachera y el dicho cacique de Fusagasugá pidió licencia a este testigo para ir al llamado del dicho cacique de Ubaque a la dicha borrachera'' \opcitp[ , f1398v]{Casilimas2001} }
\end{quotation} 

The multitudinous aspect of this ceremony impressed the Spanish witnesses who arrived there, some claim to be Luis de Peralta, who described in detail the multitude of people summoned, and of the rituals performed by them:

\begin{quotation}
[Luis de Peralta] found the  Cacique and a lot of Indians, and he thinks that there could be more than ten thousand Indians in all... [unreadable] a roadway was in front of the the Cacique Ubaque's door, about ten or twelve steps wide and very long, along which he saw a huge quantity of Indians forming squadrons, with boys behind them, all wearing net masks, lion faces, \textit{totumo} masks, tin-beaded masks, leather masks, and all wearing the devil's image. They came playing whistles, flutes, jingle bells and other instruments, some came whistling, others crying, or howling, others singing painful and sad sangs, and the ones who wore the masks caused tears to be shed by the masks themselves. They came wearing shapeless dresses and badges that the witness was unable to describe.  That day, he saw  more [Indians] arrive by the same roadway, all with the same dresses, playing [music], crying, howling and groaning like lions or tigers, with too many differences that to name that he is not able to describe them all. He thinks that these actions were summons to the Devil, since the participants wore a lot of pointed hoods, painted smocks, and other kind of dresses that when seen, frightened [people]. The caciques and capitanes  [already] announced were with the Cacique Ubaque, which had all met for the \textit{borrachera} and offerings that the  Cacique Ubaque performed...\footnote{``...[Luis de Peralta] halló al dicho cacique y a muy gran cantidad de indios, que le pareció a este testigo /(1410r.) que habría en todos más de diez mil indios de que [ilegible: - -vada] una carrera que estaba delante de la puerta del dicho cacique de Ubaque, de anchor de más de diez o doce pasos e muy larga, por la cual vido que venían muy gran cantidad de indios en escuadrones con muchachos tras ellos, todos enmascarados con máscaras de redes y caras de leones y máscaras de tutumos y máscaras de estaño y cuentas y máscaras de cueros y todos puestos de la visión del demonio, los cuales venían tañendo con pitos y flautas y cascabeles y otros instrumentos, e unos venían silvando y otros llorando y otros aullando y otros cantando cantos tristes y dolorosos, y los que traían las máscaras traían lágrimas derramadas por ellas, y que venían de disformes vestiduras e insignias que este testigo no lo sabrá contar. E que hoy dicho día este testigo vido venir a mucha más cantidad por la misma carrera con las mesmas vestiduras, tañendo y llorando y aullando y gimiendo como leones o tigueres, de tantas diferencias que no sabrá este testigo contarlo, e que le pareció a este testigo que todo era invocación y llamamiento de los demonios; e que asimesmo traían en las cabezas muchas corozas y sayos pintados y otras formas de vestiduras que ponía espanto verlas. Y que esta/(1410v.)ban con el dicho cacique de Ubaque los caciques y capitanes declarados en la cabeza de la información, los cuales todos se habían juntado a la borrachera y obsequias que el dicho cacique de Ubaque hacía.'' \opcitp[ , f1410r-f1410v]{Casilimas2001}}
\end{quotation}

The raison-d'être for such a ceremomy, according to the witness interrogated in the later trial, was that Ubaque wished to celebrate his own funeral while alive.  However, the elements shown in the description of the ceremony points to the fact that such a ceremony was the performance  of  a ritual associated to the reproduction of society, showing  the death of the cacique as a catalyst to resurrect the social dynamics of the Muisca society \citep{Correa2004}. Thus, the interpretation of the Spaniards that this ceremony is a funeral could be incomplete. How could an apparent funeral attract thousands of people, coming from several places of the territory? The Spaniards were astonished with the massive response to Ubaque; it is clear that an event like this was not witnessed by the Europeans in all the twenty-five years they colonized  New Granada. What did persuade to almost thirty muisca leaders to challenge the Spanish control against idolatry? It is very interesting that the records indicate chiefs considered enemies of Ubaque, were also summoned, as in the case of Capitán Riguativa, father of the Cacique Hontivon. Surprisingly, he accepted the summon and prepared a delegation of about twenty indians, who actively participated in the ceremony.

\subsection{Astronomical aspects of the Ubaque ceremony.}

From this, it is obvious that such a ceremony was not only a funeral, but it was a ritual moment of great importance for the Muisca society. I think its performance was a last salvo by a desperate culture  in the process of changing due to an alien power that disrupted its ancient order of beliefs, and wanted to reconnect to its roots by means of the ritual.

It is very understandable that the performing of this ceremony was an act of resistance  to the European invasion and the cultural disintegration that the Muiscas were subject to. But  now a question arises,  why this moment?  Why not before, when the Spanish power was not entirely entrenched? According to several witnesses, a ceremony of this magnitude had not been seen in the territory since before the Spaniards arrived:

\begin{quotation}
[The witness was] asked about how many \textit{borracheras} and offerings he has seen like the one done by the Cacique Ubaque and in how many he had participated. He said that since he was born, he only has seen a \textit{borrachera} like this one that the Cacique Ubaque did, the one done by the Cacique Bogotá before the Christians came.\footnote{``Preguntado cuántas borracheras y obsequias ha visto este que declara que se han hecho como la que hizo el dicho cacique de Ubaque y en cuántas se ha hallado. Dijo que desde que nació no ha visto más de otra borrachera como esta que hizo el cacique de Ubaque, que la hizo el cacique de Bogotá antes que los cristianos vinieran.'' \opcitp[ , f1431r]{Casilimas2001} }
\end{quotation}

It seems that ceremonies of such magnitude were not common, and the ones offered by Ubaque (and before by Bogotá) took place on special dates specified by the religious use of the calendar. Many religions used to schedule their ceremonies to specific moments on the calendars. So, it is plausible that the Ubaque's ceremony is not an exception. The same Cacique Ubaque gives us clues to point out that such a ceremony was comparable to the European's Easter:

\begin{quotation}
[Ubaque was] asked  why he does such meeting and \textit{borrachera}, if it is to honor the dead. He said that when God made the Indians, he left them this Easter as he did to the Christians, so they enjoy themselves like the Christians do.\footnote{``Preguntado que para qué efecto hace la dicha junta y borrachera y si es por honras de muertos. Dijo que cuando Dios hizo a los indios les dejó esta Pascua como a los /(1416r.) cristianos la suya, e que se holgaban como se huelgan los cristianos.'' \opcitp[ , f1416r]{Casilimas2001}}

\end{quotation} 

It is interesting to see that he refers to the ceremony as  ``Easter'', meaning that  this ceremony had a similar nature as the Christian Easter, in the sense that it is attached to a timekeeping system.

If this celebration is associated with a calendar system, we must firstly consider whether the time of the celebration's performance could be marked by special astronomical circumstances. After examining the position in the sky of Sun, Moon, and the planets, specially Jupiter and Saturn, it is evidenced that such a time indeed had astronomical significance.

The Real Audiencia in Santafé de Bogotá was informed  about the Muis\-ca ceremony on December 20 of 1563. The dispatched commission arrived to the Ubaque's land seven days later. That day, most of the interrogated witnesses  confessed to being summoned about eleven to twenty days ago, in the first or second week of December. The 1563 December solstice occurred on the 12th day\footnote{This historical episode happened 19 years before the reform of the Julian calendar by Pope Gregory XIII, so the real date of the solstice was shifted several days before of the expected date, about December 21.}, so it shows that the ceremony was started around the days of the December solstice.  It seems very probable that such a solstice may have an important role in the performing of the ceremony. 

 Solstices seem to be associated to other religious ceremonies in the Muisca Culture; for example,  chronicler Fray Pedro Simón (1629) informs us about another ceremony known as the \textit{Huan}, which attributes the solstice to the notion of fertility. The December solstice  coincided with the dry season in most of the territory of Colombia, and  it is in this season that  the preparation of the fields begins, in order to sow in the first months of the next year. This suggests that fertility was bound to the solstice.

During the December solstice, the sun reaches its maximum southern raising and setting  points. The arrival of the sun into a determined point in the geography during this time could serve as a calendrical marker for any person aware of this phenomenon \citep{Aveni1980}.  

On the other hand, the moon has  more ``erratic'' behavior, when compared to the sun.  Due to the inclination of its orbit, its raising points varies considerably each month, changing dramatically from northeast to the  southwest in a short time. In the case of the Ubaque's ceremony, its raising point at the moment of the full moon is very peculiar: the full moon occurred on December 29, and for that date, the moon's rising azimuth was very near (69.6\grados{}) of the raising point of the sun during the June solstice (66.4\grados{}).  The possibility exists of a moon-sun association at that moment,  each  in a ``mirror position'', around the east cardinal point. 

\begin{figure}[ht]
\begin{center}
\includegraphics[width=13cm]{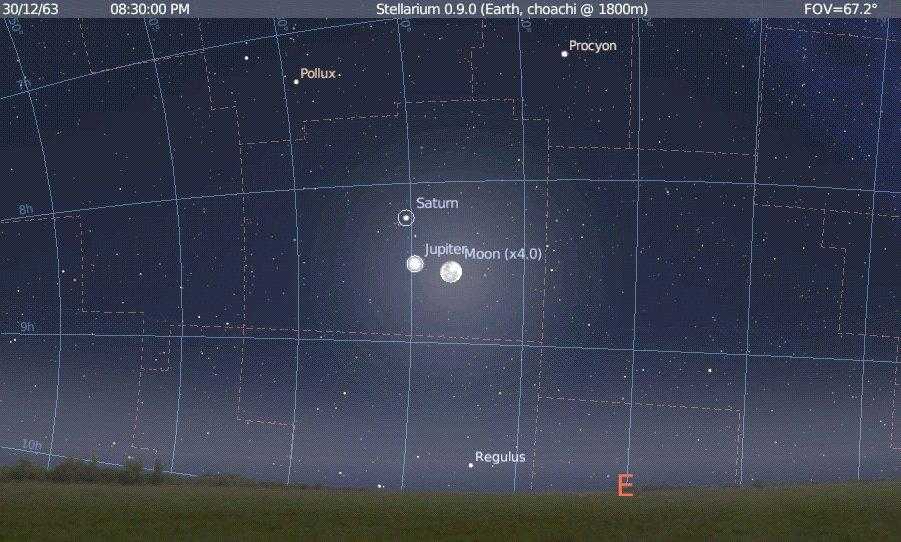}
\caption[Simulation made in \textit{Stellarium} \citep{Chereau2005} showing the Great Conjunction of 30 December 1563.]{Simulation made in \textit{Stellarium} \citep{Chereau2005} showing the Great Conjunction of 30 December 1563. The positions in the sky of Jupiter, Saturn and the Moon are shown as seen from the Ubaque's region, at 20:30, local time.}
\label{fig-conj30-12-1563}
\end{center}
\end{figure}

The moon and sun are not, however, the only celestial bodies in a special configuration that month. Surprisingly, the planets
Jupiter and Saturn were in a continuous conjunction during the years 1562 to 1564, having a mean angular separation of 4.5\grados{} during the month of December of 1563 (figure \ref{fig-conj30-12-1563}). Their appearance was very obvious, like two nearby bright stars in the sky, and in the same way as the moon, they rise on the horizon at a point near where the sun rises during the June solstice, about 70\grados{} of azimuth. The moon joined into this planetary conjunction on December 30, having an angular separation of 3.7\grados{} with Jupiter and 3.5\grados{} with Saturn,  but the bodies were below horizon at that moment (about 14:25h local time). On that same same day, when the three celestial bodies rose, the moon was displaced to the east,  having a separation of 3.1\grados{} and 5.7\grados{} from Jupiter and Saturn, respectively (table \ref{conj30-12-1563}).

\begin{table}
\begin{center}
\begin{tabular}[h]{l|ccc}
        & Moon & Jupiter \\\hline
Jupiter & 3.7\grados{} &         &              \\
Saturn  & 3.5\grados{} &  4.1\grados{}   &                      
\end{tabular}             
\caption{Angular separation between Moon, Jupiter and Saturn  during the conjunction of 30 December 1563}
\label{conj30-12-1563}
\end{center}
\end{table}

In Europe, such astronomical phenomena attracted the attention  of a well known astronomers of the time, Tycho Brahe \citep{Moore2000}, 
because it presented several astronomical conditions which can be summarized as: (a) sun in  December  solstice, (b) full moon in the vicinity (on the eastern horizon) of the June solstice, (c) planets Jupiter and Saturn in conjunction, and (d) such a conjunction is also in the vicinity of the June solstice. According to this, such a moment offered a good set of astronomical circumstances to the Muisca priests, favorable  enough to  give a cosmic purpose to their religious practices. 

A conjunction of Jupiter and Saturn is commonly known in astronomy as ``The Great Conjuntion'', and it is not an isolated event in time, but it obeys a cycle of about twenty years (19.86), when the conjunction occurs in a different place of the sky, returning approximately to the same point after three conjunctions, or sixty years \citep{Etz2000}. The cultural significance of such a cycle has been studied by \cite{Aston1970} for the Western Culture, and by \cite{Sullivan1988} for the Andean traditions. 

\section{The memory of a previous ceremony in other ethnohistorical sources}

This ethnohistorical account is very valuable because it allows one possibility to hypotesize a close relationship between the astronomical cycle of Great Conjunctions and the structure of the Calendar. However, in order to test the validity of this hypothesis, it is necessary to check whether this ceremony was an isolated event, or if on the contrary, it belonged to an already established ritual system. Using the clue of the astronomical nature of this ceremony, we could expect the performance of such  similar ceremonies in times associated with such conjunctions.  Although the ethnohistorical available sources  are sparse on in the astronomical topic, they are not, however, entirely void. 

The works of chroniclers  Fray Pedro Simón, Juan de Castellanos, Lucas Fernadez de Piedrahita, and Juan Rodriguez Freyle, written some years after the conquest of the Muisca territory, gives us some insight about the events that happened prior to the European invasion. In these,  the political landscape of the Muiscas is described as a period of conflict, caused by the expansionist policies of  a lineage of rulers known as the \textit{Zipas}, who gained political control over the southern cacicazgos of the territory, and who defied the authority of the \textit{Zaque} of Tunja, ruler of the northern territories. The Zipas, using the name of their home cacicazgo, are also referred to as the Cacique Bogotá. Despite the description of these wars as the product of a medieval misinterpretation  of  native conflicts,  aimed to justify the European occupation \citep{Correa2004}, they  provide enough historical evidence to  derive conclusions about the calendar.

In the \textit{El Carnero} \citep{Freile1636},  an episode relates how the Bogotá's forces  face  the army of the Guatavi\-ta's Cacique, loyal to the Zaque ruler, and against whom Bogotá  rebeled:

\begin{quotation}
The two armies were face-to-face, but, then there appeared signs of a break in the conflict. The night before battle, the priests, Xeques and Mohanes got together and addressed  the chiefs and main army heads, arguing that now had came the time to make sacrifice to the gods, to offer them gold and incense, and particularly, run over the land and visit the sanctuary's lakes and conduct other rites and ceremonies; so, to be understood better, \textit{they convinced the chiefs that the Jubilee year had arrived}, and it would be right to pacify the gods before the battle occurred, and in order to do it, it should be fair to allow a truce of twenty days or more...\footnote{``...Afrontados los dos campos, dieron luego muestras de venir al rompimiento de la batalla: la noche antes del día que pretendían darse la batalla se juntaron sus sacerdotes, jeques y mohanes, y trataron con los señores y cabezas principales de sus ejércitos, diciendo cómo era llegado el tiempo en que debían sacrificar a sus dioses, ofreciéndoles oro e inciensos, y particularmente correr la tierra y visitar las lagunas de los santuarios, y hacer otros ritos y ceremonias; y para que se entienda mejor, los persuadieron que era llegado el año del jubileo, y que sería muy justo cumpliesen con sus dioses primero que se diese la batalla y que para podello hacer, sería bueno asentasen treguas por veinte días o más...''\citep[chap.~IV:f10r]{Freile1636}}${}^{,}$\footnote{ The \textit{emphasis} is mine.}
\end{quotation}

What is interesting in this event, is the reason that motivated the Xeques to stop the war and impose a truce: the arrival of the Jubilee year; in other words, it was the end of a century and the beggining of the next. This suggest that the priesthood excerced an established timekeeping tradition,  fact that was already noted by Simón, in his description of the main religious chief of the Muisca, the \textit{Cacique Sogamoso}: 

\begin{quotation}
...having him spent many days in the darkness of sun, moon, stars and clouds, birds and animals, it came by experience and conjectures to know these events before they came, as the good and right astrology does, or by the fortune of being the Cacique Idacansas by who it is said this great sorcerer was initiated, and the pacts he had with th devil, to whom he used to speak, reached to know these revolutions and motions of time as a master achieves and much more in philosophy.
\footnote{``...que habiendo con el gastado muchos dias en las obscuridades del sol, luna, estrellas y nubes, aves y animales, vino por experiencia y conjeturas a sacar estos sucesos antes que vinieran, como lo hace la buena y acertada astrología, ó por ventura por ser por ser el Cacique Iducansas, en quien dicen comenzó este grande hechizero, y por pactos que tenía con el demonio, con quien de ordinario hablaba, vino a alcanzar estas revoluciones y mundanzas de tiempos como de un maestro que alcanza esto y mucho más en filosofía.'' \cite[p.317]{Simon1625}.}
\end{quotation}

\noindent which would confirm that such tradition was an important component of priesthood's crafts. Returning to \textit{El Carnero}, the story tells us how the waring sides agreed to a truce and, similarly in the Ubaque's account, they all participated massively in the ceremony thus coordinated by the Xeques:

\begin{quotation}
 ...The first ceremony they did was, form rings of both of  factions combined,  dancing men and women  with their musical instruments, as if  never existed any resentment between them or signs of war. In the prairie lying between the two rivers that divided the fields, they showed each other very joyfully, eating and drinking together in big \textit{borracheras} that lasted all day and night, and those who could bed the most people was a hero, a sentiment that remains until today. The feast and \textit{borracheras} lasted three complete days, and on the fourth day, the Xeques and Mohanes got together and agreed to begin the land run the next day, which was the greatest ceremony and sacrifice they could do for their god...
\footnote{ ``...La primera ceremonia que hicieron fue salir de ambos campos muy largos corros de hombres y mujeres danzando con sus instrumentos músicos, y como si entre ellos no hubiese habido rencor ni rastro de guerra. En aquella llanada que había entre los dos ríos que dividían los campos, con mucha fiesta y regocijo se mostraban los unos con los otros, convidándose, comiendo y bebiendo juntos en grandes borracheras que hicieron que duraban de día y de noche, a donde el que más incestos y fornicios hacía era más santo, vicio que hasta hoy les dura. Por tres días continuos dura esta fiesta y borracheras, y al cuarto día se juntaron los jeques y moanes, y acordaron que al siguiente día se comenzase a correr la tierra, que era la [fol. l0v] mayor ceremonia y sacrificio que hacían a su dios...'' \citep[chap.~IV:f9r]{Freile1636}} 

...at the dawn, the rumble of voices, trumpets, bagpipes and sea shell trumpets  was heard in the high mountains,\label{conch-guatavita} showing that the Guatavita's faction was the first to bare, and from the Bogotá's side  they went out in a great hurry to get to their assigned places, as decided by the Xeques and Mohanes. People covered the mountains and valleys, all running, as if to win a race.\footnote{``...al romper de el alba se oyeron grandes vocerías en las cordrilleras altas con muchas tronpetillas, gaitas y fotutos, que demostraban cómo el campo de Guatavita era el primero que había sa- [fol. 11r] lido a la fiesta, con lo cual, en el de Bogotá no quedó hombre con hombre, porque salieron con gran priesa a ganar los puestos que les tocaban y estaban repartidos por los jeques y moanes; cubrían las gentes los montes y valles corriendo todos, como quien pretende ganar el palio...'' \citep[chap.~IV:f10v]{Freile1636}}
\end{quotation}

The value of this story, initially pointed out by \citet{Lopez2004} \citetext{priv. comm.}, probably is  key to the understanding the Muisca calendar. The date given to this event  by \cite{Freile1636} was the year of 1538, the same year of the Spanish conquest, this being a clue that could help to correlate the Christian and Muisca calendars. 

Unfortunately, according to the other chroniclers, this year is inconsistent with the probable time of the hostilities between the Bogota and Guatavita. In general, the accounts describe the events occurring under the rule of the Zipas  \textit{Saguanmachica}, \textit{Nemequene}, \textit{Tisquesusa} and \textit{Sagipa},   the advent of the Spaniards being during the Tisquesusa's rule. \citet{Piedrahita1666}, according to \textit{the computation of moons that the natives do}\footnote{``...seg\'un el c\'omputo de lunas que hacen los naturales...'' \citep[II:~chap.~1]{Piedrahita1666}} traces back the reign of Saguanmachica to 1470, Nemequene to 1494, and Tisquesusa to 1514. Sagipa would have taken the power in 1538, after the death of Tisquesusa, who was killed during the Spanish occupation. Chroniclers such as \citet{Simon1625}, \citet{Castellanos1601}, and \citet{Piedrahita1666},  all agree that to describe the enmity between the Bogot\'a and Guatavita goes back to the times of  Saguanmachica, who  rebeled against the Guatavita's control. The chroniclers described also how Nemequene subsequently defeated Guatavita and proceeded with an expansionist campaign against the Za\-que of Tunja, conquering powerful cacicazgos loyal to the Zaque, such as the Cacicazgo of Turmequ\'e. Nemequene was, however, killed during a battle against the Zaque forces, which caused the retreat of the Bogot\'a advance. Afterwards,  Tisquesusa,  Nemequene's nephew  became the new Zipa, and prepared to renew the hostilities, although these plans were thwarted by the arrival of the Spaniards.

\begin{figure}[t]
 \centering
 \includegraphics[width=12cm]{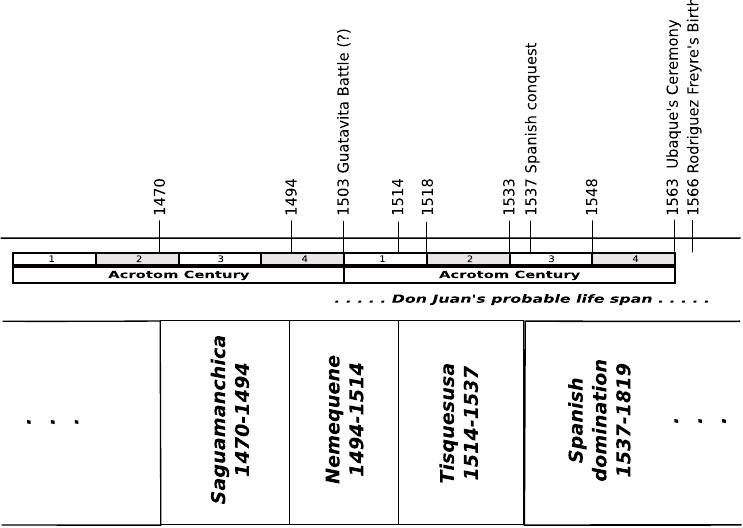}
 % muiscatimeline.pdf: 356x253 pixel, 72dpi, 12.56x8.93 cm, bb=0 0 356 253
 \caption{Time line of the Muisca's last history}
 \label{muiscatimeline}
\end{figure}

Freyre condensed this sequence of events into a very unlikely two-years timespan, may be because of a poor interpretation of the information. The chronicler admits having heard this story during his youth from a first-hand witness, Don Juan, heir of the Cacicazgo of Guatavita, who in the times of the Spanish invasion was \textit{fasting to succeed his uncle's lordship}\footnote{``...en el ayuno para la sucesión del señorío de su tío...'' \citep{Freile1636}}. Since Don Juan was  old  when he met Freyle\footnote{Juan Rodríguez Freyle was born in Santafé de Bogotá in 1566.}, one could assume that  his childhood was spent during the last years of the Nemequene rule, and may have been a direct or indirect witness of the battles against Bogotá, and perhaps participated in the ceremony (see figure \ref{muiscatimeline}). Therefore, by means of the Don Juan's testimony, Freyle describes the performance of a massive ceremony on the eve of the conquest, which we could link to the Ubaque one, and that had a religious-calendar nature. 

\section{The Duquesne's Acrotom Century and its \\ matching with the Great Conjunction cycles}
According to what was exposed in  chapter \ref{apuntes-duquesne}, Father Duquesne argues that the Muisca's Acrotom  Century had a span of approximately sixty years, which resembles the timespan of the cycle of Great Conjunctions, fact already noticed by researchers like \cite{Morales2003}.  Hence, there is a  possibility that the Muiscas did use the timespan of three consecutive conjunctions (sixty solar years), assimilating it as the Acrotom Century, and then splitting it in quarters of fifteen years (see page \pageref{muisca-centuries}), each one marking the time to perform a religious celebrations. So, we could try to fit the  Duquesne's model to the  Muisca's chronology outlined, starting with the assumption  that the  Ubaque's ceremony of 1563 celebrated the end of a century. Thus, if such a century should have started sixty years before, in 1503, it should be divided in quarters corresponding to the years 1503, 1518, 1533, and 1548,  these dates being the possible candidates for the perfomance of public ceremonies. 

\begin{figure}[t]
\begin{center}
\includegraphics[width=13cm]{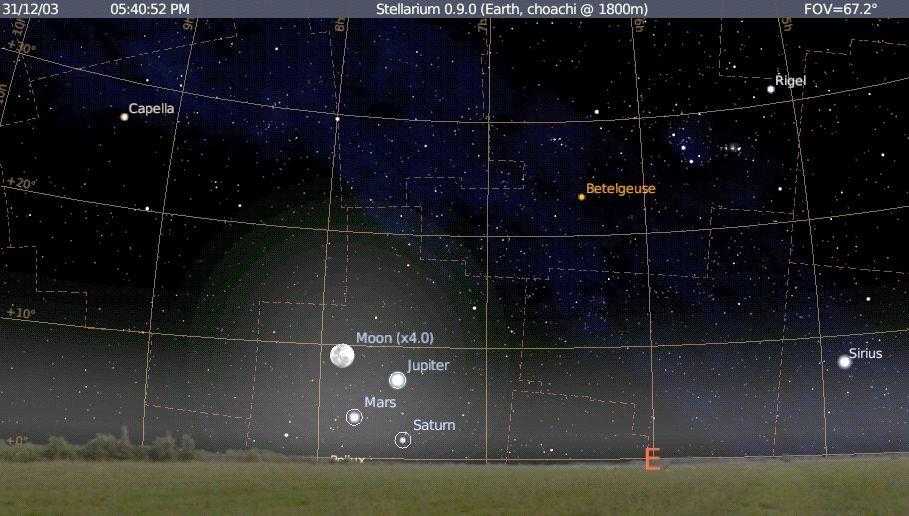}
\caption[Simulation made in \textit{Stellarium} \citep{Chereau2005} showing the Great Conjunction of 31 December 1503.]{Simulation made in \textit{Stellarium} \citep{Chereau2005} showing the Great Conjunction of 31 December 1503. The positions of Jupiter, Saturn, Mars and the Moon are shown as seen from the Ubaque's region, at 17:40, local time (daylight removed). }
\label{fig-conj31-12-1503}
\end{center}
\end{figure}

So far, no historical record of ceremonies in 1548 exists, a fact which agrees with the declaration of the Ubaque witnesses, who only remember a similar ceremony before the ``arrival of the Christians''.  Astronomically, both 1533 and 1503 are significant: the year 1533 marks the middle of that century, showing in December 31 the planets Jupiter and Saturn located in opposite places of the sky\footnote{Jupiter, $\alpha$: 18h 53m 5s, $\delta$: -23\grados{} 2' 3''; Saturn, $\alpha$: 7h 35m 57s, $\delta$: 21\grados{} 46' 39''}, having a rising azimut of 113\grados{} and 68\grados{} respectively, their rising points being near to the sunrise points during the solstices of June and December. In December 31 at 11:07am, the moon was in conjunction with Saturn, having an angular separation of 2.7\grados{}. In the year 1503, following the Great Conjunction cycle,  Jupiter and Saturn were near in the sky, but interestingly, the planet Mars and the moon adjoined  the conjunction during the month of December (figure \ref{fig-conj31-12-1503}). Again, on December 31, all  three planets and the satellite had a mean angular separation of about 5.85\grados{} (see table \ref{conj31-12-1503}).

\begin{table}
\begin{center}
\begin{tabular}[h]{l|cccc}
        & Moon & Jupiter & Saturn \\\hline
Jupiter & 5.3\grados{} &         &              \\
Saturn  & 9.2\grados{} &  5.3\grados{}   &              \\
Mars    & 5.5\grados{} &  5.0\grados{}   &  4.8\grados{}         
\end{tabular}             
\caption{Angular separation between Moon, Jupiter, Saturn and Mars during the conjunction of 31 December 1503}
\label{conj31-12-1503}
\end{center}
\end{table}

If the battle against Guatavita really happened during the rule of Nemequene, it seems probable that such a ``Jubilee year'', described by Freyle, could point to the Great Conjunction of 1503. In consequence, the Xeques' imposition of a truce was clearly justified in terms of the religious use of their calendar, forcing the sides to stop the war and attend together the ceremonies. Similarly, a (Muisca) century later, the same reason motivated the Muiscas to celebrate the end of the century, defying the Spanish control against the ``native pagan traditions''. If this is the case, we have evidence in the case of the Ubaque's ceremony and the Guatavita's battle that the Muiscas paid significant attention to the planetary phenomena, and its role determined  the numerical elements of their calendar. 

If the Acrotom Century is effectively bound to the span covered by three Great Conjunctions, a question arises: Why did they  not celebrate all the conjunctions each twenty solar years (246 moons), and instead imposed ceremonies each fifteen solar years (185 moons), thus covering  the span of three conjunctions? I believe the Muisca priests considered it important to fix the century in the series 1563, 1503, 1443, etcetera, because only during those Great Conjunctions the participating planets rose in a region of the horizon near to the sunrise point of the June solstice, which would have been significant to the native priests. The importance of the solstice's direction in the Muisca's use of the geographical space has been reported elsewhere \citep{Dolmatoff1986,Izquierdo2000,Morales2003}, and seems to be associated to the mythical route of Bochica across the territory~\citep{Morales2003}.

Note that the hypothesis presented in this chapter comes from evidence that, although precise, is yet incomplete. However, this gives us motivations to search for previously unrecognized clues in the historical sources, that lead us to complement our understanding of this issue. Hopefully, under Belmonte's classification of the archaeoastronomical work (see section \ref{belmote-classification}), I consider these ideas strictly as  \textit{serious speculations}, and await for future data to contribute to their validation.
\newpage

\begin{figure}[h]
 \begin{center}
          \includegraphics[width=15cm]{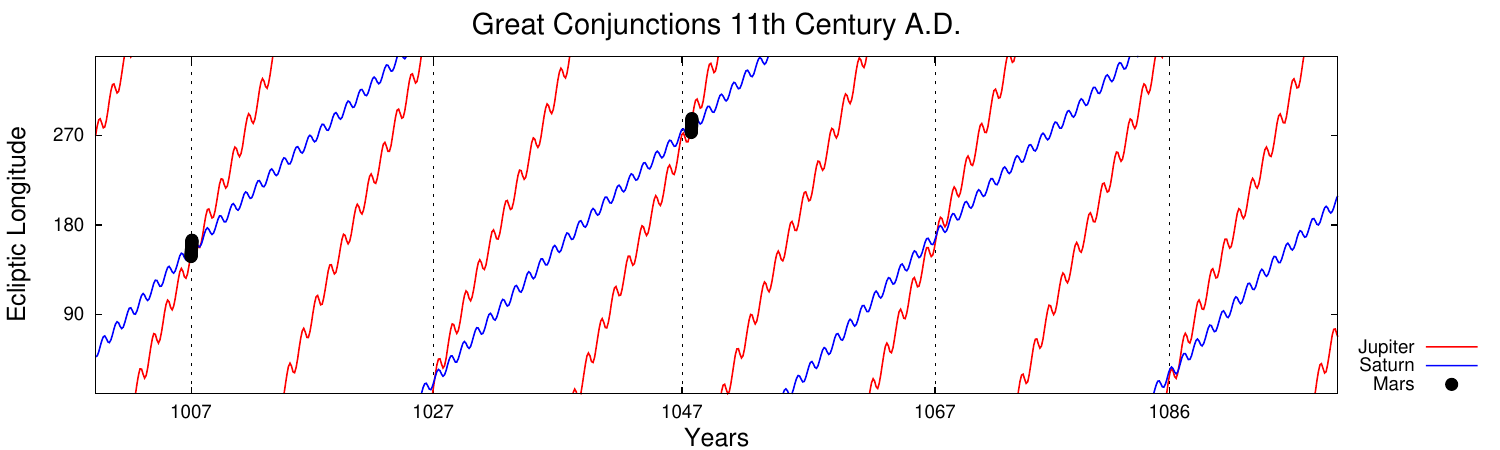}
          \includegraphics[width=15cm]{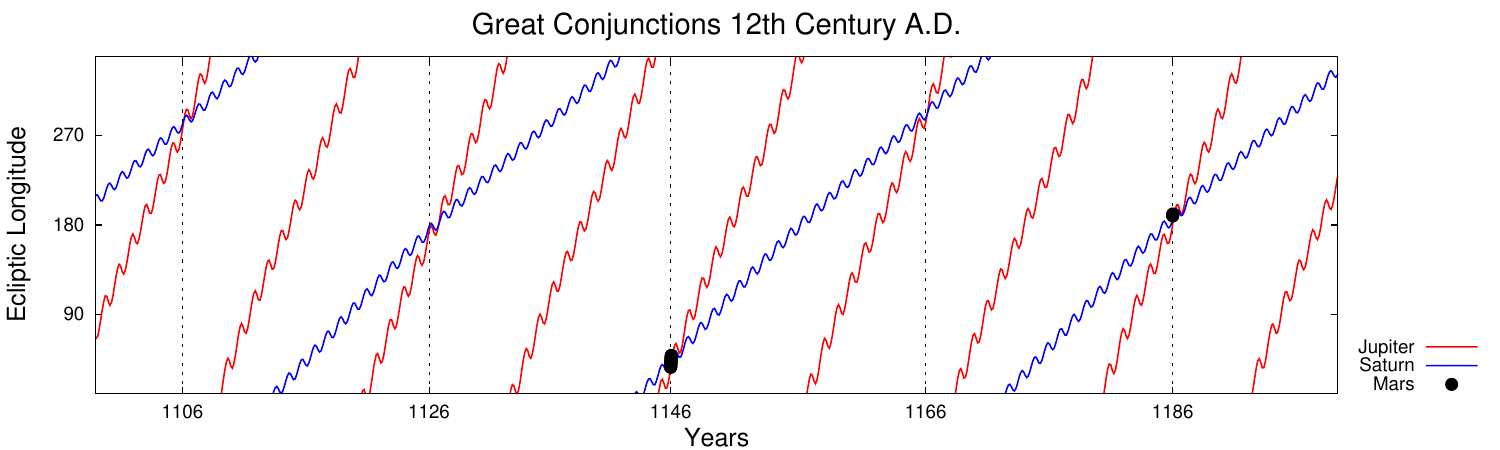}
          \includegraphics[width=15cm]{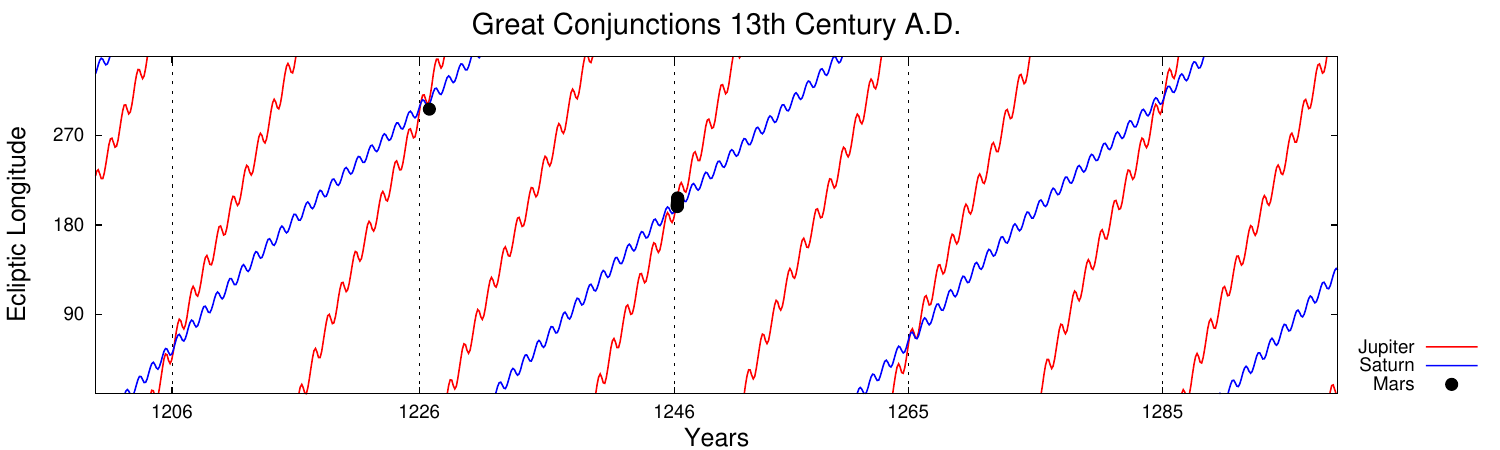}
\end{center}
\caption[Great Conjunctions between 1000 A.D. and 1600 A.D.]{Great Conjunctions between 1000 A.D. and 1600 A.D.. The $x$-axis shows the years of each conjunction. Planet's locations ($y$-axis) are in ecliptic coordinates (longitude $\lambda$, expressed in degrees). Basically,  any intersection of the %blue and red 
ondulated lines corresponds to a close encounter of Jupiter and Saturn somewhere in the sky.  The black spots correspond to the Mars' presence  when it participates also in a Great Conjunction. The vertical dotted lines mark intervals of twenty solar years. \textit{(continued in figure \ref{conjgraph2})} }
\label{conjgraph1}
\end{figure}

\begin{figure}[h]
 \begin{center}

          \includegraphics[width=15cm]{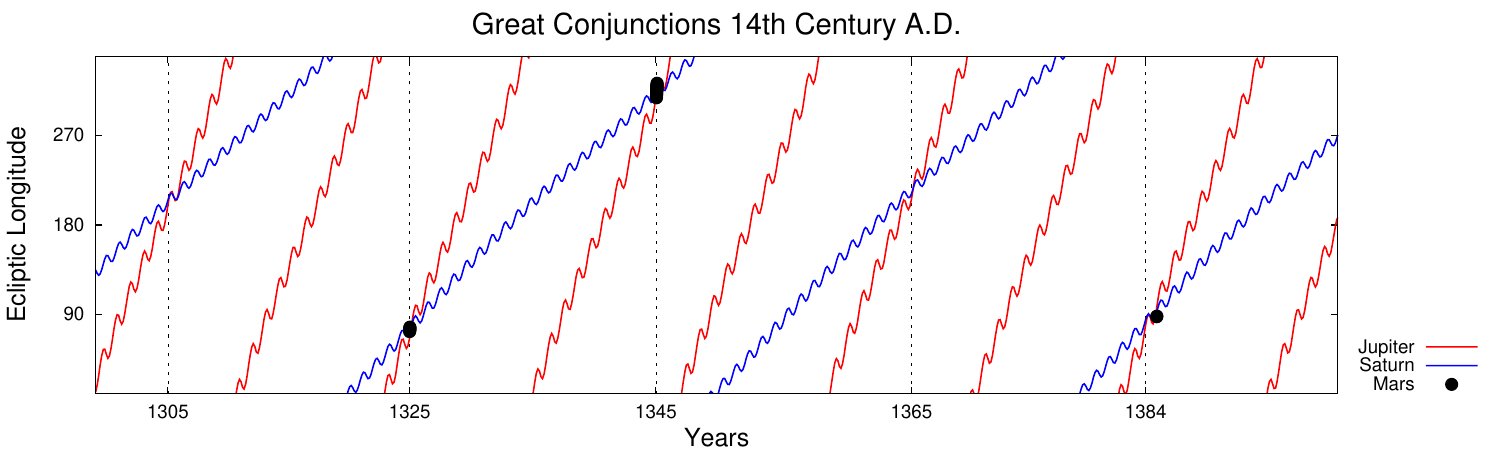}
          \includegraphics[width=15cm]{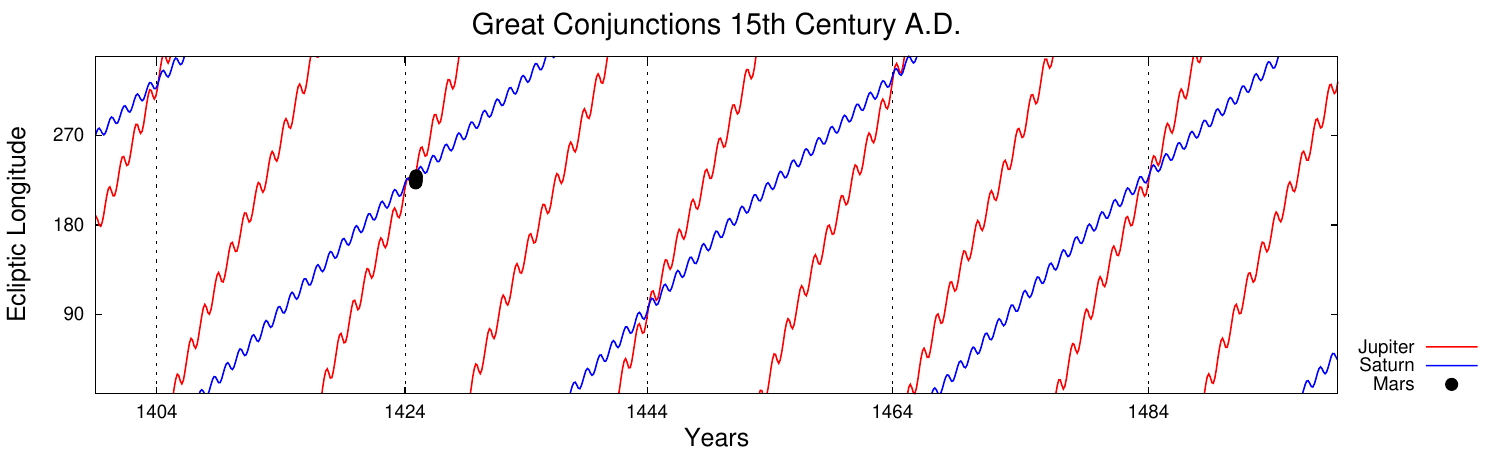}
          \includegraphics[width=15cm]{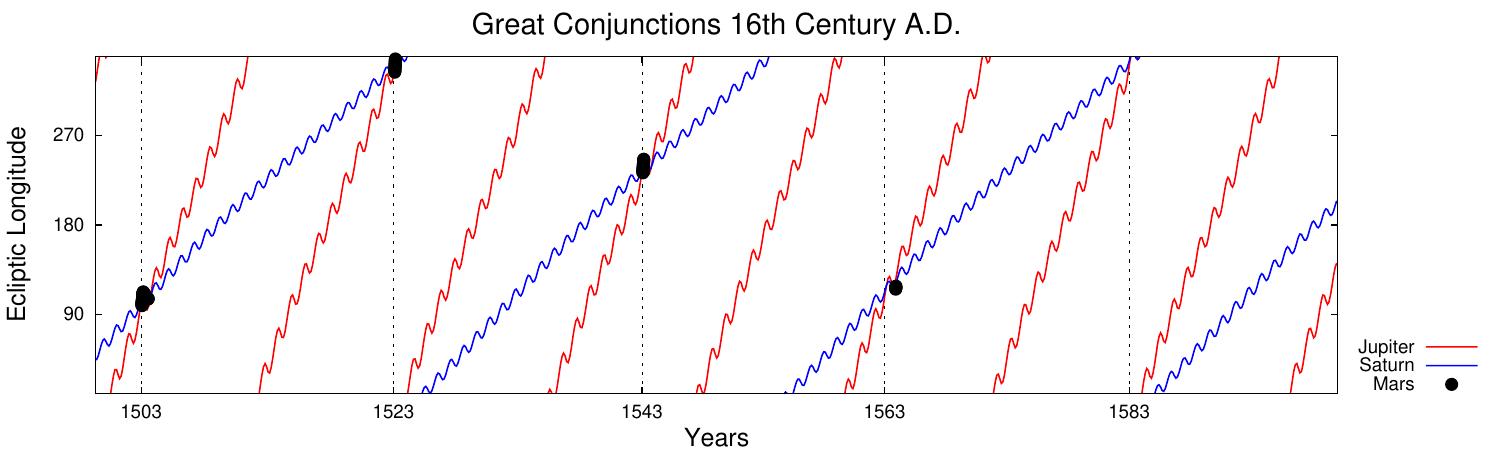}
\end{center}
\caption{Great Conjunctions between 1000 A.D. and 1600 A.D. \textit{(continuation of figure \ref{conjgraph1})}. }
\label{conjgraph2}

\end{figure}

\chapter{Archaeological artifacts associated to the calendar}\label{stones-chapter}

\section{The Duquesne's calendar stone}
\begin{figure}[h]
 \centering
 \includegraphics[width=9cm]{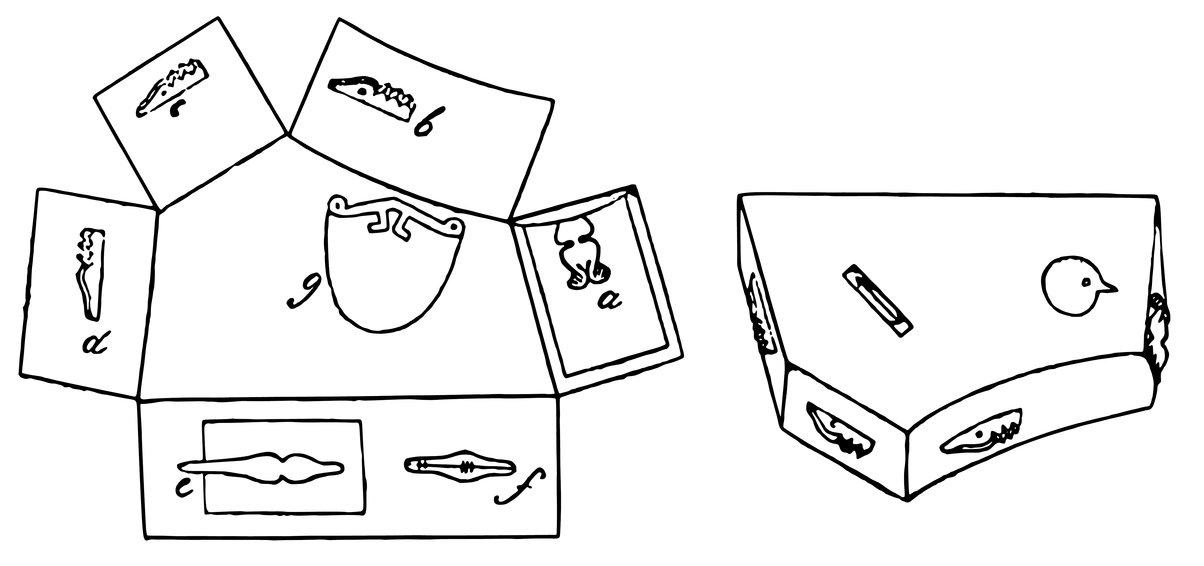}
 % duquesne-stone.pdf: 2310x1086 pixel, 72dpi, 81.49x38.31 cm, bb=0 0 2310 1086
 \caption{Stone described by Duquesne.}
 \label{duquesne-stone}
\end{figure}

Father Duquesne, in the final parts of both  the Astronomical Ring and the Dissertation, describes a small artifact of stone that he obtained from his indigenous informants and presumably contained calendar information \citep{Humboldt1878,Zerda1882}.  This artifact was made on black stone and was irregularly pentagonal in shape, having nine carved figures distributed along its sides, most of them are described as iconographic variations of the toad and the snake (figure \ref{duquesne-stone}).

According to the clergyman's interpretation, this artifact was a representation of the sequence of 185 moons which composed the first Astronomical Revolution of the Acrotom Century.  In an ingenious interpretation of its iconographic contents (see pages \pageref{stone-dissertation} and \pageref{stone-ring}), Duquesne concludes that the five sides of the stone correspond to the five Acrotom years of such a Revolution. Furthermore, each one of the nine carved figures, are interpreted as the nine zocam years by a 185 moon span\footnote{See the table of the appendix \ref{muisca-centuries-table}, for the moon number 185, and note that such moon occurs during the nineth Zocam year from the start of the century.}. A set of four  such  stones --- according Duquesne--- should yield enough information to describe an entire Acrotom Century \citep{Humboldt1878}.    

\begin{figure}[t]
 \centering
 \includegraphics[height=4.5cm]{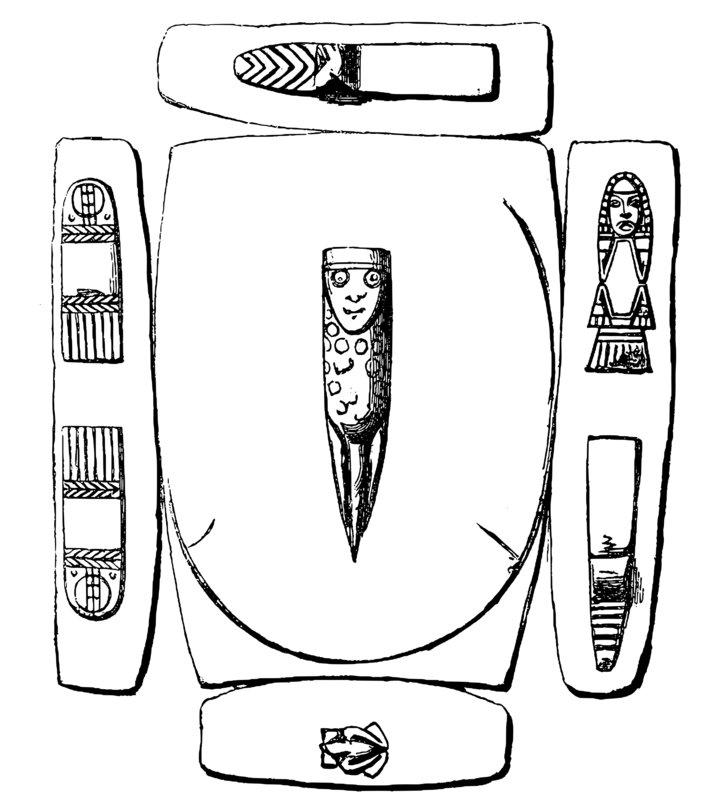}(a)
 \includegraphics[height=4.5cm]{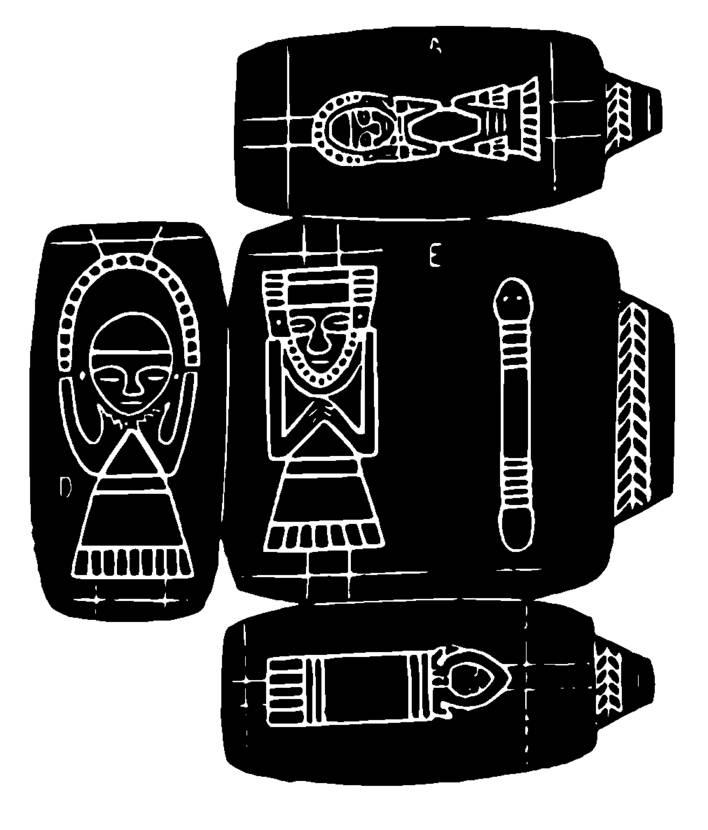}(b)
 \includegraphics[height=4.5cm]{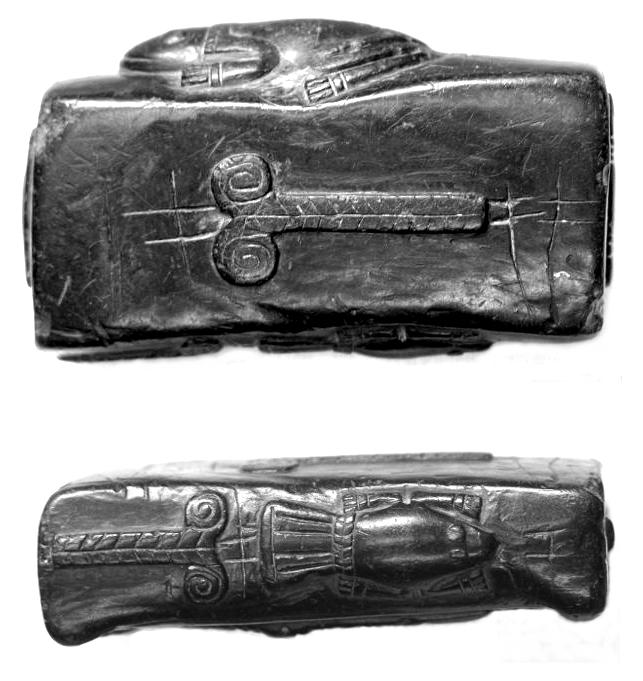}(c)
 \caption[Muisca carved stones.]{Muisca carved stones. (a) Stone described by \cite{Acosta1848} (b) Stone described by \cite{Zerda1882} (c) Stone of the Museo Nacional de Colombia's collection.}
 \label{stones}
\end{figure}

This interpretation of this stone artifact, has lead other scholars after Duquesne, (i.e. Acosta and Zerda, 1882), to publish similar descriptions of new stones found in both private and museum collections (figure \ref{stones}). Interpreted under the lens of Duquesne's work, they are referred to as `morphographic books'  containing calendar information \citep{Zerda1882}. 

However,  two problematic aspects of  Duquesne's approach arise: the first, is the fact that the clergyman's assignement of meanings to the iconographic motifs  is  subjective and difficult to corroborate. Essentially this is a case of \textit{Endearing speculation} according Belmonte's classification of the archaeoastronomical research (see page \pageref{belmote-classification}). The second  --- and more convincing-- aspect  is that similar stone artifacts as described  by Duquesne, Acosta and Zerda, have been identified as moulds used by goldsmiths to  mass-produce gold figurines \citep{Long1989}. This has minimized the credibility of the calendar-based interpretation of these stones, and consequently, the Muisca stone industry has been considered with skepticism regarding its calendrical and/or astronomical contents.

\section{The Choachí Stone}\label{choachi-section}

\begin{figure}[t]
 \centering
 \includegraphics[width=13cm]{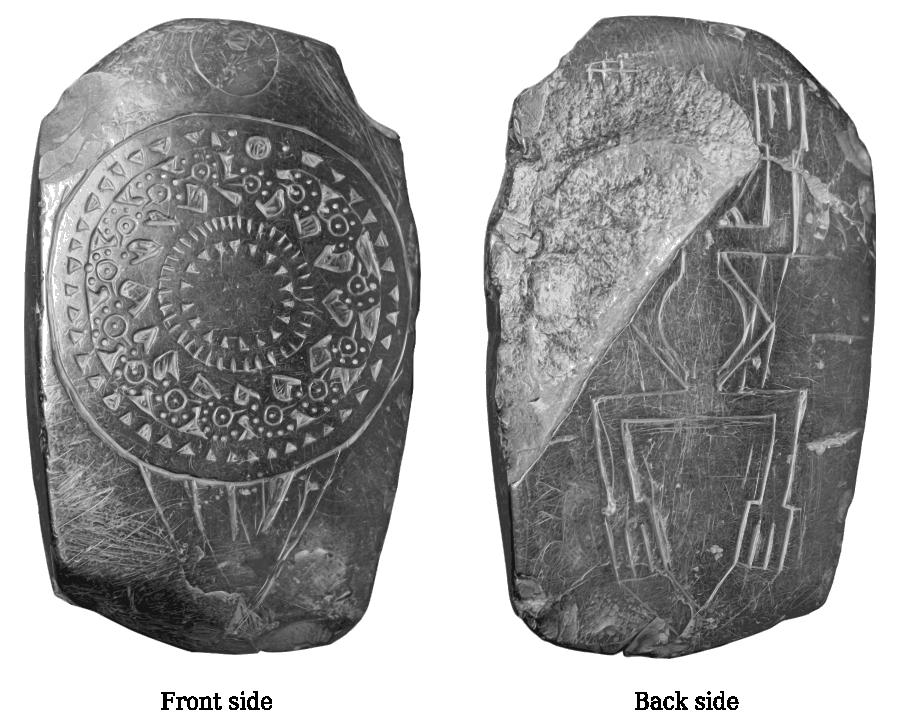}
 % piedra_choachi.png: 900x700 pixel, 72dpi, 31.75x24.69 cm, bb=0 0 900 700
 \caption{Muisca carved stone from Choachí (Cundinamarca). Museo Nacional de Colombia  (ICAN42-VIII-3920)}
 \label{choachi-stone}
\end{figure}

 Despite the apparent flaws in this part of the Duquesne's work, it is important to remark that not all the Muisca's lithic industry can be associated to goldsmith's tools, therefore it can be reconsidered, working under a less naive and more objective examination of its iconographic contents. If Duquesne elaborated his interpretation of calendar stones based on ancient memories taken from his informers, this hypothesis must be supported by the existence of further objects in the available collections, with a clearly different use than as goldsmith tools and , which may provide  more  objetive calendrical information.

A highly plausible case of one of these artifacts is an archaeological piece belonging to the Museo Nacional de Colombia,  (catalog number ICAN42-VIII-3920). It is a black (lidite) stone slab-shaped object whose dimensions measure $13\times7\times1.5$ cm  (figure \ref{choachi-stone}). Its two sides are polished bearing numerous figure carvings. It was acquired by the Instituto Colombiano de Antropología e Historia during the %Museum in the %1940's decade,
first half of the 20 century. Unfortunately, there is little information regarding its origin , except that is was found in  Choachí, a village located at twenty kilometers to the southeast of Bogotá, possibly from a grave. Its back side bears a  large fracture distorting about one third of its surface, which may be the result of careless violent extraction from its  archaeological context. The front side is, however, undamaged and depicts an impressive set of carved figures ordered in  concentric rings  (see figure \ref{study}). 

Unlike the stone described by Duquesne, the figures on this artifact provide more precise information, by depicting raw numerical values, which facilitates a safer and less  subjective analysis of its contents and their relationship to the  calendar model shown in this study. 

\section{Iconographic structure of the Choachí Stone}

\begin{figure}[t]
\centering
\begin{tabular}{cc}
\includegraphics[width=6cm]{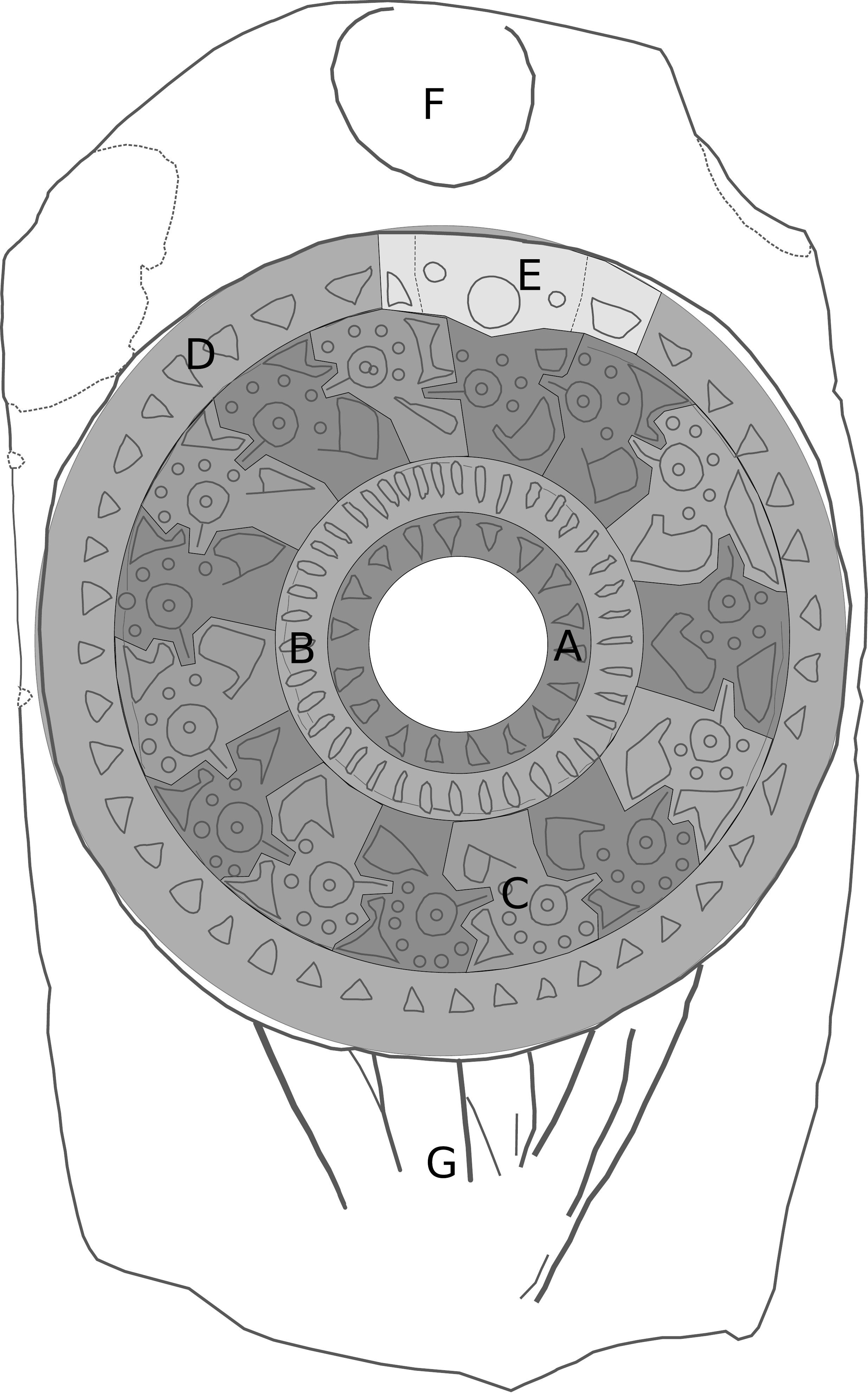} & \includegraphics[width=6cm]{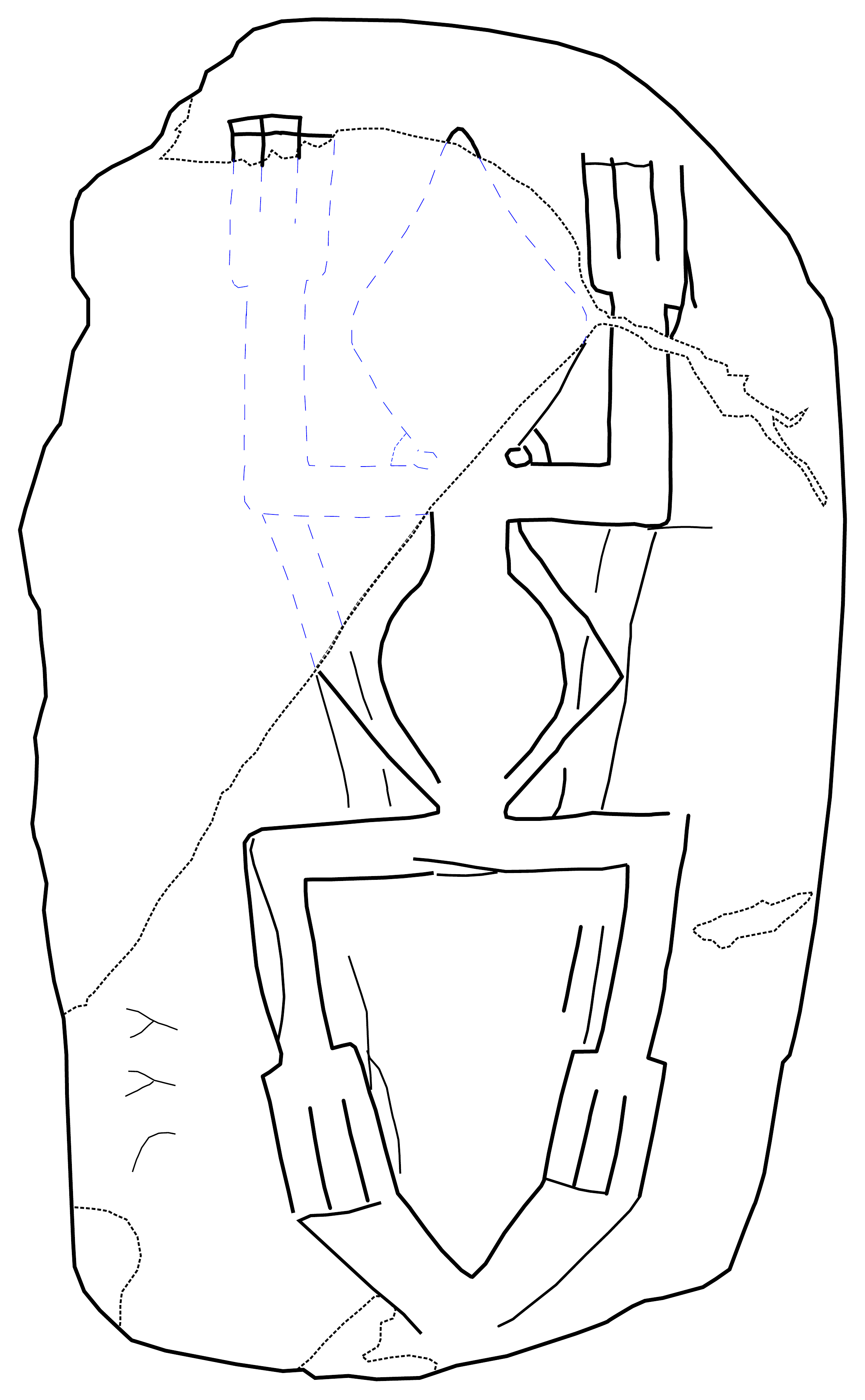}\\
Front Side & Back Side
\end{tabular}
 % choachi_study.pdf: 1185x1901 pixel, 72dpi, 41.80x67.06 cm, bb=0 0 1185 1901
 \caption{Structure of figures of the Choachí Stone.} 
\label{study}
\end{figure}

As illustrated in figure \ref{study}, the main iconographic components of this piece are in four concentric rings. In addition,  some apparently isolated figures exist outside the rings. By labeling the rings with a series of letters,  their contents are described:

\begin{description}
 \item[First ring \textit{(A)}] Shows a series of 18 small triangles whose apex points towards the center.
 \item[Second ring \textit{(B)}] Has a sequence of 40 lines, oriented towards the center.
 \item[Third ring \textit{(C)}] Shows a very complex set of 15 figures, each  composed of a bird head like figure, surrounded by two irregularly shaped figures, the inner ones resemble a flag  shape, the outer ones resemble triangles with a curved ending, and in some cases losing its triangular shape, becoming trapezoidal in form.
 \item[Fourth ring \textit{(D)} and \textit{(E)}] Shows two sets of figures, a sequence of 37 triangles, similar to the ones of the first ring, and \textit{(E)} depicts five figures that are, from right to left, a trapezium, a dot, a circle, a dot, and a triangle whose apex points away from the center. 
\end{description}

Outside these rings,  a  semicircle \textit{(F)} and  five lines \textit{(G)} starting from the bottom part of the main circle, extending themselves towards a center point of the bottom side of the stone. In the undamaged portion of the back side, there is a carved figure resembling a toad,   with  a \textsc{v} shaped carving at its feet that converges to the central bottom side of the stone.

\subsection{Hypothesis about the Choachí Stone's Nature}

\subsubsection{Acrotom year association}

In an initial observation, the 37 triangles of the fourth ring (D) appear to match  the number of lunar months of the Acrotom year.  This coincidence attracted my attention when  I first examined this piece.

\subsubsection{Zocam Centuries and Ata Cycle Association}

Although less evident, than the fourth ring, the first and second ring (A) and (B) show numbers that permit associations with the Zocam Century, the Extended Zocam Century and/or the Ata Cycle. One possibility is to consider that the 40 lines of the second ring represent the number of Zocam years of a Zocam Century, 400 if each line had a value 10. Alternatively, if each line represents a set of 20 Zocam years, the ring correspond to 800 years, an Extended Zocam year.

Another possible connection can be seen if the set of 40 lines is considered to correspond with the 18 triangles of the first ring: $40 \times 18=720$. Note that this value corresponds to the moon just before the last moon of the Acrotom Century. Note also that such a number can be divided into the span of two Ata Cycles $(160 \times 2 = 320)$ plus one Zocam Century: $320 + 400 = 720$. Therefore, 18 and 40 could be associated and expressed in terms of Ata Cycles and Zocam centuries. It is very unclear to infer  how to fit this in the system shown by Duquesne, however, I consider it may be a significant clue.

Another possible association of the number 18 is the same that is found in the mesoamerican calendars, where such a number results from the integer  division of the number of days of a solar year (365) over months of 20 days. Albeit contemporary groups liguistically akin to the Muiscas, such as the Kogi, are known to use this formula to conform their calendar \citep{Dolmatoff1949}. However, there is no clear evidence suggesting the use  of months defined this way by the Muiscas, who, according the historical accounts, had preferred a 30 day month (see section~\ref{lunarcycle}). 

\subsubsection{The numerical properties of the third ring's figures}
These properties were first reported by \cite{Izquierdo2006}, who noticed that the bird head figures of the third ring may express additional numerical values. The bird-head motif is commonly found on ceremonial pottery and spindle whorls. It is also a common motif  in most prehispanic cultures of the Intermediate Area \citep{Falchetti1993,Warwick1997}.  Ethnographic analysis of another linguistically akin contemporary group known as the U'wa \citep{Osborn1985,Osborn1995} 
suggests that the image of the bird had an  association to the seasonal movements of the sun, and hence time.  

\begin{figure}[t]
\begin{center}
\begin{small}
\begin{tabular}{cccc}
 \includegraphics[height=1.5cm]{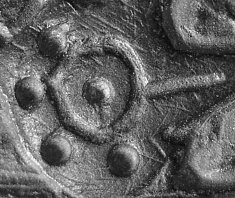} & %
 \includegraphics[height=1.5cm]{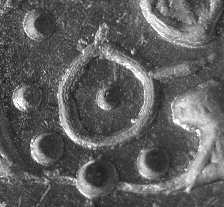} & %
 \includegraphics[height=1.5cm]{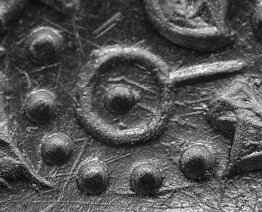} & %
 \includegraphics[height=1.5cm]{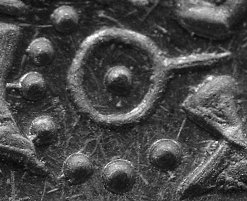} \\
\textit{Muihica} & \textit{Hisca} & \textit{Ta} & \textit{Cuhupcua}\\
4,14 & 5,15 & 6,16 & 7, 17 
\end{tabular}\\
\vspace{0.5cm}            
\includegraphics[width=7cm]{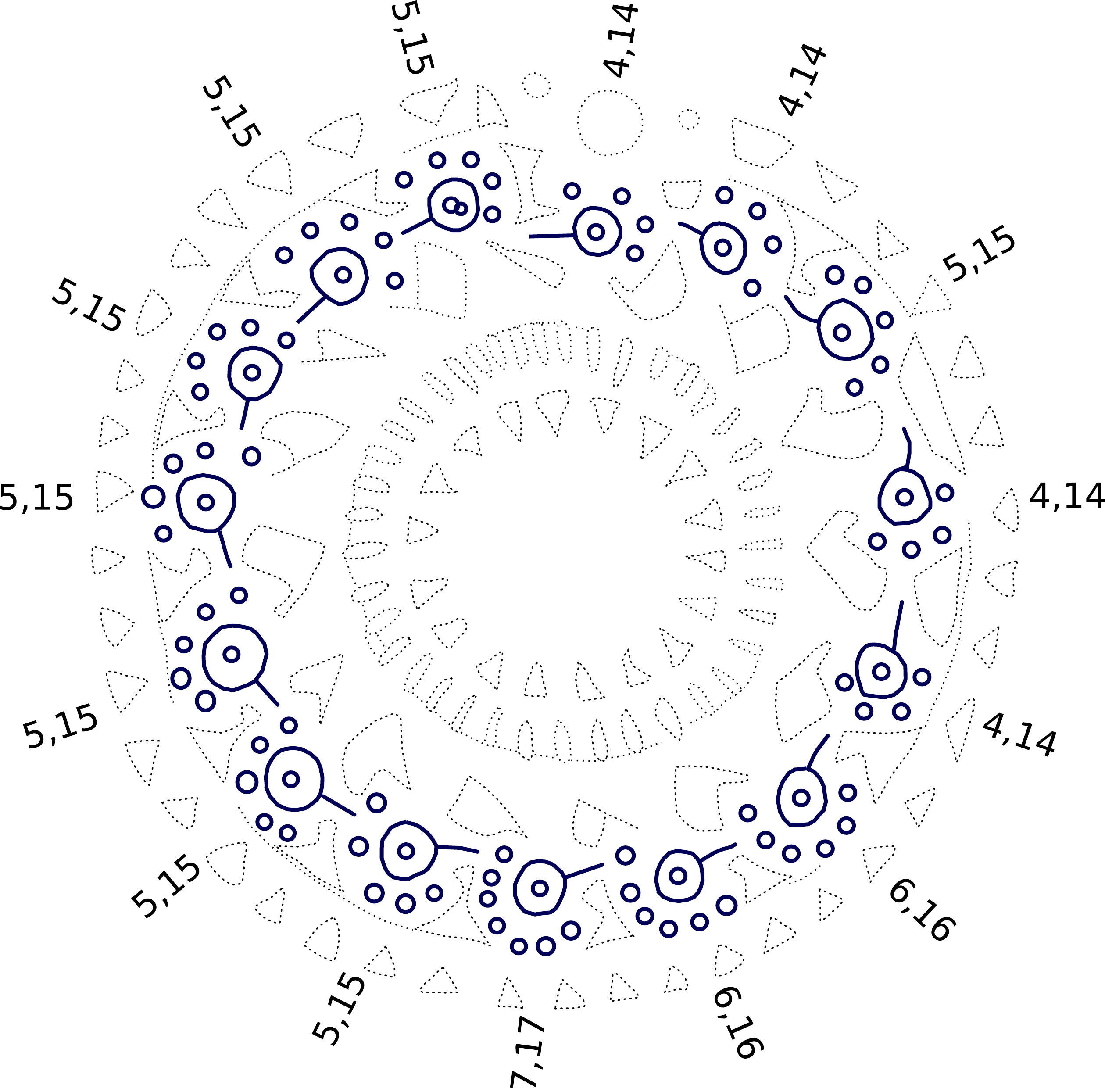}\\
\end{small}
\end{center}
\caption{Numerical values for the bird-head figures}
\label{choachi-values}
\end{figure}

In the case of the Choachí stone,  a graphical component of such motif, the feather crest, seems to be reused as a container of numerical information. Note  the number of feathers (illustrated as dots)  for each of the fifteen  bird-headed figures of the Choachí stone is intentionally set, and is not a product of any aesthetic goal, representing a sequence composed of the values 4, 5, 6, 7 (figure \ref{choachi-values}). When read counterclockwise,  the numerical values of these fifteen bird-head figures,  yields the following sequence: 5, 5, 5, 5, 5, 5, 5, 7, 6, 6, 4, 4, 5, 4, 4. 

What is the nature of such a sequence? What information is it providing? One may to consider that  such  sequence of natural numbers is being used to represent a real number by means of its average, in a similar fashion to the listings of number of days in the Maya codices. Note the average of the sequence is 5, and may suggest a special purpose of the number 5. Furthermore, considering the possibility that the circle-with-a-dot-and-a-line represents the head of each figure would have a similar use of the \textit{quihicha} prefix, which would add ten to each value, the sequence could be restated as:   15, 15, 15, 15, 15, 15, 15, 17, 16, 16,  14, 15, 14, 14, 14; whose average is 15 . This makes it even more interesting because it equals, the number of bird-head figures depicted in the third ring. 

\subsubsection{Synodic Lunar Period association of the third ring's figures} 
Subsets of the fifteen bird-heads figures whose sequence numbers can lead to values of astronomical significance: (14, 15, 15, 15), (17, 14, 14, 14), (16, 15, 14, 14). This gives each one a mean value of 14.75, which is approximately the value of half lunar synodic period \citep{Izquierdo2006}. 

\subsubsection{Ata Cycle association of the third ring's figures}
During the four years of an Ata cycle, the names of the years that finishes each year (before the supplemental series for the years 2,3,4),  follows the following sequence: \textit{Quihicha Cuhupcua, Quihicha Muihica, Quihicha Hisca, Quihicha Ta}; which could be rewritten as 17,14,15,16. Note  these are the same numbers represented in the third ring (figure \ref{choachi-values}), which could support the association of the sequence of 15 bird-head numbers to the Ata Cycle.  

\subsubsection{An hypothesis regarding the use of the Choachí Stone}

Based on the previous observations, and upon another interesting detail of this piece, its apparent ergonomics,  the following hypothetical use could be proposed: the stone's shape and the way the figures  were carved, makes this stone a suitable calculating tool, such that the user could have held the stone in  hand, and using his thumb finger, could touch  the different figures secuentially thus executing a given algorithm, until  a numerical result is obtained. The Duquesne's description the performing of finger, calendrical computations (see chapter \ref{apuntes-duquesne}) could support the ``pocket-calculator'' hypothesis for the Choachí Stone. However, if this was really the use of this stone, new questions must be considered: What was this algorithm? What was it the expected outcame?, moreover, what data was needed to start the algorithm? 

Even though this interpretation is somewhat speculative, it is clear that such numbers are not randomly chosen, rather they are intentional, serving the role of providing its owner with a tool to facilitate a particular numerical computation. However, until additional archaeological and ethnohistorical evidence surfaces,  this hypothesis should to be regarded as a plausible model, but only within the range of  \textit{Serious/Endearing speculation}.

\section{Other artifacts to be considered by future research}

An important detail of the Choachí stone is the probable reimplementation of the bird-head motif as a template to express numbers. Fortunately, there exist many archaeological artifacts depicting such a motif and deserves the attention of  researchers. The analysis of which, could shed light on our understanding of the arithmetic and the calendar of the Muiscas. 

\subsection{The ceremonial conch of the Archaeological Museum of Sogamoso}
\begin{figure}[t]
 \centering
 \includegraphics[width=10cm]{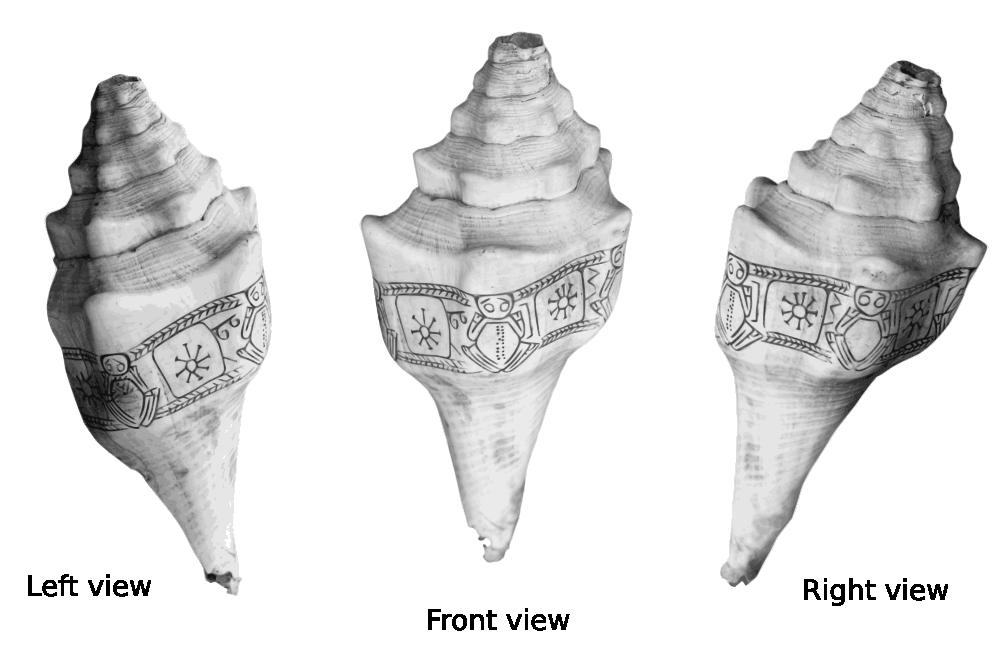}
 % concha.png: 1000x650 pixel, 45dpi, 56.43x36.68 cm, bb=0 0 1600 1040
 \caption{Conch shell bearing carved figures. From Museo Arqueológico de Sogamoso}
 \label{sogamoso-conch}
\end{figure}

This strombus conch shell was used as musical instrument (figure \ref{sogamoso-conch}), and is part  of the Archaeological Museum of Sogamoso's collection\footnote{Catalogue number \textsc{s068vi}.}. It was found by the archaeologist Eliecer Silva in the village of Socorro (department of Santander),  and corresponds to the Guane; a culture which was a neighbor  linguistically and culturally very similar to the Muiscas. The Guane also maintained close trade relations with the Muisca \citep{Lleras1987}. Strombus conchs were highly prized by the Muiscas, who used them to manufacture trumpets \textit{(fotutos)}, and were played during the religious ceremonies. Such artifact probably arrived in Guane territory through trade between the two cultures. 

\begin{figure}[t]
\centering
 \includegraphics[width=12cm]{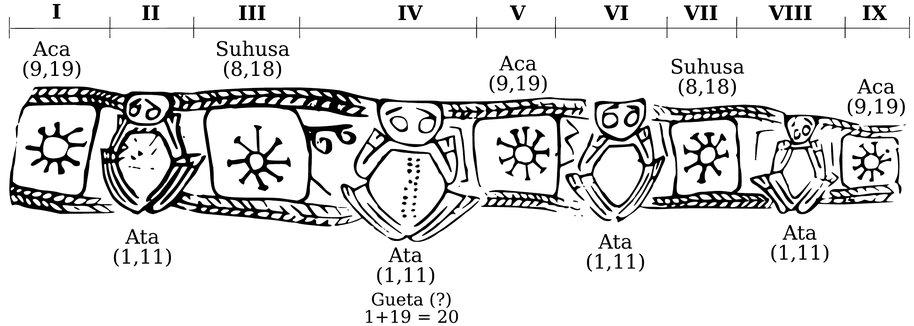}
 % shell-snakeskin-values.pdf: 1155x363 pixel, 72dpi, 40.75x12.81 cm, bb=0 0 1155 363
 \caption{Numerical values of the conch of the Archaeological Museum of Sogamoso}
 \label{conch-values}
\end{figure}

This curious artifact bears a carved strip of nine red colored figures along its surface, depicting a sequence of four  toad figures,  five rounded squares, a circle surrounded with several radial lines in the center of each square.  These figures  have been illustrated with roman numbers \textsc{i} to \textsc{xi} (figure \ref{conch-values}). 

As with the case of the dots around the bird-head figures depicted in the Choachí Stone, the  lines around the circles  suggest the intention to express numerical values, being  a possible variation  of the  bird-head theme of the Choachí Stone. With regards to the conch, the numbers 8 \textit{(Suhusa)} and 9 \textit{(Aca)} are alternated in the five circles of the strip (see sections labeled as \textsc{i, iii, v, vii, ix}, in the figure \ref{conch-values}). 

A connection with the Muisca calendar can be proposed, on the basis of these numbers and Duquesne's etymological association for the number one and the toad. Therefore, it could be speculated that the four toads represent the number 1 \textit{(Ata)} (see sections labeled as \textsc{ii, iv, vi, viii}, in the figure \ref{conch-values}). An association with the calendar can be made when these numbers are compared with the name of the first month of each Acrotom year (including the supplemental series) of the Cycle of Ata (\textit{Ata, Suhusa, Aca, Ubchihica/Gueta}, see appendix \ref{muisca-centuries-table}), According to Duquesne, these names, were used by the Muiscas to coordinate the  Cycle of Ata (see page \pageref{cycle-of-ata}).  It is interesting to note that the central figure of the strip, the toad labeled as \textsc{iv} in the figure \ref{conch-values}, bears two rows with 19 dots on its leading to the (possibility encoded)  value of 20 \textit{(Gueta)} based on the premise that: $1\,\textit{(the toad)} + 19 = 20$. Consequently this conch could be a symbolic expression of the Cycle of Ata, incorporated in the decoration of a ceremonial intrument. Furthermore it has been noted that conch trumpets were used in the context of ceremonies associated to calendar events  as described in the accounts of Spanish chroniclers (see chapter~\ref{ubaque-chapter}). 

Although hypothesis may sound attractive,  there remains some aspects not understood in this piece. For instance, the meaning of the groups of straight and curved lines drawn to in the space between of the figures \textsc{iii} to \textsc{ix}, and the symbol inside the toad labeled as \textsc{ii}, which today is obscure, depicting only tree dots in the upper part of the chest of the animal.  Duquesne not only associated the toad with the one, but with  nine and twenty, so why is the number nine not represented by a toad?. Until more evidence is made available, the interpretation of this piece has to be regarded as an \textit{Endearing speculation}.

\subsection{Spindle whorls}
\begin{figure}[t]
\centering
{\footnotesize
 \begin{tabular}{cccc}
\includegraphics[width=3cm]{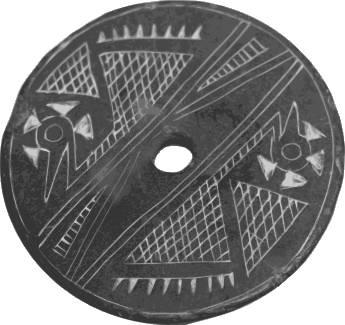} (4)&%
\includegraphics[width=2.2cm]{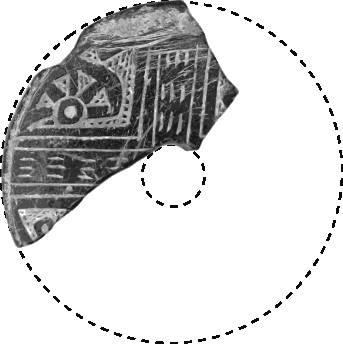} (4)&%
\includegraphics[width=2.3cm]{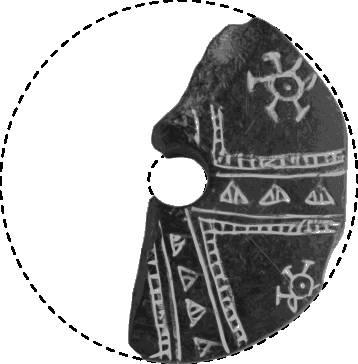} (4)&%
\includegraphics[width=2.0cm]{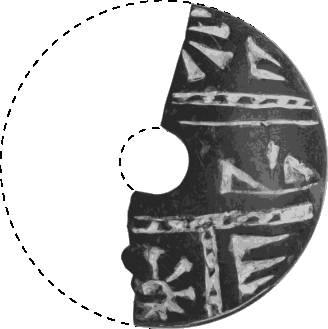} (4?)\\%
\includegraphics[width=3.3cm]{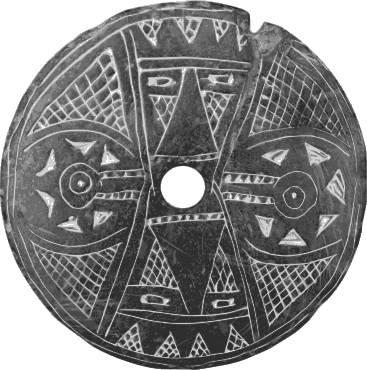} (5)&%
\includegraphics[width=2.5cm]{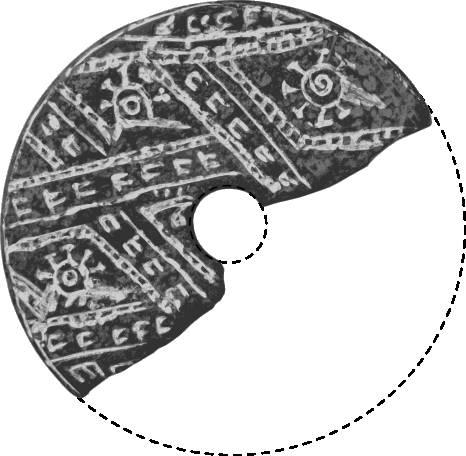} (5)&%
\includegraphics[width=3cm]{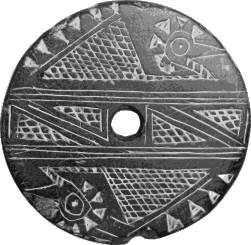} (6)&%
\includegraphics[width=3cm]{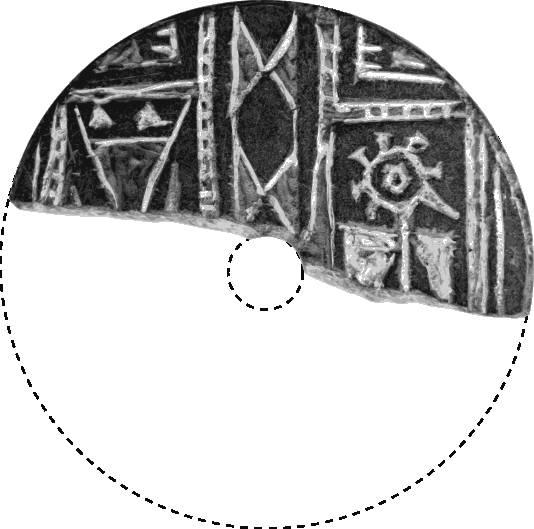} (6)\\%
\includegraphics[width=2.0cm]{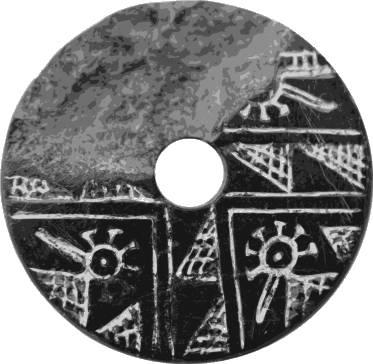} (6)&%
\includegraphics[width=2.5cm]{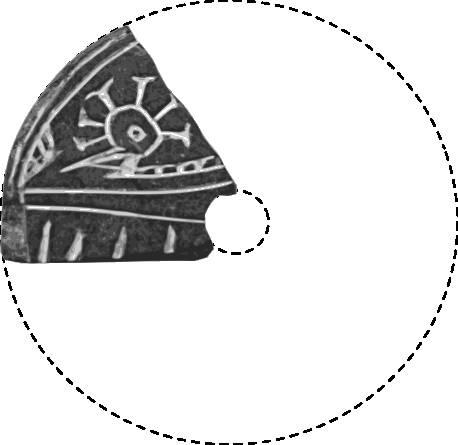} (6)&%
\includegraphics[width=3cm]{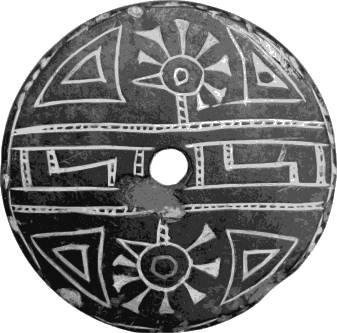} (7)&%
\includegraphics[width=2.5cm]{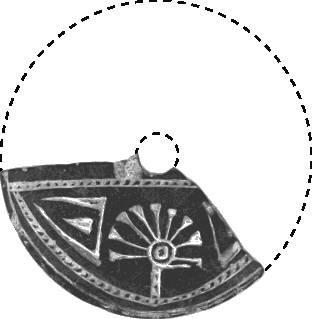} (8)
\end{tabular} 
}
 \caption[Spindle whorls of the Archaeological Museum of Sogamoso]{Spindle whorls of the Archaeological Museum of Sogamoso depicting the bird-head theme. The numerical value expressed in each bird head image is indicated in parenthesis at the bottom right of each example.}
 \label{spindle-whorls}
\end{figure}

Another example of artifacts worth mentioning are the stone spindle whorls of the Muiscas.  In the case of the Kogi culture, the Earth is seen as  ``the spindle whorl of the Mother Goddess'', whose spindle,  which is associated with the Orion constellation is the ``central post'' of the world, providing a canon for the construction of the Kogi temple that includes several elements of the Kogi calendar \citep{Kelley2004}. As the Muisca group shared several traits with the Kogi, the study of the spindle whorls iconography eventually gets relevance for understanding their calendar. 

A work aiming to classify these artifacts according their shape, size and decoration was undertaken by \cite{Silva1985}. A remarkable feature of Muisca spindle whorls is that unlike other pre-columbian cultures, who manufactured them out of clay,  the Muisca preferred to make theirs out of stone. Their decoration  is highly geometrical, portraying a complex combination of motifs conformed  by such geometrical primitives as straight and curve lines, spirals, circles, and triangles, representing both abstract and zoomorphic themes.  A preliminary classification of such decoration was made by \citep{Garcia1971}. A common decorative theme  is the bird head, which apparently displays intentional variations in the number of feathers on its head. Figure \ref{spindle-whorls} shows some  examples of spindle whorls  from the Archaeological Museum of Sogamoso's collection which bear this theme.  It can be noted that the number of feathers on each bird's head ranges from 4 to 8, additionally, some style variations are noteworthy. For example  the beak is omitted in some cases. However, it would be premature  to argue the presence of  calendar traits interpreted from  numerical contents of  spindle whorls. The arithmetic significance  of bird head theme, seems to be only a small component of a larger pictorical system of representation,  where it interacts with additional pictorical themes possibly related to diverse elements of the Muisca life.  The complete analysis of the semantics of these  decorations would require a whole new dissertation to itself, and lies beyond the scope of this project.

\section{Nota bene}
The aim of this chapter is not to assert the true meaning of the images appearing on the described artifacts, but to draw the attention of researchers to the possibility of a potential numerical representation schema in these objects.  Such interpretation applied in conjuction with Duquesne's calendar model could yield clues in each other's interpretation. This project is intended to drive the examination of more artifacts, in order to confirm or refute interpretations regarding the existence of a numerical notation system as depicted in artifacts from this region, probably associated the Muisca calendar, as it has been suggested by the Choachí Stone this far.

\chapter{Concluding remarks}

It could be said that the primary conclusion of this work is  demonstrating that Father Duquesne's descriptive model is not product of his ``imagination'' as criticized by Vicente Restrepo. Rather, it is the product of an  exercise in reflection regarding an authentic and legitimate timekeeping system, kept by oral tradition along three hundred years since the Spanish conquest by anonymous Muisca elders, a system that for Duquesne (and even, currently for us) was strange and curious.  He intended to understand and describe such a system under the logic of the western calendars, and perhaps this made his work confuse and misunderstood by the next generation of scholars.

After completing the ardous task of compiling the data used in chapter \ref{apuntes-duquesne}, I began to recognize that the `calendar'  Duquesne interpreted was really  part of  a complex system of time cycles devised by the Xeques, which was based on a combination of astronomical cycles and  arithmetic properties of  `propitious' numbers. The  master of the esoteric use of such numbers via their arithmetic properties and their application to manage religious activities in function of the natural phenomena, which was likely a key component of Muisca religion. On one hand, they used the prime number 37, in order to condensate many  astronomical cycles; on the other, the numbers 20, and its factor 5, represented culturally determined values and were likely selected for anthropomorphic reasons. Derived from these are two constituent time cycles: one is the Acrotom Century, with its astronomical association with the Sun, Moon and planets;  the other, the  Cycle of Ata, its roots in the  numeric-mystical traits of the Muisca religion. The aritmetically derived results from  their propitious numbers will bring forth a mythical, temporal frame, where their civilizator hero can be located and hence, the foundations of their own culture lie. It seems likely that other cycles existed, based on other key numbers, which currently remain unknown. It is also important to say that not all these astro-arithmetical cycles had a direct agricultural application, and possibly the `calendar' of Duquesne was  a non agricultural component of the Muisca timekeeping system, aimed at measuring long time spans. This may explain why this system is unsuitable for agriculture, as previously noted by  many other researchers \citep{Rozo1997}.

It can be argued that numbers contained further concepts beyond  simple  quantities for the Muisca, and embraced ideas related to the cyclical and  celestial manifestations (as asterisms in the sky) of time. Thus,   numbers are used to express the quantities, days, months, and years\footnote{See section \ref{cycle-of-ata} how the different years of the Cycle of Ata are named by a given numerical name.} as well as constellations. So numbers and all their arithmetic, astronomical and calendrical implications, can be understood as ideological artifacts manipulated by the Xeques, as devices to warrant their religious power.

This esoteric treatment of  numbers, and their astronomical interpretations, are not, however, unique to the Musicas. Many prehispanic cultures of Mesoamerica based their calendar systems on elaborate numerical systems, used their properties to record large time spans, and operated over a mystical framework  of their religion. 
This resembles the way the Muiscas designed their calendar, and in a sense, resemblances between the two systems are a difficult issue to deal with.  The presence of twenty as a base of the numbering system, and the use of concurrent time series (or ``calendar wheels'' in  Mesoamerican gargon), are indicators that  knowledge of  Mesoamerican arithmetic and timekeeping methods were known beyond the boundaries of the cultural realm of Mesoamerica. When describing aspects of the ancient Tairona culture\footnote{Contemporary to the Muiscas and Chibcha speakers also. Located on the northern coast of Colombia, their descendants are the current Kogi culture.}, Tom Zuidema,  addresses the issue in a direct way:  ``While these relations \textsl{[the ones with the Andes]} are rather general, connections with Mesoamerica are more specific. The Kogi calendar of eighteen months of twenty days each and their vigesimal system of counting, known also to the Tairona and the Muisca, could have been derived only from Mesoamerica, although applied to a non-Mesoamerican world. The Mesoamerican model may have reached more cultures in northern Colombia than we realize at present'' \citep{Zuidema1992}.

Similarly, although less evident, similar elements akin to Andean Cultures are illustrated in the Muisca calendar. For instance, the 30 day lunar month was used in the Inca calendar \citep{Zuidema1977,Bauer1995}  as well as the practice of intercalating one month each three lunar years in order to synchronize the sun and moon as argued by \cite{Ziolkowski1992} as a component of Inca astronomy. Furthermore, the significance of the movements of the planet Jupiter and Saturn in the Andean astronomy has been already suggested by \cite{Sullivan1988}.  Similarly, the possible identification of  `dark constellations' by the Muiscas (see section \ref{muisca-constellations}) in the celestial background of the Milky Way  is  a trait clearly identified in both past and present Andean traditions \citep{Urton1981a}. 

As a primary conjecture, it can be argued that when the Muiscas developed their own calendrical system, they  eclectically incorporated concepts  from the astronomical and arithmetic traditions of the main cultural centers of America,  developed before the emergence of Muisca Culture in the east-central highlands of  Colombia during the 9th century AD.  This is clearly demonstrated by in the  insistence,  to use lunar months (Andes) as base units of the `calendar rounds' (Mesoamerica), which compose the Cycle of Ata and the Acrotom Century of the Muisca system.  Such an eclectic approach, could have been useful in terms of dealing with some lunar periods. For example, \cite{Zuidema1977} argued  that the Incas used a lunar year of 328 days based on the sidereal\footnote{The duration of the Moon's return to the same point in the sky, relative to the background stars.} period of the Moon ($27 \frac{1}{3}$ days), opposed to the sinodical one\footnote{The time between two identical lunar phases.}. Perhaps the Muiscas would have been aware of such a system of calculation, because it is notewhorthy to see that by combining this period  with the Mesoamerican twenty, by multiplication yields a timespan of 6560 days, which falls roughly into the fifth day of the moon number 222 of the Acrotom Century (see table of appendix \ref{muisca-centuries-table}). Interestingly, this month  marks the end of the Acrotom year  six of that century. Another equally interesting association is that the next month (the 223,  the first of the seventh Acrotom year), will start on   6585.3213, which matches the Cycle of Saros\footnote{Cycle wherein two identical eclipses are produced.}, of  18 solar years and 11 days, exactly. This  makes such a cycle very easy to track within the Muisca framework, every six Acrotom years. It is difficult to determine if the Muisca were actually aware of these astronomical cycles. However, any Muisca priest, dealing experimentally with lunar phenomena, would  soon discover that the eclectic choice of Mesoamerican and Andean calendrical concepts would permit the tabulation such phenomena in a simple and elegant way.

The mechanisms, which allowed the Muiscas to profit from the calendrical concepts of Mesoamerica and the Andes and  develop their own by extension, represent an unexplored and uncharted territory.  Such  issues  involve the detailed study of the trade networks operating in the prehispanic America from north to south, transferring knowledge and ideas, as well as material goods.

%%%%%%%%%%%%%%%%%%%%%%%%%%%%%%%%%%%%%%
%% APPENDIX

\appendix

%% DUQUESNE'S WORK
%% ASCII ENCODING: ISO8859-1

\newcommand{\ra}{$^\textrm{\footnotesize a}$}

%% DISERTACION ESP
\chapter{Duquesne's works}\label{duquesne-work}

\section{Dissertation}
\newpage
\selectlanguage{spanish}
\begin{center}
\textsc{disertación sobre el calendario de los muyscas, indios naturales de este nuevo reino de granada. dedicada al s. d. d. josé celestino de mutis, director general de la expedición botanica por s. m.} 

\textsc{por }

 \textsc{el d. d. josé domingo duquesne de la madrid, cura de la iglesia de gachancipá de los mismos indios. año de 1795} 
\end{center}

\begin{center}
\textit{Calendario de los Muyscas, Indios naturales del Nuevo Reino de Granada.}                              \end{center}

\sl

 Una de las cosas que han hecho más honor a las artes y ciencias es el estudio de las antigüedades. Por este medio se han penetrado algunos secretos escondidos, se han descifrado varios misterios, y se ha ilustrado en una gran parte la historia. No contentos los doctos anticuarios con las lápidas sepulcrales é inscripciones de sus países, han procurado desenterrar a Menfis, y los viajes a Egipto han enriquecido al orbe literario con sus descubrimientos.

 La América no puede hacer ostentación de estas magnificas antigüedades. Por más que se haya pretendido que Sesostris extendió sus conquistas hasta estos remotos fines de la tierra, no encontraremos en ella los trofeos de sus victorias que dejó en el Asia. No hallaremos obeliscos con que adornar nuestras ciudades, laberintos, ruinas de edificios, medias columnas, pirámides, cuyos lados tengamos que medir, para describir sus fachadas, ni otros monumentos que en si mismos conservan, a pesar de su ruina, no sé que aire de magnificencia en cuyo prólijo registro se emplea con gusto la curiosidad.

 Los fragmentos históricos de estas partes son tan sencillos como sus primeros pobladores. Pero, aunque no se encuentre entre éstas gentes el fausto de los antiguos Egipcios, se ven sus misterios. No hallamos monumentos Faraónicos, pero si algunos pequeños trozos de los fundamentos sobre que se edificaron. Hablo de los hieroglíficos que se han encontrado entre los Indios. Esta palabra griega quiere decir: imágenes ó figuras sagradas. Dióse este nombre a aquellas de que se servían los Egipcios para representar los dogmas de su teología, o de su ciencia moral y política, que se veían esculpidas sobre piedras, pirámides, etc.

 Las pinturas de los Indios son puramente simbólicas; se insistió poco sobre ellas en aquellos tiempos en que pudieron haberse examinado. Nada penetramos de los caracteres de los Egipcios, y los que tenemos de los Indios no pueden explicarse. Así estas dos naciones que poseyeron, o, por decir, cultivaron  más bien que otras los símbolos y caracteres primitivos de que nació el uso de las letras, se han hecho igualmente célebres e ininteligibles, sirviendo ya más estos sus monumentos para atormentar los ingenios que para adelantar la erudición.

 Como quiera que sea, la antigua América no ha dejado de hacer alguna ostentación de sus pinturas simbólicas entre los eruditos. Pero la nación de los Muyscas, Indios del Nuevo Reino de Granada, no ha podido entrar hasta ahora a la parte de esta pequeña gloria. El Padre Torquemada se queja de la negligencia de las primeras personas de letras que entraron en esta tierra. El señor Piedrahita abiertamente pronuncia que ignoraron estos indios los hieroglíficos, y los quipus de los Peruanos; lo cual es falso, como se convence de muchos fragmentos que nos han quedado de su antigua superstición.

 Tengo pues el honor de servir a la historia con un nuevo descubrimiento, y de exponer el año y siglo de los Muyscas; interpretando los signos que lo contienen, y que hemos hallado por propia investigación. Esta interpretación esta fundada en el conocimiento de sus costumbres, de su historia, de su idolatría y de su lengua. Esta ultima, así como me ha sido de mucho auxilio, me ha dado también mucho trabajo, porque ya no se habla este idioma, y me ha sido necesario sacarlo de entre los cartapacios en que se halla reducido al método de la lengua latina, con quien no tiene analogía, para restituirlo a su verdadero principio, formándolo como de nuevo sobre el genio de las lenguas orientales para investigar las raíces y deducir las etimologías.

 Los Muyscas contaban por los dedos. Solo tienen nombres propios para diez, y para el número veinte. A saber: \textit{Ata, Bosa, Mica, Muyhica, Hisca, Ta, Cuhupcua, Suhuza, Aca, Ubchihica, Gueta}. En concluyendo con una vuelta de las manos, pasaban a los pies, repitiendo los mismos nombres, a que anteponían la palabra \textit{Quihicha}, que quiere decir el pié, \textit{Quihicha ata}, once: \textit{Quihicha bosa}, doce, etc.

El número 20, expresado por la dicción \textit{gueta} (casa y sementera), en que se encerraban todos los bienes y felicidad de esta nación, fenecía todas sus cuentas. Y así en terminando con un 20, pasaban a contar otro, uniéndolo con el primero hasta formar un veinte de veintes.

De modo que, así como los matemáticos han dado al círculo 360 grados, por la facilidad con que este número se subdivide en otros menores para formar cualquiera cálculo, así ellos dividían sus cuentas en cuatro partes tomadas de la misma naturaleza, partiéndolas de cinco en cinco. Y así sus números más privilegiados eran: 5, 10, 15, 20, de los cuales se servían en el arreglo de todos sus negocios.

 La luna era el objeto de sus observaciones y de sus cultos. Este astro, de que no apartaban los ojos, les dio el modelo de sus casas, cercados, templos, labranzas, en una palabra de todas sus cosas. Fijaban en el suelo un palo de que hacían centro, y con una cuerda trazaban el círculo. Este palo, y la cuerda, si se consideran bien los caracteres o símbolos que hemos descrito en la tabla, se conocerá que son los principales elementos sobre que se hallan formados. Los diferentes significados que tienen estas voces numerales en su lengua, todos son alusivos a las fases de la luna, a las labores de sus sementeras, y a las supersticiones de su idolatría, y así nos conducen derechamente a la formación de un calendario.

 Tenían los indios colocados en las manos mentalmente estos símbolos, a manera que los músicos los signos del sistema de Aretino. Y así, con solo dar una vuelta a los dedos, sabían el estado de la luna y el gobierno de sus cosas v de sus sementeras.

 El año constaba de veinte lunas, y el siglo de veinte años; comenzaban a contar el mes desde la oposición o plenilunio figurado en \textit{Ubchihica}, que significa luna brillante; contando siete días en los dedos comenzando por \textit{Ata,} que se sigue a \textit{Ubchihica,} hallaban la cuadratura en \textit{Cuhupcua}; contando de allí siete encontraban la próxima inmersión de la luna en \textit{Muyhica}, que significa cosa negra, y al día siguiente la conjunción simbolizada en \textit{Hisca,} que en su concepto era una unión de la luna con el sol, que representaba las nupcias de estos dos astros, que era el dogma capital de su creencia, y el objeto de sus mas execrables cultos; contando después ocho días hallaban la otra cuadratura en \textit{Mica}, que significa cosa varia, como queriendo significar la perpetua variación de sus fases. El primer aspecto de la primera faz la señalaban en \textit{Cuhupcua}, y como en este símbolo caía la cuadratura le daban dos orejas, y le llamaban sordo por otros motivos de superstición.

Estos mismos símbolos servían a contar los años, y contenían una doctrina general, en orden a la siembra, \textit{Ata}, pues, y \textit{Aca}, representaban las aguas en el Sapo. El mas frecuente graznido de este animal les sirvió de señal para conocer que se acercaba el tiempo de sembrar.

\begin{description}
 
 \item \textit{Bosa}: una sementera que hacían alrededor de la principal para defender el centro de los daños.

\item  \textit{Mica}: buscar, hallar, escoger cosas menudas: significa la elección que debían hacer de las semillas para la siembra.

\item  \textit{Muyhica:} cosa negra: representa el tiempo tempestuoso y oscuro. Su raíz significa crecer las plantas, porque con el beneficio de las aguas toma cuerpo la sementera.

\item  \textit{Hisca}: cosa verde: con las lluvias aparece el campo hermoso y alegre. También significa holgarse. Las plantas mas crecidas los alegraban con la esperanza de los frutos.

\item  \textit{Ta:} Sementera: al sexto mes de la siembra corresponde la cosecha.

\item  \textit{Cuhupcua}: sus graneros tienen la figura de caracol o de oreja. \textit{Cuhutana}, que tiene la misma raíz, significa los rincones de la casa donde depositan los granos: alude a la cosecha.

\item  \textit{Suhuza}: cola, rabo: mes que viene al fin de las siembras. Tiene alusión al palo de sus calzadas, donde hacían sus solemnidades verificada la cosecha.

\item \textit{Ubchihica }puede aludir a sus convites.

\item  \textit{Gueta: }casa, y sementera. Esta marcado con un sapo tendido, que entre ellos era el símbolo de la felicidad. 

\end{description}

 Los indios miraban estos avisos como otros tantos oráculos: enseñaban a sus hijos con tesón esta doctrina de sus mayores, y, no contentos con estas precauciones, para no perder el gobierno del año lo señalaban con la sangre de muchas víctimas.

 No decían jamás esta palabra: \textit{Zocam} (el año) solo, sino con el número que le correspondía, \textit{Zocam Ata}, \textit{Zocam Bosa}, etc. Lo mismo ejecutaban con la palabra \textit{Suna} (la calzada), en donde hacían en cada siembra y cosecha sus mogigangas y sacrificios, \textit{Suna Ata,} \textit{Suna Bosa}, una calzada, dos calzadas. Y de este modo estos lugares eran como un libro donde se iban registrando las cuentas.

 Veinte lunas, pues, hacían el año. Terminadas estas, contaban otras veinte, y así sucesivamente, rodando en un círculo continuo hasta concluir un veinte de veintes. La intercalación de una luna, que es necesario hacer después de la luna trigésima sexta, para que el año lunar corresponda al año solar. Y se guarde la regularidad de las estaciones la ejecutaban con suma facilidad. Porque, como tenían en las manos todo el calendario, sembraban dos sementeras seguidas con un signo de por medio, y la tercera con dos. Como  sobre este principio rueda toda su astronomía, idolatría, política, economía, u lo que ahora nos es más interesante, su iconografía, es necesario expresarlo con mayor individuación.

 Distribuyamos pues los signos muyscos en los dedos, y esta tabla digita nos dará todas las combinaciones. Supongamos que \textit{Ata}, que esta en el primer dedo, corresponde a enero, y que es un mes apto para sembrar. Corridos los dedos corresponde la segunda sementera en Mica, interceptando a \textit{Bosa}, que esta en medio de \textit{Ata}, y \textit{Mica}. De suerte que esta sementera se hace en la luna decima tercia respecto de \textit{Ata}.

 Corriendo ahora los dedos desde \textit{Mica}, corresponde la sementera en \textit{Hisca,} interceptando a \textit{Muyhica}, que esta en medio de \textit{Mica}, é \textit{Hisca}. De modo que se hace la sementera en la luna décima tercia respecto de Mica.

 Corramos últimamente los dedos desde Hisca, y se hará la sementera en \textit{Suhuza}, interceptando dos signos: \textit{Ta} y \textit{Cuhupcua}, que están en medio de \textit{Hisca} y \textit{Suhuza}; esto es en la luna décima cuarta respecto de Hisca.

 Esta luna \textit{Cuhupcua} (que en su lengua quiere decir sorda) es la que se intercala, por que es la décima séptima al año segundo muysco, cuyo número, añadido a las veinte lunas del año primero, produce 37, con lo que queda igualado el año lunar con el solar, y \textit{Suhuza} viene a ser un verdadero enero.

 Esta intercalación, que se verifica perpetuamente, dejando pasar como inoficiosa o como sorda la luna 37, nos hace concebir que dentro de los dos años vulgares, de veinte lunas cada uno, hay otro año astronómico oculto que consta de 37 lunas, de modo que la luna 38 sera un verdadero enero. Los Indios, sin penetrar la teórica de esta proposición, que ha sido embarazosa en otras naciones más cultas, por esta luna que ha sido necesario añadir al fin de cada tres años lunares por ser los doce anteriores de doce lunas, y el tercero de trece, tenían suma facilidad en la práctica de su intercalación, siguiendo el método propuesto, conservándose así el año astronómico, sin que el pueblo notase diferencia alguna en sus años vulgares de veinte lunas cada uno.

 El año vulgar de veinte lunas servía para las treguas en la guerra, como consta de su historia, para las compras y ventas, y otros negocios de la sociedad. Pero el año astronómico é intercalar de 37 lunas, que se contaba por tres sementeras, servía principalmente a la agricultura y a la religión; y así llevaban su cuenta con mucha prolijidad los xeques, y mayores a quienes correspondía, notando sus épocas con sacrificios mas particulares, y grabándolas también en piedras, por medio de símbolos y figuras, como se ve en un pentágono que tengo en mi poder y voy a explicar al fin de este papel.

 El siglo pues de los Muyscas constaba de veinte años intercalares de 37 lunas cada uno,  que corresponden a 60 años nuestros, y le componían de cuatro revoluciones contadas de cinco en cinco, cada una de las cuales constaba de diez años muyscos, y quince nuestros, hasta completar los veinte, en que el signo Ata vuelve a tomar el turno de donde comenzó la vez primera. La primera revolución se cerraba en \textit{Hisca}, la segunda en \textit{Ubchihica}, la tercera en \textit{Quihicha Hisca} y la cuarta en \textit{Gueta}.

 La inteligencia de estos calculos es tan necesaria para penetrar su historia antigua, y descifrar sus símbolos y figuras, que sin ella no pueden comprenderse, y así nos ha sido indispensable formar una tabla cronológico-muisca, en que fácilmente se percibe toda la economía de su siglo, que ponemos al fin con la debida explicación.

 La semana era de tres días, y estaba señalada con un mercado que hacían cada primer día de ella en Turmequé, de los mas ricos y opulentos,  como se puede ver en el Padre Zamora.

 Dividían el día \textit{Sua}, y la noche \textit{Za}, desde el oriente al medio día \textit{Suamena}, la mañana; desde el medio día al ocaso, \textit{Suameca}, la tarde; del ocaso al fin de los crepúsculos (hacían la comida), \textit{Zasca}, prima noche: de la media noche (se levantan al mayor trabajo) a la aurora, \textit{Cagui}. De la aurora (almuerzan) al oriente (\textit{Así esta.})

 El fundador de los Muyscas no quizo dejar el calendario, por fácil que fuese su ejecución, al arbitrio del pueblo. Mandó que se consultase a sus jefes, y esta providencia paso con el tiempo a superstición. Llegaron a persuadirse que obtenían estos el imperio de las estrellas, y que eran dueños absolutos de los tiempos favorables o adversos y aún de todas las miserias y calamidades que afligen al hombre. Nada pues se hacia sin su consejo, y sin que recibiesen por él muchos donativos, y así no hubo pueblo en donde se vendiesen más caros los almanaques.

 Tenían, a más de eso, el cuidado de señalar las revoluciones del año con las cosas más notables. No había siembra ni cosecha sin sacrificio. Tenían en cada pueblo una calzada ancha y nivelada que salía del cercado, o casa del cacique, y corría como por media legua, rematando en un palo labrado en figura de una gavia de que prendían al miserable cautivo que ofrecían al sol y a la luna para obtener una cosecha abundante.

 Venían en mogiganga los indios, repartidos en diferentes cuadrillas, adornados de muchas joyas, lunas y medias lunas de oro: disfrazados unos con pieles de osos, tigres y leones; enmascarados otros con máscaras de oro, y lágrimas bien retratadas, a los cuales seguían otros con mucha gritería y risadas, bailando y brincando con descompasados movimientos: otros traían unas grandes y largas colas, que iban pisando los que los seguían, y llegando al término de la calzada disparaban todos sus flechas y tiraderas al infeliz cautivo matándole con larga muerte, y, recibiendo su sangre en diferentes vasijas, terminaban la barbara función con sus acostumbradas borracheras.

 Nuestros historiadores se admiran mucho del fausto y de la extravagancia de estas procesiones, pero nos dieron una idea muy diminuta, refiriendo por mayor sus cuadrillas. En lo poco que describieron se conoce que esta mogiganga era un símbolo de su calendario, y, si las hubiesen dibujado todas, nos ayudarían a formar el concepto de sus signos, y de los caracteres que les atribuyan.

 Pero la víctima destinada a solemnizar las cuatro lunas intercalares que partían el siglo, estaba señalada con muchas circunstancias. Era este un miserable mancebo, que precisamente había de ser natural de cierto pueblo, sito en los llanos que llamamos hoy de San Juan. Horadábanle las orejas, le criaban desde mediano en el templo del sol; en llegando a diez años nuestros, le sacaban para pasearle, en memoria de las peregrinaciones del Bochica su fundador, a quien se figuraban colocado en el sol, y continuando, en un matrimonio feliz con la luna, una lucidísima descendencia. Vendíanle en precio muy alto, y era depositado en el templo del sol hasta cumplir quince años nuestros, en cuya precisa edad hacían el bárbaro sacrificio, sacándole vivo el corazón y las entrañas para ofrecerlas al sol.

 A este mozo le llamaban \textit{Guesa}, esto es sin casa, por lo dicho. Llamabanle tambien \textit{Quihica,} que quiere decir puerta, con la misma alusión que los Romanos llamaron Jano al principio del año. Significa también boca, porque llevaba la voz de su nación para hablar de cerca a la luna intercalar y sorda que no oía desde acá abajo sus lamentos. Esta gente ilusa se figuraba que sus víctimas le hablaban por ellos dentro de su misma casa, y por eso hacían muchos sacrificios de loros, pericos y guacamayos; y solían matar hasta doscientos en cada vez de estos animales. mas no llegaban a las aras sin haber aprendido la lengua. Pero, por muchos sacrificios que hiciesen, la luna intercalar y sorda proseguía de la misma suerte en todos sus turnos, sin que se alterase el calendario. Los pericos y guacamayos hacían desde luego en tanto número una terrible algazara. \textit{Et sequitur cursus surda Diana suos.}

 Las muchas precauciones que tomo el legislador para el gobierno del año hicieron a los Muyscas demasiadamente atentos a su observancia. Mirábanle como un invento divino, y a su autor como un Dios que habitaba en las mismas estrellas. Colocaron pues al Bochica en el sol, ya su mujer Chia en la luna, para que continuasen desde allí una protección benéfica sobre su descendencia.

 A éste su Bochica daban dos compañeros, o hermanos, a que simbolizaban de un cuerpo con tres cabezas, porque decían que tenían un corazón y una alma. Entre tanto el Bochica les dirigía desde el sol sus sementeras. (Véase una imagen de Endimion, de quien afirma Plinio que paso una gran parte de su vida en la contemplación de la luna. De donde nació la fabula de que estaba enamorado de ella.)

 Tuvo también su lugar entre los astros el Sapo, para acompañar al Escorpión, y a los demás animales de los Egipcios. Jamás ha dado esta sabandija mayor brinco del charco al cielo, y nunca bajó el hombre más del cielo al cieno, y de la altura de los astros, a quienes domina por su sabiduría, a la bajeza de la más profunda ignorancia en que es dominado de todas las pasiones. Por este pequeño rasgo se conoce la uniformidad de los progresos de la idolatría en todas las naciones del mundo.

 No contentos con haber divinizado a su legislador formaron otra divinidad de uno de sus héroes sobre el mismo calendario. Fue este el portentoso Tomagata, uno de sus mas antiguos Zaques. En vez de tejer su historia, haremos su retrato. Tenia un ojo solo, porque era tuerto; pero este defecto lo suplían las orejas, porque tenia cuatro, y una cola muy larga a manera de león, o tigre, que le arrastraba por el  suelo. Fué fortuna de la miserable nación que fuese impotente, porque no se multiplicasen los monstruos. El sol lo había despojado de la potencia generativa la noche anterior a su matrimonio, para que le heredase su hermano Tutasua. Fué lastima que no fuese cojo, porque era, decían, tan ligero que todas las noches hacia diez viajes de ida, y otros tantos de vuelta, a Sogamoso, que dista ocho leguas de Tunja, visitando todas sus hermitas. Vivió cien años, y los Muyscas pretendieron hacerle vivir muchos mas. Sus facultades se median por sus defectos, pues tenia del sol el poder de convertir en culebra, tigre, lagarto, etc., a cualquiera que lo irritase. Los indios le llamaban el cacique rabón. Su nombre Tomagata, significa fuego que hierve. Ellos pasaron al cielo astrológico este espantoso cometa, y yo, según las circunstancias de su historia, creo que le señalarían mas bien por eunuco de la Virgen Espigadora que por compañero de Sagitario. \label{eunuco-virgen}

 Tal fué el cielo de los Muyscas, lleno de animales como el de los Egipcios. En él vemos introducidos al Bochica y a Chía sus fundadores, como en aquel a Osiris é Isis: las trasformaciones de aquellos en el carnero, en el toro, y en otros animales celestes, se ven igualmente imitadas, entre estas gentes, en las trasformaciones de Tomagata, é que aludían las de sus cuadrillas. Se ve también una gran conformidad entre los signos de los Egipcios y los símbolos de los Indios. No pretendemos que los caracteres de que hoy usamos en la astronomía sean los mismos originales que inventaron los antiguos: pero todos conocen que retienen alguna semejanza de los elementos sobre que se formaron. Como también que los Egipcios no fueron sus primeros inventores, habiéndose propagado desde el valle de Senaar, junto con los primeros conocimientos astronómicos. Pero los Egipcios y los Indios. que son descendientes de Can en la más probable opinión, como aquellos, cultivaron la escritura simbólica. con mas aplicación que otras naciones, hasta hacerla propia.

 \textit{Ata}: es un sapo en acción de brincar. que caracteriza bien la entrada del año. \textit{Aca}: es otro sapo de cuya cola se empieza a formar otro; símbolo de aquella luna en que observaban la generación de estos animales, cuyos frecuentes graznidos anunciaban las próximas aguas, y eran la señal de acercarse sus siembras. Por donde se conoce la alusión que hace al signo de Piscis. \textit{Gueta}: es un sapo tendido; significa la abundancia y la felicidad. A otros signos dieron facciones humanas, de donde parece ha llegado hasta nosotros el uso de pintar el sol y la luna con ojos y narices. \textit{Bosa:} representa unas narices. \textit{Mica:} dos ojos abiertos. \textit{Muyhica} dos ojos cerrados. \textit{Cuhupcua:} dos orejas. \textit{Ubchihica}: una oreja. Verisímilmente quisieron dar a entender las diversas fases de la luna, y abusaron después por erradas aplicaciones. \textit{Cuhupcua:} tiene también la idea de una canasta, para significar la cosecha. \textit{Ta, Suhuza:} figuran el palo y la cuerda con que formaban el círculo de sus casas y de sus labranzas. \textit{Hisca: }la unión de dos figuras. Era símbolo de la fecundidad; y se conoce la alusión que hace a Géminis. En sus significados, que son varios, se nota también la conformidad con los antíguos, y que esta doctrina de los tiempos, la recibieron los Indios como las demás naciones al tiempo de la dispersión de las gentes.

\label{stone-dissertation}

 Hemos visto el calendario muysca en los dedos; también le gravaban en piedras por medio de sus figuras simbólicas. Mantengo en mi poder una que lo expresa según mi modo de pensar, y tengo el honor de servir a la historia con este nuevo descubrimiento. En este reino ninguno ha pensado hasta ahora en trabajar sobre la iconografía de los Muyscas, y así estos pequeños rasgos son los primeros elementos de este genero en que tanto se interesa la historia. El Sapo es indubitablemente el símbolo de la primera luna del año y del siglo. Pusieronle los Indios entre sus divinidades, y le dibujaban de distintas maneras. En acción de brincar correspondía al primer signo, \textit{Ata}, y así se halla grabado en varias piedras. He notado en otras que esta grabado con rabo o cola, lo que me ha hecho pensar que en esta acción caracteriza a \textit{Quihicha ata}, esto es al número 12. Porque, continuando el brinco para denotar los meses futuros, señala con la cola los que deja detrás. Símbolo que en otros animales usaron los antiguos, y que representaban estos mismos en las cuadrillas de sus procesiones, de que hemos hablado. Observando varias piedras con la debida atención, he notado que figuran también el cuerpo del sapo sin patas, lo que me representa el signo de \textit{Gueta,} o también un signo en quietud, sin que influya en las operaciones del campo. Algunas veces la cabeza del sapo se ve unida a la cabeza de hombre: otras el cuerpo sin patas trasformado en ídolo: esto es con una vestidura o túnica propia de hombre: y asimismo el sapo de cola y sin patas de que hemos hablado.

 Supuesto este corto número de observaciones porque carecemos de otros monumentos sobre que hacerlas, explicaré la piedra, que se ve dibujada en la figura 1. Es un pentágono, señalado con las letras a. b. c. d.

\textit{ a.} es un sapo sobre un plano en acción de brincar. \textit{b.} es una especie de dedo señalado de tres líneas gruesas. \textit{c.} es lo mismo, pero se debe notar que esta fuera del centro o línea que siguen los otros. \textit{d.} es lo mismo conservando el centro del primero. \textit{e.} es el cuerpo de un sapo, con cola, y sin patas, sobre un piano. \textit{f.} es una culebrilla, \textit{g.} es un círculo en el plano de la piedra en cuyo segmento se ve la figura H. Y. es el reverso del plano de la misma piedra. L. es un círculo con dos segmentos formados por una cuerda y un radio. M. es una culebra, etc.

\begin{center}
\textsc{interpretación}
\end{center}

Esta simbolizada en esta piedra la primera revolución del siglo muysca, que comienza en \textit{Ata}, y acaba en \textit{Hisca}, el cual incluye nueve años y cinco lunas muyscas. Los Indios, que para todo usan del círculo, aquí prefieren el pentágono para significar que hablan de cinco años intercalares.

 \textit{a.} El sapo en acción de brincar: principio del año y del siglo. \textit{b.} Esta especie de dedo señala en las tres lineas gruesas tres años. Omitiendo pues el dedo \textit{c.}, que está a un lado, cuento en el dedo \textit{d.} otros tres años, que, juntos con los del dedo \textit{b.,} producen seis. Lo cual denota la intercalación de \textit{Quihicha ata}, que sucede puntualmente a los seis años muyscas, como se ve en la tabla; y es de mucha consideración entre los Indios, por pertenecer al sapo que regla todo el calendario.

 \textit{e.} es el cuerpo de un sapo de cola y sin patas. Símbolo de \textit{Quihicha ata}, y por carecer de patas figura muy propia para expresar su intercalación. Porque el mes intercalar no se computa para la sementera, y así lo imaginaban sin acción y sin movimiento. Se ve sobre un plano, como también el sapo Ata, lo que conduce a significar que en una y otra parte se habla del sapo.

\textit{f.} Esta culebrilla representa el signo \textit{Suhuza}, que es el que se intercala después de \textit{Quihicha ata} a los dos años muyscas representados en las dos líneas gruesas que tiene en el dorso. Lo que corresponde al año octavo, como se ve en la tabla.

 Como concluimos con los lados del pentágono pasamos al plano \textit{i. }La culebra \textit{m} es una reproducción de S\textit{uhuza} y como esta tendida sobre una especie de triángulo símbolo de Hisca, significa que se intercala inmediatamente después de \textit{Suhuza} al segundo año, lo que esta figurado igualmente en las dos líneas gruesas que tiene en el dorso.

 Como el fin principal de esta piedra cronológica es señalar la intercalación del signo de \textit{Hisca}, por ser el término de la primera revolución del siglo muysca, para mayor claridad están contados estos años en los tres dedos; conviene a saber: \textit{b, c, d.}, que juntos producen nueve años, que son los que dan puntualmente esta notable intercalación, que sucede a los nueve años y cinco meses como se ve en la tabla.

\textit{ g.} es un templo cerrado. \textit{h.} es una cerradura que hasta el día de hoy usan algunos indios, y llaman candado cormo. Los agujeros de las dos orejas sirven a las estacas que le ponen, y los dos ganchos interiores a asegurar la puerta. Significa la primera revolución del siglo, cerrada en Hisca, y para que continuase el tiempo era necesario en su imaginación que el Guesa abriese la puerta con el sacrificio de que hemos hablado, y cuyas circunstancias eran simbólicas, relativas a estas revoluciones del siglo.

 La culebra, por otra parte, ha sido un símbolo del tiempo en todas las naciones. Esta primera revolución de siglo estaba consagrada principalmente a las nupcias del sol y la luna, simbolizadas en el triángulo, no sólo según los Indios, sino según otras naciones.

\begin{center}
 \textsc{Explicación de la tabla de los años muyscas.} %(Lam 1a Fig. 3a) 
\end{center}

 El círculo interior representa las veinte lunas del año muysco vulgar, cuyos signos todos se intercalan en el espacio del siglo.

 El círculo segundo expresa los años muyscas a que corresponde la intercalación de cada signo.

   El círculo tercero expresa el orden de esta intercalación; ejemplo: deseo saber en que año muysco se intercala el signo \textit{Mica}. Veo en la tabla el número 3 en el círculo interior, hallo en el segundo que le corresponde el número 30\footnote{There is a typesetting error in the Acosta's text. Instead 30 it should be 36.}, y este es el año que se busca; veo en el siguiente círculo que le corresponde el número 19, y así la intercalación de \textit{Mica} es en orden la décima nona del siglo.

 La intercalación de Gueta (20) es la ultima del año muysco 37. Esto es después de un siglo vulgar muysco de años 20 lunas, y más 17 años, de suerte que, terminando el siglo, o revolución astronómica de 20 años intercalares de 37 lunas cada uno, les faltan tres años vulgares para completar dos siglos vulgares. En llegando pues a este caso no hacían más cuenta de aquellos tres años vulgares de que no necesitaban para la labranza, ni para la religión, ni para la historia, y empezaban en Ata (a que había llegado el turno) un año vulgar, nuevo principio de un siglo nuevo en todo semejante al primero que hemos descrito. 

\begin{center}
 \includegraphics[width=12cm]{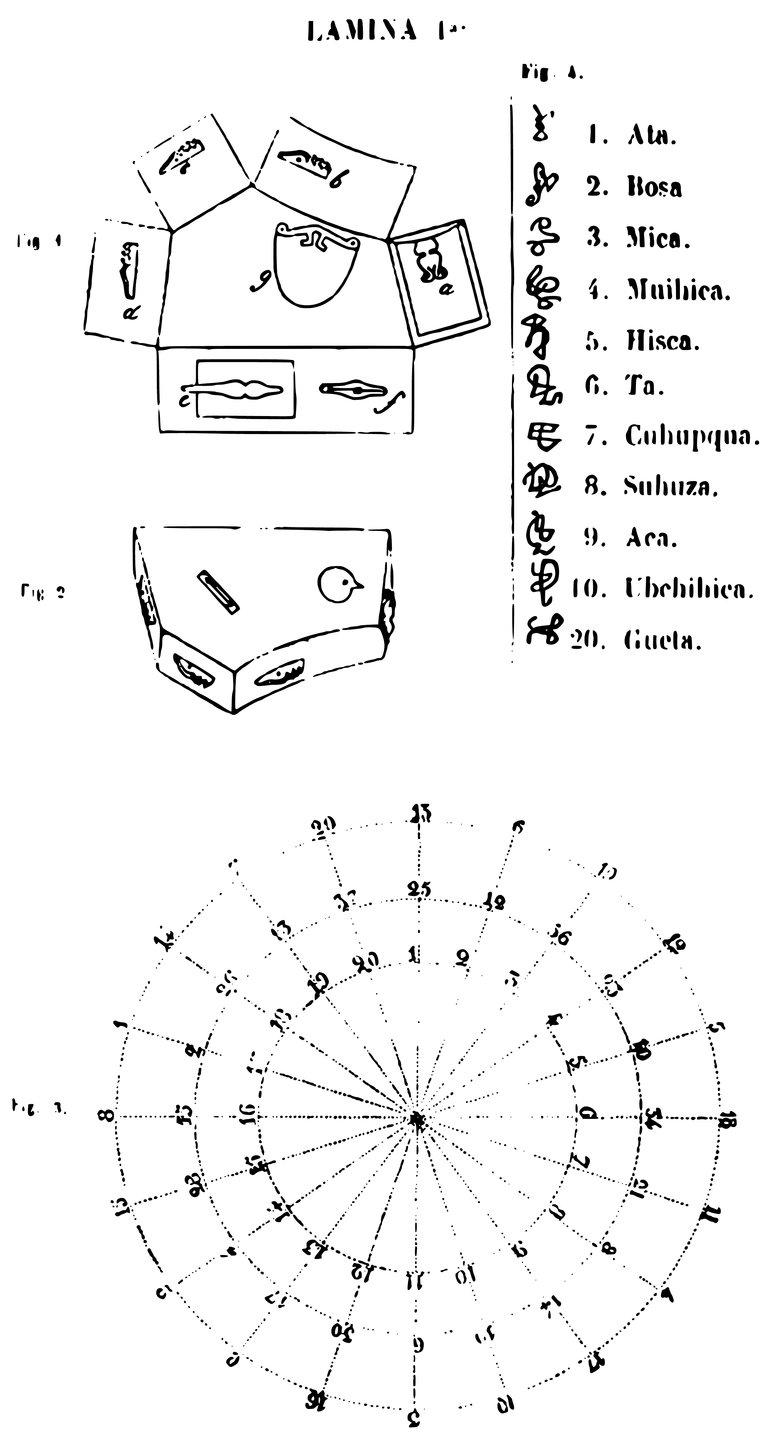}
 % duquesne-stone.pdf: 2310x1086 pixel, 72dpi, 81.49x38.31 cm, bb=0 0 2310 1086
\end{center}

%%%%%%%%%%%%%%%%%%%%%%%%%%%%%%%%%%%%%%%%%%%%%%%%%%%%%%%%%%%%%%%%%%%%%%%%%%%%%%%%%%%%%%%%%%%%%%%%%%%%
%% ANILLO ESP
\newpage

\section{Astronomical Ring}
\newpage
\begin{center}
\textsc{disertacion sobre el origen del calendario jeroglífico de los moscas}
\end{center}

\begin{center}
(Por el Doctor José Domingo Duquesne. 1795).
\end{center}

El calendario de los moscas es una pieza original; los indios atribuían esta invención al Bochica su fundador, y sirve de prueba el haber fundado sobre esta misma idea toda su religión y todas sus costumbres, mirándole no sólo como una tabla astronómica para el computo de los tiempos sino como un cuerpo de legislación de ritos y ceremonias para el gobierno de la nación. En efecto, el Bochica puso en planta su calendario en las tierras de que se posesionó, pero traía la idea de otra parte, y debemos pensar que la había recibido en aquella misma escuela en que cursaron juntos todos los hombres.

Nada hay tan natural como creer que los hijos de Noé extendidos en las vastas llanuras de Senaar, convinieron entre sí en algunos reglamentos cómodos para medir el tiempo arreglando por ellos las operaciones de la labranza y los negocios de la sociedad. La medida mas sencilla de que pudieron servirse fué la luna, así por sus revoluciones periódicas como por la notable diversidad de sus fases, a las cuales podían ligar sus diferentes ideas, para sus juntas, sacrificios, estaciones, teniendo todos en el cielo un libro público por donde gobernarse con la mayor seguridad.

Yo no pienso en detenerme en probar una verdad que ya otros han establecido con sólidos fundamentos; pero que habiendo hallado los padres de todas las naciones reunidos por bastante tiempo bajo un jefe, en unos mismos intereses y ceremonias, es constante que de esta misma fuente bebieron todos aquellos primeros elementos en que después se han visto convenir todos los pueblos del orbe.

Si conviniéramos en esta idea, daríamos una salida fácil a muchas cuestiones embarazosas que han atormentado los ingenios que se han querido fatigar voluntariamente. No habría necesidad de hacer paralelos entre los egipcios o los fenicios y los indios para buscar en aquellos como en su fuente los símbolos de que usaron éstos para hacerlos sus descendientes. No habría que buscar las libaciones, purificaciones y sacrificios ya entre estas gentes o las otras, para pretender que los indios habían copiado de otros estas o semejantes costumbres.

Conviniendo en ciertos usos que fueron comunes a los primeros maestros del orbe, nos desembarazaríamos de estas dificultades. Entre tanto no pienso, como he dicho en detenerme en probar una verdad tan bien establecida, pero me serviré de ella para ilustrar el argumento que tratamos, por la cual se conocerá el fondo o política de estos primeros habitantes.

A la verdad todos recibieron de Noé las primeras lecciones de astronomía. Todos se conformaron en el
gobierno del año, y al principio fue para todos el curso de la luna la regla general de los tiempos. Esta luna intercalar, tan necesaria para arreglar el movimiento de este astro con el sol, fue conocida de todas las naciones. y aunque se pueda pretender que debieron este conocimiento a sus propias posteriores observaciones, los calendarios de América prueban que no les fue desconocida a sus primeros pobladores, los cuales fueron al mismo tiempo mas cuidadosos en conservar estos primeros elementos que los otros tal vez despreciaron.

Si cotejamos el calendario de los moscas con los que usaron las naciones del mundo antiguo, hallaremos que este es una pieza original que en nada se les parece sin embargo de haber tenido todos un mismo principio. La idea no puede ser mas singular, a excepción de la luna intercalar que es un principio común sobre que ruedan todos. ¡Qué artificio! ¡Qué invención tan nueva y rara conservar sin añadir ni quitar ninguna luna, la adición de aquella luna tan extraordinaria que se les hizo dificultosa de entender con el tiempo a los otros, y que hubieron de abandonarla o confundirla con vergonzosa ignorancia aun aquellos pueblos que han sido reputados por mas políticos y sagaces! ¡Cuantas veces perdieron el hilo! ¡Cuantas pretendieron hallarlo de nuevo! Entre tanto los moscas usaron de las reglas de sus mayores por tantos siglos, sin tener que variar jamás el primer método que recibieron. Siempre lograron sus cosechas, siempre tuvieron bien conocidas sus respectivas estaciones, siempre conservaron el orden de su cronología en aquella parte que necesitaron, y aún con toda la larga carrera de sus años se hubiera hallado arreglada tal vez de algún modo, si se hubiesen descubierto sus quipus que ocultaron, y con que llevaban una cuenta tan sencilla como segura en todos los negocios que consideraban de alguna importancia.

Si echamos una ojeada sobre el mundo antiguo, apenas se encontrará en toda la antigüedad cosa de significación más variada que el año. Casi no se puede creer que haya habido gentes que hayan tenido por año una luna, si no fuese igualmente cierto que hubo quienes le tuvieron de un día solo. Los egipcios, según Plinio, tenían el año de una luna; los arcades de cuatro. Los indios habían ya logrado, por su plan de medir el tiempo, muchas cosechas, cuando los caldeos, no se sabe cómo se gobernaban, ignorándose de qué manera computaban los padres de la astronomía. Pasando de aquellos tiempos más oscuros a otros más conocidos, los romanos desde su primera fundación nos presentan un calendario de diez meses muy artificioso; no hay cosa más especiosa que su primera fachada: las calendas, las nonas, los idus, son unos nombres de mucha idea, pero que importa si le dejaron en la parte sustancial tan imperfecto, que el pueblo mismo, según dice Macrobio, añadía al fin de los diez meses tantos números de días cuantos eran menester para que el principio del año coincidiese con la primavera, sin dar a este tiempo nombre especial de mes, lo que puso a Numa en la necesidad de ajustarlo. Con todo, o por defecto de calculo o por política, quedo tan defectuoso, que fue preciso que Julio César basase el año sobre un plan enteramente nuevo, que es el que nosotros hemos adoptado.

En orden a los hebreos no se sabe qué género de año usaron en los tiempos antiquísimos; creen algunos que hayan sido solares. Según los diferentes estados del miserable pueblo, unas veces siguieron a los egipcios, otras a los caldeos y a los persas, y finalmente a los griegos. Estos se gobernaban para el año por el sol, y por la luna para los meses. En los libros de Moisés, solo un mes, que es Abib, se designa con su propio nombre, los demás tienen el de primero, segundo, etc.

Cuando este pueblo se hallo en su libertad, uso de años de doce lunas y, al tercero de trece. Este plan nos recuerda la primera forma de aquellos arios antiguos del tiempo de Noé, cuya tradición es muy natural que guardasen. Como quiera que sea, los moscas, entre todas las gentes, no tuvieron alteraciones ni variaciones en el gobierno del año; su fundador lo arreglo sobre el pie que recibieron de los hijos de Noé todos los hombres cuando, la tierra era de un solo labio; y cuando tuvo una lengua distinta le acomodo según sus ideas y el genio de su idioma, en los términos en que lo hemos explicado, dándole tanta regularidad, y tomando tantas precauciones, que aseguró su perpetuidad por largos siglos entre sus hijos; estos contribuyeron, por su parte, a su conservación; pero, convertida la observancia en superstición, y entendiendo siniestramente la doctrina de su legislador, mancharon con feos borrones una de las piezas mas finas y memorables que nos han quedado de aquella respetable antigüedad.

Los jeroglíficos tuvieron a mi ver el mismo principio que los calendarios; los egipcios cultivaron con tanto empeño los símbolos, que han pasado por inventores de ellos, entre muchos eruditos. Los monumentos faraónicos que ellos levantaron en los tiempos de su mayor opulencia, contribuyeron en una gran parte a ponerlos en posesión de esta gloria. Pero si atendemos al uso que hicieron los indios de los caracteres y pinturas simbólicas, nos veremos precisados a darles un origen más antiguo, y le habremos de buscar entre los primeros hombres; éstos, así como escogieron la luna para medir por ella los tiempos, señalaron también sus fases con ciertas figuras alusivas a las facciones humanas, las cuales fueron, en mi modo de pensar, el fundamento no solo de todos los símbolos sino de todas las letras. Así como con el curso de la escritura fueron declinando los rasgos de la pluma o del estilo hacia unas u otras formas, de la misma manera vario el pincel las lineas de los jeroglíficos, según el genio y gusto de las gentes.

Si se cotejan los jeroglíficos y cifras de los moscas ya con los signos de los meses de los egipcios, ya con varias letras asiáticas, se hallara toda la conformidad y analogía que es necesaria para establecer este pensamiento; Libra es una oreja, como la de ubchihica; Leo y Pisis, son el mica y muihica de los indios, y asi de otros. La medheoris, yet, thet, sirlaco y muchas otras letras asiáticas tienen una grande afinidad con estos caracteres.

Yo no decido sobre estas materias, someto mi juicio al de los eruditos, y me contento solo con este pequeño descubrimiento, que si merece la aprobación de los doctos, podrá contribuir de algún modo o dar alguna mayor luz a varios lugares oscuros de la historia.

\begin{center}
{\rm Papel Periódico Ilustrado. Año III. Pagina: 278 a 280.}
\end{center}

\newpage

\begin{center}
\textsc{anillo astronómico de los moscas}

Por el doctor Don José Domingo Duquesne de la Madrid.

1795

\end{center}

\begin{center}
\textsc{i - de los jeroglíficos}
\end{center}

Hieroglíphicos, según las dos palabras griegas que componen esta dicción quiere decir imagen o figura grabada. Diose este nombre a aquellas de que se servían los egipcios para representar los dogmas de su teología o las máximas de su ciencia moral y política que se veían esculpidas sobre piedras, pirámides, etc. No hubo rasgo o carácter entre estos indios gentiles, que no mirase alguno de estos objetos, y principalmente al primero; y así se cegó voluntariamente Walton para despreciar los símbolos mexicanos, conociéndose que no los entendió, en el mismo ejemplo que puso para rebatirlos.

Las pinturas de los indios algunas veces eran una pura escritura, explicaban sus pensamientos con imágenes, y faltando imágenes, con caracteres. En el padre García se puede ver un ejemplo de la confesión de los peruanos. Pero nuestros autores no explicaron los dichos caracteres ni nos han dado una lamina de ellos, en lo cual ha perdido mucho la historia. Los egipcios nos dejaron sus representaciones jeroglíficas en sus pirámides, los indios en sus diferentes piedras, pero los caracteres de éstos no se entendieron y se perdió el antiguo alfabeto de aquéllos, y de esta suerte estas dos naciones que cultivaron tanto los símbolos, se han hecho igualmente célebres y oscuras, y así sus monumentos sirven mas para atormentar los ingenios que para adelantar la erudición.

Pero todo esto que se admiró en los peruanos fue común a los moscas. Aunque muchas de las figuras simbólicas están tan enterradas, como las noticias de sus historias, no deja de haber algunas descubiertas por donde pueda cualquiera certificarse de esta verdad. Todavía se encuentran algunas piedras con animales grabados y distinguidas con lineas, ángulos, triángulos, etc. Se conservan algunas pinturas de colores en piedras expuestas al sol y al aire. que no ha podido borrar el tiempo, y entre otras, una muy particular de que habla el padre Zamora por estas palabras, tratando de una imagen que se halla en el pueblo de Guasca: ``Tiene barbas, sandalias y un libro en la mano; y a sus pies cinco renglones de caracteres tan incógnitos que no se ha podido entender su significado. Están a su lado dos compañeros con el mismo género de vestido''.

No es mi asunto la iconología de los moscas, si hubiese de hablar de sus diferentes imágenes, ya las que representan sus dioses, ya las que simbolizan a sus héroes, y son lo mismo que los manes de los antiguos, ya las que eran únicamente un voto u ofrenda que contenía la petición que hacían a sus númenes, me desviaría de mi principal argumento. Pero me ha sido preciso recordar o afirmar estas noticias para que a su vista se hagan memorables los jeroglíficos que voy a dar, y en cuya explicación se conocerá mejor el genio, las costumbres, la idolatría y el gobierno civil y político de los moscas.

\begin{center}
\textsc{ii - de los números}
\end{center}

Los moscas contaban por los dedos, esto significa el verbo \textit{zegitisuca}, cuya raiz es \textit{igiti}, el dedo. Solo tienen nombres propios para diez, en concluyendo con ellos pasaban de las manos a los pies, anteponiendo a cada uno la palabra \textit{quihicha}, que significa el pie: \textit{quihicha ata}, el uno del pie u once; \textit{quihicha bosa}, doce, etc.

El número 20, expresado por la palabra \textit{gueta}, casa y sementera, en que encerraban todos los bienes y felicidades de esta nación, era un total completo que cerraba todas las cuentas. Concluido un 20 pasaban a otro, que parece de la misma especie, al que llamaban \textit{gueta bosa, }dos veintes; \textit{gueta mica}, tres veintes, y de esta manera unían los unos con los otros hasta completar un veinte de veintes, y así en lo demás. Estos mismos números empleaban en sus medidas: \textit{iana}, el palmo, que era la menor, y \textit{pcuacua}, la brazada que era la mayor, y para el suelo se gobernaban por el paso, \textit{gata}, al que regulaban por la misma numeración, encerrándose en esto toda la aritmética que necesitaban para su comercio, agricultura, casas, labranzas, etc.

Los años pasados, meses, etc., los contaban para atrás como se ve en la tabla. El año presente corresponde a ata, el segundo a bosa, el tercero a mica, etc. y con el número 20 llenaban grandes espacios de tiempo en el uso de su cronología. Los números cardinales: lo primero, lo segundo, etc., \textit{quihina}, corresponden en la tabla a los numerales.

Cuando denunciaban la guerra asistían por 20 días seguidos en el campo, cantando y alegrándose con la esperanza de la victoria, y si perdían la batalla permanecían en el mismo campo otros 20 días llorando y lamentando su negra y desdichada fortuna. Se dice que el zipa Nemequene y el zaque Michua ajustaron una solemne tregua de 20 lunas; y que el misterioso sueño de su memorable Bochica duro en su fantasía veinte veces cinco veintes de años, todo lo cual nos confirma en la idea de su modo de contar que llevamos insinuado, para que no se juzgue que procedemos voluntariamente en estas imputaciones.

No solo tenían los moscas nombres para sus números, sino también guarismos para expresarlos. Descubrimiento también nuevo y correlativo a los símbolos. Tengo el gusto de servir al lector con una tabla de estas cifras que verosímilmente son las mismas que usaban los peruanos y de que solo nos había quedado una generalísima y confusa noticia que nos suministra el padre García por estas palabras: ``Suelen pintar los indios del Perú la confesión de todos sus pecados, pintando cada uno de los diez mandamientos por cierto modo, y luego allí haciendo ciertas señales como cifras que significan los pecados que han hecho contra aquel mandamiento''.

Este cierto modo y estas ciertas señales son las que hemos descubierto. Son, pues, estas cifras unos compendios de los símbolos, de los cuales cada uno vale uno de los números a que corresponde. El método que tenían de llevar estas cuentas es el siguiente; representaban, por ejemplo, diez sujetos que debían una determinada cantidad, en diez lineas, y al pie de la linea ponían la cantidad pagada, por donde se conocía lo que quedaba restando, hasta enterar todo el débito que quedaba señalado con el carácter correspondiente a \textit{gueta}, que en esta su aritmética es (según parece) no solo 20, sino una cifra igual a cualesquiera totales, Estos caracteres, en mi modo de pensar, ilustran mucho la historia como veremos adelante.

\begin{center}

\textsc{iii - origen de los numeros}

\end{center}

El círculo fue la figura mas usada de los moscas; daban esta figura a los cercados y palacios de los zipas y zaques, a sus casas particulares, a sus labranzas, a sus templos, en una palabra, a todas sus cosas. Fijaban en la tierra un palo, de que hacian centro, y con una cuerda describían alrededor el círculo.

Este parece haber sido el origen de los números: como entre ellos la casa y la labranza hacían todos sus bienes, el círculo con que describían uno y otro, fue la medida más propia para expresarlas. \textit{Abstascua }significa dar esta vuelta y sus dos raíces tienen una significación muy apropiada. \textit{Abos}. alrededor, y \textit{bta}, volver a otra cosa, y así de estas dos palabras formaron los dos primeros números (ata-bosa), porque concluyendo con una vuelta de los dedos, tenían que volver a otra, y como en las mismas manos tenían una imagen o representación del círculo, pareció esto lo mas natural y propio para explicarlo. Pero así estos números como los demás tienen otras significaciones muy acomodadas a todos los objetos a que los aplicaban.

Los símbolos tienen los mismos nombres de los números, pero aplicados a los meses; su representación es la siguiente:

\begin{description}
\item[1] \includegraphics[width=1cm]{ata-zerda.pdf} Ata. \textit{Los bienes --- otra cosa.}

\textit{Ata}: Un sapo en acción de brincar, que caracteriza la entrada del año.

\item[2] \includegraphics[width=1cm]{bosa-zerda.pdf} Bosa: \textit{Alrededor.}

\textit{Bosa:} Unas narices y las dos ventanas.

\item[3] \includegraphics[width=1cm]{mica-zerda.pdf} Mica. \textit{Parar, hallar, abrir, buscar, coger, cosa varia.}

\textit{Mica:} Dos ojos abiertos y las narices.

\item[4] \includegraphics[width=1cm]{muihica-zerda.pdf} Muihica. \textit{Piedra de la casa, cosa negra, crecer.}

\textit{Muihica:} Dos ojos cerrados.

\item[5] \includegraphics[width=1cm]{hisca-zerda.pdf} Hisca. \textit{Cosa verde. Alegría, echarse uno sobre otro, medicina.}

\textit{Hisca:} Unión de 2 fig.: símbolo de fecundidad.

\item[6] \includegraphics[width=1cm]{ta-zerda.pdf} Ta. \textit{Labranza, cosecha.}

\textit{Ta:} El palo y la cuerda: con que formaban el círculo de sus casas y de sus labranzas.

\item[7]\includegraphics[width=1cm]{cuhupcua-zerda.pdf} Cuhupcua. \textit{Sordo.}

\textit{Cuhupcua:} Las dos orejas tapadas.

\item[8]\includegraphics[width=1cm]{suhusa-zerda.pdf} Suhusa. \textit{No tirar de otra cosa. La raíz significa tender, extender.}

\textit{Suhusa:} El palo y la cuerda.

\item[9] \includegraphics[width=1cm]{aca-zerda.pdf} Aca. \textit{Los bienes.}

\textit{Aca:} sapo de cuya cola principia a formarse otro.

\item[10] \includegraphics[width=1cm]{ubchichica-zerda.pdf} Ubchihica. \textit{Luna resplandeciente, casa pintada, Pintar.}

\textit{Ubchihica:} Oreja, significa las fases de la luna.

\item[20]\includegraphics[width=1cm]{gueta-zerda.pdf} Gueta. \textit{Casa y sementera, tocar.}

\textit{Gueta:} Un sapo extendido o echado.

\end{description}

En estos nombres puede notarse la alusión que tienen los números con las letras de algunos orientales (que entre ellos son numerales), que son significativas y tomadas en una gran parte de los bienes, casas y sentidos del hombre.

Estas figuras son dirigidas a expresar los meses y el año, cuya artificiosa distribución esta simbolizada en estos aspectos, como vamos a exponer con la mayor claridad.

\begin{center}

\textsc{iv - año de los moscas}

\end{center}

Sin embargo de estar asistidos de las dos mayores luces, el sol y la luna, entramos en una provincia muy oscura. Habiendo puesto Dios estas dos grandes lumbreras en el cielo para que todos los hombres pudiesen computar por ellas los días, las noches, los meses y los años, entramos con ellas a registrar las profundas y lóbregas cavernas de la idolatría de esta nación y sus mas remotas antigüedades. No nos ha quedado otro medio, porque los autores que escribieron en el tiempo de su descubrimiento, no nos dieron noticia alguna fija en orden al año. Retrocederemos, pues, a aquella edad, caminando con tiento sobre algunas huellas y vestigios que han quedado aun estampados sobre sus labranzas, y el método que guardaban algunos sobre el tiempo de las siembras, junto con otras noticias ya generales, ya particulares, que nos pondrán en estado de conocer su antiguo año con la mayor seguridad.

En efecto, el año que voy a dar de los moscas, es una pieza completa y estoy perfectamente asegurado en orden a todo lo que voy a exponer. Me veo precisado a repetir esta advertencia, porque pudiéndose colegir de aquí algunas importantes verdades que ilustren la historia antigua, no quiero que se piense que me burlo con pensamientos ingeniosos en una materia en que interesa tanto la erudición. Bien que, creo que el lector erudito que se tome el trabajo de combinar bajo una idea las noticias históricas que nos han quedado de esta nación, si por otra parte ha tratado con alguna frecuencia (y no superficialmente) a los indios; si ha penetrado su genio y su carácter misterioso y enfático, conocerá la solidez de los fundamentos sobre que establecemos esta interpretación. En fin, el lector juzgara como gustare: yo estoy certificado de esta verdad.

Tenían los moscas su calendario descrito en las manos, teniendo en cada dedo colocados mentalmente sus signos, según el orden de sus números, a manera que los músicos tienen en la mano las cuerdas del sistema de Aretino. Este era el libro, o si se quiere, la tabla pública de su calendario.

Los signos que tienen facciones humanas representan los diferentes aspectos de la luna, que sucesivamente se varían en cada mes.

\textit{Mica }y \textit{Muihica} figuran las dos Zisigias; \textit{Mica}, en los ojos abiertos, el plenilunio y Ia oposición; \textit{Muihica} en los ojos cerrados, la conjunción o girante.

\textit{Bosa} y \textit{Cuhupcua}, las primeras fases de la luna, y \textit{Ubchihica} las cuadraturas.

El mes y la luna tienen un solo y mismo nombre en su lengua: Chia, y es la palabra que representaba todo lo hermoso, lo brillante, lo honorifico, porque la luz de este astro era entre ellos el símbolo de la belleza y de la virtud.

Esta distribución de fases en la mano esta llena de artificio; pero como no sabemos a punto fijo el uso que hicieran de ella los moscas, nos abstenemos por ahora de explicarla, contentándonos con insinuar lo que significa respecto a lo que nos consta que ellos practicaron.

\textit{Cuhupcua} es la neomenia. No es esta una neomenia tan puntual y precisa como la de los hebreos, observada desde los lugares mas altos y anunciada con trompetas, es una primera fase vulgar y universalmente conocida y como el novilunio en unos meses es mas breve, en otros mas tardío, se colocó en el día cuarto para que hubiese diferencia, a no ser que queramos que \textit{Bosa} represente la neomenia de los unos y \textit{Cuhupcua} la de los otros.

Contando desde \textit{Muihica} hallaban al número 7 la primera cuadratura, en \textit{Ubchihica} después de la neomenia; y contando desde Mica encontraban al número 8 la segunda después de la oposición, en el mismo signo. La media oreja que lo caracteriza es un símbolo muy natural de la cuadratura.

Finalmente \textit{Bosa} les daba la ultima fase de la luna colocada al tercer dedo antes de \textit{Muihica}.

Esta distribución vulgar y acomodada de las fases de la luna, les daba dividido un mes en diferentes términos, de que podían servirse para el arreglo de los negocios públicos, y a la verdad \textit{Ubchihica} estaba destinado tal vez para sus juntas o cosas de religión (significa también sentarse mucho en compañía), puede aludir a sus cacerías, y así le podemos considerar como señalado con una de aquellas letras que los romanos llamaron nundinales.

Si parece artificiosa esta tabla o digamos mejor esta mano astronómica, de que no hemos explicado mas que los primeros aspectos, lo es mucho mas el año, pues tenían dos especies de año; el uno vulgar de veinte lunas y el otro de treinta y siete, al que por ahora llamaremos astronómico, pero de tal suerte mezclados entre si, que no sólo no se turbaban ni confundían, sino se ayudaban sirviendo el uno a dirigir. conservar y facilitar el uso y la perpetuidad del otro.

Ya hemos dicho muchas veces que los moscas miraban como sagrado el número 20. No podían menos que ajustar por él el año, porque de otra suerte se hubieran confundido en todas sus cuentas. Los plazos para los pagos en su comercio, las convenciones solemnes entre sus jefes, el orden de los sucesos y la cronología de su nación, todo se debía gobernar por este número; \textit{Gueta }era el símbolo de la felicidad, y entre esta gente supersticiosa hubieran sido menguados e infelices los años que no se hubiesen sellado con este carácter; era, pues, inexcusable entre ellos el año de veinte lunas. Pero en este supuesto, ¡Cuántos inconvenientes! Los equinoccios se hubieran mudado: esto les hubiera sido de poco cuidado a los indios, porque como distamos unos cuatro grados y medio de la linea, tenemos los días y las noches perfectamente iguales. Pero los dos inviernos y los dos veranos que logramos en la zona tórrida y que solo consisten en que llueve o no llueve, se hubieran trocado; corriendo sus signos indiferentemente por un círculo perpetuo pasarían por todas las estaciones; y se hubiera trastornado también el orden de las siembras y jamás hubieran podido hallar un punto fijo para gobernar sus labranzas, negocio importante para toda gente, pero mucho mas para ellos que hacían su primer ídolo de la sementera, por cuya causa inventaron los otros años; pudiéndose decir de los moscas que Ceres fue la madre de todos sus dioses.

No hubo, pues, otro medio que ocultar un año particular en el año vulgar de veinte lunas, pero con artificio tan fino y delicado que su uso se facilitase, aun respecto del rudo pueblo, y que para este fin se correspondiesen entre si con tanta regularidad que en ninguna manera pudiesen confundirse.

Para explicarlo mejor debo suponer que un año lunar de doce lunaciones compone solamente la suma de días y seis horas, y así, es once días mas corto que el solar que consta de 365; es necesario, pues, añadir una lunación entera al tercer año lunar, de suerte que tenga trece lunaciones, y de esta manera vienen a coincidir el lunar y el solar pasada la luna 37\ra{}, pues 37 meses lunares componen la misma suma de días que 36 solares. Este computo es muy antiguo y le han conocido casi todas y las naciones, y así sobre este principio ruedan todos los calendarios.

En este supuesto se entenderá el año oculto de los moscas, porque pasadas las 20 lunas de un año, al siguiente, llegando a la 17\ra{} en que les competía sembrar, según el mes por donde habían comenzado, la intercalaban, es decir, la dejaban pasar como inoficiosa, y sembraban en la siguiente que era la 18\ra{}, y de este modo, aunque corrían perpetuamente sin intermisión el círculo de 20 lunas, cada bienio intercalaban del modo expresado la luna, a la que correspondía el número diez y siete.

Este método hubiera sido muy expuesto y confuso si no se hubiera establecido bajo un pie fijo y con reglas fáciles y generales. Tenía, pues, destinados para este objeto, cuatro signos: el primero \textit{Ata}, y los tres últimos del año \textit{Suhusa}, al cual llamaban por esto, la cola; \textit{Aca }y \textit{Gueta}, que eran los que fenecían el año de veinte lunas y a los que por esta razón podemos llamar terminales. Comenzando, pues, por \textit{Ata}, tocaba el número 17 del siguiente a \textit{Cuhupcua}, a quien por esta intercalación llamaban con toda propiedad e1 sordo, y se hacia la siembra en el siguiente \textit{Suhusa} 18. Pasado el año de veinte lunas, al siguiente tocaba el número 17 a \textit{Muihica}, el ciego, y pasaba la siembra y representación del primer mes a \textit{Aca}. En e1 bienio de éste correspondía el 17 a \textit{Hisca}, y entraba \textit{Gueta} en el lugar y en los oficios de \textit{Aca}; en su bienio era 17, \textit{Ta}, y volvía el turno al primero, \textit{Ata}.

Aunque \textit{Gueta}, 20, es a quien toca el expresado oficio en la tabla que hemos formado de este año. en la restante explicación por la mayor facilidad usamos del carácter \textit{Ubchihica}, que es lo mismo para este efecto.

Al año de veinte lunas llamaban Zocam, que según sus raíces quiere decir asir o aferrar de arriba, término muy significativo, que expresa la unión del uno con el otro, esto es, del año de veinte lunas que pasó con el de las otras veinte que sigue para seguir su vuelta en un continuado círculo; palabra enérgica que corresponde a la \textit{xiuhmolpili} con que los mexicanos conocían aquella su célebre rueda de cincuenta y dos en cincuenta y dos años, que en aquella lengua quiere decir atadura de los años.

Para el otro año que estaba como embebido en este, o no tenía voz con qué significarlo o le conocían por el de \textit{Cuhupcua}, en cuya virtud, para un año tan desconocido me permitirá el lector usar de término nuevo, y le llamaremos \textit{Acrótomos}; esta palabra griega significa una cosa cuyas extremidades o partes últimas están cortadas, y así me parece que señala con toda propiedad el año de que vamos hablando.

Con lo que hemos expuesto se conoce la idea del calendario, pero para que se pueda formar un entero
concepto de la correspondencia de todos sus meses, hemos formado la tabla que lo contiene, y que para mayor claridad e individuación explicaremos en el número siguiente con todas sus circunstancias.

\begin{center}
\textsc{v - del año acrotomos}

(Año astronómico o intercalar)
\end{center}

El año acrótomos se gobierna con tres signos terminales que están como partidos, para este intento, y que sin embargo conservan las representaciones de su número en el año que va continuado de 20 lunas en el modo siguiente: Sea \textit{Ata} por suposición correspondiente a enero como apto para las siembras; este signo tiene dos oficios en este caso: ser primer mes en el año de veinte lunas que comienza, cuyo carácter no pierde jamás; y ser primero del año acrótomos, cuyo carácter pierde en llegando a su término que es la luna 37. Señala, pues, las sementeras que se deben hacer de doce en doce lunas, y que son otros tantos meses enero del año acrótomos, en el modo siguiente: (a \textit{Ata}, y a los demás terminales les corresponde la siembra primera de su respectivo año) conviene a saber, 2\ra{} siembra \textit{Mica}, 3\ra{} \textit{Hisca}, 4\ra{} \textit{Cuhupcua}, inoficioso intercalar, y así sigue otro turno; entre tanto, \textit{Ata} dejando este oficio a \textit{Suhusa} continua gobernando su año de 20 lunas, que no se ha concluído, \textit{Suhusa} en este círculo perpetuo y en esta situación es el mes l8; pero en el turno del año acrótomos le toca ser primero por exclusión de \textit{Cuhupcua}; en éste, pues, señala las sementeras siguientes: 2\ra{} \textit{Ubchihica}, 3\ra{}  Bosa, 4\ra{} número 37, \textit{Muihica }se elude por inoficioso; concluye su turno \textit{Suhusa}, y en su lugar y con la representación de enero entra \textit{Aca} que es número 19 en el año de 20 lunas; pero primero en el año acrótomos señala las sementeras siguientes: 2\ra{} \textit{Ata,} 3\ra{} \textit{Mica,} 4\ra{} \textit{Hisca} intercalar número 37; concluído este turno entra \textit{Ubchihica}, último mes del año de 20 lunas y primero en este turno del año acrótomos; sus sementeras son 2\ra{} \textit{Bosa}, 3\ra{} \textit{Muihica}, 4\ra{} la número 37 se elude y vuelve el turno a Ata. Este es el sistema del año de los indios muiscas.

\begin{center}

---

\end{center}

\textsf{En la memoria que Duquesne presento al célebre botánico Mutis y que fue el resumen de sus estudios, dice:}

Esta intercalación que se verifica perpetuamente, dejando pasar como inoficiosa o como sorda la luna 37, nos hace concebir que dentro de los dos años vulgares de veinte lunas cada uno, hay otro año astronómico oculto que consta de 37 lunas, de modo que la luna 38 seria un verdadero enero. Los indios sin penetrar la teoría de esta proposición, que ha sido embarazosa en otras naciones mas cultas, por esta luna que ha sido necesario añadir al fin de cada tres años lunares, por ser los dos anteriores de doce lunas y el tercero de trece, tenían suma facilidad en la práctica de su intercalación, siguiendo el método propuesto, conservando así el año astronómico, sin que el pueblo notase diferencia alguna en sus años vulgares de veinte lunas cada uno.

El año vulgar de veinte lunas servia para las treguas en la guerra, como consta de su historia, para las compras y ventas y otros negocios de la sociedad. Pero el año astronómico e intercalar de 37 lunas, que se contaba por tres sementeras, servía principalmente a la agricultura y a la religión, y así llevaban su cuenta con mucha prolijidad los jeques (sacerdotes) y mayores a quienes correspondía, notando sus épocas con sacrificios mas particulares, y grabándolas también en piedra por medio de símbolos o figuras, como se ve en un pentágono que tengo en mi poder y voy a explicar al fin de este papel.

El siglo, pues, de los muiscas constaba de veinte años intercalares de 37 lunas cada uno, que corresponden a sesenta años nuestros y lo componían de cuatro revoluciones contadas de cinco en cinco, cada una de las cuales constaba de diez años muiscas, y quince nuestros, hasta completar los veinte, en que el signo \textit{Ata} vuelve a tomar el turno de donde comenzó la vez primera. La primera revolución se cerraba en \textit{Hisca}, la segunda en \textit{Ubchihica, }la tercera en \textit{Quihicha Hisca} y la cuarta en Gueta.%»

\begin{center}
\textsc{vi -- de las semanas, dias, etc. }
\end{center}

Al día artificial le llamaban \textit{Sua}, esto es, un sol midiéndole de levante a poniente. A la noche llamaban \textit{Za }del ocaso hasta su levante. Al día lo dividían en dos partes, del levante al punto de la meridiana, la mañana, \textit{Suamena}; y desde ésta al ocaso la tarde \textit{Suameca}. Del ocaso hasta el término de una hora u hora y media, la prima noche, \textit{Zasca}; y desde la una de la noche al levante del sol, la madrugada \textit{Cagui}. Estas eran sus horas, distinguidos los dos puntos de madrugada y prima noche, con dos comidas que eran y son las únicas que hacen.

La semana era de tres días, y se conoce que usaban de ella porque cada tres días hacían en Turmequé, lugar perteneciente al Guatavita, un mercado. De suerte que así como los mexicanos hacían sus mercados de cinco en cinco días para dividir por este número los días de la luna, como lo afirma el padre Torquemada, por una analogía semejante hacían el mas rico los moscas de tres en tres días, por relación a la memoria de que hemos hablado.

\begin{center}
{\rm (Papel Periódico Ilustrado. Año III. Pagina: 298 a 303)}
\end{center}

\newpage

\begin{center}
\textsc{sacrificio de los moscas y significados o alusiones de los nombres de sus víctimas}

Por el doctor José Domingo Duquesne -- 1795
\end{center}

Tenían los moscas el cuidado de señalar las revoluciones del año con las cosas más notables. No había
siembra ni cosecha sin sacrificio. Tenían en cada pueblo una calzada ancha y nivelada que salía del cercado o casa del cacique, y corría como por media legua, rematando en un palo labrado en figura de una gavia, de que pendían al miserable cautivo que ofrecían al sol y a la luna para obtener una cosecha abundante.

Venían en mojiganga los indios, repartidos en diferentes cuadrillas, adornados de muchas joyas, lunas y
medias lunas de oro; disfrazados unos con pieles de osos, tigres y leones; enmascarados otros con máscaras de oro, y lagrimas bien retratadas, a los cuales seguían otros con mucha gritería y risotadas, bailando y brincando con descompasados movimientos; otros traían unas grandes y largas colas, que iban pisando los que los seguían, y llegando al término de la calzada disparaban todos sus flechas y tiraderas al infeliz cautivo, matándole con larga muerte, y recibiendo su sangre en diferentes vasijas, terminaban la barbara función con sus acostumbradas borracheras.

Nuestros historiadores se admiran mucho del fausto y de la extravagancia de estas procesiones, pero nos dieron una idea muy diminuta, refiriendo de una manera general sus cuadrillas. En lo poco que describieron se conoce que esta mojiganga o procesión era un símbolo de su calendario, y, si las hubieran dibujado todas, nos ayudarían a formar el concepto de sus signos, y de los caracteres que les atribuían.

Pero la víctima destinada a solemnizar las cuatro lunas intercalares que partían el siglo, estaba señalada con muchas circunstancias. Era esta un miserable mancebo, que precisamente había de ser natural de cierto pueblo, sito en los llanos que llamamos hoy de San Juan. Horadábanle las orejas, le criaban desde mediano en el templo del sol; en llegando a diez años nuestros, le sacaban para pasearle, en memoria de las peregrinaciones del Bochica su fundador, a quien se figuraban colocado en el sol, y continuando en un matrimonio feliz con la luna y una lucidísima descendencia. Vendíanle en precio muy alto, y era depositado en el templo del sol hasta cumplir quince años nuestros, en cuya precisa edad hacian el bárbaro sacrificio, sacándole el corazón y las entrañas para ofrecérselas al sol.

A este mozo le llamaban \textit{Guesa}, y también \textit{Quihica}: \textit{Guesa}, que quiere decir mancebo, de aquí sale el verbo \textit{guesansuca}, ir creciendo en edad, pero esta palabra tiene mucho énfasis porque denota precisamente una edad que no puede llegar a 20 años; y así para la juventud más crecida o mayor, tienen otro término que es Guas-gua-cha. Esto es así, y se reconoce más buscando las raíces de que esta formada, que son \textit{gue} y \textit{za}, partícula negativa que quiere decir literalmente sin veinte; por la historia se conoce la imposición de este nombre.

Mas, así como \textit{gue} es la raiz de veinte (\textit{Gueta}) también lo es de esta palabra casa, y quiere decir sin casa,lo que encierra la otra circunstancia que según la historia debía tener este miserable mozo, pues que lo criaban en un templo del sol, y en comprándolo debían colocarlo en otro templo, y por esta razón no tenia casa. Querían los indios que no tuviese el menor comercio, y menos impuro, porque en este caso le desechaban, y a esto miran directamente estas significaciones.

Mas también puede significar casa oscura, porque \textit{gue} es la casa, y \textit{zu} la noche, y es modo que se conforma con sus frases. Lo que hace relación a la conjunción de la luna.

\textit{Quihica} es la boca, y por alusión la puerta, pero en su lenguaje no tiene otro término que éste para significar la puerta de la casa, o de cualquier otra cosa. Por tanto, la primera y obvia significación de estas dicciones, \textit{gueza quihica}, es el mancebo, que es boca, o el mancebo que es puerta; según, pues, las explicaciones de arriba, puede decir, \textit{la puerta de la casa oscura}, o d\textit{e la noche oscura, }o \textit{de la juventud.}

Se debe suponer que en esta lengua, como en todas las orientales, cada palabra es una definición, y las compuertas encierran muchos sentidos, y son muy enfáticas en todas sus alusiones. \textit{Cuhucuaque} tiene la misma raíz que \textit{Cuhupcua}, o por decirlo mejor, es la misma voz tomada adverbialmente, y significa señal. \textit{Muihica}, se compone de \textit{mui} y de \textit{hica:} \textit{mui} es el palacio, la casa grande: \textit{hica}, la palabra, y es lo mismo que decir en nuestra lengua la palabra de la casa grande o del palacio. Mas, esta raíz, \textit{mui}, es raíz del verbo \textit{muisca}, que significa tender o extender, el participio del presente que también es muisca, significa el hombre y esta palabra que denota toda la nación, por lo cual los españoles (aun sin entender todavía la lengua de este pueblo) atendiendo a su prodigiosa multitud, corrompieron el vocablo y les llamaron moscas. Los indios tuvieron en cuenta el mismo fin cuando aplicaron a toda su nación este verbo, esto es, gente extendida. Es muy enérgica esta voz, porque alude al barro de que Dios formo el primer hombre, y de aquí nació también la palabra \textit{muisquien}, que significa la naturaleza, y \textit{muihica}, cosa negra, por el color del barro y el de la gente de la misma nación. Del mismo verbo nace la palabra \textit{muiso}, mudada la partícula \textit{sca} en \textit{so}, como acostumbran estos indios, significa cosa tendida o arrastrada, y aplicaron esta voz a significar la culebra, y los indios de hoy (1790), llaman en español a este animal rastra, conservando el significado de su lengua.

Por esto se vendrá en conocimiento de los misterios que encierran la lunación designada con esta voz \textit{muihica}, que hace relación a las casas de los indios, a sus personas, a su nación, a su lengua, a las culebras y a la oscuridad de la conjunción, porque en si encierra con un énfasis particular todos estos significados, y unida a la voz \textit{guesa}, la puerta de la casa o de la noche oscura, o la boca de la inocencia, de la juventud, etc.

Esta víctima (\textit{el guesa}) según la historia, estaba dedicada al sol; véase aquí, pues, una representación del Jano de las otras naciones, No era otro este dios multifirme que el sol que gobierna el tiempo, que con una cara mira a lo pasado y con otra a lo futuro. Se han encontrado también medallas que tenían cuatro caras, aludiendo a las cuatro estaciones. La etimología de puerta (\textit{quihica}) es bien conocida, pero se debe saber que la imposición de este nombre no era sólo porque mirase los tiempos, sino porque le consideraban como una puerta por donde entraban todos sus ruegos a los dioses. Así Ovidio, gran comentador de la teología pagana, dice:

\begin{center}
\ldots Cur quamvis aliorum numina placem,\\
¿Jane tibiprimum thura merumqle fero?\\
Ut possis aditum per me qui limina servo\\
ad quos cumque voles. inquit, habere deos.
\end{center}

No es tiempo de examinar el origen de la fábula en los indios, asunto que tocaremos en disertación separada, y que es igualmente original porque hasta ahora no la ha tratado ningún erudito. Pero no es menester mucha penetración para conocer que las mismas causas que han movido a los romanos para estas invenciones, intervinieron entre los indios, Testo señala esta para aquellas gentes: \textit{quod fuerit omnium primus }(dice)\textit{ a quor erum omnium putabant initium ideo ei suplicabant velut paretis}:. Los indios, según la historia, habían colocado en el sol a su padre y fundador Bochica, y así les fue fácil tratarlo con los mismos respetos; pero el verdadero motivo de los indios estaba en la persuación en que estaban de la sordera de la luna; estaba su casa cerrada y era necesario abrir la puerta y esta miserable victima (\textit{el guesa}) era en su concepto la puerta por donde entraban sin estorbo sus ruegos.

A mas de esto era la boca de la nación que hablaba de cerca a la luna sorda y así no podía desentenderse de sus gemidos cuando llegasen hasta el cielo, porque gritando ellos desde acá abajo no los oía. Este era el modo de discurrir de esta gente ilusa, y es bien claro porque no hacían sacrificios de otros animales que de loros, papagayos y pericos, y éstos no llegaban a las aras hasta que hubiesen aprendido su lengua, porque sus victimas habían de tener voz para gritar de cerca a la luna. Volaban, en su opinión, los pericos y los loros después de muertos, aún más arriba que cuando vivos. Pero si no basto el águila a Júpiter, fue necesario que esta ave arrebatase a Ganimedes: este es el infeliz y desdichado guesa, esto es, la señal de sus años y la víctima que se hacía a \textit{Cuhupcua }cada quince años nuestros, es decir, cada cinco acrótomos; hasta que señalado cada uno de los caracteres con una victima particular concluyese en el círculo del tiempo, una vuelta entera hasta veinte que hacen 60 años nuestros, que era su edad privilegiada.

Es verdad que el señor Piedrahita habla tan generalmente y pone tan vagas todas las circunstancias de este sacrificio, que no lo pudo determinar. Parece una solemnidad incierta y voluntaria, pero no es así. No digo yo un sacrificio de tan prolijas circunstancias, que efectivamente se conoce ser el compendio y la cifra misteriosa de todas sus supersticiones: pero aun las cosas de menos consideración, todo, grande o pequeño, es nivelado por sus reglas entre los indios. Todos saben que éstos han sido la gente de las ceremonias, y que no hay ninguna que no tenga su peso y su medida determinados.

En esta consideración, habiendo omitido nuestros autores todas aquellas circunstancias que desestimaron por parecerles ridículas y extravagantes o porque no eran necesarias para llenar los intentos y objetos de sus historias, nos vemos en la necesidad de ilustrar esta parte, haciéndonos prolijos, contra nuestro genio, para satisfacer cumplidamente al lector, y para que no juzgue que trabajamos sobre nuestras voluntarias imaginaciones.

El Bochica (de cuyos caracteres trataremos separadamente) fue el fundador, legislador y padre de los moscas. Sea que viniese por la Groenlandia a pasar por el Istmo de Panamá, por larguísimos rodeos y giros interminables, o sea porque navegase desde el cabo de San Vicente, como parece natural, es verdad averiguada que llegó por los llanos que hoy llamamos de San Juan a tomar posesión de Sogamoso, y desde allí, de todos los hermosos países que habito la extendida y númerosa nación de los moscas. Vivió largo tiempo (aunque no el que quisiesen los indios), como se puede creer de los primeros pobladores, habiendo sido los maestros de sus hijos a quienes comunicaron los primeros elementos de las artes, en las que después se ejercitaron, las leyes fundamentales de sus gobiernos y el uso de los tiempos para el arreglo de las labranzas. A lo menos los fundamentos para pensar que el Bochica tuvo tiempo para sembrar en sus nietos estos conocimientos. Hallándose viejo, sea que quisiese despedirse de sus hijos y que quisiese repartirles por si mismo las tierras en que se establecieron; o sea que los que se encaminaron a Bacatá, necesitasen de sus consejos y dirección para facilitar el paso al rió Funza, que anegaba los mejores y mas extendidos campos; ya porque se detenían en algunos estorbos formados por el inmediato diluvio; ya porque los mismos campos bajos facilitasen la salida de las aguas, es de suponerse que él intervino en la dirección de estas obras, y que visito muchos de aquellos lugares. Restituido a Sogamoso, murió dejando por heredero de la suprema autoridad, que le competía como a cabeza y padre de toda la nación, a su primogénito.

Es muy regular que un anciano de tanto mérito, fuese el oráculo de los moscas cuando vivo, y que lo fuese de sus lagrimas y de sus deseos después de muerto. Pero los indios no conocieron los limites del respeto y de los obsequios debidos al padre y pasaron a los cultos propios de la divinidad. Estos hombres quisieron perpetuar la memoria de la venida del Bochica haciendo una calzada o carrera desde la boca de los Llanos a Sogamoso, que tendrá como cien leguas de longitud, muy ancha y con sus valladares o pretiles por una y otra parte, aunque ya maltratada y oscurecida con la paja y barrizal que se han criado en ella, por la cual dicen que subió Bochica desde los Llanos al Nuevo Reino.

De aquí tomaron idea para hacer otras calzadas semejantes, como la de Bacatá, y en los lugares mas señalados con los vestigios del Bochica, hasta que paso a ser adorno general de todos sus pueblos, y entrada de los templos y casas de sus caciques, en donde se ejecutaban las danzas, procesiones y sacrificios.

Por alusión a estas tradiciones o por la mayor opulencia de los caciques se hicieron con el tiempo más
célebres algunos lugares, y como adoraban al Bochica, colocado en el sol, pretendieron ennoblecerlos con templos más suntuosos que eran como los santuarios de su mayor veneración.

El templo de Sogamoso, dedicado al sol, era el centro de su religión y el mas privilegiado. Seguían a éste los de Bacatá, Guachetá y Guatavita; pero el principal era aquel que tenían en el pueblo en donde comenzaba la antigua y espaciosa calzada que servia de memoria perpetua de su establecimiento en este lugar y de las hazanas gloriosas de su héroe.

Esta es en pocas palabras la historia primitiva de los moscas, desenredada de las fabulas de los indios que de estos pasajes históricos hicieron innumerables misterios. Entre ellos se debe reparar el misterioso sueño del Bochica que estuvo durmiendo en Sogamoso veinte veces cinco veintes de años, cuya portentosa época quisieron conservar a esfuerzos de las mayores crueldades.

He aquí puesta en claro la historia del guesa con todas sus circunstancias. El 20 y el 5 veces 20 es misterioso en toda la secuela del sacrificio; por eso dividieron los años terminales de cinco en cinco, acabando de contar en el que habían comenzado y de esta manera cada cuatro acrótomos salía la víctima del templo a recordar a los indios que se acercaba esta estupenda solemnidad, pues que así acostumbraban estarse avisando sucesivamente de varios modos de lo que tienen mas presente y nunca se les olvida. Era sacrificada la victima al fin del quinto acrótomos, y ese mismo día era entregado a los sátrapas del templo de los llanos el sucesor de esta desgraciada, y de esta manera señalado cada quinquenio con una victima en cuatro actos detestables y crueles, se concluía la lastimosa tragedia, y se contaba uno de los señalados 20 que no tenia cuando acabarse, y empezaba otra escena.

Los indios tuvieron la crueldad de imponer a sus víctimas los nombres y la representación de sus dioses: las historias de México están llenas de estas narraciones; por lo que mira a los moscas, ya que no tuvieron al guesa por el Bochica, le tenían por hijo suyo, por tanto el paseo que le hacían dar por las poblaciones no carecía de misterio, era este una especie de peregrinación que llevaba sus representantes de los viajes del Bochica. Verosímilmente le tocaba a cada templo de los memorables y dedicados al sol por esta circunstancia, la compra de la víctima en cada turno. Los templos eran cuatro, y servía esta distribución de nueva serial para el gobierno de los años, porque empezaba nueva edad cada vez que se concluían las cuatro estaciones; el que conozca los libros de cuentas que usan los indios hallara verosimilitud en este computo. Es de advertir que a las calzadas o camellones les daban el nombre que a los años; suna ata, suna bosa, etc., un camellón, dos camellones. etc., porque como éstos eran los teatros de sus procesiones y sacrificios eran también el libro en que se iban registrando.

Aun cuando caminaban con el muchacho (el guesa) buscando en todas partes quien lo comprara, esto era una formula, pues no lo podían vender sino en el lugar de su destino. Y así, aunque por veneración compareciesen haciendo ostentación de hacerse dueños de una victima, entre ellos tan preciosa, se cuidaban mucho los mercaderes de ponerle unos precios excesivamente crecidos, como que no podían venderla, y esta cruel ficción había pasado a costumbre y a misterio. Así nacen las cosas entre el pueblo, y así crecen entre supersticiones. Lo cierto es que aunque fuese cara para que no se hiciese común, ya tenia su precio señalado y fijo, del que no podía pasarse, y yo aseguro que en cada pueblo estaba tan bien repartido este precio entre las personas a quienes tocaba comprarlo, porque esto es lo conforme con su genio y con su política.

Esta víctima, que era un sacrificio publico de la nación, bastaba para fijar los años, no solo entre los magistrados y sacerdotes de los templos que debían llevar una cuenta exacta del calendario sino en todos los pueblos: pero como este sacrificio se hacia en una sola parte, a la que no podían concurrir todos, es de presumirse que tendrían en los demás otra ceremonia igualmente cruel para solemnizar esta memorable revolución de sus años. El padre Zamora asegura que ademas del sacrificio del guesa, el demonio les había persuadido que no había otro mas grato a los dioses que el de algunos mancebos que no llegasen a veinte años; así, aunque en las demás partes no fuesen tan costosas, ni tan circunstanciadas éstas víctimas, estaban señaladas con la edad, que era el misterio principal en semejantes ofrendas.

\begin{center}
{\rm (Papel Periódico Ilustrado. Año III. Pagina: 313 a 315) }                                             
\end{center}

\newpage
\begin{center}
\textsc{explicación de los símbolos del siglo}

%(Figura 48)

Por el doctor José Domingo Duquesne
\end{center}

\label{stone-ring}

Tengo en mi poder un manuscrito, que según parece contiene una de estas revoluciones del tiempo. Los amantes de la bella literatura gustaran de leer en estos caracteres tan extraños, y mucho mas cuando se puede mirar como un pedazo del alfabeto chibcha, con cuyas notas se podrán imponer y aun adelantar en otros semejantes; de suerte que estos fragmentos que suelen encontrarse, no serán en adelante un mueble vano, sino un adorno importante de un gabinete de historia.

Es una piedra chica, especie de jaspe negro, tersa, y su figura un pentágono. El primer lado es más largo que los otros, tiene de relieve la figura de un sapo o rana, con más cola y sin patas, \textit{e}, sobre un plano limitado por cuatro líneas; mas adelante se encuentra una linea gruesa, \textit{f,} en forma de una culebrilla que en el dorso tiene dos lineas paralelas de iguales dimensiones. El segundo tiene grabado un sapo, \textit{a}, en actitud de brincar sobre un plano limitado por cuatro lineas como el anterior. El tercer lado contiene una linea gruesa en forma de un dedo, \textit{b}, señalado con tres líneas gruesas transversales y en medio del dorso se levanta una prominencia casi como una nariz, señalada por los lados con dos puntos opuestos, uno en cada lado. El cuarto contiene otra linea gruesa, \textit{c}, como la anterior; esta figura se distingue en que no ocupa el centro, esta hacia un lado del plano. El quinto, \textit{d}, es como los dos anteriores, conservando el centro como la primera.

Uno de los planos tiene un círculo, \textit{g}, cortado por un segmento, en el cual hay una especie de dibujo con doble línea en angulo obtuso, \textit{h}. En el otro plano se ve un círculo menor, \textit{k}, que tiene en el centro un punto, y esta cortado, con un segmento, de cuyos extremos parten dos líneas que se unen hacia fuera formando un angulo; en el mismo plano se encuentra la figura \textit{m}, que es una culebrilla; en su dorso se ven dos lineas paralelas; hacia la cabeza tiene un angulo agudo puesto de lado y a la cola un triangulo partido por una linea mas gruesa tirada descuidadamente.

\textit{Interpretación}. -- Esta simbolizada en esta piedra la primera revolución del siglo muisca, que comienza en Ata, y acaba en Hisca, el cual incluye nueve años y cinco lunas muiscas. Los indios, que para todo usan el círculo, aquí prefieren el pentágono, para significar que hablan de cinco años intercalares.

\textit{a}. El sapo en acción de brincar es el signo del principio del año y del siglo.

\textit{b}. Esta especie de dedo señala en las tres lineas gruesas, tres años.

Omitiendo, pues, el dedo \textit{c}, que esta a un lado, cuento en el dedo \textit{d}, otros tres años, que, juntos con los del dedo \textit{b}, producen seis. Lo cual denota la intercalación de \textit{Quihicha ata}, que sucede puntualmente a los seis años muiscas, como se ve en la tabla; y es de mucha consideración entre los indios por pertenecer al sapo que arregla todo el calendario.

\textit{e}. Es el cuerpo de un sapo de cola y sin patas, es símbolo de Quihicha ata; y por carecer de patas es figura muy propia para expresar su intercalación, porque el mes intercalar no se computa para la sementera, y así lo imaginaban sin acción y sin movimiento. Se ve sobre un plano, como también el sapo \textit{Ata}, para significar que en una y otra parte se habla del sapo.

\textit{f}. Esta culebrilla representa el signo Suhuza, que es el que se intercala después de\textit{ Quihicha ata}, a los dos años muiscas representados en las dos líneas gruesas que tiene el dorso. Lo que corresponde al año octavo. como se ve en la tabla.

Como concluimos con los lados del pentágono pasemos al plano.

La culebra \textit{m} es una reproducción de Suhuza, y como esta tendida sobre una especie de triangulo, símbolo de Hisca, significa que se intercala inmediatamente después de Suhuza al segundo año, lo que esta figurado igualmente en las dos lineas gruesas que tiene el dorso.

Como el fin principal de esta piedra cronológica es señalar la intercalación del signo de Hisca, por ser el término de la primera revolución del signo muisca, para mayor claridad están contados estos años en los tres dedos, conviene a saber: \textit{b, c, d,} que juntos producen nueve años, que son los que dan puntualmente esta notable intercalación, que sucede a los nueve años y cinco meses como se ve en la tabla.

\textit{g}. Es un templo cerrado; \textit{h}, es la cerradura que hasta el dia de hoy usan los indios en sus puertas, y llaman candado \textit{cormo}. Los agujeros de las dos orejas sirven a la estacas que le ponen, y los dos ganchos interiores para asegurar la puerta. Significa la primera revolución del siglo, cerrada en \textit{Hisca}, y para que continuase el tiempo, era necesario en su imaginación (es decir, en opinión de los indios) que el guesa abriese la puerta con el sacrificio de que hemos hablado y cuyas circunstancias eran simbólicas, relativas a esta revolución del siglo.

El círculo menor, \textit{k}, con los radios que están en el otro plano, figuran a \textit{Cuhupcua}, esto es, la luna intercalar 3 y sorda, y la unión y conjunción particular del sol con la luna que veneran tan misteriosamente y a la que se dirigía esta revolución.

La culebra \textit{m}, es símbolo del tiempo. El angulo es número cinco como el de los romanos: le usaban los indios para explicar cinco, porque contaban por los dedos levantando el dedo indice y el dedo medio en alto, como todavía practican; esta figura y las líneas del dorso de la culebra, que es una representación de \textit{Suhuza}, significan que se deben tomar los terminales cinco veces. como ya hemos explicado.

La culebra. por otra parte, ha sido el símbolo del tiempo en todas las naciones. Esta primera revolución del siglo estaba consagrada principalmente a las nupcias del sol y la luna, simbolizadas en el triangulo, no sólo según los indios, sino según otras naciones.

\begin{center}
\textsc{explicación de la tabla de los años moscas}

Por el doctor José Domingo Duquesne                                   
\end{center}

El círculo interior representa las 20 lunas del año muisca vulgar, cuyos signos todos se intercalan en el espacio del siglo.

El círculo segundo expresa los años muiscas a que corresponde la intercalación de cada signo.

El círculo tercero expresa el orden de esta intercalación.

Ejemplo: Deseo saber en qué año muisca se intercala el signo mica. Veo en la tabla en el número tres en el círculo interior, hallo en el segundo que le corresponde el número 36, y este es el año que se busca; veo en el siguiente círculo que le corresponde el número 19, y así la intercalación de mica es en orden la décima nona del siglo.

La intercalación de \textit{gueta} (20) es la ultima del año muisca treinta y siete, esto es, después de un siglo vulgar muisca de años de 20 lunas y mas diez y siete años, de suerte que, terminado el siglo, o revolución astronómica de 20 años intercalares de 37 lunas, cada uno, les faltan tres vulgares, para completar dos siglos vulgares. En llegando, pues, a este caso, no hacían mas cuenta de aquellos tres años vulgares de que no necesitaban para la labranza, ni para la religión, ni para la historia, y empezaban en \textit{ata} (a que había llegado el turno) un año vulgar, nuevo, principio de un siglo nuevo en todo semejante al primero que hemos descrito.

\begin{center}
{\rm (Papel Periódico Ilustrado. Año III. Páginas 315 a 318)}                                                       
\end{center}

\begin{center}
 \includegraphics[width=9cm]{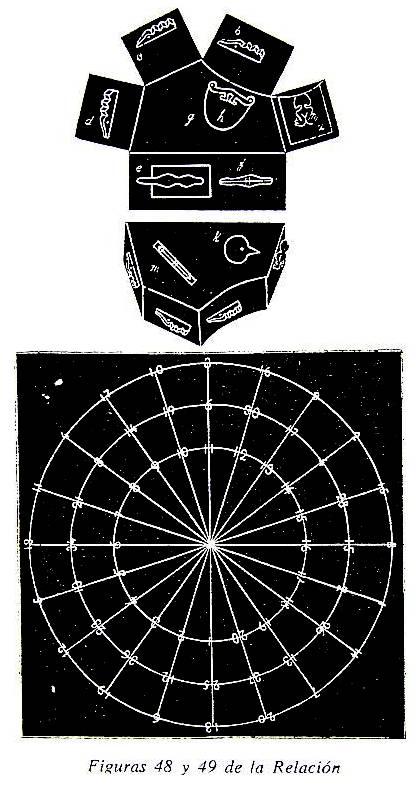}
 % zerda-figuras.png: 420x798 pixel, 72dpi, 14.81x28.15 cm, bb=0 0 420 798
\end{center}

\rm 
\selectlanguage{english}

%%%%%%%%%%%%%%%%%%%%%%%%%%%%%%%%%%%%%%%%%%%%%%%%%%%%
%% CALENDAR TABLES

\chapter{Muisca centuries table}\label{muisca-centuries-table}
\section{Acrotom and Extended Zocam Centuries}

This is a table with the sequence of lunar months in the span of an Acrotom 740 century and 2 Zocam centuries or 800 moons. The symbols of the fourth column have the following meaning:

\begin{itemize}
\item  (\revolucion{n}): Number $n$ of Revolution of Acrotom century (185 moons)
\item  (\sementera{}): Harvest time (12 or 13 moons)
%\item  (\metonico{}):  Metonic cycle of 19 years
\item  (\siglozocam{n}): Number $n$ of Zocam century (400 moons)
\item  (\sigloacrot{n}): Number $n$ of Acrotom century (740 moons)
\item  (\bxogonoa{n}): Number $n$ of Bxogonoa (3700 moons)
\item  (\zocam{n}): Number $n$ of Zocam year
\item  (\acrot{n}): Number $n$ of  Acrotom year   
\end{itemize}
\newpage
\begin{footnotesize}
\begin{longtable}{|r|rrr|c||r|r|}\hline
\multicolumn{7}{|c|}{\textbf{Intercalary year of \textsc{ Ata      }}}\\\hline
{\leftmoon} & \multicolumn{1}{|c|}{Z} & \multicolumn{1}{|c|}{A} & \multicolumn{1}{|c|}{R}& \ding{85} &\multicolumn{1}{|c|}{Days}& \multicolumn{1}{|c|}{Years} \\   \hline
\textbf{    1}&  Ata      $_{1}$ & 1 &  1 & \sementera{} &      29.5306 &      0.0809 \\  
\textbf{    2}&  Bosa     $_{2}$ & 2 &  2 &  &      59.0612 &      0.1617 \\  
\textbf{    3}&  Mica     $_{3}$ & 3 &  3 &  &      88.5918 &      0.2426 \\  
\textbf{    4}&  Muihica  $_{4}$ & 4 &  4 &  &     118.1224 &      0.3234 \\  
\textbf{    5}&  Hisca    $_{5}$ & 5 &  5 &  &     147.6529 &      0.4043 \\  
\textbf{    6}&  Ta       $_{6}$ & 6 &  6 &  &     177.1835 &      0.4851 \\
\hline\hline
\end{longtable}\newpage
\begin{longtable}{|r|rrr|c||r|r|}\hline
\multicolumn{7}{|c|}{\textbf{Intercalary year of \textsc{ Suhusa   }}}\\\hline
{\leftmoon} & \multicolumn{1}{|c|}{Z} & \multicolumn{1}{|c|}{A} & \multicolumn{1}{|c|}{R}& \ding{85} &\multicolumn{1}{|c|}{Days}& \multicolumn{1}{|c|}{Years} \\   \hline
\textbf{   38}& Q.Suhusa   $_{18}$ & 1 &  1 & \sementera{} &    1122.1624 &      3.0724 \\  
\textbf{   39}& Q.Aca      $_{19}$ & 2 &  2 &  &    1151.6930 &      3.1532 \\  
\textbf{   40}& Gueta    $_{20}$ & 3 &  3 & \zocam{2} &    1181.2236 &      3.2341 \\  
\textbf{   41}&  Ata      $_{1}$ & 4 &  4 &  &    1210.7543 &      3.3149 \\  
\textbf{   42}&  Bosa     $_{2}$ & 5 &  5 &  &    1240.2849 &      3.3958 \\  
\textbf{   43}&  Mica     $_{3}$ & 6 &  6 &  &    1269.8156 &      3.4766 \\
\hline\hline
\end{longtable}\newpage
\begin{longtable}{|r|rrr|c||r|r|}\hline
\multicolumn{7}{|c|}{\textbf{Intercalary year of \textsc{ Aca      }}}\\\hline
{\leftmoon} & \multicolumn{1}{|c|}{Z} & \multicolumn{1}{|c|}{A} & \multicolumn{1}{|c|}{R}& \ding{85} &\multicolumn{1}{|c|}{Days}& \multicolumn{1}{|c|}{Years} \\   \hline
\textbf{   79}& Q.Aca      $_{19}$ & 1 &  5 &  &    2332.9175 &      6.3873 \\  
\textbf{   80}& Gueta    $_{20}$ & 2 &  6 & \zocam{4} &    2362.4480 &      6.4682 \\  
\textbf{   81}&  Ata      $_{1}$ & 3 &  7 &  &    2391.9785 &      6.5490 \\  
\textbf{   82}&  Bosa     $_{2}$ & 4 &  8 &  &    2421.5090 &      6.6299 \\  
\textbf{   83}&  Mica     $_{3}$ & 5 &  9 &  &    2451.0396 &      6.7107 \\  
\textbf{   84}&  Muihica  $_{4}$ & 6 & 10 &  &    2480.5701 &      6.7916 \\
\hline\hline
\end{longtable}\newpage
\begin{longtable}{|r|rrr|c||r|r|}\hline
\multicolumn{7}{|c|}{\textbf{Intercalary year of \textsc{ Ubchihica}}}\\\hline
{\leftmoon} & \multicolumn{1}{|c|}{Z} & \multicolumn{1}{|c|}{A} & \multicolumn{1}{|c|}{R}& \ding{85} &\multicolumn{1}{|c|}{Days}& \multicolumn{1}{|c|}{Years} \\   \hline
\textbf{  120}& Gueta    $_{20}$ & 1 &  9 & \zocam{6} &    3543.6687 &      9.7022 \\  
\textbf{  121}&  Ata      $_{1}$ & 2 & 10 &  &    3573.1992 &      9.7831 \\  
\textbf{  122}&  Bosa     $_{2}$ & 3 & 11 &  &    3602.7297 &      9.8639 \\  
\textbf{  123}&  Mica     $_{3}$ & 4 & 12 &  &    3632.2603 &      9.9448 \\  
\textbf{  124}&  Muihica  $_{4}$ & 5 &  1 & \sementera{} &    3661.7908 &     10.0256 \\  
\textbf{  125}&  Hisca    $_{5}$ & 6 &  2 &  &    3691.3213 &     10.1065 \\
\hline\hline
\end{longtable}\newpage
\begin{longtable}{|r|rrr|c||r|r|}\hline
\multicolumn{7}{|c|}{\textbf{Intercalary year of \textsc{ Ata      }}}\\\hline
{\leftmoon} & \multicolumn{1}{|c|}{Z} & \multicolumn{1}{|c|}{A} & \multicolumn{1}{|c|}{R}& \ding{85} &\multicolumn{1}{|c|}{Days}& \multicolumn{1}{|c|}{Years} \\   \hline
\textbf{  161}&  Ata      $_{1}$ & 1 &  1 & \sementera{} &    4754.4253 &     13.0172 \\  
\textbf{  162}&  Bosa     $_{2}$ & 2 &  2 &  &    4783.9561 &     13.0980 \\  
\textbf{  163}&  Mica     $_{3}$ & 3 &  3 &  &    4813.4868 &     13.1789 \\  
\textbf{  164}&  Muihica  $_{4}$ & 4 &  4 &  &    4843.0176 &     13.2597 \\  
\textbf{  165}&  Hisca    $_{5}$ & 5 &  5 &  &    4872.5483 &     13.3406 \\  
\textbf{  166}&  Ta       $_{6}$ & 6 &  6 &  &    4902.0791 &     13.4214 \\
\hline\hline
\end{longtable}\newpage
\begin{longtable}{|r|rrr|c||r|r|}\hline
\multicolumn{7}{|c|}{\textbf{Intercalary year of \textsc{ Suhusa   }}}\\\hline
{\leftmoon} & \multicolumn{1}{|c|}{Z} & \multicolumn{1}{|c|}{A} & \multicolumn{1}{|c|}{R}& \ding{85} &\multicolumn{1}{|c|}{Days}& \multicolumn{1}{|c|}{Years} \\   \hline
\textbf{  198}& Q.Suhusa   $_{18}$ & 1 &  1 & \sementera{} &    5847.0635 &     16.0087 \\  
\textbf{  199}& Q.Aca      $_{19}$ & 2 &  2 &  &    5876.5942 &     16.0896 \\  
\textbf{  200}& Gueta    $_{20}$ & 3 &  3 & \zocam{10} &    5906.1250 &     16.1704 \\  
\textbf{  201}&  Ata      $_{1}$ & 4 &  4 &  &    5935.6558 &     16.2513 \\  
\textbf{  202}&  Bosa     $_{2}$ & 5 &  5 &  &    5965.1865 &     16.3321 \\  
\textbf{  203}&  Mica     $_{3}$ & 6 &  6 &  &    5994.7173 &     16.4130 \\
\hline\hline
\end{longtable}\newpage
\begin{longtable}{|r|rrr|c||r|r|}\hline
\multicolumn{7}{|c|}{\textbf{Intercalary year of \textsc{ Aca      }}}\\\hline
{\leftmoon} & \multicolumn{1}{|c|}{Z} & \multicolumn{1}{|c|}{A} & \multicolumn{1}{|c|}{R}& \ding{85} &\multicolumn{1}{|c|}{Days}& \multicolumn{1}{|c|}{Years} \\   \hline
\textbf{  239}& Q.Aca      $_{19}$ & 1 &  5 &  &    7057.8247 &     19.3237 \\  
\textbf{  240}& Gueta    $_{20}$ & 2 &  6 & \zocam{12} &    7087.3555 &     19.4045 \\  
\textbf{  241}&  Ata      $_{1}$ & 3 &  7 &  &    7116.8862 &     19.4854 \\  
\textbf{  242}&  Bosa     $_{2}$ & 4 &  8 &  &    7146.4170 &     19.5662 \\  
\textbf{  243}&  Mica     $_{3}$ & 5 &  9 &  &    7175.9478 &     19.6471 \\  
\textbf{  244}&  Muihica  $_{4}$ & 6 & 10 &  &    7205.4785 &     19.7279 \\
\hline\hline
\end{longtable}\newpage
\begin{longtable}{|r|rrr|c||r|r|}\hline
\multicolumn{7}{|c|}{\textbf{Intercalary year of \textsc{ Ubchihica}}}\\\hline
{\leftmoon} & \multicolumn{1}{|c|}{Z} & \multicolumn{1}{|c|}{A} & \multicolumn{1}{|c|}{R}& \ding{85} &\multicolumn{1}{|c|}{Days}& \multicolumn{1}{|c|}{Years} \\   \hline
\textbf{  280}& Gueta    $_{20}$ & 1 &  9 & \zocam{14} &    8268.5850 &     22.6386 \\  
\textbf{  281}&  Ata      $_{1}$ & 2 & 10 &  &    8298.1152 &     22.7195 \\  
\textbf{  282}&  Bosa     $_{2}$ & 3 & 11 &  &    8327.6455 &     22.8003 \\  
\textbf{  283}&  Mica     $_{3}$ & 4 & 12 &  &    8357.1758 &     22.8812 \\  
\textbf{  284}&  Muihica  $_{4}$ & 5 &  1 & \sementera{} &    8386.7061 &     22.9620 \\  
\textbf{  285}&  Hisca    $_{5}$ & 6 &  2 &  &    8416.2363 &     23.0429 \\
\hline\hline
\end{longtable}\newpage
\begin{longtable}{|r|rrr|c||r|r|}\hline
\multicolumn{7}{|c|}{\textbf{Intercalary year of \textsc{ Ata      }}}\\\hline
{\leftmoon} & \multicolumn{1}{|c|}{Z} & \multicolumn{1}{|c|}{A} & \multicolumn{1}{|c|}{R}& \ding{85} &\multicolumn{1}{|c|}{Days}& \multicolumn{1}{|c|}{Years} \\   \hline
\textbf{  321}&  Ata      $_{1}$ & 1 &  1 & \sementera{} &    9479.3262 &     25.9535 \\  
\textbf{  322}&  Bosa     $_{2}$ & 2 &  2 &  &    9508.8564 &     26.0344 \\  
\textbf{  323}&  Mica     $_{3}$ & 3 &  3 &  &    9538.3867 &     26.1152 \\  
\textbf{  324}&  Muihica  $_{4}$ & 4 &  4 &  &    9567.9170 &     26.1961 \\  
\textbf{  325}&  Hisca    $_{5}$ & 5 &  5 &  &    9597.4473 &     26.2769 \\  
\textbf{  326}&  Ta       $_{6}$ & 6 &  6 &  &    9626.9775 &     26.3578 \\
\hline\hline
\end{longtable}\newpage
\begin{longtable}{|r|rrr|c||r|r|}\hline
\multicolumn{7}{|c|}{\textbf{Intercalary year of \textsc{ Suhusa   }}}\\\hline
{\leftmoon} & \multicolumn{1}{|c|}{Z} & \multicolumn{1}{|c|}{A} & \multicolumn{1}{|c|}{R}& \ding{85} &\multicolumn{1}{|c|}{Days}& \multicolumn{1}{|c|}{Years} \\   \hline
\textbf{  358}& Q.Suhusa   $_{18}$ & 1 &  1 & \sementera{} &   10571.9463 &     28.9450 \\  
\textbf{  359}& Q.Aca      $_{19}$ & 2 &  2 &  &   10601.4766 &     29.0259 \\  
\textbf{  360}& Gueta    $_{20}$ & 3 &  3 & \zocam{18} &   10631.0068 &     29.1067 \\  
\textbf{  361}&  Ata      $_{1}$ & 4 &  4 &  &   10660.5371 &     29.1876 \\  
\textbf{  362}&  Bosa     $_{2}$ & 5 &  5 &  &   10690.0674 &     29.2684 \\  
\textbf{  363}&  Mica     $_{3}$ & 6 &  6 &  &   10719.5977 &     29.3493 \\
\hline\hline
\end{longtable}\newpage
\begin{longtable}{|r|rrr|c||r|r|}\hline
\multicolumn{7}{|c|}{\textbf{Intercalary year of \textsc{ Aca      }}}\\\hline
{\leftmoon} & \multicolumn{1}{|c|}{Z} & \multicolumn{1}{|c|}{A} & \multicolumn{1}{|c|}{R}& \ding{85} &\multicolumn{1}{|c|}{Days}& \multicolumn{1}{|c|}{Years} \\   \hline
\textbf{  399}& Q.Aca      $_{19}$ & 1 &  5 &  &   11782.6875 &     32.2599 \\  
\textbf{  400}& Gueta    $_{20}$ & 2 &  6 & \zocam{20},\siglozocam{1} &   11812.2178 &     32.3408 \\  
\textbf{  401}&  Ata      $_{1}$ & 3 &  7 &  &   11841.7480 &     32.4216 \\  
\textbf{  402}&  Bosa     $_{2}$ & 4 &  8 &  &   11871.2783 &     32.5025 \\  
\textbf{  403}&  Mica     $_{3}$ & 5 &  9 &  &   11900.8086 &     32.5833 \\  
\textbf{  404}&  Muihica  $_{4}$ & 6 & 10 &  &   11930.3389 &     32.6642 \\
\hline\hline
\end{longtable}\newpage
\begin{longtable}{|r|rrr|c||r|r|}\hline
\multicolumn{7}{|c|}{\textbf{Intercalary year of \textsc{ Ubchihica}}}\\\hline
{\leftmoon} & \multicolumn{1}{|c|}{Z} & \multicolumn{1}{|c|}{A} & \multicolumn{1}{|c|}{R}& \ding{85} &\multicolumn{1}{|c|}{Days}& \multicolumn{1}{|c|}{Years} \\   \hline
\textbf{  440}& Gueta    $_{20}$ & 1 &  9 & \zocam{22} &   12993.4287 &     35.5748 \\  
\textbf{  441}&  Ata      $_{1}$ & 2 & 10 &  &   13022.9590 &     35.6557 \\  
\textbf{  442}&  Bosa     $_{2}$ & 3 & 11 &  &   13052.4893 &     35.7365 \\  
\textbf{  443}&  Mica     $_{3}$ & 4 & 12 &  &   13082.0195 &     35.8174 \\  
\textbf{  444}&  Muihica  $_{4}$ & 5 & 13 & \acrot{12} &   13111.5498 &     35.8982 \\  
\textbf{  445}&  Hisca    $_{5}$ & 6 &  1 & \sementera{} &   13141.0801 &     35.9791 \\
\hline\hline
\end{longtable}\newpage
\begin{longtable}{|r|rrr|c||r|r|}\hline
\multicolumn{7}{|c|}{\textbf{Intercalary year of \textsc{ Ata      }}}\\\hline
{\leftmoon} & \multicolumn{1}{|c|}{Z} & \multicolumn{1}{|c|}{A} & \multicolumn{1}{|c|}{R}& \ding{85} &\multicolumn{1}{|c|}{Days}& \multicolumn{1}{|c|}{Years} \\   \hline
\textbf{  481}&  Ata      $_{1}$ & 1 & 13 & \acrot{13} &   14204.1699 &     38.8897 \\  
\textbf{  482}&  Bosa     $_{2}$ & 2 &  1 & \sementera{} &   14233.7002 &     38.9705 \\  
\textbf{  483}&  Mica     $_{3}$ & 3 &  2 &  &   14263.2305 &     39.0514 \\  
\textbf{  484}&  Muihica  $_{4}$ & 4 &  3 &  &   14292.7607 &     39.1322 \\  
\textbf{  485}&  Hisca    $_{5}$ & 5 &  4 &  &   14322.2910 &     39.2131 \\  
\textbf{  486}&  Ta       $_{6}$ & 6 &  5 &  &   14351.8213 &     39.2940 \\
\hline\hline
\end{longtable}\newpage
\begin{longtable}{|r|rrr|c||r|r|}\hline
\multicolumn{7}{|c|}{\textbf{Intercalary year of \textsc{ Suhusa   }}}\\\hline
{\leftmoon} & \multicolumn{1}{|c|}{Z} & \multicolumn{1}{|c|}{A} & \multicolumn{1}{|c|}{R}& \ding{85} &\multicolumn{1}{|c|}{Days}& \multicolumn{1}{|c|}{Years} \\   \hline
\textbf{  518}& Q.Suhusa   $_{18}$ & 1 & 13 & \acrot{14} &   15296.7900 &     41.8812 \\  
\textbf{  519}& Q.Aca      $_{19}$ & 2 &  1 & \sementera{} &   15326.3203 &     41.9620 \\  
\textbf{  520}& Gueta    $_{20}$ & 3 &  2 & \zocam{26} &   15355.8506 &     42.0429 \\  
\textbf{  521}&  Ata      $_{1}$ & 4 &  3 &  &   15385.3809 &     42.1237 \\  
\textbf{  522}&  Bosa     $_{2}$ & 5 &  4 &  &   15414.9111 &     42.2046 \\  
\textbf{  523}&  Mica     $_{3}$ & 6 &  5 &  &   15444.4414 &     42.2854 \\
\hline\hline
\end{longtable}\newpage
\begin{longtable}{|r|rrr|c||r|r|}\hline
\multicolumn{7}{|c|}{\textbf{Intercalary year of \textsc{ Aca      }}}\\\hline
{\leftmoon} & \multicolumn{1}{|c|}{Z} & \multicolumn{1}{|c|}{A} & \multicolumn{1}{|c|}{R}& \ding{85} &\multicolumn{1}{|c|}{Days}& \multicolumn{1}{|c|}{Years} \\   \hline
\textbf{  559}& Q.Aca      $_{19}$ & 1 &  4 &  &   16507.5352 &     45.1961 \\  
\textbf{  560}& Gueta    $_{20}$ & 2 &  5 & \zocam{28} &   16537.0664 &     45.2770 \\  
\textbf{  561}&  Ata      $_{1}$ & 3 &  6 &  &   16566.5977 &     45.3578 \\  
\textbf{  562}&  Bosa     $_{2}$ & 4 &  7 &  &   16596.1289 &     45.4387 \\  
\textbf{  563}&  Mica     $_{3}$ & 5 &  8 &  &   16625.6602 &     45.5195 \\  
\textbf{  564}&  Muihica  $_{4}$ & 6 &  9 &  &   16655.1914 &     45.6004 \\
\hline\hline
\end{longtable}\newpage
\begin{longtable}{|r|rrr|c||r|r|}\hline
\multicolumn{7}{|c|}{\textbf{Intercalary year of \textsc{ Ubchihica}}}\\\hline
{\leftmoon} & \multicolumn{1}{|c|}{Z} & \multicolumn{1}{|c|}{A} & \multicolumn{1}{|c|}{R}& \ding{85} &\multicolumn{1}{|c|}{Days}& \multicolumn{1}{|c|}{Years} \\   \hline
\textbf{  600}& Gueta    $_{20}$ & 1 &  8 & \zocam{30} &   17718.3164 &     48.5111 \\  
\textbf{  601}&  Ata      $_{1}$ & 2 &  9 &  &   17747.8477 &     48.5920 \\  
\textbf{  602}&  Bosa     $_{2}$ & 3 & 10 &  &   17777.3789 &     48.6728 \\  
\textbf{  603}&  Mica     $_{3}$ & 4 & 11 &  &   17806.9102 &     48.7537 \\  
\textbf{  604}&  Muihica  $_{4}$ & 5 & 12 &  &   17836.4414 &     48.8345 \\  
\textbf{  605}&  Hisca    $_{5}$ & 6 &  1 & \sementera{} &   17865.9727 &     48.9154 \\
\hline\hline
\end{longtable}\newpage
\begin{longtable}{|r|rrr|c||r|r|}\hline
\multicolumn{7}{|c|}{\textbf{Intercalary year of \textsc{ Ata      }}}\\\hline
{\leftmoon} & \multicolumn{1}{|c|}{Z} & \multicolumn{1}{|c|}{A} & \multicolumn{1}{|c|}{R}& \ding{85} &\multicolumn{1}{|c|}{Days}& \multicolumn{1}{|c|}{Years} \\   \hline
\textbf{  641}&  Ata      $_{1}$ & 1 & 12 &  &   18929.0977 &     51.8261 \\  
\textbf{  642}&  Bosa     $_{2}$ & 2 &  1 & \sementera{} &   18958.6289 &     51.9070 \\  
\textbf{  643}&  Mica     $_{3}$ & 3 &  2 &  &   18988.1602 &     51.9878 \\  
\textbf{  644}&  Muihica  $_{4}$ & 4 &  3 &  &   19017.6914 &     52.0687 \\  
\textbf{  645}&  Hisca    $_{5}$ & 5 &  4 &  &   19047.2227 &     52.1495 \\  
\textbf{  646}&  Ta       $_{6}$ & 6 &  5 &  &   19076.7539 &     52.2304 \\
\hline\hline
\end{longtable}\newpage
\begin{longtable}{|r|rrr|c||r|r|}\hline
\multicolumn{7}{|c|}{\textbf{Intercalary year of \textsc{ Suhusa   }}}\\\hline
{\leftmoon} & \multicolumn{1}{|c|}{Z} & \multicolumn{1}{|c|}{A} & \multicolumn{1}{|c|}{R}& \ding{85} &\multicolumn{1}{|c|}{Days}& \multicolumn{1}{|c|}{Years} \\   \hline
\textbf{  678}& Q.Suhusa   $_{18}$ & 1 & 12 &  &   20021.7539 &     54.8177 \\  
\textbf{  679}& Q.Aca      $_{19}$ & 2 &  1 & \sementera{} &   20051.2852 &     54.8986 \\  
\textbf{  680}& Gueta    $_{20}$ & 3 &  2 & \zocam{34} &   20080.8164 &     54.9794 \\  
\textbf{  681}&  Ata      $_{1}$ & 4 &  3 &  &   20110.3477 &     55.0603 \\  
\textbf{  682}&  Bosa     $_{2}$ & 5 &  4 &  &   20139.8789 &     55.1411 \\  
\textbf{  683}&  Mica     $_{3}$ & 6 &  5 &  &   20169.4102 &     55.2220 \\
\hline\hline
\end{longtable}\newpage
\begin{longtable}{|r|rrr|c||r|r|}\hline
\multicolumn{7}{|c|}{\textbf{Intercalary year of \textsc{ Aca      }}}\\\hline
{\leftmoon} & \multicolumn{1}{|c|}{Z} & \multicolumn{1}{|c|}{A} & \multicolumn{1}{|c|}{R}& \ding{85} &\multicolumn{1}{|c|}{Days}& \multicolumn{1}{|c|}{Years} \\   \hline
\textbf{  719}& Q.Aca      $_{19}$ & 1 &  4 &  &   21232.5352 &     58.1327 \\  
\textbf{  720}& Gueta    $_{20}$ & 2 &  5 & \zocam{36} &   21262.0664 &     58.2136 \\  
\textbf{  721}&  Ata      $_{1}$ & 3 &  6 &  &   21291.5977 &     58.2944 \\  
\textbf{  722}&  Bosa     $_{2}$ & 4 &  7 &  &   21321.1289 &     58.3753 \\  
\textbf{  723}&  Mica     $_{3}$ & 5 &  8 &  &   21350.6602 &     58.4561 \\  
\textbf{  724}&  Muihica  $_{4}$ & 6 &  9 &  &   21380.1914 &     58.5370 \\
\hline\hline
\end{longtable}\newpage
\begin{longtable}{|r|rrr|c||r|r|}\hline
\multicolumn{7}{|c|}{\textbf{Intercalary year of \textsc{ Ubchihica}}}\\\hline
{\leftmoon} & \multicolumn{1}{|c|}{Z} & \multicolumn{1}{|c|}{A} & \multicolumn{1}{|c|}{R}& \ding{85} &\multicolumn{1}{|c|}{Days}& \multicolumn{1}{|c|}{Years} \\   \hline
\textbf{  760}& Gueta    $_{20}$ & 1 &  8 & \zocam{38} &   22443.3164 &     61.4477 \\  
\textbf{  761}&  Ata      $_{1}$ & 2 &  9 &  &   22472.8477 &     61.5286 \\  
\textbf{  762}&  Bosa     $_{2}$ & 3 & 10 &  &   22502.3789 &     61.6094 \\  
\textbf{  763}&  Mica     $_{3}$ & 4 & 11 &  &   22531.9102 &     61.6903 \\  
\textbf{  764}&  Muihica  $_{4}$ & 5 & 12 &  &   22561.4414 &     61.7711 \\  
\textbf{  765}&  Hisca    $_{5}$ & 6 &  1 & \sementera{} &   22590.9727 &     61.8520 \\
\hline\hline
\end{longtable}\end{footnotesize}
\newpage

\section{Bxogonoa}\newpage
\begin{footnotesize}\begin{longtable}{|r|rrr|c||r|r|}\hline
\multicolumn{7}{|c|}{\textbf{Intercalary year of \textsc{ Ata      }}}\\\hline
{\leftmoon} & \multicolumn{1}{|c|}{Z} & \multicolumn{1}{|c|}{A} & \multicolumn{1}{|c|}{R}& \ding{85} &\multicolumn{1}{|c|}{Days}& \multicolumn{1}{|c|}{Years} \\   \hline
\textbf{ 3681}&  Ata      $_{1}$ & 1 &  6 &  &  108704.0938 &    297.6217 \\  
\textbf{ 3682}&  Bosa     $_{2}$ & 2 &  7 &  &  108733.6250 &    297.7025 \\  
\textbf{ 3683}&  Mica     $_{3}$ & 3 &  8 &  &  108763.1562 &    297.7834 \\  
\textbf{ 3684}&  Muihica  $_{4}$ & 4 &  9 &  &  108792.6875 &    297.8643 \\  
\textbf{ 3685}&  Hisca    $_{5}$ & 5 & 10 &  &  108822.2188 &    297.9451 \\  
\textbf{ 3686}&  Ta       $_{6}$ & 6 & 11 &  &  108851.7500 &    298.0260 \\
\hline\hline
\end{longtable}               \end{footnotesize}
\newpage
\section{Bochica's coming}\newpage
\begin{footnotesize}\begin{longtable}{|r|rrr|c||r|r|}\hline
\multicolumn{7}{|c|}{\textbf{Intercalary year of \textsc{ Aca      }}}\\\hline
{\leftmoon} & \multicolumn{1}{|c|}{Z} & \multicolumn{1}{|c|}{A} & \multicolumn{1}{|c|}{R}& \ding{85} &\multicolumn{1}{|c|}{Days}& \multicolumn{1}{|c|}{Years} \\   \hline
\textbf{14799}& Q.Aca      $_{19}$ & 1 & 12 &  &  437032.5312 &   1196.5544 \\  
\textbf{14800}& Gueta    $_{20}$ & 2 & 13 & \zocam{740},\acrot{400},\revolucion{},\siglozocam{37},\sigloacrot{20},\bxogonoa{4} &  437062.0625 &   1196.6353 \\  
\textbf{14801}&  Ata      $_{1}$ & 3 &  1 & \sementera{} &  437091.5938 &   1196.7162 \\  
\textbf{14802}&  Bosa     $_{2}$ & 4 &  2 &  &  437121.1250 &   1196.7970 \\  
\textbf{14803}&  Mica     $_{3}$ & 5 &  3 &  &  437150.6562 &   1196.8778 \\  
\textbf{14804}&  Muihica  $_{4}$ & 6 &  4 &  &  437180.1875 &   1196.9587 \\
\hline\hline
\end{longtable}               \end{footnotesize}
\newpage

\section{29600 Moon cycle}\newpage
\begin{footnotesize}\begin{longtable}{|r|rrr|c||r|r|}\hline
\multicolumn{7}{|c|}{\textbf{Intercalary year of \textsc{ Ubchihica}}}\\\hline
{\leftmoon} & \multicolumn{1}{|c|}{Z} & \multicolumn{1}{|c|}{A} & \multicolumn{1}{|c|}{R}& \ding{85} &\multicolumn{1}{|c|}{Days}& \multicolumn{1}{|c|}{Years} \\   \hline
\textbf{29560}& Gueta    $_{20}$ & 1 & 10 & \zocam{1478} &  872574.3750 &   2389.0276 \\  
\textbf{29561}&  Ata      $_{1}$ & 2 & 11 &  &  872603.8750 &   2389.1082 \\  
\textbf{29562}&  Bosa     $_{2}$ & 3 & 12 &  &  872633.3750 &   2389.1890 \\  
\textbf{29563}&  Mica     $_{3}$ & 4 & 13 & \acrot{799} &  872662.8750 &   2389.2698 \\  
\textbf{29564}&  Muihica  $_{4}$ & 5 &  1 & \sementera{} &  872692.3750 &   2389.3506 \\  
\textbf{29565}&  Hisca    $_{5}$ & 6 &  2 &  &  872721.8750 &   2389.4314 \\
\hline\hline
\end{longtable}               \end{footnotesize}
\newpage

\section{Bochica's dream}\newpage
\begin{footnotesize}\begin{longtable}{|r|rrr|c||r|r|}\hline
\multicolumn{7}{|c|}{\textbf{Intercalary year of \textsc{ Aca      }}}\\\hline
{\leftmoon} & \multicolumn{1}{|c|}{Z} & \multicolumn{1}{|c|}{A} & \multicolumn{1}{|c|}{R}& \ding{85} &\multicolumn{1}{|c|}{Days}& \multicolumn{1}{|c|}{Years} \\   \hline
\textbf{73999}& Q.Aca      $_{19}$ & 1 & 12 &  & 2183525.0000 &   5978.2886 \\  
\textbf{74000}& Gueta    $_{20}$ & 2 & 13 & \zocam{3700},\acrot{2000},\revolucion{},\siglozocam{185},\sigloacrot{100},\bxogonoa{20} & 2183554.5000 &   5978.3691 \\  
\textbf{74001}&  Ata      $_{1}$ & 3 &  1 & \sementera{} & 2183584.0000 &   5978.4497 \\  
\textbf{74002}&  Bosa     $_{2}$ & 4 &  2 &  & 2183613.5000 &   5978.5308 \\  
\textbf{74003}&  Mica     $_{3}$ & 5 &  3 &  & 2183643.0000 &   5978.6113 \\  
\textbf{74004}&  Muihica  $_{4}$ & 6 &  4 &  & 2183672.5000 &   5978.6924 \\
\hline\hline
\end{longtable}
\end{footnotesize}
\newpage

\backmatter
\bibliography{muiscacalendar}

\end{document}